\documentclass[fleqn,usenatbib]{mnras}

\usepackage{newtxtext,newtxmath}
\usepackage{subcaption}

\usepackage[T1]{fontenc}

\DeclareRobustCommand{\VAN}[3]{#2}
\let\VANthebibliography\thebibliography
\def\thebibliography{\DeclareRobustCommand{\VAN}[3]{##3}\VANthebibliography}

\usepackage{graphicx}	
\usepackage{amsmath}	
\usepackage{threeparttable}

\title[Gas-Phase Metallicity Gradients at $\rm z\sim 7-10$]{Metal Mayhem at $\rm z \sim 7\text{--}10$: Diversity and Evolution of Gas-Phase Metallicity Gradients}

\author[M. Koller et al.]{Maria Koller,$^{1,2}$\thanks{E-mail:mk2264@cam.ac.uk}
Roberto Maiolino,$^{1,2,3}$
Hannah Übler,$^{4}$
Qiao Duan,$^{1,2}$
Jan Scholtz,$^{1,2}$
\newauthor
Santiago Arribas,$^{5}$
William M. Baker,$^{6}$
Stefano Carniani,$^{7}$
Stephane Charlot,$^{8}$ 
Mirko Curti,$^{9}$
\newauthor
Luca Graziani,$^{10,11}$
Gareth Jones, $^{1,2}$
William McClymont,$^{1,2}$
Michele Perna,$^{5}$
Bruno Rodríguez Del Pino,$^{5}$
\newauthor
Sandro Tacchella,$^{1,2}$
Alessandra Venditti,$^{12,13}$\thanks{Cosmic Frontier Center Prize Fellow}
Giacomo Venturi,$^{7}$
Joris Witstok$^{14,15}$
\\
$^{1}$Kavli Institute for Cosmology, University of Cambridge, Madingley Road, Cambridge, CB3 OHA, UK\\
$^{2}$Cavendish Laboratory, University of Cambridge, 19 JJ Thomson Avenue, Cambridge CB3 0HE, UK\\
$^{3}$Department of Physics and Astronomy, University College London, Gower Street, London WC1E 6BT, UK \\
$^{4}$Max-Planck-Institut für extraterrestrische Physik (MPE), Gießenbachstraße 1, 85748 Garching, Germany \\
$^{5}$Centro de Astrobiolog\'ia (CAB), CSIC–INTA, Cra. de Ajalvir km.~4, 28850- Torrej\'on de Ardoz, Madrid, Spain \\
$^{6}$DARK, Niels Bohr Institute, University of Copenhagen, Jagtvej 155A, DK-2200 Copenhagen, Denmark \\
$^{7}$Scuola Normale Superiore, Piazza dei Cavalieri 7, I-56126 Pisa, Italy \\
$^{8}$Sorbonne Universit\'e, CNRS, UMR 7095, Institut d'Astrophysique de Paris, 98 bis bd Arago, 75014 Paris, France \\
$^{9}$INAF - Osservatorio di Astrofisica e Scienza dello Spazio, Via Piero Gobetti 93/3, 40129, Bologna, Italy\\
$^{10}$Dipartimento di Fisica, "Sapienza" Universit$\grave{a}$ di Roma, Piazzale Aldo Moro 2, 00185 Roma, Italy \\
$^{11}$INFN, Sezione Roma1, Dipartimento di Fisica, ``Sapienza'' Universit$\grave{a}$ di Roma, Piazzale Aldo Moro 2, 00185, Roma, Italy \\
$^{12}$Department of Astronomy, University of Texas, Austin, TX 78712, USA \\
$^{13}$Cosmic Frontier Center, The University of Texas at Austin, Austin, TX 78712, USA \\
$^{14}$Cosmic Dawn Center (DAWN), Copenhagen, Denmark \\
$^{15}$Niels Bohr Institute, University of Copenhagen, Jagtvej 128, DK-2200, Copenhagen, Denmark \\
}

\date{Accepted XXX. Received YYY; in original form ZZZ}

\pubyear{\the\year{}}

\begin{document}
\label{firstpage}
\pagerange{\pageref{firstpage}--\pageref{lastpage}}
\maketitle

\begin{abstract}
We present a JWST/NIRSpec Integral Field Unit (IFU) study of metallicity gradients in seven low-metallicity systems at $z=7.2-9.5$. The main sample spans stellar masses of $\rm \log(M_*/M_{\odot}) \sim 7.8-9.5$, star formation rates (SFRs) of $\rm \log(\text{SFR} / M_{\odot} \text{yr}^{-1}) \sim 0.5-2.5$, and gas-phase metallicities of $4\%-13 \%~Z_\odot$. Within our sample, we also identify three low-metallicity satellite galaxies associated with two of our sources, providing a rare view of early-epoch interactions. The three satellites exhibit even more primordial properties, with metallicity $3\% -4\% ~Z_\odot$ and low star-formation activity ($\rm \log(\text{SFR} / M_{\odot} \text{yr}^{-1}) \sim -0.5$ to $-0.9$). We find that our galaxies, and especially the satellites, are significantly offset from the local Fundamental Metallicity Relation (FMR), with deviations reaching $\Delta \text{FMR} \approx -1.0$ dex. This indicates that these galaxies are likely experiencing strong accretion of pristine gas. Overall, we observe a large scatter in radial metallicity gradients, ranging from positive to negative with an average metallicity gradient of $\rm -0.03 \pm 0.04 \ dex \ kpc^{-1}$. Flat gradients are found in systems with confirmed satellites, suggesting that tidal interactions and mergers drive the radial mixing necessary to homogenise the interstellar medium. The (tentative) presence of an AGN in two of our sources suggests that strong feedback may also be responsible for the observed flat gradients. Conversely, the detection of a positive gradient in one source points toward a direct funnelling of metal-poor gas inflow into the central region of the galaxy. These results show that galaxies in the first billion years grow through diverse, episodic processes, suggesting that early evolution is characterised by structural variety rather than a single, predictable path. 
\end{abstract}




\section{Introduction}

Galaxy evolution is governed by the regulation of gas content, metallicity, and stellar mass, shaped by gas inflows from the intergalactic medium (IGM), feedback-driven outflows, and internal recycling \citep{lilly_gas_2013,madau_cosmic_2014,peroux_cosmic_2020}, that is, the so-called cosmic baryon cycle \citep{tumlinson_circumgalactic_2017}. Inflowing, metal-poor gas dilutes the ISM and its metal content \citep{koeppen_evolution_1994, keres_how_2005}, which then cools inside the galaxy, fuelling star formation \citep{mckee_theory_2007,kennicutt_star_2012}. Stellar feedback mechanisms such as stellar winds and supernovae can enrich the interstellar medium (ISM) and circumgalactic medium (CGM) \citep{lilly_gas_2013,peng_haloes_2014,tumlinson_circumgalactic_2017}, while both stellar and AGN feedback influence the galaxy by heating or expelling gas, suppressing star formation \citep{fabian_observational_2012, ubler_why_2014, bourne_recent_2023}. Expelled gas can also restart the cycle, if it manages to cool, by raining back down as recycled metal-enriched gas onto the galaxy, what is often referred to as a 'galactic fountain' \citep{oppenheimer_mass_2008, oppenheimer_feedback_2010, brook_hierarchical_2011, brook_hierarchical_2012, ubler_why_2014}. Therefore, one of the most important tracers of the gas cycle is the gas-phase metallicity, which is linked to the time-integrated production of chemical elements; hence, it reflects a galaxy’s star formation history, assembly history and the underlying processes governing galaxy evolution \citep[see review by][]{maiolino_re_2019}. 

Metallicity is known to correlate strongly with the stellar mass, giving rise to the so-called mass-metallicity relation  \citep[MZR; see e.g.,][for a review]{tremonti_origin_2004, kewley_metallicity_2008, sanchez_mass-metallicity_2017, maiolino_re_2019}. In addition, a secondary dependence on the SFR has been found, leading to what is known as the fundamental metallicity relation \citep[FMR;][]{ellison_clues_2008, mannucci_fundamental_2010, lara-lopez_fundamental_2010, cresci_fundamental_2019, curti_mass-metallicity_2020, curti_jades_2024, pistis_comparative_2024, sarkar_unveiling_2025}. The dependence on stellar mass has often been interpreted as an indirect proxy for the gravitational potential, thereby reflecting the capability to retain metals \citep[e.g.][]{tremonti_origin_2004}. Nonetheless, recent studies have shown that there is also a direct relation between metallicity and stellar mass, probably resulting from both tracing the time-integrated production of stars and metals \citep{baker_stellar_2023}, while other studies have clarified that the primary dependence on stellar mass or gravitational potential strongly depends on the galaxy scales that are probed \citep{koller_both_2026}. The anticorrelation with SFR is often interpreted in terms of low-metallicity gas accretion (from the halo or from the IGM), which both dilutes the metallicity and fosters star formation \citep[e.g.][]{bothwell_fundamental_2013}. For a given stellar mass, the MZR evolves towards higher metallicities with cosmic time \citep[e.g.][]{maiolino_amaze_2008,troncoso_metallicity_2014,curti_chemical_2023,curti_jades_2024,nakajima_jwst_2023, jain_uniform_2026}. However, this is thought to be mostly dominated by a non-evolving FMR \citep[e.g.][]{mannucci_fundamental_2010,cresci_fundamental_2019} (suggesting that galaxies at high redshift follow the same smooth, secular evolution in quasi equilibrium between gas accretion, star formation and feedback) combined with a strong SFR evolution. Yet, recently various studies have shown evidence for galaxies departing from the FMR at z$>$3 \citep[e.g.][]{tacchella_jwst_2023, heintz_dilution_2023,nakajima_jwst_2023, curti_jades_2024, stanton_jwst_2026, scholte_jwst_2025,pollock_novel_2026, laseter_investigation_2025, nishigaki_dreamsii_2025}. A possible explanation for this scenario is that high-z galaxies have bursty star formation histories (SFHs), likely driven by strong, pristine inflowing gas that lowers metallicity and boosts star formation, i.e., they appear significantly offset from the local FMR. On top of that, it has been discovered that rapid accretion of pristine gas in the early Universe can cause a compaction effect; strong star formation is recovered within compact metal-poor systems \citep[e.g.][]{tacchella_jwst_2023, langeroodi_ultraviolet_2023}. However, a possible contribution from enriched outflows cannot be neglected \citep[e.g.][]{curti_jades_2024, laseter_investigation_2025}. Lastly, it can also be argued that the redshift evolution is actually a mass evolution: more massive galaxies at high-redshift have been observed to fall close to or on the local FMR \citep[e.g.][]{stanton_jwst_2026, faisst_alpinecristaljwst_2026, rowland_rebels-ifu_2026}. This could indicate an observational bias: we do not probe the same mass range in the early Universe as we do for nearby galaxies, which could add to the apparent redshift evolution of the FMR offset. 

Studying the chemical enrichment on spatially resolved scales allows us to gain insights into the different physical processes driving galaxy evolution. By studying radial gas-phase metallicity gradients, we can gain important insights into the distribution of metals within these galaxies. Negative gradients are characterised as radially decreasing, meaning that the metallicity is highest in the centre of the galaxy and decreases outwards. These are usually interpreted to result from inside-out galaxy formation, where stars in the inner part of the galaxy form earlier than the outskirts and therefore have more time to chemically enrich the central region \citep[e.g.][]{samland_modeling_1997, prantzos_chemo-spectrophotometric_2000, dave_galaxy_2011, pilkington_metallicity_2012,gibson_constraining_2013, hemler_gas-phase_2021, tissera_evolution_2022, venturi_gas-phase_2024, baker_core_2025, deepak_global_2025,ibrahim_impact_2025, garcia_metallicity_2025, li_13_2025}. Flat gradients are theorised to stem from efficient radial mixing of gas, which causes the metals to redistribute inside the galaxy. This is believed to be caused by supernova (SN) winds or metal-enriched galactic outflows that are re-accreted onto the outer regions of the galaxy, so-called galactic fountains, stemming from intense stellar feedback \citep{gibson_constraining_2013, ma_why_2017}. Galaxy mergers and interactions are also expected to flatten metallicity gradients \citep{rupke_gas-phase_2010, rich_integral_2012, torres-flores_star-forming_2014}. Finally, positive gradients, where the metallicity radially increases from the centre to the outskirts of the galaxy, are potentially produced by the accretion of pristine (metal-poor) gas towards the centre \citep[e.g.][]{cresci_gas_2010, ceverino_gas_2016}. Strong enriched outflows can also cause inverted gradients \citep[e.g.][]{rodriguez_del_pino_ga-nifs_2024}. In conclusion, tracing radial metallicity gradients enables us to investigate inside-out growth, gas inflows, and mixing throughout cosmic time. 

In the local Universe, intensive surveys using integral field spectroscopy (IFS), such as MaNGA \citep{bundy_overview_2015}, SAMI \citep{croom_sydney-aao_2012}, and CALIFA \citep{sanchez_califa_2012}, have shown that local galaxies tend to have negative metallicity gradients, where the centre is more metal-enriched than the outskirts, with typical values of around $-0.05 \pm 0.05$ dex kpc$^{-1}$ \citep[e.g.][]{zaritsky_h_1994,magrini_metal_2010,rupke_gas-phase_2010, kewley_metallicity_2010, bresolin_abundance_2011,stanghellini_radial_2014,sanchez_characteristic_2014, ho_metallicity_2015, berg_chaos_2015,belfiore_sdss_2017}. Furthermore, spatially resolved scaling relations such as the resolved mass-metallicity relation \citep[rMZR; e.g.][]{rosales-ortega_new_2012, sanchez_mass-metallicity_2013, barrera-ballesteros_galaxy_2016, sanchez_almeida_fundamental_2019, baker_metallicitys_2022}, resolved star formation main sequence \citep[e.g.][]{cano-diaz_spatially_2016, baker_almaquest_2022}, and the resolved fundamental metallicity relation \citep[rFMR; e.g.][]{baker_metallicitys_2022, koller_magpi_2024} have been established for the local Universe.

IFS instruments such as MUSE \citep{bacon_muse_2010}, SINFONI \citep{eisenhauer_sinfoni_2003}, and KMOS \citep{sharples_first_2013} at the Very Large Telescope (VLT) have enabled spatially resolved metallicity studies at cosmic noon, that is, at intermediate redshifts of $\rm z \sim1-4$. These observations benefit from gravitational lensing, which provides the necessary magnification to resolve high-redshift targets at sub-kiloparsec scales. Observations have found varying results, ranging from negative, flat, to positive gradients \citep[e.g.][]{cresci_gas_2010, yuan_metallicity_2011, queyrel_massiv_2012, swinbank_properties_2012, jones_origin_2013, jones_grism_2015,stott_relationship_2014, troncoso_metallicity_2014, leethochawalit_keck_2016, wuyts_evolution_2016, carton_first_2018,forster_schreiber_sinszc-sinf_2018,curti_klever_2020,simons_clear_2021,wang_early_2022, ju_msa-3d_2025}. However, the majority of studies agree that metallicity gradients at cosmic noon are more or less flat ($\lesssim |0.1|$ dex kpc$^{-1}$, \citealt{curti_klever_2020}), suggesting that the distribution of metals is likely driven by efficient radial gas mixing. Additionally, observations have also confirmed the existence of a resolved star formation main sequence (rSFMS) and rMZR in the local Universe at these redshifts \citep{jones_resolved_2010, jones_origin_2013, yuan_metallicity_2011}, revealing that even at cosmic noon, the epoch of peak star formation in the Universe, galaxies already maintained a tightly regulated cycle of gas processing, star formation, and chemical enrichment. 

Recent advances, mainly via JWST NIRSpec \citep{jakobsen_near-infrared_2022}, have made it possible to study high-redshift ($\rm z \gtrsim 4$)galaxies on a spatially resolved basis, including deriving gas-phase metallicity gradients \citep[e.g.][]{venturi_gas-phase_2024, tripodi_spatially_2024, arribas_ga-nifs_2024,marconcini_ga-nifs_2025, fujimoto_alpine-cristal-jwst_2025, ivey_exploring_2026, lee_alpine-cristal-jwst_2026}. The rapid growth in publicly available data within this fast-moving field now allows for unprecedented quantification of early galaxy evolution, with essentially all investigations reaching past $\rm z >6$ being published within the last two years. For example, \citet{venturi_gas-phase_2024} analysed three systems at redshifts $\rm z = 6-8$, which feature multiple spatial components, and found that their gas-phase metallicity gradients cover a large scatter and are mostly flat or flat within the uncertainties. \citet{fujimoto_alpine-cristal-jwst_2025} revealed the largest high-z sample of metallicity gradients to date, with 18 galaxies spanning $z=4-6$ revealing mostly flattened gradients, implying that efficient chemical mixing via gas inflows, outflows and mergers is taking place at these epochs. \citet{lee_alpine-cristal-jwst_2026} re-analysed these same galaxies, combining several more metallicity diagnostics and found that on average the gradients are slightly positive at this epoch. Additionally, they also find that dynamical maturity of disks (quantified via $\rm V_{rot}/\sigma$) plays a crucial role in defining the increasingly negative metallicity gradient trend we observe towards lower redshifts. On the contrary, \citet{li_13_2025} utilised JWST/NIRISS observations of over 400 galaxies, median-stacking them into two redshift bins from the ASPIRE survey ($\rm z\sim5-7$) and FRESCO survey ($\rm z\sim7-9$) and found overwhelmingly negative gradients at these epochs. They conclude that these negative gradients likely stem from rapid growth in inside-out mode, which was supported by continuous replenishment from cold gas accretion. One should, however, be cautious about the metallicity gradients inferred from stacks, given the different sizes of the galaxies involved in the process. Additionally, another caution is that there is an intrinsic difficulty when interpreting gradients in high-redshift galaxies. These galaxies are often irregular or undergoing mergers, making the interpretation via averaged annuli difficult. The compactness of these early galaxies also adds a significant uncertainty in inferring metallicity gradients. Nonetheless, a large scatter has been observed among results at high redshifts. A possible explanation for this observed scatter was given by \citet{asada_glimpse-ddt_2026}, where they attribute this finding to possibly a variety of evolutionary pathways for metal enrichment, specifically at the low-mass ($\rm M_* \lesssim 10^{6} M_{\odot}$) end of the MZR. In this regime, galaxies can either experience an 'overshoot' enrichment characterised by rapid first chemical enrichment, or they undergo a delayed enrichment accompanied by a smooth transition from a predominantly Population III to Population II stellar population. Moreover, these recent spatially resolved investigations have also found evidence for an rMZR and rFMR \citep[e.g.][]{gillman_resolved_2022,marconcini_ga-nifs_2025, marconcini_ga-nifs_2024, fujimoto_alpine-cristal-jwst_2025} in the early Universe. Specifically, \citet{fujimoto_alpine-cristal-jwst_2025} found a stronger dependence of the rFMR on $\rm \Sigma_{SFR}$ at $\rm z\sim 4-6$ than in the local Universe, i.e. the spread of metallicities as a function of $\rm \Sigma_{SFR}$ is much larger than at $\rm z \sim 0$. This indicates that short-timescale processes, defined by inflows of pristine gas and SFR-driven outflows, strongly regulate metallicity at these early epochs. 

It is also known that the gradient slope correlates with stellar mass. For the local Universe, it has been observed that galaxies with higher stellar masses exhibit steeper, more negative gradients, while low-mass galaxies show mostly flattened gradients, i.e. there exists an inverse correlation between stellar mass and metallicity gradient with a turnover point commonly assumed at $\rm \log(M_*/M_{\odot}) \sim 10$, above which there is a slightly positive correlation where galaxies with stellar masses $\rm \log(M_*/M_{\odot}) \gtrsim 10$ tend to exhibit more flattened gradients \citep{belfiore_sdss_2017, mingozzi_sdss_2020,poetrodjojo_sami_2021,khoram_direct-method_2025,li_negative_2025}. Physically, inside-out galaxy formation is the likely explanation for higher mass galaxies exhibiting more negative gradients. The fact that galaxies at the highest stellar masses tend to show a flattening of metallicity gradient could be explained by their central regions already reaching a stage of saturation. \citet{li_13_2025} investigated the redshift evolution of the stellar mass vs. gradient slope relation and found that between redshifts $z \sim 0.1 - 3.5$  there is only a weakly negative relation. In contrast, at high redshifts of $z \sim 5-9$, stellar mass and gradient slope are positively correlated; lower stellar masses exhibit steeper negative gradients, while galaxies with higher stellar masses tend to be flatter or even positive. At higher redshift, the positive correlation between stellar mass and metallicity gradient could reflect the dynamical nature of galaxy growth in the early Universe: higher-mass galaxies formed earlier, consisting mainly of a star-forming bulge, and then gradually built up their outer disks over time. This accumulation of metals in the outskirts of galaxies causes flattening of metallicity gradients towards higher stellar masses.

The redshift evolution of gas-phase metallicity gradients has been extensively studied through various simulations. However, results vary significantly depending on the specific models used, particularly the type of stellar feedback implemented \citep{garcia_metallicity_2025, kim_agora_2026}. Some simulations suggest that gradients should become more positive with increasing redshift \citep{mott_abundance_2013} while others predict that gradients flatten with decreasing redshift \citep{gibson_constraining_2013, taylor_metallicity_2017, hemler_gas-phase_2021}. On the contrary, other models \citep{ma_why_2017, bellardini_3d_2021, bellardini_3d_2022, sharda_physics_2021, sun_physical_2025, sun_galaxy_2026, graf_spatial_2025} found that gradients should become more negative towards the local Universe. \citet{tissera_evolution_2022} found a very weak positive redshift evolution, which is mainly driven by an increase in the scatter of gradients at higher redshifts rather than an increase in the fraction of galaxies with positive gradients. Nonetheless, there exist only a few theoretical studies that investigate the evolution of gradients beyond $\rm z>3$. Specifically, \citet{garcia_metallicity_2026} reveal that, across all employed simulations (EAGLE, Illustris, IllustrisTNG, and SIMBA; for redshifts $z=0-8$), metallicity gradients become increasingly negative with increasing redshift, a trend accompanied by substantial scatter. However, these models all implemented smooth stellar feedback, i.e., non-bursty, more gradual star formation over time. \citet{garcia_metallicity_2025} compared two classes of star formation feedback: bursty (FIRE-2, SPICE Bursty, and Thesan Zoom) and smooth (SPICE Smooth and Thesan Box). Here, bursty feedback refers to star formation histories (SFHs) that show distinct, strong peaks of sSFR associated with short timescales, while smooth star formation feedback correlates to SFHs that still show time variations, but generally less and not as strong, i.e. the SFH looks smoother. They found that simulations employing bursty feedback mechanisms mostly overlap with recent observational evidence of flat gradients at high redshifts ($\rm z>4$), suggesting smooth feedback models may not provide the gas-mixing required to reproduce the observations. Lastly, \citet{sun_galaxy_2026} used FIRE-2 cosmological hydrodynamic zoom-in simulations to investigate the evolution of metallicity gradients at $\rm z\sim 5-10$ and found that gradients become more negative with redshift and their scatter increases.

\begin{table*}

    \begin{threeparttable}[b]
        \centering
        \caption{Basic properties of our sample.}
        \begin{tabular}{c|c|c|c|c|c|c}
        \hline
        Name & RA [deg] & DEC [deg] & z & $\rm \mu$ &Scale [kpc arcsec$^{-1}$] \tnote{(a)} & Reference\\
        \hline
        SMACS0723\_4590 & 110.8593 & -73.4492 & 8.45 & 3.74 & 2.477 & \citet{curti_chemical_2023} \\
        JADES\_8013 & 53.1645 & -27.8022 & 8.48 & 1.0 & 4.738 & \citet{curti_jades_2024, curtis-lake_jades_2026} \\
        JADES\_10058975 & 53.1124 & -27.7746 & 9.43 & 1.0 & 4.415 &  \citet{curti_jades_2025, curtis-lake_jades_2026} \\
        RX2129\_11022 & 322.4003 & 0.0832 & 8.15 & 3.29 & 2.68 & \citet{langeroodi_evolution_2023} \\
        RX2129\_11027 & 322.4216 & 0.0917 & 9.51 & 19.2 & 1.003 & \citet{williams_magnified_2023} \\
        Abell\_Z7885 & 3.5960 & -30.3858 & 7.89 & 2.12 & 3.407 & \citet{heintz_dilution_2023} \\
        SXDF\_NB1006-2 & 34.7356 & -5.3330 & 7.21 & 1.0 & 5.245 & \citet{inoue_detection_2016, langeroodi_evolution_2023}\\
        \hline
        \end{tabular}
        \begin{tablenotes}
           \item \textit{Note.} $^{(a)}$ Scales are corrected for the lensing magnification factor. 
         \end{tablenotes}
        \label{tab:data}
    \end{threeparttable}

\end{table*}

It is important to stress that there are very few high-z ($\rm z>6$) observations of gas-phase metallicity gradients using IFU data \citep[e.g.][]{arribas_ga-nifs_2024, venturi_gas-phase_2024, ivey_exploring_2026}, and none of them reaches out to $\rm z>8$. In this paper, we aim to further explore the cosmic evolution of the gas-phase metallicity using a sample of seven low-metallicity galaxies at redshifts $z = 7.2-9.5$, observed with the JWST/NIRSpec IFU. These are some of the highest redshift galaxies for which metallicity gradients have been determined. We focus our analysis on deriving gas-phase metallicity gradients via the $\rm [OIII]\lambda 5007/H \beta$ and $\rm [NeIII]\lambda3869/[OII]\lambda\lambda 3727,29$ strong line ratio diagnostics. By combining our NIRSpec-IFS data with publicly available NIRCam data, we derive updated stellar mass measurements, allowing us to study the galaxies' position on the MZR. We also include recent results from observations and simulations, which allow us to analyse the redshift evolution of gas-phase metallicity gradients from $z=0$ to $z=10$. 

This paper is organised as follows. In section~\ref{sec:data-reduction-analysis}, we describe the NIRSpec-IFS observations and the reduction (section~\ref{sec:data}), spectral fitting (section~\ref{sec:analysis-EL}), and spectral energy distribution (SED) fitting (section~\ref{sec:sedfitting}) procedures. Within section~\ref{sec:analysis-EL}, we also describe the methods for deriving gas-phase metallicities and the corresponding gradients. We present the results in section~\ref{sec:results} and discuss the physical implications of our observations in section~\ref{sec:discussion}. Lastly, we summarise our main findings in section~\ref{sec:conclusions}. 

Throughout this paper, we assume a \cite{chabrier_galactic_2003} initial mass function (IMF) and adopt a flat $\mathrm{\Lambda CDM}$ cosmology with $\mathrm{H_0 = 67.4 \ km s^{-1} Mpc^{-1}}$, $\mathrm{\Omega_m = 0.315}$, and $\mathrm{\Omega_{b} = 0.0492}$ \citep{planck_collaboration_planck_2020}. We assume a solar metallicity value of $\rm 12+log(O/H) = 8.69$ \citep{asplund_chemical_2009}. 

\section{Observations, Data Reduction, and Analysis}\label{sec:data-reduction-analysis}

\subsection{Data}\label{sec:data}

\subsubsection{NIRSpec Data}\label{sec:data-nirspec}

The focus of this work lies on the JWST/NIRSpec-IFS observations of seven low-metallicity galaxies carried out between October 2023 and August 2024 as part of the GO program ID 2957 (PIs: H.~Übler, R.~Maiolino). These galaxies were pre-selected based on existing emission line measurements to have $\rm 12+log(O/H)<7.5$ (see reference column in Table~\ref{tab:data}). Previous metallicity values, which were used to select this sample, were derived using NIRSpec/MSA data, and can thus give different results to our integrated IFU measurements shown in section~\ref{sec:results_MZR}, depending on the MSA coverage of the galaxy. The seven sources were observed using a medium cycle pattern of 12 dither positions (10 dither positions for SXDF\_NB1006-2) and a total integration time of $\sim$ 4.9h per source ($\sim$ 4.1h for SXDF\_NB1006-2). The observations were performed with the PRISM/CLEAR disperser-filter combination covering a wavelength range of 0.6 -- 5.3$\rm \mu m$ ($\rm R \sim 30-300$). For our targets at $z=7.2-9.5$, this covers line emission from Ly$\alpha$ to [O~III]$\lambda5007$.

Raw data files were downloaded from the Barbara A.~Mikulski Archive for Space Telescopes (MAST) and subsequently processed with the {\it JWST} Science Calibration pipeline\footnote{\url{https://jwst-pipeline.readthedocs.io/en/stable/jwst/introduction.html}} version 1.15.0 under the Calibration Reference Data System (CRDS) context jwst\_1281.pmap. To increase data quality, we added several steps to the standard reduction steps. They are described by \cite{perna_ga-nifs_2023}, and we briefly summarise them here. Count-rate frames were corrected for $1/f$ noise through a polynomial fit. During Stage 2, we masked regions affected by bad pixels, cosmic ray hits, and failed open MSA shutters. 
We removed remaining outliers in individual exposures following \cite{deugenio_fast-rotator_2024}, rejecting pixels for which the normalised derivative in wavelength direction was higher than the 99.9\textsuperscript{th} percentile (see \citealt{deugenio_fast-rotator_2024} for details). The final cubes were combined using the `drizzle' method with a pixel scale of $0.05''$. We use spaxels away from the central source and free of line emission and galaxy continuum emission to perform a background subtraction.

No target acquisition was included in our observing setup. Due to NIRSpec IFU observations relying on blind target pointing for their science exposures, significant astrometry offsets have been reported. These are usually contained within scales of $\sim 0.1-0.3^{\prime \prime}$ \citep[e.g.][]{arribas_ga-nifs_2024,jones_ga-nifs_2024,lamperti_ga-nifs_2024, ubler_ga-nifs_2024,fujimoto_alpine-cristal-jwst_2025, fujimoto_primordial_2025, parlanti_ga-nifs_2025, jones_ga-nifs_2026}. To align our NIRSpec IFU and NIRCam data, we correct the astrometry by aligning the centroid of a pseudo F444W image created from the IFU cube with the centroid of the F444W image. We find typical offsets of $\rm \sim 0.13-0.18\arcsec$, which are consistent with the aforementioned previous studies. 

We obtain integrated spectra by selecting a circular aperture on the IFS data which are of sizes $\rm 5-10$ pixel (corresponding to roughly $\rm 0.5-1 \arcsec$ diameters).

As reported by \cite{ubler_ga-nifs_2023}, the flux uncertainties provided in the `ERR' extension of the IFU data cubes tend to underestimate the noise relative to the r.m.s. measured directly from the spectra. Nevertheless, the `ERR' extension contains valuable information about outliers and the correlated noise between spectral channels. This specifically occurs when integrating several spaxels \citep[e.g.][]{jones_ga-nifs_2026, venturi_ga-nifs_2025}; thus, we need to correct the errors for integrated spectra when we compute the galaxies' global values and metallicity gradients via annuli. We rescale the errors so that their median value matches the $\rm \sigma$-clipped r.m.s. measured in emission-line–free regions. The scaling factor results in about 2 to 4, depending on the dataset.

The JWST NIRSPec IFU observations span a total nominal field of view (FOV) of $\sim 3^{\prime\prime} \times 3^{\prime\prime}$. An overview of the seven sources is given in Table~\ref{tab:data}.  

\subsubsection{NIRCam Data}\label{sec:data-nircam}

We use JWST/NIRCam imaging data from the Dawn JWST Archive (DJA) mosaic release v7. The DJA is an online repository that provides reduced images, photometric catalogues, and spectroscopic data derived from publicly available JWST observations. Details of the data processing and reduction are described in \citet{valentino_atlas_2023} and \citet{brammer_grizli_2023}.

NIRCam photometry for six of our galaxies is extracted from the following DJA mosaic fields: \texttt{smacs0723-grizli-v7.4}, \texttt{gds-grizli-v7.2}, \texttt{rxj2129-grizli-v7.0}, and \texttt{abell2744clu-grizli-v7.2}. The relevant comes from the following programs: SMACS0723 (PID: 2736, PI: K. Pontoppidan, \citealt{pontoppidan_jwst_2022}; PID: 4043, PI: C. Witten, \citealt{witten_not_2025}), Goods-South (GDS) (PID: 1180, PI: D. Eisenstein, \citet{eisenstein_overview_2026}; PID: 1210, PI: N. Lützgendorf, \citet{eisenstein_overview_2026}; PID: 1895, PI: P. Oesch, \citealt{oesch_jwst_2023}; PID: 1963, PI: C. Williams, \citealt{williams_jems_2023}; PID: 2079, PI: S. Finkelstein, \citealt{bagley_next_2024}; PID: 2514, PI: C. Williams, \citealt{williams_panoramic_2025}; PID: 3215, PI: D. Eisenstein, \citealt{eisenstein_jades_2025}), RXJ2129 (PID: 2767, PI: P. Kelly, \citealt{williams_magnified_2023}), Abell2744 (PID: 1324, PI: T. Treu, \citealt{merlin_early_2022}; PID: 2561, PI: I. Labbe, \citealt{bezanson_jwst_2024}; PID: 2756, PI: W. Chen, \citealt{paris_glass-jwst_2023}; PID:3516, PI: K. Suess \citealt{naidu_all_2024}; PID:3990, PI:, \citealt{morishita_beacon_2025}; PID: 4111, PI:, \citealt{suess_medium_2024}). No photometric data for SXDF\_NB1006-2 are currently available in the DJA. The archive provides reduced and flux-calibrated images in photometric units of 10 nJy, which we use to measure aperture photometry and derive stellar masses through SED fitting (see Section~\ref{sec:sedfitting}).  

Aperture photometry from the NIRCam data is obtained through the following steps:

\begin{enumerate}
\item Each NIRCam mosaic field is cropped around the target galaxy and aligned with the NIRSpec data. We match the brightest pixel of the F444W filter to the brightest pixel of a pseudo F444W filter image created from the IFU cube (see Sect.~\ref{sec:data-nirspec}).
\item A median background, using pixels free of source mission, subtraction is applied to each cutout.
\item All additional filters used in the aperture photometry are point spread function (PSF)-matched to the $\rm F444W$ resolution.
\end{enumerate}

We obtain PSF-models from the \texttt{STPSF} Python package \citep{perrin_updated_2014}. Finally, we extract a circular aperture centered on the galaxy and compute the total flux and associated uncertainty.

\subsection{Data Analysis}\label{sec:analysis}

\subsubsection{Emission Line Fitting}\label{sec:analysis-EL}

In this section, we describe the emission line fitting analysis of the NIRSpec R100 IFU data. In summary, we obtain emission line fluxes using Gaussian fitting by modelling the spectra, both on an integrated basis and spatially resolved (spaxel-by-spaxel). This analysis consists of measuring the following rest-frame optical and near-UV emission lines available in the observed spectral range: $\rm [NeV]\lambda 3436$, $\rm [OII]\lambda\lambda 3727,29$, $\rm [NeIII] 3869$, $\rm H\delta$, $\rm [OIII]\lambda 4364$, $\rm H\gamma$, $\rm H\beta$, and $\rm [OIII]\lambda 5007$. Hereafter, we will refer to $\rm [OII]\lambda 3728 = [OII]\lambda 3727 + [OII]\lambda 3729$. 

We additionally include the $\rm [HeI]\lambda 3889$ emission line solely to aid in more accurately fitting the emission lines of interest mentioned above, as some of these lines are partially blended. For measuring emission line fluxes, we group the following emission lines:

\begin{enumerate}
    \item $\rm [NeV]\lambda 3436$
    \item $\rm [OII]\lambda 3728$
    \item $\rm [NeIII]\lambda 3869,3968$, $\rm [HeI]\lambda 3889$, $\rm H\epsilon$, $\rm H\delta$
    \item $\rm [OIII]\lambda 4364$, $\rm H\gamma$
    \item $\rm H\beta$, $\rm [OIII]\lambda 4960,5007$
\end{enumerate}

In PRISM spectra, whose spectral resolution is $\rm \sigma_{res}\sim 1250~km~s^{-1}$ at $\rm \sim 3~ \mu m$ to $\rm \sigma_{res}\sim 450~km~s^{-1}$ at $\rm \sim 5~ \mu m$, all emission lines in our targets are spectrally unresolved, therefore, we use a single Gaussian function per emission line. For practical simplicity, each line complex detailed above is fitted separately: we tie the redshift and FWHM of the emission lines within each group, but allow each amplitude to vary. The doublet $\rm [OII]\lambda 3728$ is always blended at the resolution of our PRISM spectra,  so we fitted it with a single Gaussian. We fixed the flux ratios of $\rm [OIII]\lambda 5007/4960$ to their theoretical value of 2.99 \citep{dimitrijevic_flux_2007}, and of $\rm [NeIII]\lambda 3968/3869$ to 0.301 \citep{osterbrock_astrophysics_2006,jones_ga-nifs_2026}. 

Additionally, we model and subtract the underlying continuum before applying our Gaussian fits. We use only the regions surrounding the emission lines of interest and free of any residual artefacts for fitting the continuum. For galaxies whose spectra extend redward of the $\rm [OIII]\lambda 5007$ line, we perform two separate continuum fits, to fit the slope of the conitnuum more accurately accross the entire spectrum. The first fit covers the regions surrounding all emission lines blueward of $\rm H\beta$, while the second fit includes the regions immediately before $\rm H\beta$ and after $\rm [OIII]\lambda 5007$. The first fit is used for continuum subtraction of the emission lines blueward of $\rm H\beta$, whereas the second fit is used exclusively for the continuum subtraction of $\rm H\beta$ and $\rm [OIII]\lambda 5007$. For galaxies whose spectra terminate shortly after $\rm [OIII]\lambda 5007$, corresponding to the highest-redshift galaxies in our sample (JADES\_8013, JADES\_10058975, and RX2129\_11027), we derive only a single continuum fit. All continua are modelled using a second-order polynomial. 

Furthermore, in order to improve the S/N at the spaxel level for visual purposes, we apply the Gaussian fitting routine to a spatially smoothed version of the NIRSpec IFU cubes. The smoothing is performed using a $3 \times 3$ spaxel median filter, where each spaxel is recalculated as the median between its own flux and that of its eight immediate neighbours. We present smoothed 2D maps in Appendices~\ref{app:spec_flux} and ~\ref{app:metal_grads}; however, we use the un-smoothed cubes to measure metallicity gradients and only rely on the smoothed data cubes for better visualisation of the 2D maps.

As for the metallicity, since we work with line flux ratios that are close in wavelength, we do not apply any correction for dust attenuation, and we do not PSF-match the line maps. Metallicities are computed via the $\mathrm{R3} = \log ([\mathrm{O\,III}]\lambda5007 / \mathrm{H}\beta)$ and $\mathrm{Ne3O2} = \log ([\mathrm{Ne\,III}]\lambda3869 / [\mathrm{O\,II}]\lambda3728)$ ratios using newly-defined calibrations based on high-z galaxies \citep{isobe_jades_2026}. We select the emission-line ratios used for metallicity estimation in an adaptive manner, depending on the signal-to-noise (S/N) of the relevant lines. The R3 ratio is employed only when both the $\mathrm{H}\beta$ and $[\mathrm{O\,III}]\lambda5007$ lines have $\mathrm{S/N}>3$, but it is known that this curve can result in ambiguous values as it contains a low- and high-metallicity branch. Thus, the Ne3O2 ratio is additionally included when the $[\mathrm{Ne\,III}]\lambda3869$ and $[\mathrm{O\,II}]\lambda3728$ lines also meet the same $\mathrm{S/N}>3$ criterion. This approach ensures that each spectrum contributes only line ratios based on reliably detected emission lines, avoiding the use of ratios involving low-S/N measurements while still making use of all available high-quality data. 

For a given set of calibrations, each diagnostic ratio $r_d$ is modelled as a polynomial function of $x = 12+\log(\mathrm{O/H}) - 8.0$, with coefficients and intrinsic scatters taken from the calibration. For each galaxy, we construct the vector of observed log ratios $\mathbf{r}_\mathrm{obs}$ and associated uncertainties $\boldsymbol{\sigma}_\mathrm{obs}$ from the available diagnostics (R3, and Ne3O2 when detected). At a trial metallicity $x$ we evaluate the model ratios $\mathbf{r}_\mathrm{mod}(x)$ and compute 
\begin{equation}
    \chi^2(x) = \sum_d \frac{\left[r_{\mathrm{obs},d} - r_{\mathrm{mod},d}(x)\right]^2}{\sigma_{\mathrm{obs},d}^2 + \sigma_{\mathrm{int},d}^2},
\end{equation}
where $\sigma_{\mathrm{int},d}$ is the intrinsic scatter of diagnostic $d$ in the calibration. We adopt a uniform prior in $\rm x$, and sample the posterior $P(x \,|\, \mathbf{r}_\mathrm{obs}) \propto \exp[-\chi^2(x)/2]$ with the \textsc{emcee} affine-invariant MCMC sampler \citep{foreman-mackey_emcee_2013}. We quote the posterior median of $12+\log(\mathrm{O/H})$ as our fiducial metallicity, and the upper and lower $1\sigma$ uncertainties correspond to the difference between the median and the 84th and 16th percentiles of the posterior, respectively. For R3-only metallicities, we break the branch degeneracy by adopting the lower-branch solution where multiple solutions exist.

We derive Star Formation Rates (SFRs) by converting $\rm H\beta$ to $\rm H\alpha$ assuming an intrinsic ratio of 2.86 (Case B recombination, $\rm T_e = 10,000$ K; \citealt{osterbrock_astrophysics_2006}). While $\rm H\gamma$ is blended with the $[\mathrm{OIII}]\lambda 4363$ auroral line in our PRISM spectra, we calculate dust attenuation via the $\rm H\delta/H\beta$ ratio. We adopt an intrinsic $\rm H\delta/H\beta$ of 0.26 ($\rm T_e = 15,000$ K, $\rm n_e = 100\text{ cm}^{-3}$; \citealt{reddy_aurora_2026}) and the \citet{calzetti_dust_2000} attenuation law.

This correction is applied only where both lines exceed $\rm S/N > 3$ in the integrated spectra; for ratios exceeding the intrinsic value, we set $\rm A_V = 0$ (no dust correction). Due to insufficient $\rm S/N$ at the spaxel level, we apply these global $\rm A_V$ values to both integrated and spatially-resolved SFRs. Under these criteria, only SXDF\_NB1006-2 yielded a positive correction ($\rm A_V = 1.6 \pm 0.5$); all other galaxies are not corrected for dust attenuation.

SFRs are computed via the conversion law from \citet{kennicutt_star_2012}, and we convert $\rm \log SFR$ from a Kroupa to a Chabrier IMF by subtracting a constant factor of $\sim0.027$ dex \citep{madau_cosmic_2014}:
\begin{equation}
    \mathrm{\log{(SFR  /M_{\odot} yr^{-1})} = \log{(L_{H\alpha}} /erg \ s^{-1}) - 41.27} - 0.027.
\end{equation}

This relation is inferred by simply assuming a constant star formation over time and solar metallicity, which are conditions that certainly do not apply for the galaxies in our sample. Deriving more accurate SFRs would require inferring star formation histories and iterative processes. Additionally, other assumptions might be violated, like that all ionising photons interact with the gas, and not with dust or simply escape \citep{tacchella_h_2022}. As we do not need an accurate measurement of the SFR in this analysis, we decided to adopt the simplified approach of using the simple linear relation given above.

\subsubsection{SED fitting}\label{sec:sedfitting}
In this section, we describe our \texttt{Prospector} \citep{johnson_stellar_2021} model setup used to infer galaxy properties, including stellar mass, metallicity, star formation history (SFH) and star formation rate (SFR), gas-phase metallicity and ionization parameter, and dust attenuation.

\texttt{Prospector} builds upon the Flexible Stellar Population Synthesis (\texttt{FSPS}) code \citep{conroy_propagation_2009, conroy_propagation_2010}, accessed via \texttt{python-fsps} \citep{johnson_stellar_2021}. Our models adopt the MIST stellar isochrones \citep{choi_mesa_2016, dotter_mesa_2016}, the MILES stellar spectral library \citep{sanchez-blazquez_medium-resolution_2006}, and assume a \citet{chabrier_galactic_2003} initial mass function. 

We fit galaxies’ spectroscopic and photometric data simultaneously. We utilise the following NIRCam filters, when available, for our SED fitting: F090W, F115W, F150W, F200W, F277W, F356W, and F444W. However, for galaxy JADES\_100558975, not all of the aforementioned wide filters are available, and we therefore rely on the following set of filters: F182M, F210M, F277W, F357W, F444W. During the fitting, we mask all data bluer than the Lyman-break wavelength ($1216$ \AA) and fix all galaxy redshifts to their spectroscopic values. The stellar mass follows a prior informed by the observed stellar mass function, as introduced in the \texttt{Prospector-$\beta$} model of \citet{wang_inferring_2023}. In this prior, lower-mass galaxies have higher prior probabilities, helping to avoid spurious high-redshift, high-mass solutions. The stellar metallicity follows a truncated Gaussian prior with mean $\mathcal{N}\left(0.5;\ -1.0\right)$, truncated to $\left[-2.0;\ 0.0\right]$. For the SFH, we adopt the non-parametric continuity model of \citet{leja_how_2019} with 8 age bins. The logarithm of the SFR ratios between adjacent bins is fitted and follows a Student’s $t$ prior, with $\sigma = 0.3$ and $\nu = 2.0$, and an expected value centred at $0$. Nebular emission is modelled using \texttt{Cloudy} (v13.03; \citealt{ferland_2013_2013, ferland_2017_2017}) as implemented in \texttt{FSPS} \citep{byler_nebular_2017}. The $\log_{10}$ of the gas-phase metallicity and the $\log_{10}$ ionisation parameter $U$ are treated as free parameters, with uniform priors over the ranges $[-3.0,0.0]$ and $[-4.0,-1.0]$, respectively.

Dust attenuation follows the two-component model of \citet{charlot_simple_2000}, where the diffuse component optical depth $\tau_{\mathrm{dust,2}}$ follows $\mathcal{N}(0.3, 1)$ truncated to $[0, 4]$, and the birth-cloud to diffuse ratio $\tau_{\mathrm{dust,1}}/\tau_{\mathrm{dust,2}} \sim \mathcal{N}(1, 0.3)$ truncated to $[0, 2]$. The attenuation curve slope $n$ is drawn from a uniform prior over $[-1.2, 0.4]$ following \citet{noll_analysis_2009}. Lastly, we include a single “jitter” parameter that rescales the spectral uncertainties to account for residual calibration imperfections and to achieve an adequate fit.

\section{Results}\label{sec:results} 
\begin{table*}
    \caption{Main derived properties of the individual galaxies and satellites. Properties are corrected for lensing magnification.}
    \label{tab:data_results}
    \begin{threeparttable}[b]
        \centering
        \renewcommand{\arraystretch}{1.5}
        \begin{tabular}{c|c|c|c|c|c|c}
        \hline
        Name & $z$ & $\rm \log M_{*}/M_{\odot}$ &
        $\rm \log SFR_{H\beta}$ [$\rm M_{\odot}$ yr$^{-1}$] &
        $\rm 12+\log(O/H)$ &
        $\nabla_r \log(Z)\ [\mathrm{dex\ kpc^{-1}}]$ &
        $\rm \Delta FMR$\\
        \hline
        SMACS0723\_4590   & 8.45 & $9.17^{+0.02}_{-0.02}$ & $0.90 \pm 0.02$ & $7.38^{+0.07}_{-0.07}$ & $0.07^{+0.10}_{-0.10}$ & $-0.96 \pm 0.07$ \\
        SMACS0723\_4590-S1  & 8.45 & $6.77^{+0.05}_{-0.04}$ & $-0.52 \pm 0.05$ & $7.12^{+0.09}_{-0.08}$ & --                & $-0.72 \pm 0.09$ \\
        SMACS0723\_4590-S2                 & 8.45 & $7.29^{+0.04}_{-0.04}$ & $-0.66 \pm 0.09$ & $7.12^{+0.13}_{-0.11}$ & --                & $-0.89 \pm 0.12$ \\
        JADES\_8013 \tnote{(a)}        & 8.48 & $7.81^{+0.04}_{-0.04}$ & $0.62 \pm 0.04$ & $7.71^{+0.10}_{-0.13}$ & $-0.20^{+0.38}_{-0.39}$ & $-0.25 \pm 0.12$ \\
        JADES\_10058975 \tnote{(a)}   & 9.44 & $8.92^{+0.06}_{-0.05}$ & $1.12 \pm 0.03$ & $7.80^{+0.07}_{-0.07}$ & $-0.03^{+0.16}_{-0.15}$ & $-0.48 \pm 0.07$ \\
        RX2129\_11022      & 8.15 & $8.39^{+0.04}_{-0.05}$ & $0.57 \pm 0.03$ & $7.18^{+0.07}_{-0.06}$ & $-0.13^{+0.15}_{-0.14}$ & $-0.97 \pm 0.07$ \\
        RX2129\_11027      & 9.51 & $8.08^{+0.02}_{-0.02}$ & $0.11 \pm 0.03$ & $7.72^{+0.08}_{-0.08}$ & $ 0.04^{+0.10}_{-0.10}$ & $-0.41 \pm 0.08$ \\
        RX2129\_11027-S1 & 9.51 & $6.83^{+0.19}_{-0.10}$ & $-0.85 \pm 0.07$ & $7.26^{+0.16}_{-0.12}$ & --                & $-0.65 \pm 0.15$ \\
        Abell\_Z7885       & 7.89 & $8.99^{+0.03}_{-0.03}$ & $0.98 \pm 0.02$ & $7.75^{+0.08}_{-0.10}$ & $ -0.08^{+0.09}_{-0.08}$ & $-0.51 \pm 0.09$ \\
        SXDF\_NB1006-2    & 7.21 & $9.47^{+0.01}_{-0.01}$ & $2.52 \pm 0.15$ & $7.78^{+0.06}_{-0.06}$ & $ 0.14^{+0.05}_{-0.06}$ & $-0.36 \pm 0.07$ \\        
        \hline
        \end{tabular}
        \begin{tablenotes}
           \item \textit{Note.} $^{(a)}$ Galaxies JADES\_10058975 and JADES\_8013 correspond to JADES DR5 IDs 265801 and 110748, respectively.
         \end{tablenotes}
    \end{threeparttable}
\end{table*}

\begin{figure*}
    \centering
    \includegraphics[width=\linewidth]{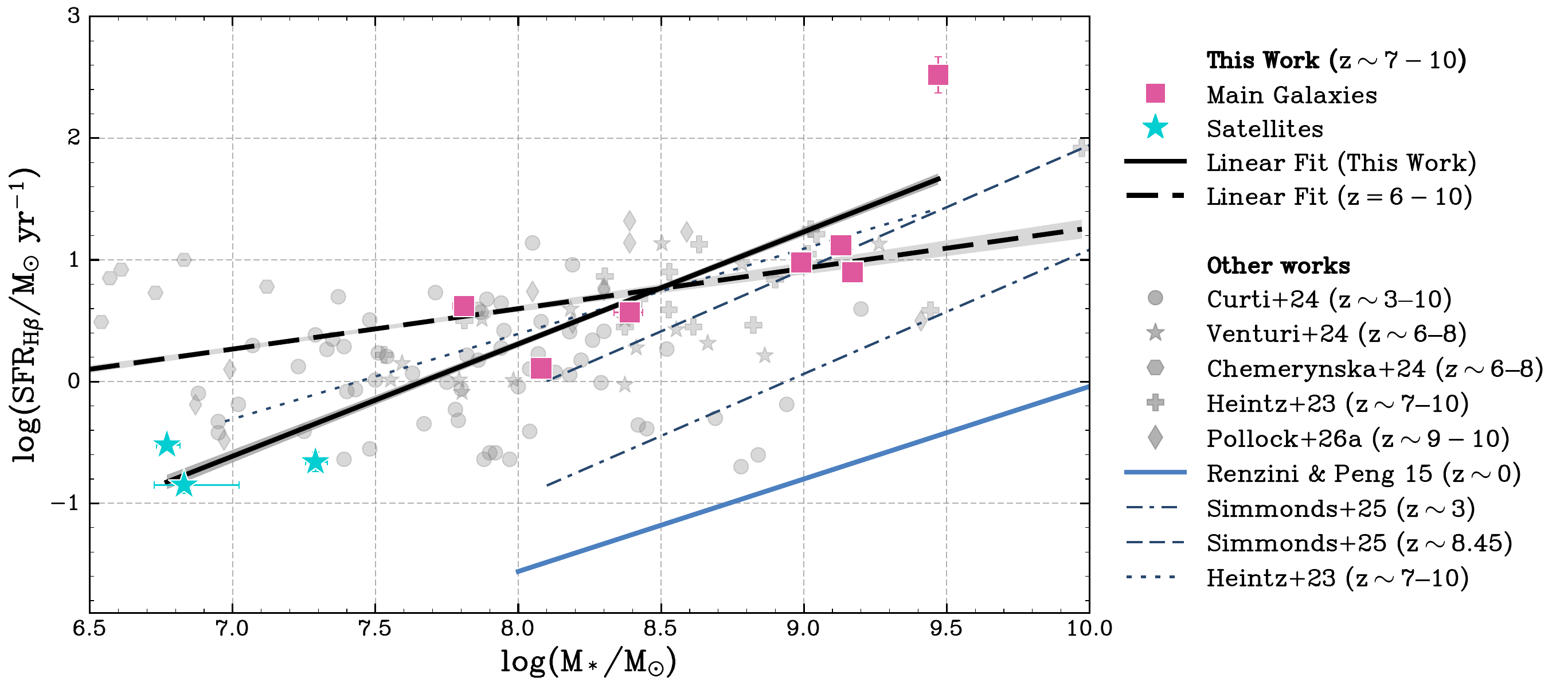}
    \caption{Star Formation Main Sequence (SFMS) of our seven main galaxies (pink squares) and three detected satellites (blue stars). In differently shaped grey points, we show recent JWST observations spanning similar redshifts to our sample of $\rm z\sim 3-10$ from \citet{curti_chemical_2023} (circles), \citet{venturi_gas-phase_2024} (stars), \citet{chemerynska_extreme_2024} (hexagons), \citet{heintz_dilution_2023} (plus), and \citet{pollock_novel_2026} (diamonds). We display the SFMS at $\rm z=0$ by \citet{renzini_objective_2015}, at $\rm z\sim 3$ and $\rm z=8.45$ (i.e. the average redshift within our sample) using the derived fit from \citet{simmonds_bursting_2025}, and at $\rm z=7-10$ \citep{heintz_dilution_2023}. Our own derived linear fits are shown as solid black lines together with their $\rm 1\sigma$ confidence interval as a grey shaded area.}
    \label{fig:SFMS}
\end{figure*}

\subsection{Integrated Scaling Relations}

We start by investigating the global properties of our galaxies and how they are located on the known scaling relations. We obtained global properties of each galaxy by applying a circular aperture, covering roughly 5-10 pixels in radius, depending on the extent of each galaxy, to the IFU data and taking the sum of the fluxes of each spaxel to derive integrated spectra. We then computed gas-phase metallicities and total SFRs as described in section~\ref{sec:analysis-EL}. Global results can be found in Table~\ref{tab:data_results}.

Integrated galaxy spectra, NIRCam imaging cutouts, and 2D flux maps can be seen in Figures~\ref{fig:spec_flux_SMACS0723_4590} to \ref{fig:spec_flux_SXDF_NB1006-2} of Appendix~\ref{app:spec_flux}. 

\subsubsection{The Star Formation Main Sequence (SFMS)}

We infer SFRs from the integrated galaxy spectra to explore the location of our galaxies on the Star Formation Main Sequence (SFMS), as shown in Figure~\ref{fig:SFMS}. We compare our data to those for galaxies in the local Universe \citep{renzini_objective_2015} and at high redshifts \citep{heintz_dilution_2023}. We also plot the relations from \citet{simmonds_bursting_2025} at $\rm z=3$
and at $\rm z \sim 8.45$, which is the median redshift of our sample. The galaxies of our sample are situated among the individual data points from other metallicity studies at high redshift \citep{heintz_dilution_2023, curti_jades_2024, chemerynska_extreme_2024, venturi_gas-phase_2024, pollock_novel_2026}.

We also fit an error-weighted linear fit for the SFMS and find the following relation fitted to our data points: $$\rm \log(SFR \ /M_{\odot} yr^{-1}) = (0.92 \pm 0.04) \times \log(M_*/M_{\odot}) - (7.04\pm 0.29).$$ Furthermore, we perform another linear fit combining our data points and those of studies at similar redshift \citep[$\rm z\sim 6-10$;][]{heintz_dilution_2023, curti_jades_2024, chemerynska_extreme_2024, venturi_gas-phase_2024, pollock_novel_2026}: $$\rm \log(SFR \ /M_{\odot} yr^{-1}) = (0.33 \pm 0.03) \times \log(M_*/M_{\odot}) - (2.06\pm 0.20).$$ These two fits are strikingly different: our sample is clearly biased to very low $\rm SFRs$ at the low mass end (i.e. the satellites introduced in Section~\ref{sec:sats}), while our main galaxies align with results covering similar redshifts.

As mentioned, on average our galaxies are aligned with galaxies in previous metallicity studies, but are slightly above ($\rm \sim 0.42$ dex) the SFMS determined at a similar redshift by \citet{simmonds_bursting_2025}, indicating a slight bias towards higher SFR; this is not unexpected given that high-z samples exploring the gas metallicity require the detection of nebular emission lines with good S/N, hence typically translating in higher than average SFR. We note that SXDF\_NB1006-2 is, in particular, notably offset from our own relation and those we plot for comparison. This is as expected, as this galaxy has already previously been identified as undergoing an intense starburst \citep[see][and section~\ref{sec:discuss-grads} for details]{inoue_detection_2016, ren_updated_2023, ren_rioja_2025}. 

\begin{figure*}
    \centering
    \includegraphics[width=\linewidth]{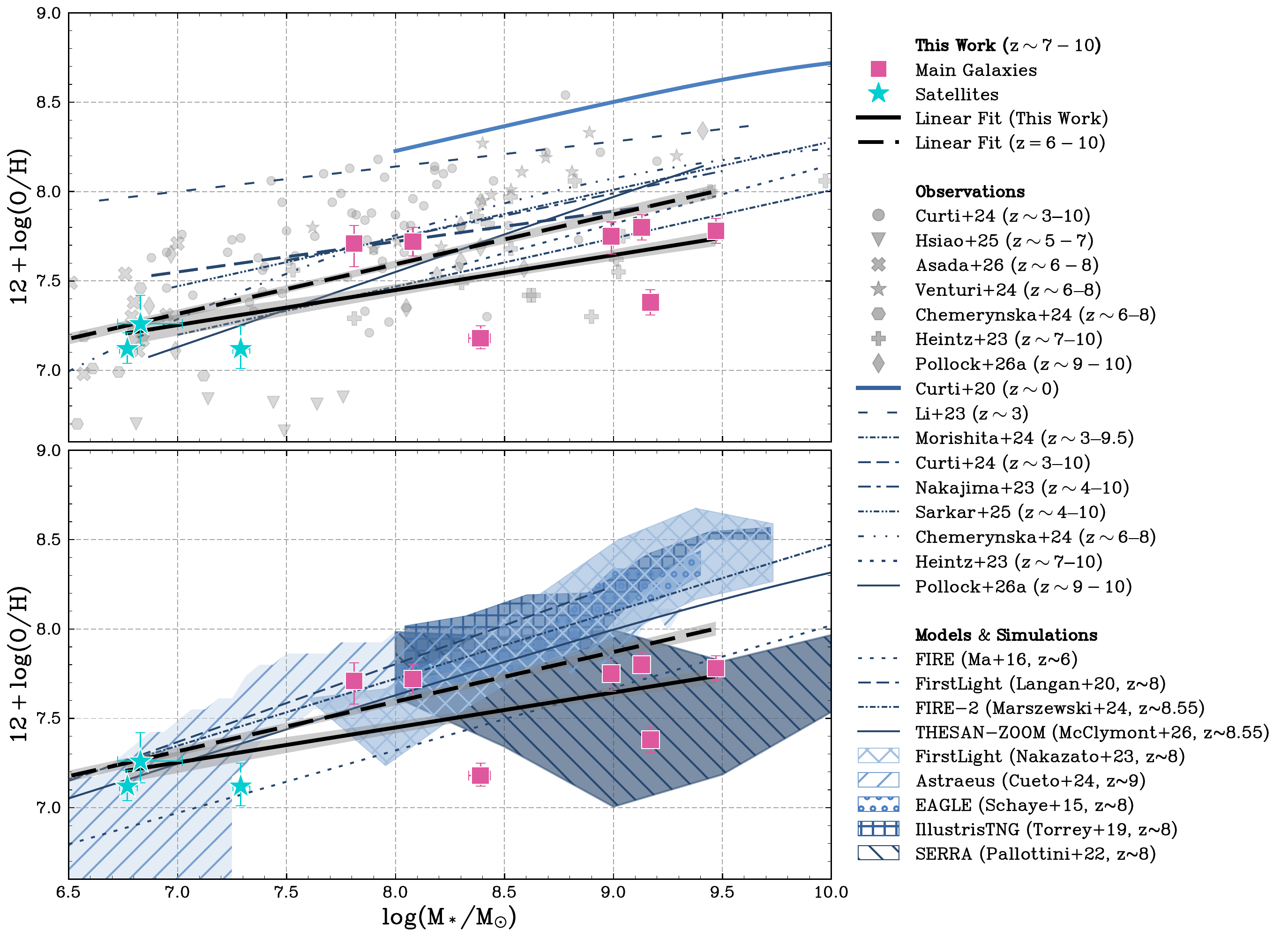}
    \caption{Mass-Metallicity Relation (MZR) of our seven main galaxies (pink squares) and three detected satellites (blue stars). Our own derived linear fits are shown as solid black lines together with their $\rm 1\sigma$ confidence interval as a grey shaded area. \textit{Top:} MZR compared to other observations. In differently shaped grey points, we show recent JWST observations spanning similar redshifts to our sample of $\rm z\sim 3-10$ from \citet{curti_chemical_2023} (circles), \citet{hsiao_sapphires_2025} (triangles), \citet{venturi_gas-phase_2024} (stars), \citet{chemerynska_extreme_2024} (hexagons), \citet{heintz_dilution_2023} (plus), \citet{asada_glimpse-ddt_2026} (crosses), and \citet{pollock_novel_2026} (diamonds). We also compare to previous fits for the local Universe \citep{curti_mass-metallicity_2020}, dwarf galaxies at $\rm z \sim 3$ \citep{li_mass-metallicity_2023}, as well as various best-fits using JWST observations spanning $\rm z=3-10$ \citep{morishita_diverse_2024, curti_chemical_2023, nakajima_jwst_2023, sarkar_unveiling_2025, chemerynska_extreme_2024, heintz_dilution_2023, pollock_novel_2026}. \textit{Bottom:} MZR compared to simulations and models at similar redshifts. We display the high-z predictions from the FIRE \citep{ma_origin_2016}, FirstLight \citep{langan_weak_2020, nakazato_simulations_2023}, FIRE-2 \citet{marszewski_high-redshift_2024}, THESAN-ZOOM \citep{mcclymont_thesan-zoom_2026}, EAGLE \citep{schaye_eagle_2015}, IllustrisTNG \citep{torrey_evolution_2019}, Astraeus assuming an evolving IMF \citep{cueto_astraeus_2024} and SERRA \citep{pallottini_survey_2022} cosmological simulations as differently dashed/dotted lines as well as hatched areas.}
    \label{fig:MZR}
\end{figure*}

\begin{figure*}
    \centering
    \includegraphics[width=\linewidth]{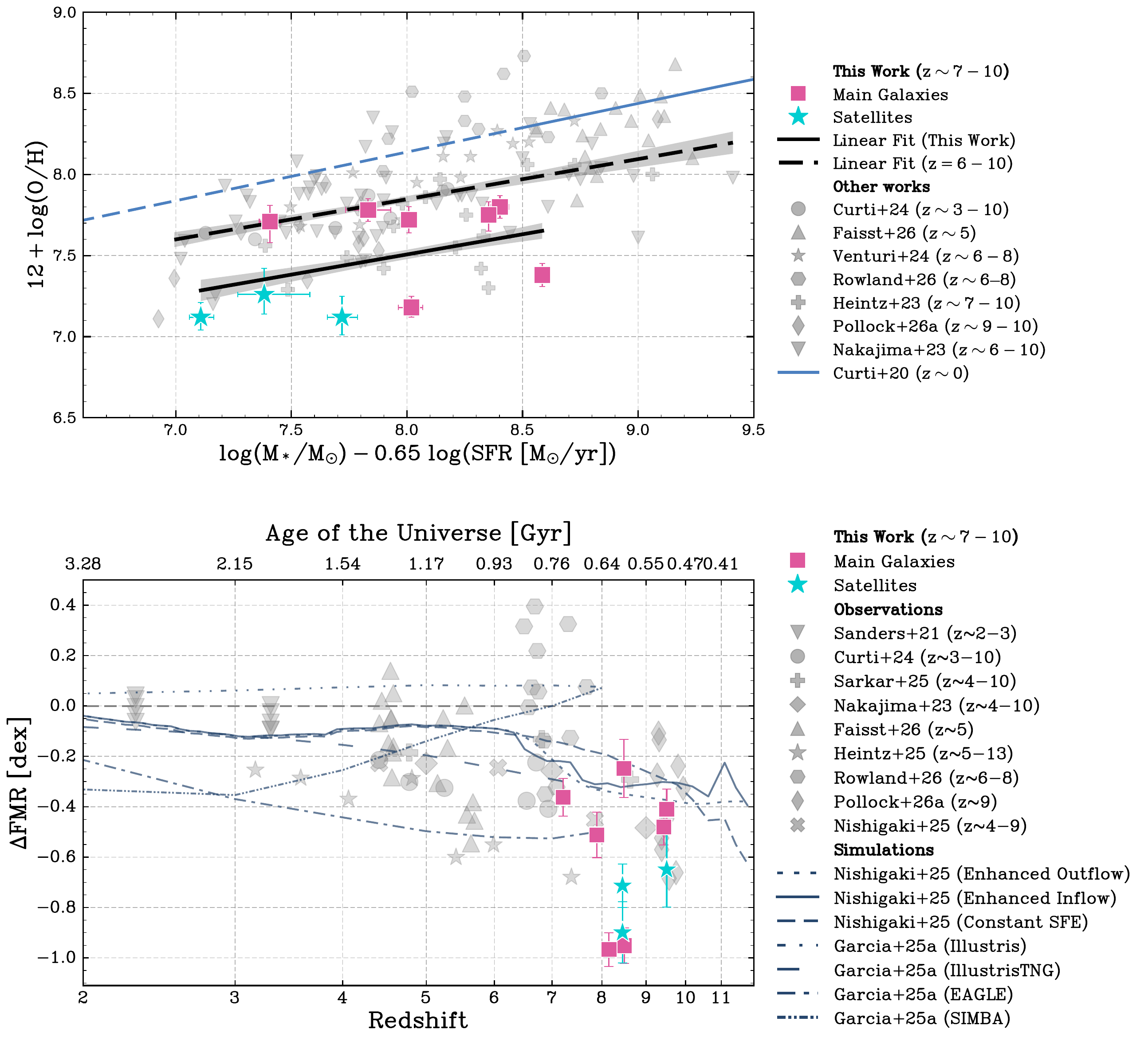}

    \caption{The fundamental metallicity relation at $\rm z=7-10$. Results for the main galaxies and satellites are shown as pink squares and blue stars, respectively. \textit{Top:} $\rm \mu_{0.65}$ vs. metallicity relation (see equation~\ref{eq:FMR_alpha}).  The local FMR by \citet{curti_mass-metallicity_2020} is shown as a solid dark blue line, which is extrapolated to lower $\rm \mu_{\alpha}$ indicated as a dashed line. Differently shaped grey scatter points indicate results covering similar redshifts of $\rm z\sim 3-13$ from various other publications \citep{curti_jades_2024, faisst_alpinecristaljwst_2026, venturi_gas-phase_2024, rowland_rebels-ifu_2026, heintz_dilution_2023, pollock_novel_2026, nishigaki_dreamsii_2025, heintz_jwst-primal_2025}. Our own linear fit is shown as a solid black line together with its $\rm 1\sigma$ confidence interval as the shaded grey region. Our results are clearly below the local results, but occupy their own space that is parallel to the local relation. \textit{Bottom:} Offset in metallicity from the local FMR \citep{curti_mass-metallicity_2020} plotted against the age of the Universe and redshift. We compare to other observations covering redshifts from the local Universe up to $\rm z\sim 9$ \citep{sanders_mosdef_2021,faisst_alpinecristaljwst_2026,rowland_rebels-ifu_2026,sarkar_unveiling_2025,curti_jades_2024,pollock_novel_2026, nishigaki_dreamsii_2025} and the predictions from \citet{nishigaki_dreamsii_2025}. We observe a clear trend of the FMR offset increasing with increasing redshift.}
    \label{fig:FMR}
\end{figure*}

\subsubsection{The Mass-Metallicity Relation (MZR)}\label{sec:results_MZR}

 Figure~\ref{fig:MZR} shows the location of the seven galaxies in our sample on the MZR (pink squares). We measure gas-phase metallicities within the range of $\rm 12+\log(O/H)\sim 7.3-7.9$, corresponding to roughly $4\%-15\%$ the solar metallicity. In the top panel, we compare our observations to other results at $\rm z \sim 3-10$ \citep{heintz_dilution_2023, nakajima_jwst_2023, curti_jades_2024, venturi_gas-phase_2024, morishita_diverse_2024, chemerynska_extreme_2024, pollock_novel_2026, sarkar_unveiling_2025, hsiao_sapphires_2025, asada_glimpse-ddt_2026}, at $\rm z=3$ \citep{li_mass-metallicity_2023}, and the local Universe \citep{curti_mass-metallicity_2020}. Furthermore, in the bottom panel of Figure~\ref{fig:MZR}, we compare our observations to the FIRE \citep{ma_origin_2016}, FirstLight \citep{langan_weak_2020, nakazato_simulations_2023}, FIRE-2 \citet{marszewski_high-redshift_2024}, THESAN-ZOOM \citep{mcclymont_thesan-zoom_2026}, EAGLE \citep{schaye_eagle_2015}, Astraeus \citep{cueto_astraeus_2024}, IllustrisTNG \citep{torrey_evolution_2019}, Astraeus assuming an evolving IMF \citep{cueto_astraeus_2024} and SERRA \citep{pallottini_survey_2022} cosmological simulations. 

Overall, we find that our samples lie significantly below ($\rm \sim 1.1$ dex) the relation for local galaxies and that at $\rm z=3$ ($\sim 0.67$ dex), confirming that galaxies in the early Universe are less enriched. However, it is worth noting again that our sample is pre-selected to be metal-poor and is thus likely to be even more metal-poor than other results included at similar redshifts. Furthermore, our data points are either situated on or below other relations at similar redshifts.

The solid black line in Figure~\ref{fig:MZR} represents an error-weighted linear fit to our data, both in mass and metallicity, resulting in the following relation: 
$$\rm 12+\log(O/H) = (0.19 \pm 0.03) \times \log(M_*/M_{\odot}) + (5.90 \pm 0.26).$$
Our resulting fit is also situated significantly below any other relation at a similar redshift, which is expected, given that our sample was selected to include some of the most metal-poor galaxies at z$>$7 known at the time of proposal submission. Additionally, we derive another linear fit by incoorporating our own results and those at similar redshifts \citep[$\rm z\sim 6-10$;][]{heintz_dilution_2023, curti_jades_2024, chemerynska_extreme_2024,venturi_gas-phase_2024, pollock_novel_2026, hsiao_sapphires_2025, asada_glimpse-ddt_2026}, resulting in the following fit (indicated as a dashed black line in Figure~\ref{fig:MZR}): $$\rm 12+\log(O/H) = (0.28 \pm 0.02) \times \log(M_*/M_{\odot}) + (5.36 \pm 0.15).$$ Our two fits meet at the very low mass ends, where EMPGs are located, but they diverge towards higher masses; our main galaxies are clearly more metal-poor than others at similar redshift.

We are also in good agreement with some of the cosmological simulations included in the bottom panel of Figure~\ref{fig:MZR}. Our four highest mass galaxies are in well agreement with the results from the SERRA simulations at $\rm z\sim 8$ \citep{pallottini_survey_2022}, and our three other sources are siutated among the simulations from the FirstLight \citep[$\rm z\sim 8$,][]{nakazato_simulations_2023} and Astraeus \citep[$\rm z \sim 9$,][]{cueto_astraeus_2024} simulations. RX2129\_11022 notably does not overlap with any of the included simulation results and is also the most metal-poor galaxy among our sample.

\subsubsection{The Fundamental Metallicity Relation (FMR)}

\begin{figure*}
    \centering
    \includegraphics[width=\linewidth]{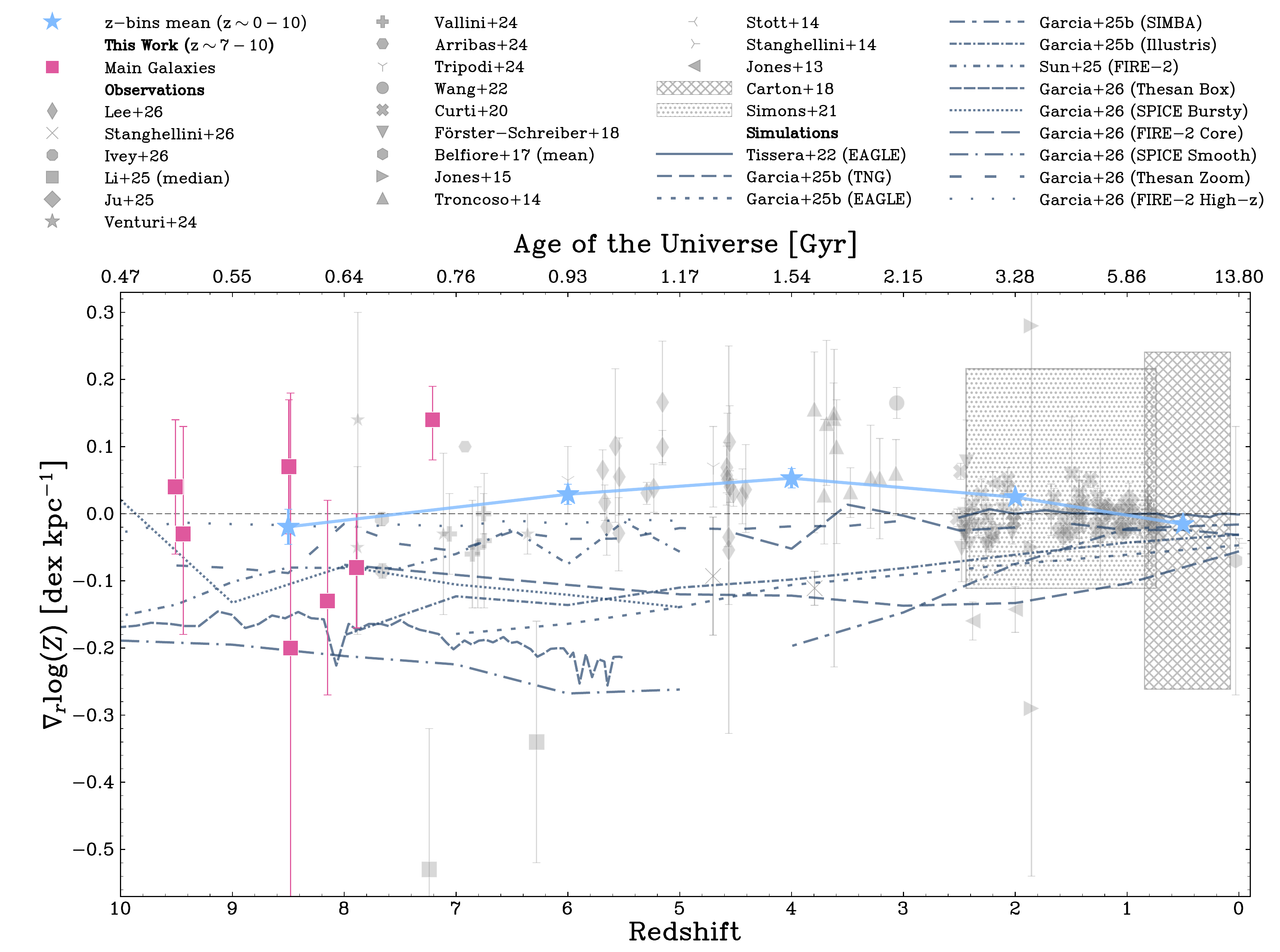}
    \caption{Gas-phase metallicity gradients ($\rm dex / kpc$) as a function of redshift and age of the Universe (Gyr). The sources from our sample are indicated as pink squares. We provide redshift-binned means (large blue stars) by combining observational results from the literature and our results. All other grey symbols represent a compilation of results from other observations spanning all of cosmic time. Most notably, we present results for the local Universe ($\rm z \lesssim 1$) by \citet{stanghellini_radial_2014,stott_relationship_2014, belfiore_sdss_2017, carton_first_2018} and cosmic noon ($\rm 1 \leq z \lesssim 3$) by \citet{jones_origin_2013,jones_grism_2015, forster_schreiber_sinszc-sinf_2018, curti_klever_2020,simons_clear_2021, wang_early_2022, ju_msa-3d_2025}. We use hatched regions to depict larger samples. Past the cosmic noon, observations start to become sparse with redshifts  $\rm 3 \leq z \lesssim 6$ still having a good number of data points \citep[e.g.][]{troncoso_metallicity_2014, tripodi_spatially_2024,fujimoto_alpine-cristal-jwst_2025, stanghellini_direct_2026, lee_alpine-cristal-jwst_2026}. However, there are only a handful of observations towards even higher redshift simply due to the lack of available, and with good enough S/N, data \citep[e.g.][]{arribas_ga-nifs_2024,vallini_spatially_2024,venturi_gas-phase_2024,li_13_2025, ivey_exploring_2026}. In this redshift regime, most previous observational studies have only encompassed a few galaxies and are unable to cover larger samples, and do not reach beyond $\rm z\sim 8$. In addition to observations, we also conduct a comparison with predictions from recent cosmological simulations, shown as differently dashed and dotted dark blue lines: EAGLE \citep{tissera_evolution_2022}, which extends from the local Universe to the cosmic noon, and FIRE-2 \citep{sun_galaxy_2026}, covering the epoch of reionisation. \citet{garcia_metallicity_2025} and \citet{garcia_metallicity_2026} used several different simulations to cover almost all of cosmic time using two different star-formation feedback modes: bursty (SPICE Bursty, Thesan Zoom, FIRE-2) and smooth (EAGLE, Illustris, TNG, SIMBA, Thesan Box, SPICE Smooth).}
    \label{fig:grads_redshift}
\end{figure*}

Lastly, in terms of global scaling relations, we also investigate the FMR in detail. First, we look at the parametrisation introduced by \citet{mannucci_fundamental_2010}, where the SFR is included as a secondary relation within the MZR to reduce its scatter:
\begin{equation}\label{eq:FMR_alpha}
    \rm \mu_{\alpha} = \log(M_*/M_{\odot})-\alpha \times \log(SFR  /M_{\odot} \ yr^{-1}),
\end{equation}

where $\rm \alpha$ is a constant between 0 and 1, which is defined so that it minimises the metallicity scatter at a given $\rm \mu_{\alpha}$. Previously, \citet{mannucci_fundamental_2010} and then \citet{sanders_mosdef_2021} found that the FMR does not evolve between $\rm z=0$ and $\rm z=3$; however, recent JWST observations at redshifts $\rm z>3$ \citep[e.g.][]{heintz_dilution_2023, nakajima_jwst_2023, curti_jades_2024, pollock_novel_2026}, and simulations \citep[e.g.][]{garcia_does_2024, garcia_does_2025} have found deviations from the locally defined relation in the early Universe.  

In the top panel of Figure~\ref{fig:FMR}, we show the results from our main sample as pink squares, adopting an $\rm \alpha = 0.65$ parameter \citep[for high sSFR galaxies; see Appendix A of][]{curti_mass-metallicity_2020}, and compare it to previous JWST observations \citep{curti_jades_2024, venturi_gas-phase_2024, faisst_alpinecristaljwst_2026, rowland_rebels-ifu_2026} and the local FMR as a solid dark blue line, whose extrapolation towards $\rm \mu_{\alpha}$ values below what probed locally in \citet{curti_mass-metallicity_2020} is indicated as a dashed line. Clearly, the galaxies in our sample are situated below the local relation; however, they still tentatively follow a linear relation with $\mu_\alpha$, fitted as: 
$$\rm 12+\log(O/H) = (0.24 \pm 0.07) \times \mu_{\alpha} + (5.58 \pm 0.52).$$ As done for the SFMS and MZR, we also provide an additional fit by looking at our results and those at similar redshift \citep[$\rm z\sim 6-10$,][]{heintz_dilution_2023, nakajima_jwst_2023,curti_jades_2024, venturi_gas-phase_2024, pollock_novel_2026, rowland_rebels-ifu_2026}: $$\rm 12+\log(O/H) = (0.25 \pm 0.04) \times \mu_{\alpha} + (5.85 \pm 0.35).$$

Comparing these two linear fits gives us great insight into our sample: the two fits are almost perfectly parallel to each other but significantly offset along the y-axis. This suggests that our sample is even metal poorer than many other galaxies measured at similar redshifts. 

\begin{figure*}
    \centering
    \includegraphics[width=\linewidth]{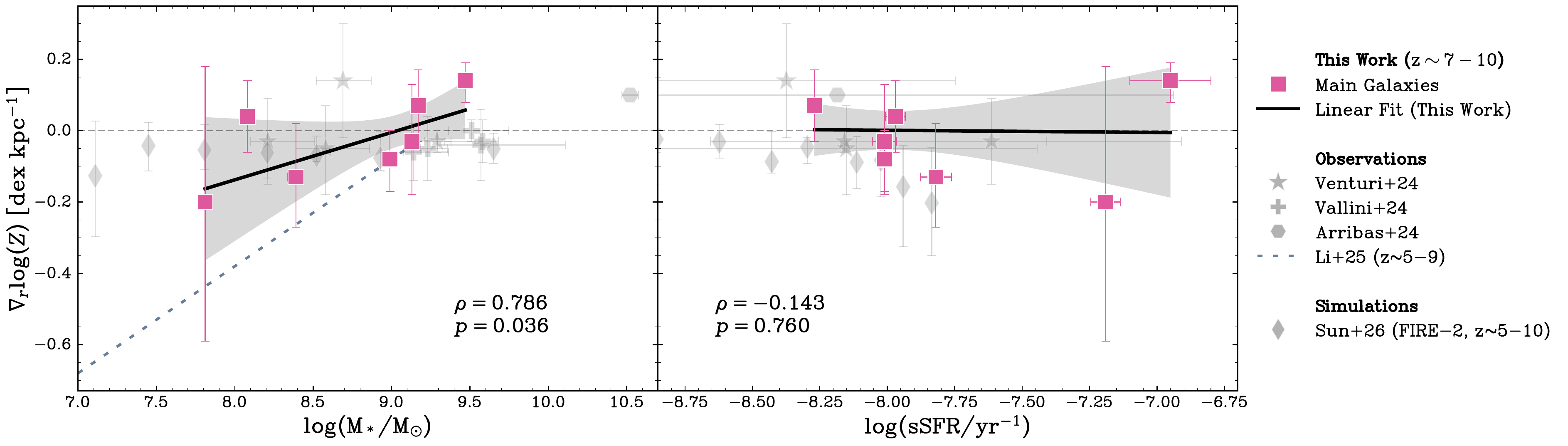}
    \caption{Gas-Phase Metallicity Gradients vs. Stellar Mass (left) and sSFR (right). We show our results as pink squares and compare them to observations covering similar redshifts \citep{vallini_spatially_2024, arribas_ga-nifs_2024, venturi_gas-phase_2024} using differently shaped symbols. Furthermore, we also compare our linear relation to that obtained by \citet{li_13_2025} for galaxies at $\rm 5<z<9$ as well as the predictions from the FIRE-2 cosmological simulation by \citet{sun_galaxy_2026}. Our linear fit is shown as a solid black line with the $\rm 1\sigma$ confidence interval as a grey band. Among our results, we see a clear positive dependence on the stellar mass: lower-mass sources exhibit more flat to slightly negative gradients, while higher mass galaxies tend to have more positive gradients. The scatter and gradients' slope uncertainties also decrease with increasing stellar mass. In contrast, a negative correlation is observed between sSFR and metallicity gradient. Both subplots include the Spearman rank coefficient and the associated p-value. According to these values, the stellar-mass relation shows a moderate but statistically non-significant trend, whereas no significant correlation is found with sSFR.}
    \label{fig:grads_mass_sSFR}
\end{figure*}

Furthermore, we examine the galaxies' offset from the local FMR, i.e., the difference between the measured metallicity and that inferred at a given $\rm \mu_{\alpha}$ assuming the local FMR parameterisation from \citet{curti_mass-metallicity_2020}. Our results are shown in the top plot of Figure~\ref{fig:FMR} together with other observational results obtained across cosmic time \citep{sanders_mosdef_2021, curti_jades_2024, pollock_novel_2026, sarkar_unveiling_2025, rowland_rebels-ifu_2026, fujimoto_alpine-cristal-jwst_2025}.
We compare with predictions from the ChemicalUniverseMachine model \citep{nishigaki_dreamsii_2025} for three different scenarios: enhanced outflow defined by a decrease in metal distribution fraction from $\rm z\sim 6$ to $\rm z\sim 8$, corresponding to a stronger metal loss with redshift (solid line), enhanced inflow which is incoorporated by increasing $\rm M_{H_{2}}$ beyond the standard extrapolation from $\rm z\sim 6$ to $\rm z\sim 8$ (dotted line), and constant star formation efficiency (SFE; dashed line). Additionally, we include predictions from several simulations presented in \citet{garcia_does_2025}. As shown, the FMR offset increases with redshift, consistent with the overall trend established by previous observations and simulations. However, some of our sources (SMACS0723\_4590, RX2129\_11022, and the three satellites) are significantly more offset at a given redshift, even well below the theoretical models. This likely indicates that fundamentally different physical properties govern galaxies in the early Universe, suggesting a non-equilibrium state that is characterised by inhomogeneous enrichment and persistent inflow of pristine gas \citep[e.g.][]{kim_agora_2026,li_insights_2025,asada_glimpse-ddt_2026}.  Additionally, the possibility of metal-enriched outflows cannot be excluded \citep{nishigaki_dreamsii_2025}.

\subsection{Gas-Phase Metallicity Gradients}

We derive radial gas-phase metallicity gradients by measuring emission line flux ratios from the average spectra of concentric annuli centred around the peak of the stellar continuum, measured between $\rm H\gamma$ and $\rm H\beta$ ($\rm \sim 4393 - 4810$ \AA  $\ $rest-frame) in the NIRSpec IFU data. The annuli have a width of one spaxel ($\rm 0.05 \arcsec$ corresponding to roughly $\rm \sim 0.22-0.26$ kpc), with the central annulus being defined as a circular aperture of radius one around the central spaxel, resulting in it always containing five spaxel. The amount of annuli were chosen to represent a compromise between sufficient $\rm S/N$ and achieving enough resolution. We determine gradients by fitting a linear relation to the radial profile obtained via the annuli while accounting for uncertainties in metallicity. Gradient slope errors are estimated by MCMC sampling 1000 times using the \texttt{emcee} package \citep{foreman-mackey_emcee_2013}.  

In the following two sub-subsections, we present our results for the gas-phase metallicity gradients, comparing them to other observations and simulations spanning redshifts $\rm z\sim 0-10$ and relating the gradient slope to stellar mass. We also provide a detailed analysis of each galaxy in section~\ref{sec:discuss-grads}.

\subsubsection{Evolution of Metallicity Gradients over Cosmic Time}

We present our results for the gas-phase metallicity gradients in units of $\rm dex \ kpc^{-1}$ in Figure~\ref{fig:grads_redshift}, comparing them to results from both observations and simulations across cosmic time.  Our results reveal a wide range of metallicity gradients spanning from $\rm \Delta_r \log(Z)\sim -0.20$ to $\rm \Delta_r \log(Z) \sim 0.14$ $\rm dex \ kpc^{-1}$. However, most gradients are consistent with being flat, with the exception of SXDF\_NB1006-2. Several of our sources' gradients are accompanied by large uncertainties, which are largely caused by weak emission lines, limiting us to measuring metallicities using only the $\rm R3$ ratio. We recover an average metallicity gradient of $\rm -0.03 \pm 0.04$ for our sample. We also present redshift-binned means across cosmic time by combining the shown literature data with our results (omitting the results from \citet{li_13_2025} for this, as those are based on stacks and not individual measurements, see below). 

The flat gradients we observe at high redshift are consistent with previous observations at slightly lower redshifts \citep[$\rm z \lesssim 8$, e.g.][]{vallini_spatially_2024, arribas_ga-nifs_2024, venturi_gas-phase_2024, fujimoto_alpine-cristal-jwst_2025, ivey_exploring_2026}, however, they strongly disagree with recent results from \citet{li_13_2025}, who utilised stacked NIRCam or NIRISS GRISM spectroscopy to measure metallicity gradients at $\rm z \sim 5-9$ and found strongly negative metallicity gradients. A possible explanation for this difference is that the sample presented in \citet{li_13_2025} is biased towards galaxies with strong $\rm [OIII]+H\beta$ lines (required for detection in grism spectra), which are often characterised by bursty star formation. A potential caveat in the analysis by \citet{li_13_2025} arises from stacking galaxies of varying physical sizes; this can introduce artefacts if specific galaxies contribute disproportionately to different radial bins. 

In the same figure, we also report predictions from the Astraeus \citep{hutter_astraeus_2021} and FIRE-2 \citep{hopkins_fire-2_2018} simulations reported in \citet{cueto_astraeus_2024} and \citet{sun_galaxy_2026}, respectively. Recently, \citet{garcia_metallicity_2025} extended the predictions to the highest redshifts to date via the EAGLE \citep{schaye_eagle_2015}, Illustris \citep{vogelsberger_introducing_2014}, SIMBA \citep{dave_simba_2019}, and IllustrisTNG \citep{pillepich_simulating_2018} cosmological simulations. Additionally, \citet{garcia_metallicity_2026} compared cosmological simulations implementing gradual star formation feedback to those applying bursty star-formation feedback utilising results from the Thesan Box \citep{kannan_introducing_2022, garaldi_thesan_2022, smith_thesan_2022}, SPICE \citep{bhagwat_spice_2024}, EAGLE, Illustris, IllustrisTNG, SIMBA, FIRE-2, Thesan Zoom \citep{kannan_introducing_2025} simulations and found that bursty star formation is needed for the simulations to align with recent observational results. A bursty SFH may predict flatter gradients as these can be accompanied by metal-loaded galactic outflows, which can disrupt the gas disc and any previously formed gradient \citep{tissera_evolution_2022, venturi_gas-phase_2024, garcia_metallicity_2026}. While our sample exhibits significant scatter, the observed gradients are generally consistent with simulation predictions, except for our strongly positive gradient, which is rarely produced in current models and is usually not indicated by the averaged results, as is seen in Figure~\ref{fig:grads_redshift}.

\subsubsection{Evolution of Metallicity Gradients with Stellar Mass and sSFR}

Recent observations suggest a correlation between stellar mass and metallicity gradients at high redshift \citep[e.g.][]{li_13_2025, sun_galaxy_2026}. In the early Universe, low-mass systems with $\rm \log(M_*/M_{\odot}) \lesssim 10$ typically exhibit negative gradients, while more massive galaxies often display flatter or even slightly positive (inverted) gradients \citep[e.g.][]{belfiore_sdss_2017, mingozzi_sdss_2020, li_negative_2025}. Analysis of the FIRE-2 simulations by \citet{sun_physical_2025} at $\rm z = 0.4$--$3$ supports this, showing a positive correlation between gradient and stellar mass up to $\rm \log(M_*/M_{\odot}) \approx 10$. Beyond this mass threshold, however, the trend reverses: more massive galaxies with $\rm \log(M_*/M_{\odot}) \gtrsim 10$ begin to exhibit increasingly negative gradients, consistent with the well-established profiles of massive galaxies in the local Universe. This transition reflects a shift in underlying physical drivers: the high scatter and negative gradients in low-mass systems likely stem from chaotic internal structures and bursty feedback, whereas the more stable, mature structures of high-mass galaxies eventually promote centrally concentrated enrichment patterns observed at lower redshifts.

On the left-hand side of Figure~\ref{fig:grads_mass_sSFR}, we present our results for the relationship between gradients and stellar mass (pink squares), along with a linear fit shown as a solid black line. As is apparent via the data points and the linear fit, we also tentatively confirm more positive/flat gradients at increasing stellar mass within our sample, although with large uncertainty. Moreover, as best seen in the $\rm 1\sigma$ confidence interval (shaded grey region) of our linear fit (solid black line), we observe a reduction in scatter with increasing mass, consistent with theoretical expectations.

Additionally, in the right panel of Fig.~\ref{fig:grads_mass_sSFR}, we investigate the correlation between the sSFR ($\rm sSFR = SFR/M_*$) and metallicity gradient shown in the right panel of Figure~\ref{fig:grads_mass_sSFR}. We find a slight negative correlation: higher sSFR is associated with steeper gradients. 
While recent studies \citep[e.g.][]{sun_galaxy_2026} suggest that sSFR-driven feedback is the primary regulator of chemical structure at $z>5$, our results indicate that stellar mass ($\rm \rho = 0.786$, $\rm p=0.036$) exhibits a more dominant correlation with gas-phase metallicity gradients. The lack of a significant correlation with sSFR ($\rm \rho = -0.143$, $\rm p=0.760$) and SFR ($\rm \rho = 0.429$, $\rm p=0.337$) suggests that at $z \sim 7-10$, the chemical profile is not a simple reflection of instantaneous bursty star formation. Instead, it likely points to a regime where radial mixing timescales are long relative to star-formation bursts. It needs to be stressed, however, that we are working with a very small sample size here and that in the future, when more results for high-z ($\rm z>7$) metallicity gradients exist, a better conclusion can be drawn. 

\subsection{Resolved Mass Metallicity and Fundamental Metallicity Relation}

The FMR has already been established on spatially resolved scales for the local Universe \citep[e.g.,][]{belfiore_sdss_2017, baker_metallicitys_2022}, which has aided in understanding the mechanisms and feedback driving the local ISM. However, studying this relation at earlier cosmic epochs has previously proven difficult due to limited spatial resolution. Most recently, \citet{fujimoto_alpine-cristal-jwst_2025} were able to find evidence for an rFMR at redshifts $\rm z=4-6$ with a projection parameter of $\rm \alpha=2.1$ (see Eq.~\ref{eq:FMR_alpha}), indicating a much stronger dependence on the local SFR than the local stellar mass. \citet{marconcini_ga-nifs_2024} also found evidence of an rFMR for a single source at $\rm z=9.11$, but with a weaker dependence on the SFR than what was found by \citet{fujimoto_alpine-cristal-jwst_2025} at $\rm z\sim 4-6$. 

To derive resolved stellar mass maps, we use the stellar continuum between $\rm H\gamma$ and $\rm H\beta$ as a spatial proxy for the stellar mass distribution. We normalise the flux of this continuum map so that the sum of all spaxels equals unity, and then multiply this template by the total global stellar mass. This approach assumes a spatially uniform mass-to-light ratio across the galaxy within this spectral range. We then derive stellar mass surface densities by dividing the stellar mass of each spaxel by its corresponding area. Similarly, we take our individual SFR measurements per spaxel, which were derived as described in section~\ref{sec:analysis-EL} for integrated spectra, and derive SFR surface densities $\rm \Sigma_{SFR}$ via a division by the area. 

Our results for the rFMR for each galaxy can be found in Appendix~\ref{app:metal_grads}. In addition to that, we compute the Spearman r-rank coefficient $\rm \rho$ and associated p-value between the gas-phase metallicity and SFR surface density for each galaxy, to further quantify if the anti-correlation is observed at high-z. The results can be found in the bottom right subplot of the Figures shown in Appendix~\ref{app:metal_grads}. In conclusion, RX2129\_11027 is the only source in our sample exhibiting a reasonably strong $\rm \rho$ accompanied by a small p-value, indicating an rFMR. Additionally, JADES\_10058975, Abell\_Z7885, SMACS0723\_4590, and JADES\_8013 show some weak indications of an anti-correlation between metallicity and SFR. However, their p-values are typically considered too large to constitute significant evidence.

Summarizing, our analysis suggests that a resolved FMR is generally not in place at such early times for most galaxies. This is in line with the strong deviations from the FMR, suggesting that excessive accretion of pristine gas is not promptly processed through star formation, and primarily results in dilution without accompanying enhancement of star formation.

\begin{figure}
    \centering
    
    \includegraphics[width=\linewidth]{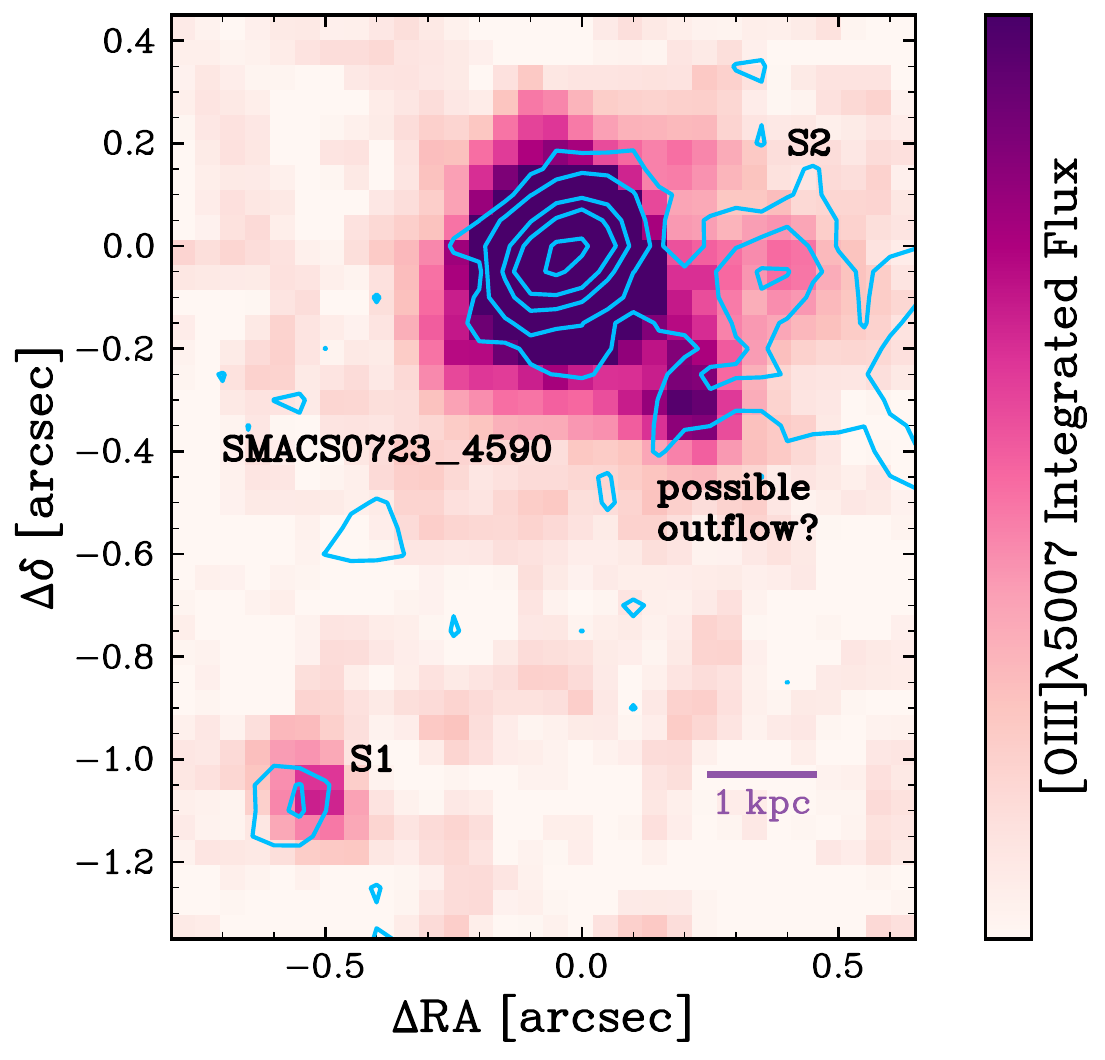}
    
    \vspace{0.6cm}
    
    \includegraphics[width=\linewidth]{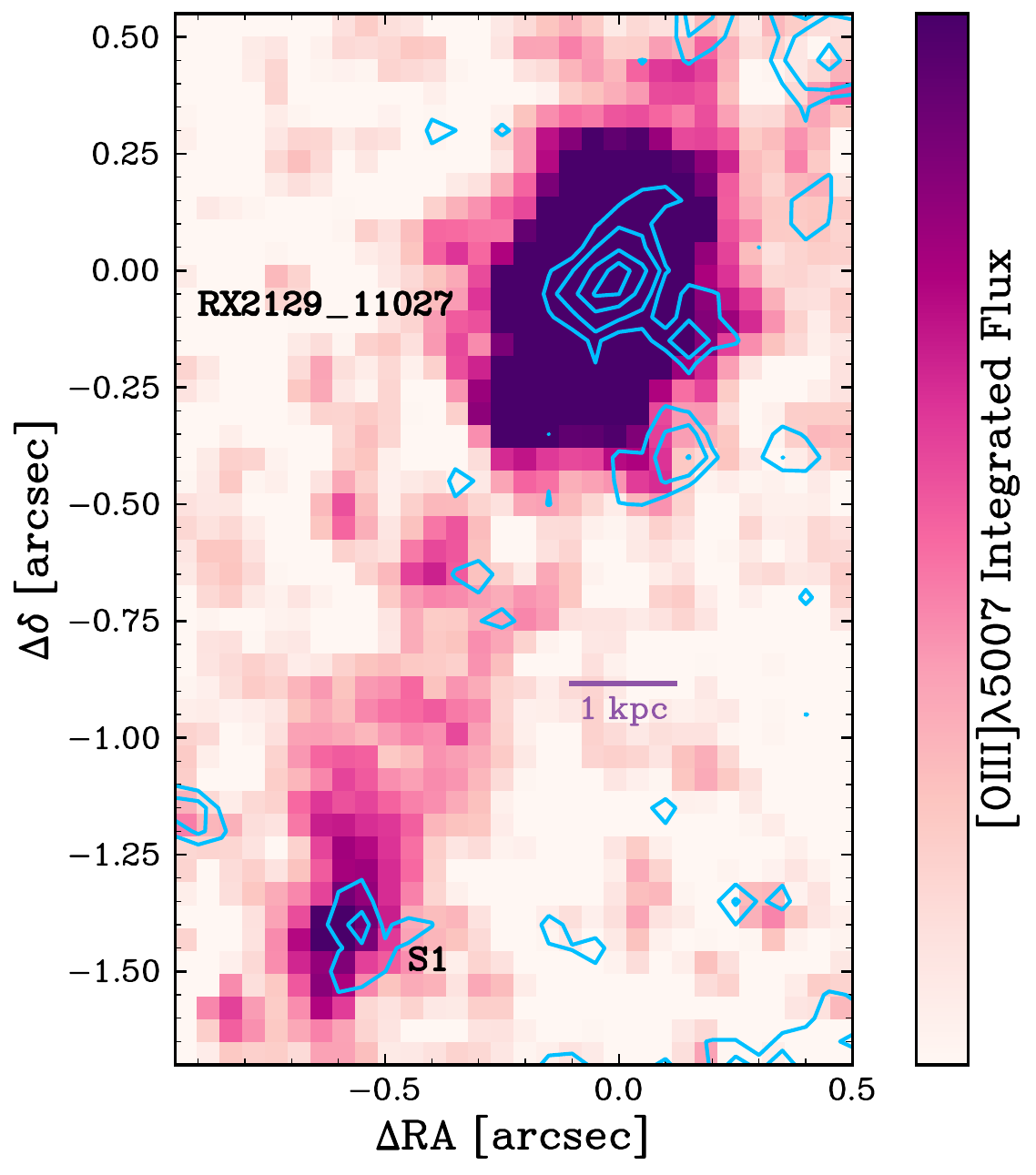}
    
    \caption{NIRSpec IFU cutouts of the two sources containing satellite galaxies. The NIRSpec IFU cubes were collapsed around the $\rm [OIII] \lambda 5007$ emission line, and the blue contour lines represent the F277W NIRCam filter. \textit{Top:} SMACS0725\_4590 with its two satellites SMACS0725\_4590-S1 and SMACS0725\_4590-S2 as well as a possible outflow. The F277W contour lines are shown for $\rm [3, 5, 10, 15, 25]\times RMS$ levels. \textit{Bottom:} RX2129\_11027 and its satellite RX2129\_11027-S1. The contour lines are shown for levels of $\rm [ 7, 8, 10, 12]\times RMS$.}
    \label{fig:satellites}
\end{figure}

\subsection{Low-Metallicity Satellite Galaxies}\label{sec:sats}

The field of view (FoV, $\rm 3 \times 3 \arcsec$ corresponding to $\rm \sim 13-16$ kpc for our sample) of the JWST IFU observations also allows us to investigate possible, line-emitting, close-by satellite candidates. Among such candidates, we confirm the presence of three, most likely, satellite galaxies surrounding two of our primary sources for which we spectroscopically confirm their redshift through the location of their $\rm H\beta$ and $\rm [OIII] \lambda 5007$ emission lines. 

We showcase the detected satellites via intensity maps of the $\rm [OIII]$ line in Figure~\ref{fig:satellites}, which were obtained by collapsing the cube around the emission line peak.
We also overlay the NIRCam F277W filter contour lines in blue, which we chose as it traces the UV stellar continuum and is not contaminated by emission line features. Additionally, we present the full 1D spectra and NIRCam image cutouts of the satellites in Figures~\ref{fig:spec_RX2129_11027-S1}, \ref{fig:spec_SMACS0723_4590-S1}, and \ref{fig:spec_SMACS0723_4590-S2} of Appendix~\ref{app:spec_flux}. 

We derive stellar masses and SFRs as done for the main galaxies (see Sections~\ref{sec:sedfitting} and~\ref{sec:analysis-EL} for details) and metallicities using only the R3 ratio due to only $\rm H\beta$ and $\rm [OIII]\lambda 5007$ being significantly detected. However, due to their compactness and weak signal, we are unable to reliably derive metallicity gradients for the satellites.

SMACS0723\_4590 is surrounded by two satellite galaxies, which we denote as SMACS0723\_4590-S1 and SMACS0723\_4590-S2. We show the main galaxy together with its two satellites in the top panel of Figure~\ref{fig:satellites}. We note that the strong F277W emission (blue contour lines) exhibited at the far right edge of the figure must be from a foreground source. The two satellites are easily identified via coincident peaks in $\rm [OIII]$ and F277W emission. SMACS0723\_4590-S1 has a stellar mass of $\rm \log(M_*/M_{\odot}) = 6.77^{+0.05}_{-0.04}$, SFR of $\rm \log(SFR_{H\beta}) = -0.52 \pm 0.05$, and a gas-phase metallicity of $\rm 12+log(O/H)=7.12^{+0.09}_{-0.08}$. We estimate a projected on-sky separation to its host galaxy of 5.82 kpc. SMACS0723\_4590-S2 is located even closer to its host galaxy with a projected separation of 2.14 kpc. We estimate a stellar mass of  $\rm \log(M_*/M_{\odot}) = 7.29^{+0.04}_{-0.04}$, SFR of $\rm \log(SFR_{H\beta}) = -0.66 \pm 0.09$, and a gas-phase metallicity of $\rm 12+log(O/H)=7.12^{+0.13}_{-0.11}$. Given the proximity of SMACS0723\_4590-S2 to the main galaxy, it is likely to be in the process of merging or interacting with SMACS0723\_4590. Immediately south of SMACS0723\_4590-S2, or south-east of SMACS0723\_4590, at a projected distance of 1.85 kpc to the host galaxy, there is also a distinct $\rm [OIII]$ feature without significant continuum,  which could be an accreting clump or possibly a metal-enriched outflow, considering that SMACS0723\_4590 is an AGN candidate (see Appendix~\ref{app:SMACS4590_BLR_outflow}). We also highlight this possible outflow in Figure~\ref{fig:satellites}. We derive a metallicity of $\rm 12+log(O/H) \sim 7.64$ and SFR of $\rm \log(SFR [M_{\odot}/yr]) \sim -1$ for this clump, if the nebular emission is due to in-situ star formation.. 


\begin{figure}
    \centering
    
    \includegraphics[width=\linewidth]{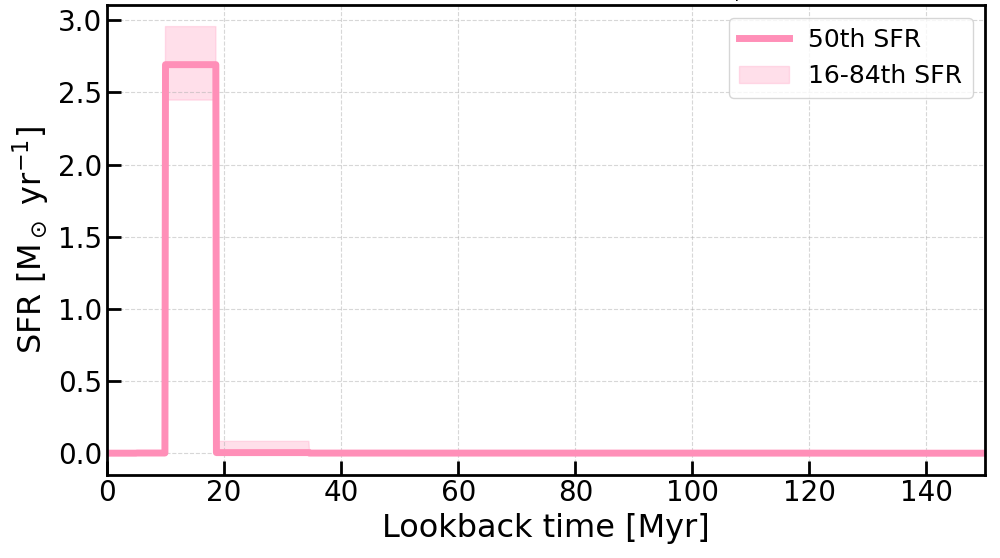}
    
    \vspace{0.15cm}
    
    \includegraphics[width=\linewidth]{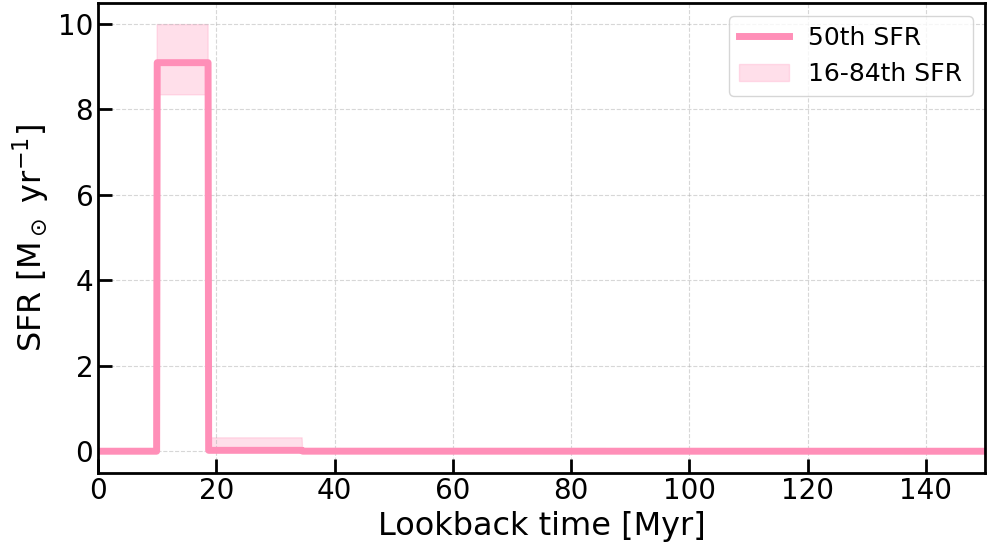}
    
    \vspace{0.15cm}
    
    \includegraphics[width=\linewidth]{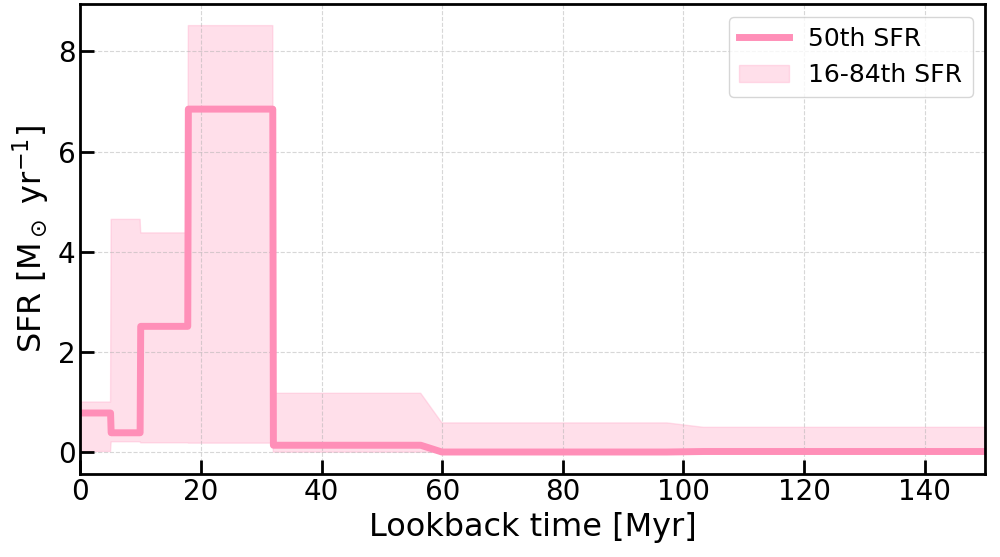}
    
    \caption{Star Formation Histories (SFHs) as measured by Prospector (see section~\ref{sec:sedfitting} for details) of the three low-metallicity satellite galaxies SMACS0723\_4590-S1 (top), SMACS0723\_4590-S2 (middle), and RX2129\_11027 (bottom).}
    \label{fig:satellites_SFHs}
\end{figure}

For RX2129\_11027, we can confirm one satellite, which we refer to as RX2129\_11027-S1, with an approximate projected separation of 6.2 kpc from its host galaxy, as shown in the bottom panel of Figure~\ref{fig:satellites}. This system is especially interesting as the satellite and host galaxy are connected via a gaseous bridge, with mainly $\rm [OIII]\lambda 5007$ and some $\rm H\beta$ detection, and is devoid of any clear detection of stellar continuum, as investigated via the NIRCam filters. We estimate a stellar mass of $\rm \log(M_*/M_{\odot}) = 6.83^{+0.19}_{-0.10}$, SFR of $\rm \log(SFR_{H\beta}) = -0.85 \pm 0.07$, and a gas-phase metallicity of $\rm 12+log(O/H)=7.26^{+0.16}_{-0.12}$ for the satellite. For the gaseous bridge, we were able to derive a metallicity of $\rm 12+\log(O/H)\sim 7.5$ (via an elliptical aperture), although with large uncertainties as the $\rm H\beta$ detection within the bridge is quite weak. This metallicity is intermediate between that of the satellite and the host galaxy, which could suggest a tidal tail between the satellite and the main galaxy.

High redshift ($\rm z >5$) observations from JWST spectroscopy in the hopes of finding metal-free galaxies consistently resulted in finding galaxies with metallicities $\gtrsim 2 \%$ solar \citep{nakajima_jwst_2023, curti_jades_2024, hsiao_sapphires_2025}, conjoining the term "metallicity floor" above which most high-z metallicity measurements lie. Our satellites are all situated close to this "metallicity floor" of $\rm 12+log(O/H) \sim 7.0$, or approximately $\rm 2\%$ of the solar metallicity, and are accompanied by very low SFRs (see Figure~\ref{fig:SFMS}) and low stellar masses. Therefore, the satellites are significantly less massive and more metal-poor than our main sample, situating them among confirmed extremely metal-poor galaxies (EMPG) at similar redshifts \citep{chemerynska_extreme_2024, hsiao_sapphires_2025, asada_glimpse-ddt_2026}, and help extend our measured MZR (see Figure~\ref{fig:MZR}) towards the low-mass end. The three satellites are strongly offset from the local FMR, as can be seen in Figure~\ref{fig:FMR}, exhibiting offsets down to $\sim -0.9$dex. The location of the satellites on the SFMS (see Fig.~\ref{fig:SFMS}) and MZR (see Fig.~\ref{fig:MZR}) suggest that they are similarly enriched, but produce significantly fewer stars compared to galaxies at similar redshifts and stellar masses. 

Additionally, via our SED modelling, we find that the satellites had undergone starbursts that started roughly $\rm \sim 20$ Myr and ended about $\rm 5-10 ~Myr$ ago, likely showing their initial assembly. The results can be seen in Figure~\ref{fig:satellites_SFHs}. We present possible implications of investigating these satellites and what their primitive chemical state reveals about the early Universe in section~\ref{sec:discuss-sats}.

\section{Discussion}\label{sec:discussion}

\subsection{Driving mechanisms of early metallicity gradients: analysis of the individual sources and overview}\label{sec:discuss-grads}

Our analysis of metallicity gradients at $z \approx 7-10$ reveals a chemically diverse landscape characterised by a variety of metallicity profiles spanning positive, flat, and negative gradients that are also accompanied by significant scatter and substantial observational uncertainties. This variation suggests that during the first billion years of cosmic time, galaxies do not follow a singular evolutionary path toward chemical maturity. In our sample, sources such as RX2129\_11022 (Fig.~\ref{fig:metal_grads_RX2129_11022}), Abell\_Z7885 (Fig.~\ref{fig:metal_grads_Abell_Z7885}), and JADES\_0813 (Fig.~\ref{fig:metal_grads_JADES_8013}) exhibit gradients that are negative but also consistent with being flat within their error bars. While JADES\_0813 shows a weak FMR offset of $-0.25$ dex, RX2129\_11022 is more extreme, with a significant offset of $-0.97$ dex that situates it among the metal-poor satellites we detect. Additionally, there is some tentative indication of a broadening of the $\rm H\beta$ line in the spectrum of RX2129\_11022 (see Fig.~\ref{fig:spec_flux_RX2129_11022}), which could suggest that this source is an AGN. The results from these sources suggest a state of stochastic growth, in which stable inside-out enrichment has not yet taken hold. Below, we explore how extreme gas accretion, satellite-driven mixing, and AGN feedback act as competing mechanisms that actively shape metallicity gradients. 


Within our sample, we have confirmed that SMACS0723\_4590 (Fig.~\ref{fig:metal_grads_SMACS0723_4590}) and RX2129\_11027 (Fig.~\ref{fig:metal_grads_RX2129_11027}) host low-metallicity satellite galaxies within a few kpc. For both sources, we measure slightly positive gradients, which are, however, also flat within their uncertainties. This could indicate that they are currently experiencing, or have experienced, interactions or mergers. In the early Universe, the fraction of mergers and interactions is higher than in the local Universe, as observed and as found via cosmological simulations and recent JWST observations \citep[e.g.][]{kohandel_velocity_2020, pallottini_survey_2022, puskas_constraining_2025, duan_galaxy_2025}.

The finding that the gradients of the galaxies with confirmed companions are slightly positive aligns with recent findings by \citet{fujimoto_alpine-cristal-jwst_2025} who analysed IFU observations of 18 galaxies from the ALPINE-CRISTAL-JWST survey at $\rm z \sim 4-6$ and found slightly positive gradients for galaxies with confirmed neighbours. They conclude that this trend can be explained by galaxies with companions residing in more massive dark-matter halos, which have likely recently accreted pristine gas, leading to more positive gradients. 

SMACS0723\_4590 represents a compelling case of a galaxy in a primordial stage of assembly. Previous observations by \citet{heintz_gas_2023} reported a high gas fraction exceeding $90\%$, which is highly consistent with our measured FMR offset of $-0.96$ dex, if ascribed to excess of gas accretion \citep[see also][]{tacchella_jwst_2023}. This massive deviation, representing a tenfold metal deficiency, aligns with the global findings of \citet{curti_chemical_2023} and indicates a system in a state of extreme chemical non-equilibrium, where the rapid accretion of pristine gas heavily outpaces enrichment. While \citet{curti_chemical_2023} suggests this system is being swamped by accretion, our spatially resolved results provide the direct physical mechanism: the observed flat-to-positive metallicity gradient demonstrates that this pristine gas has reached the galactic core and diluted the central metallicity. This is further indicated by the clearly metal-diluted centre in this galaxy, as shown in the 2D metallicity map in Figure~\ref{fig:metal_grads_SMACS0723_4590}. Furthermore, the presence of two confirmed satellite galaxies suggests that this swamping is likely driven by tidal interactions and minor mergers. These companions induce the radial mixing necessary to homogenise the ISM, effectively suppressing the formation of a stable, negative inside-out radial metallicity profile. However, this source also has indications of an AGN and a possible outflow, as indicated by Figure~\ref{fig:satellites} (see Appendix~\ref{app:SMACS4590_BLR_outflow} for more details). If this outflow can transport metal-enriched material from the centre of the galaxy towards the outskirts, or even completely out of the galaxy, it could explain the observed flattened to slightly positive gradient. 

For RX2129\_11027, we find an FMR offset of $-0.41$ dex, which is slightly lower than the $-0.6$ dex reported by \citet{williams_magnified_2023}. They also identified a half-light radius of only 16.2 pc, suggesting a highly dense star-forming core. We find that this core is connected to a nearby satellite via a gas bridge consisting mainly of $\rm [OIII]\lambda 5007$, providing a rare, spatially resolved view of an ongoing interaction. The fact that this bridge exhibits a metallicity intermediate between the host and its satellite strongly suggests tidal stripping and mixing of gas between the two systems. In such an ultra-compact environment, these tidal forces can rapidly redistribute metals across the entire galaxy scale. This ongoing redistribution likely prevents the establishment of a centralised metal enrichment, resulting in the observed flat-to-slightly positive metallicity gradient of 0.04 dex/kpc.


\citet{scholtz_jades_2025} measured specific rest-frame optical and UV emission lines to tentatively identify JADES\_10058975 (Fig.~\ref{fig:metal_grads_JADES_10058975}) as a Type-2 AGN, making it the highest-redshift candidate of its class to date. While \citet{scholtz_jades_2025} and \citet{curti_jades_2025} suggest that this galaxy may be undergoing a starburst, our measurements indicate that the measured $\rm SFR$ of this source aligns with previous JWST observations at similar redshifts and stellar masses, placing it within the star-forming main sequence at $z\sim 7-10$. We find an FMR offset of $-0.48$ dex, suggesting that the system is significantly metal-deficient compared to local expectations. This global deficiency, combined with our measured flat metallicity gradient of $-0.03$ dex/kpc, suggests two possible scenarios. First, we cannot fully rule out the impact of active AGN feedback, although JADES\_10058975 has not yet been fully confirmed to host an AGN. Such AGN activity could explain the radial redistribution and removal of enriched material, effectively homogenising the radial metallicity profile and producing a flattened gradient \citep[e.g.][]{taylor_metallicity_2017,villar_martin_agn_2024}. On the other hand, the steeply rising SFH, high $\rm SFR$, and rapid $\rm N$ enrichment, suggesting a vigorous star formation activity, reported by \citet{curti_jades_2025}, indicate that this system has recently accreted metal-poor gas, which would be better reflected by a positive metallicity gradient. Additionally, \citet{pollock_characterising_2026} reported a relatively large $\rm HI$ column density from fitting the $\rm Ly\alpha$ damping wing for this source, further suggesting the recent inflow of pristine gas. In fact, the 2D metallicity map shown in Figure~\ref{fig:metal_grads_JADES_10058975} could suggest such a positive trend, which might not be accurately reflected by the averaging of spectra within annuli.


Previous ALMA observations of SXDF\_NB1006-2 (Fig.~\ref{fig:metal_grads_SXDF_NB1006-2}) identified the galaxy as undergoing an intense starburst \citep{inoue_detection_2016, ren_updated_2023}. Additionally, a recent study found evidence of galactic outflows \citep{ren_rioja_2025}. Given these outflows are likely metal-enriched and coupled with a strong inflow of pristine gas, the observed strongly positive metallicity gradient in this source likely stems from a combination of both mechanisms: central dilution and expulsion of freshly produced heavy elements. Indeed, metal-enriched outflows have been observed for star-forming galaxies at $\rm z\sim 3-9$ \citep{rodriguez_del_pino_ga-nifs_2026}. The source's intermediate FMR offset of $-0.36$ dex (see Fig.~\ref{fig:FMR}) further suggests the presence of strong pristine inflows. The observed positive gradient in this galaxy effectively provides a spatially resolved view of the FMR's dilution mechanism; the pristine gas required to maintain this chemical equilibrium is likely concentrated in the galactic cores, simultaneously fuelling star formation and lowering central metallicities.

By combining literature results across redshifts and our newly derived gradients at $\rm z\sim 7-10$, we can recover redshift-binned means across cosmic time. This allows us to better visualise the redshift evolution: we see a slightly negative trend for the local Universe ($\rm z\sim 0-1$), positive gradients across cosmic noon and even out to $\rm z\sim 6$, and finally we arrive back to a slightly negative, but consistent with being flat, trend at the highest redshift bin ($\rm z>7$). However, it should be noted that the highest redshift bin only contains a handful of data points and will need substantially more observations in the future to give any statistically relevant conclusions about this epoch. 

Finally, we should mention that, although we infer radial metallicity gradients, most of our sources do not show a regular radial metallicity profile, but rather a complex distribution, of which our radial gradient analysis only captures the average radial trends. Additionally, the estimated uncertainty comes from applying MCMC, which should reflect the variance of the radial trends. Such complex morphologies are expected, as a consequence of high merger rates, galactic interactions, intense feedback, and rapid accretion, accompanied by dynamically unstable disks, which can collectively disrupt metal distribution. These factors result in complex measurements, which are often exacerbated by low signal-to-noise ratios in high-redshift galaxies, particularly when studying low-mass systems.

Ultimately, our observations suggest a prevailing trend toward flattened gradients; however, the substantial scatter, encompassing both positive and negative slopes, and azimuthal variations within individual galaxies, highlights the diverse processes governing chemical enrichment during the early stages of galaxy evolution. In conclusion, these consistent results of nearly flat gradients suggest that strong radial mixing processes are in place at such high redshifts and even down to cosmic noon.

\subsection{Probing the first chemical enrichment using satellite galaxies}\label{sec:discuss-sats}

Through our investigations, we identified and measured the galaxy properties of three low-mass, low-metallicity satellite galaxies. Their metallicities ( $\rm 12+log(O/H) \sim 7.1-7.3$) correspond to a stage of chemical enrichment, galaxies are likely to harbour the first population II stars, shortly after the enrichment of population III \citep{rusta_metal-polluted_2025, rusta_pristine_2026}.
All three satellites exhibit a strong offset from the local FMR, down to $\rm -1$ dex, indicating a stark accretion of pristine gas. 
On the contrary, their SFRs are low ($\rm \log(SFR / M_{\odot} \ yr^{-1}) \sim -0.7$) and consistent with the SFMS for their stellar mass, showing no indication of an active starburst, which the inflow of pristine gas would otherwise suggest. 

The combination of low metallicity, significant FMR offsets, and low star-formation activity in these satellites aligns with the metallicity relation observed in the THESAN-Zoom simulations by \citet{mcclymont_thesan-zoom_2026}. At the low-mass regime ($\rm \log(M_*/M_{\odot}) \leq 9$), the canonical anti-correlation between SFR and gas-phase metallicity is found to weaken or even invert \citep[e.g.][]{laseter_investigation_2025}. This inversion is primarily driven by the prevalence of pristine gas inflows that dilute the ISM of low-SFR galaxies. In this framework, the observed FMR offset of $\sim -7$ to $-0.9$ dex likely characterises a dilution-dominated phase, where the arrival of pristine gas has successfully lowered the global metallicity. Still, the subsequent star formation has not yet enriched the gas or restored chemical equilibrium.
Furthermore, the satellite status of these galaxies adds a layer of complexity to their chemical evolution. According to \citet{mcclymont_thesan-zoom_2026}, satellite galaxies at these redshifts often exhibit lower metal retention efficiencies and lower gas fractions compared to central galaxies of similar mass. This is partly due to the impact of metal pollution from the outflows of their neighbouring central galaxies. 

The fact that we observe little to no star formation activity within the last $\rm 10 \ Myr$ within the SFHs of the satellites (see Figure~\ref{fig:satellites_SFHs}) could also indicate that these are mini-quenched galaxies \citep[e.g.][]{looser_recently_2024}. However, the fact that we are still able to measure emission lines in these sources makes this scenario not very plausible.

The location of these satellites near the metallicity floor of $\rm \sim 2\% \ Z_{\odot}$ ($\rm 12+\log(O/H) \sim 7.0$) highlights important implications for the transition from the first stars to the first galaxies. Metal production from the first generation of stars can rapidly enrich a host halo to a baseline of $\sim 10^{-3} \ Z_{\odot}$, triggering a transition to Population II star formation \citep{klessen_first_2023}. Finding galaxies at this metallicity floor suggests that they are primordial systems in the immediate aftermath of this initial enrichment from strong pristine inflows \citep{mcclymont_thesan-zoom_2026}. However, the timescale and uniformity of this enrichment are sensitive to stellar initial mass function, and the efficiency of metal mixing \citep{wise_birth_2012, ritter_metal_2015}. Furthermore, \citet{asada_glimpse-ddt_2026} introduce two possible first enrichment scenarios: 'overshoot' or 'undershoot' enrichment, which are characterised by a rapid or slow transition from a Pop III to Pop II stellar population, respectively. Depending on how rapid this transition happens, young galaxies such as our satellites can cross the metallicity floor towards higher metallicities rather quickly. Nonetheless, the recent detection of even more pristine sources, with an estimated gas-phase metallicities below 1\% solar \citep{vanzella_pristine_2026,nakajima_ultra-faint_2026,morishita_pristine_2025,maiolino_black_2026}, some of which have been found even towards lower redshifts of $\rm z\sim 3$ \citep{cai_metal-free_2025}, suggests that this floor is permeable, representing the very first star formation in near-pristine conditions. 

\section{Conclusions}\label{sec:conclusions}

We have presented NIRSpec/PRISM IFU observations of seven low-metallicity galaxies spanning a redshift range of $z \sim 7-10$. This sample constitutes the largest systematic study of spatially resolved metallicity gradients around and beyond the Epoch of Reionisation to date. The broad wavelength coverage of the PRISM allowed for the simultaneous measurement of multiple rest-frame optical emission lines, enabling us to map gas-phase metallicities on kiloparsec scales. Beyond our primary sample, we identified and characterised three low-metallicity satellite galaxies associated with the host systems. Our main findings are summarised as follows:

\begin{itemize}

    \item \textbf{Metal-poor Sample:} Our sample is systematically shifted toward lower metallicities compared to local Mass-Metallicity Relations (MZR), appearing chemically immature for their stellar masses, $\log(M_*/M_{\odot}) \sim 7.8-9.5$. Conversely, their star formation activity remains high, with the majority of sources aligning with the $z \sim 7-10$ Star-Forming Main Sequence (SFMS), with the notable exception of the starbursting system SXDF\_NB1006-2. 
    
    \item \textbf{Low-Metallicity Satellites:} The three identified satellites exhibit metallicities are very metal-poor with $12+\log(\text{O/H}) \sim 7.1-7.3$ ($\sim 3\%-4\% Z_{\odot}$). Their combination of low star formation rates and significant FMR offsets (down to $-1$ dex) suggests they are possibly in a dilution-dominated pre-burst phase, where the accretion of pristine gas has lowered the global metallicity but has not yet significantly enhanced their star formation. These systems provide a rare spatially resolved view of the first chemical enrichment stages.
    
    \item \textbf{Gas-Phase Metallicity Gradients:} We report an average metallicity gradient of $-0.03 \pm 0.04$ dex/kpc with a significant scatter ($\sigma \approx 0.12$ dex/kpc). The prevalence of flat-to-slightly negative gradients suggests that efficient radial mixing, driven by AGN feedback, supernovae-driven outflows, and tidal interactions, is already prevalent at $z > 7$. The source with a strongly positive gradient likely reflects the direct funnelling of pristine gas into the galactic cores, which dilutes central metallicities and outpaces inside-out enrichment. We observe a tentative trend where higher-mass systems exhibit more flat-to-positive gradients, suggesting that centrally concentrated accretion may scale with stellar mass.
    
    \item \textbf{Large FMR offset:} Our sources are significantly offset from the local Fundamental Metallicity Relation (FMR), exhibiting a downward scatter in metallicity for a fixed $\rm \mu_{\alpha}$. This deviation, coupled with the absence of a spatially resolved FMR, indicates that these galaxies are in a state of non-equilibrium, whereby an excess of pristine gas accretion could not be readily processed into star formation but is diluting the metallicity. However, the emergence of a coherent (though offset) relation suggests that the fundamental regulatory mechanisms of the baryon cycle were already established within the first Gyr of cosmic time.
\end{itemize}

In conclusion, these results provide some of the first spatially resolved constraints on the baryon cycle during the early stages of chemical enrichment. The observed diversity in chemical architectures, ranging from well-mixed flat profiles to inflow-dominated positive gradients, underscores that early galaxy assembly is a stochastic process driven by rapid gas fluctuations and intense feedback.

\section*{Acknowledgements}

We are grateful to the referee for their helpful feedback. We thank Xunda Sun, Moka Nishigaki, Pratika Dayal, and Alex M. Garcia for kindly sharing their simulation results. MK thanks the University of Cambridge Harding Distinguished Postgraduate Scholars Programme, UK Science and Technology Facilities Council (STFC) Center for Doctoral Training (CDT) in Data Intensive Science, and Girton College Cambridge for a PhD studentship. RM acknowledges support from the Science and Technology Facilities Council (STFC), by the European Research Council (ERC) through Advanced Grant 695671 ``QUENCH'', by the UK Research and Innovation (UKRI) Frontier Research grant RISEandFALL. RM also acknowledges support from a Royal Society Research Professorship grant. H\"U thanks the Max Planck Society for support through the Lise Meitner Excellence Program. H\"U acknowledges funding by the European Union (ERC APEX, 101164796). SA acknowledges support from grant PID2021-127718NB-I00 funded by Spanish Ministerio de Ciencia e Innovaci\'on MCIN/AEI/10.13039/501100011033. WMB gratefully acknowledges support from DARK via the DARK fellowship. This work was supported by a research grant (VIL54489) from VILLUM FONDEN. WM thanks the Science and Technology Facilities Council (STFC) Center for Doctoral Training (CDT) in Data Intensive Science at the University of Cambridge (STFC grant number 2742968) for a PhD studentship. MP acknowledges support through the grants PID2021-127718NB-I00, PID2024-159902NA-I00, and RYC2023-044853-I, funded by the Spain Ministry of Science and Innovation/State Agency of Research MCIN/AEI/10.13039/501100011033 and El Fondo Social Europeo Plus FSE+. B.R.P acknowledges support from grant PID2024-158856NA-I00 funded by Spanish Ministerio de Ciencia e Innovación MCIN/AEI/10.13039/501100011033 and by “ERDF A way of making Europe. GV and SC acknowledge support by European Union’s HE ERC Starting Grant No. 101040227 - WINGS. AV acknowledges funding from the Cosmic Frontier Center and the University of Texas at Austin’s College of Natural Sciences. Views and opinions expressed are those of the authors only and do not necessarily reflect those of the European Union or the European Research Council Executive Agency. Neither the European Union nor the granting authority can be held responsible for them. Some of the data products presented herein were retrieved from the Dawn JWST Archive (DJA). DJA is an initiative of the Cosmic Dawn Center (DAWN), which is funded by the Danish National Research Foundation under grant DNRF140. 

\section*{Data Availability}

The NIRSpec data used in this research were obtained within the
NIRSpec-IFU GTO with programme ID 2957, publicly available at MAST. The NIRCam data used are publicly available at the DJA: \url{https://dawn-cph.github.io/dja/}.



\bibliographystyle{mnras}
\bibliography{references} 

@article{maiolino_jades_2024-1,
	title = {{JADES} - {The} diverse population of infant black holes at 4 {\textless} z {\textless} 11: {Merging}, tiny, poor, but mighty},
	volume = {691},
	copyright = {© The Authors 2024},
	issn = {0004-6361, 1432-0746},
	shorttitle = {{JADES} - {The} diverse population of infant black holes at 4 {\textless} z {\textless} 11},
	url = {https://www.aanda.org/articles/aa/abs/2024/11/aa47640-23/aa47640-23.html},
	doi = {10.1051/0004-6361/202347640},
	language = {en},
	urldate = {2026-09-15},
	journal = {Astronomy \& Astrophysics},
	publisher = {EDP Sciences},
	author = {Maiolino, Roberto and Scholtz, Jan and Curtis-Lake, Emma and Carniani, Stefano and Baker, William and Graaff, Anna de and Tacchella, Sandro and Übler, Hannah and D’Eugenio, Francesco and Witstok, Joris and Curti, Mirko and Arribas, Santiago and Bunker, Andrew J. and Charlot, Stéphane and Chevallard, Jacopo and Eisenstein, Daniel J. and Egami, Eiichi and Ji, Zhiyuan and Jones, Gareth C. and Lyu, Jianwei and Rawle, Tim and Robertson, Brant and Rujopakarn, Wiphu and Perna, Michele and Sun, Fengwu and Venturi, Giacomo and Williams, Christina C. and Willott, Chris},
	month = nov,
	year = {2024},
	pages = {A145},
}

@misc{isobe_jades_2026,
	title = {{JADES}: the mass-metallicity relation at \$z=1-10\$. {New} calibrations, extremely metal-poor galaxies, and chemical diversity},
	shorttitle = {{JADES}},
	url = {https://ui.adsabs.harvard.edu/abs/2026arXiv260611345I},
	doi = {10.48550/arXiv.2606.11345},
	urldate = {2026-09-15},
	publisher = {arXiv},
	author = {Isobe, Yuki and Curti, Mirko and Maiolino, Roberto and Duan, Qiao and McClymont, William and Puskás, Dávid and D'Eugenio, Francesco and Rinaldi, Pierluigi and Trussler, James A. A. and Scholtz, Jan and Looser, Tobias J. and Nelson, Erica and Ji, Xihan and Langeroodi, Danial and Tacchella, Sandro and Jones, Gareth C. and Juodžbalis, Ignas and Pascalau, Robert G. and Hsiao, Tiger Yu-Yang and Übler, Hannah and Baker, William M. and Bunker, Andrew J. and Carniani, Stefano and Charlot, Stéphane and Curtis-Lake, Emma and Geris, Sophia and Koller, Maria and Lyu, Jianwei and Robertson, Brant and Williams, Christina C. and Wu, Zihao},
	month = jun,
	year = {2026},
	note = {ADS Bibcode: 2026arXiv260611345I},
}

@inproceedings{perrin_updated_2014,
	title = {Updated point spread function simulations for {JWST} with {WebbPSF}},
	volume = {9143},
	url = {https://ui.adsabs.harvard.edu/abs/2014SPIE.9143E..3XP},
	doi = {10.1117/12.2056689},
	urldate = {2026-03-06},
	booktitle = {{SPIE}},
	author = {Perrin, Marshall D. and Sivaramakrishnan, Anand and Lajoie, Charles-Philippe and Elliott, Erin and Pueyo, Laurent and Ravindranath, Swara and Albert, Loïc.},
	month = aug,
	year = {2014},
	note = {ADS Bibcode: 2014SPIE.9143E..3XP},
	pages = {91433X},
}

@book{osterbrock_astrophysics_2006,
	title = {Astrophysics of gaseous nebulae and active galactic nuclei},
	url = {https://ui.adsabs.harvard.edu/abs/2006agna.book.....O},
	urldate = {2025-08-11},
	publisher = {The MIT Press},
	author = {Osterbrock, Donald E. and Ferland, Gary J.},
	month = jan,
	year = {2006},
	note = {Publication Title: Astrophysics of gaseous nebulae and active galactic nuclei
ADS Bibcode: 2006agna.book.....O},
}

@article{lee_alpine-cristal-jwst_2026,
	title = {The {ALPINE}-{CRISTAL}-{JWST} {Survey}: {Gas}-phase abundance gradients of main-sequence star-forming galaxies and their kinematics at 4 {\textless} z {\textless} 6},
	volume = {710},
	issn = {0004-6361},
	shorttitle = {The {ALPINE}-{CRISTAL}-{JWST} {Survey}},
	url = {https://ui.adsabs.harvard.edu/abs/2026A&A...710A.310L},
	doi = {10.1051/0004-6361/202557589},
	urldate = {2026-07-21},
	journal = {Astronomy and Astrophysics},
	publisher = {EDP},
	author = {Lee, L. L. and Förster Schreiber, N. M. and Fujimoto, S. and Faisst, A. L. and Herrera-Camus, R. and Genzel, R. and Tacconi, L. J. and Lutz, D. and Renzini, A. and Sanders, R. and Wisnioski, E. and Wuyts, S. and Parlanti, E. and Jones, G. and Übler, H. and Liu, D. and Chen, J. and Davies, R. I. and Tozzi, G. and Burkert, A. and Price, S. H. and Aravena, M. and Boquien, M. and Béthermin, M. and da Cunha, E. and Davies, R. L. and De Looze, I. and Dessauges-Zavadsky, M. and Ferrara, A. and Fisher, D. B. and Gillman, S. and Ginolfi, M. and Ibar, E. and Koekemoer, A. M. and Molina, J. and Naab, T. and Relaño, M. and Riechers, D. A. and Sanders, D. B. and Spilker, J. S. and Vallini, L. and Zamorani, G. and Nanni, A. and Dam, P. and Diaz-Santos, T. and Gómez-Espinoza, D. and Hadi, A. and Ikeda, R. and Posses, A. and Romano, M. and Sternberg, A. and Villanueva, V. and Wang, W.},
	month = jun,
	year = {2026},
	note = {ADS Bibcode: 2026A\&A...710A.310L},
	pages = {A310},
}

@article{stanton_jwst_2026,
	title = {The {JWST} {EXCELS} {Survey}: gas-phase metallicity evolution at 2 {\textless} z {\textless} 8},
	volume = {547},
	issn = {0035-8711},
	shorttitle = {The {JWST} {EXCELS} {Survey}},
	url = {https://ui.adsabs.harvard.edu/abs/2026MNRAS.547ag449S},
	doi = {10.1093/mnras/stag449},
	urldate = {2026-07-06},
	journal = {Monthly Notices of the Royal Astronomical Society},
	publisher = {OUP},
	author = {Stanton, T. M. and Cullen, F. and Carnall, A. C. and Scholte, D. and Arellano-Córdova, K. Z. and Shapley, A. E. and McLeod, D. J. and Donnan, C. T. and Begley, R. and Davé, R. and Dunlop, J. S. and McLure, R. J. and Rowlands, K. and Bondestam, C. and Hamadouche, M. L. and Leung, H.-H. and Stevenson, S. D. and Taylor, E.},
	month = apr,
	year = {2026},
	note = {ADS Bibcode: 2026MNRAS.547ag449S},
	pages = {stag449},
}

@article{rusta_pristine_2026,
	title = {The {Pristine} {He} {II} {Emitter} near {GN}-z11: {Constraining} the {Mass} {Distribution} of the {First} {Stars}},
	volume = {1003},
	issn = {0004-637X},
	shorttitle = {The {Pristine} {He} {II} {Emitter} near {GN}-z11},
	url = {https://ui.adsabs.harvard.edu/abs/2026ApJ..1003L..14R},
	doi = {10.3847/2041-8213/ae64e1},
	urldate = {2026-07-06},
	journal = {The Astrophysical Journal},
	publisher = {IOP},
	author = {Rusta, Elka and Salvadori, Stefania and Maiolino, Roberto and Gelli, Viola and Koutsouridou, Ioanna and Carniani, Stefano and Übler, Hannah and Marconi, Alessandro and Schaerer, Daniel},
	month = may,
	year = {2026},
	note = {ADS Bibcode: 2026ApJ..1003L..14R},
	pages = {L14},
}

@article{ren_rioja_2025,
	title = {{RIOJA}. {Young} starburst and ionized gas outflows in a z = 7.212 galaxy uncovered by {JWST} {NIRCam} and {NIRSpec} observations},
	volume = {544},
	issn = {0035-8711},
	url = {https://ui.adsabs.harvard.edu/abs/2025MNRAS.544.4722R},
	doi = {10.1093/mnras/staf1886},
	urldate = {2026-07-06},
	journal = {Monthly Notices of the Royal Astronomical Society},
	publisher = {OUP},
	author = {Ren, Yi W. and Inoue, Akio K. and Álvarez-Márquez, Javier and Hashimoto, Takuya and Colina, Luis and Sugahara, Yuma and Costantin, Luca and Mawatari, Ken and Fudamoto, Yoshinobu and Arribas, Santiago and Gómez, Alejandro Crespo and Ceverino, Daniel and Nakazato, Yurina and Hagimoto, Masato and Usui, Mitsutaka and Marques-Chaves, Rui and Matsuo, Hiroshi and Hashigaya, Takeshi and Osone, Wataru and Blanco-Prieto, Carmen and Tamura, Yoichi and Yoshida, Naoki and Bakx, Tom J. L. C. and Pereira-Santaella, Miguel},
	month = dec,
	year = {2025},
	note = {ADS Bibcode: 2025MNRAS.544.4722R},
	pages = {4722--4743},
}

@article{reddy_aurora_2026,
	title = {The {AURORA} {Survey}: {Multiple} {Balmer} and {Paschen} {Emission} {Lines} for {Individual} {Star}-forming {Galaxies} at z = 1.5─4.4. {I}. {A} {Diversity} of {Nebular} {Attenuation} {Curves} and {Evidence} for {Non}-unity {Dust} {Covering} {Fractions}},
	volume = {999},
	issn = {0004-637X},
	shorttitle = {The {AURORA} {Survey}},
	url = {https://ui.adsabs.harvard.edu/abs/2026ApJ...999...15R},
	doi = {10.3847/1538-4357/ae38da},
	urldate = {2026-07-06},
	journal = {The Astrophysical Journal},
	publisher = {IOP},
	author = {Reddy, Naveen A. and Shapley, Alice E. and Sanders, Ryan L. and Topping, Michael W. and Ellis, Richard S. and Pettini, Max and Brammer, Gabriel and Cullen, Fergus and Förster Schreiber, Natascha M. and Khostovan, Ali A. and McLeod, Derek J. and McLure, Ross J. and Narayanan, Desika and Oesch, Pascal A. and Pahl, Anthony J. and Steidel, Charles C. and Berg, Danielle A.},
	month = mar,
	year = {2026},
	note = {ADS Bibcode: 2026ApJ...999...15R},
	pages = {15},
}

@article{pollock_novel_2026,
	title = {Novel z ∼ 10 auroral line measurements extend the gradual offset of the fundamental metallicity relation deep into the first gigayear of cosmic time},
	volume = {708},
	issn = {0004-6361},
	url = {https://ui.adsabs.harvard.edu/abs/2026A&A...708A.203P},
	doi = {10.1051/0004-6361/202556032},
	urldate = {2026-07-06},
	journal = {Astronomy and Astrophysics},
	publisher = {EDP},
	author = {Pollock, Clara L. and Gottumukkala, Rashmi and Heintz, Kasper E. and Brammer, Gabriel B. and Roberts-Borsani, Guido and Oesch, Pascal A. and Witstok, Joris and Arellano-Córdova, Karla Z. and Cullen, Fergus and Scholte, Dirk and Terp, Chamilla and Rowland, Lucie and Sneppen, Albert and Ito, Kei and Valentino, Francesco and Matthee, Jorryt and Watson, Darach and Toft, Sune},
	month = apr,
	year = {2026},
	note = {ADS Bibcode: 2026A\&A...708A.203P},
	pages = {A203},
}

@article{nakajima_ultra-faint_2026,
	title = {An ultra-faint, chemically primitive galaxy forming in the reionization era},
	volume = {653},
	issn = {0028-0836},
	url = {https://ui.adsabs.harvard.edu/abs/2026Natur.653..363N},
	doi = {10.1038/s41586-026-10374-1},
	urldate = {2026-07-06},
	journal = {Nature},
	author = {Nakajima, Kimihiko and Ouchi, Masami and Harikane, Yuichi and Vanzella, Eros and Ono, Yoshiaki and Isobe, Yuki and Nishigaki, Moka and Tsujimoto, Takuji and Nakamura, Fumitaka and Xu, Yi and Umeda, Hiroya and Zhang, Yechi},
	month = may,
	year = {2026},
	note = {ADS Bibcode: 2026Natur.653..363N},
	pages = {363--367},
}

@article{maiolino_black_2026,
	title = {A black hole in a near pristine galaxy 700 {Myr} after the big bang},
	volume = {548},
	issn = {0035-8711},
	url = {https://ui.adsabs.harvard.edu/abs/2026MNRAS.548f2109M},
	doi = {10.1093/mnras/staf2109},
	urldate = {2026-07-06},
	journal = {Monthly Notices of the Royal Astronomical Society},
	publisher = {OUP},
	author = {Maiolino, Roberto and Übler, Hannah and D'Eugenio, Francesco and Scholtz, Jan and Juodžbalis, Ignas and Ji, Xihan and Perna, Michele and Bromm, Volker and Dayal, Pratika and Koudmani, Sophie and Liu, Boyuan and Schneider, Raffaella and Sijacki, Debora and Valiante, Rosa and Trinca, Alessandro and Zhang, Saiyang and Volonteri, Marta and Inayoshi, Kohei and Carniani, Stefano and Nakajima, Kimihiko and Isobe, Yuki and Witstok, Joris and Jones, Gareth C. and Tacchella, Sandro and Arribas, Santiago and Bunker, Andrew and Cataldi, Elisa and Charlot, Stephane and Curti, Giovanni Cresci Mirko and Fabian, Andrew C. and Katz, Harley and Kumari, Nimisha and Laporte, Nicolas and Mazzolari, Giovanni and Robertson, Brant and Sun, Fengwu and Rodriguez Del Pino, Bruno and Venturi, Giacomo},
	month = may,
	year = {2026},
	note = {ADS Bibcode: 2026MNRAS.548f2109M},
	pages = {staf2109},
}

@article{kim_agora_2026,
	title = {The {AGORA} {High}-resolution {Galaxy} {Simulations} {Comparison} {Project}. {X}. {Formation} and {Evolution} of {Galaxies} at the {High}-redshift {Frontier}},
	volume = {1000},
	issn = {0004-637X},
	url = {https://ui.adsabs.harvard.edu/abs/2026ApJ..1000..276K},
	doi = {10.3847/1538-4357/ae4a23},
	urldate = {2026-07-06},
	journal = {The Astrophysical Journal},
	publisher = {IOP},
	author = {Kim, Hyeonyong and Kim, Ji-Hoon and Jung, Minyong and Roca-Fàbrega, Santi and Ceverino, Daniel and Granizo, Pablo and Nagamine, Kentaro and Primack, Joel R. and Velázquez, Héctor and Barrow, Kirk S. S. and Feldmann, Robert and Fukushima, Keita and Mayer, Lucio and Oh, Boon Kiat and Powell, Johnny W. and Abel, Tom and Agertz, Oscar and Jeong, Chaerin and Lupi, Alessandro and Oku, Yuri and Quinn, Thomas R. and Revaz, Yves and Rodríguez-Cardoso, Ramón and Shimizu, Ikkoh and Teyssier, Romain and {Agora Collaboration}},
	month = apr,
	year = {2026},
	note = {ADS Bibcode: 2026ApJ..1000..276K},
	pages = {276},
}

@article{jones_ga-nifs_2026,
	title = {{GA}-{NIFS}: interstellar medium properties and tidal interactions in the evolved massive merging system {B14}-65666 at z = 7.152},
	volume = {547},
	issn = {0035-8711},
	shorttitle = {{GA}-{NIFS}},
	url = {https://ui.adsabs.harvard.edu/abs/2026MNRAS.547ag336J},
	doi = {10.1093/mnras/stag336},
	urldate = {2026-07-06},
	journal = {Monthly Notices of the Royal Astronomical Society},
	publisher = {OUP},
	author = {Jones, Gareth C. and Bowler, Rebecca A. A. and Bunker, Andrew J. and Curti, Mirko and Arribas, Santiago and Carniani, Stefano and Charlot, Stephane and Perna, Michele and Rodríguez Del Pino, Bruno and Übler, Hannah and Willott, Chris J. and Chevallard, Jacopo and Cresci, Giovanni and Parlanti, Eleonora and Scholtz, Jan and Venturi, Giacomo},
	month = apr,
	year = {2026},
	note = {ADS Bibcode: 2026MNRAS.547ag336J},
	pages = {stag336},
}

@article{jain_uniform_2026,
	title = {A {Uniform} {Analysis} of {Gas}-phase {Metallicity} {Evolution} with 1─3 {Gyr} {Time} {Sampling} over the {Past} 12 {Gyr}},
	volume = {1000},
	issn = {0004-637X},
	url = {https://ui.adsabs.harvard.edu/abs/2026ApJ..1000..109J},
	doi = {10.3847/1538-4357/ae4590},
	urldate = {2026-07-06},
	journal = {The Astrophysical Journal},
	publisher = {IOP},
	author = {Jain, Shweta and Sanders, Ryan L. and Khostovan, Ali Ahmad and Jones, Tucker and Shapley, Alice E. and Reddy, Naveen A. and Garcia, Alex M. and Torrey, Paul and Coil, Alison},
	month = mar,
	year = {2026},
	note = {ADS Bibcode: 2026ApJ..1000..109J},
	pages = {109},
}

@article{foreman-mackey_emcee_2013,
	title = {emcee: {The} {MCMC} {Hammer}},
	volume = {125},
	issn = {1538-3873},
	shorttitle = {emcee},
	url = {https://iopscience.iop.org/article/10.1086/670067},
	doi = {10.1086/670067},
	language = {en},
	number = {925},
	urldate = {2026-07-06},
	journal = {Publications of the Astronomical Society of the Pacific},
	publisher = {IOP Publishing},
	author = {Foreman-Mackey, Daniel and Hogg, David W. and Lang, Dustin and Goodman, Jonathan},
	month = feb,
	year = {2013},
	pages = {306},
}

@misc{ferland_2013_2013,
	title = {The 2013 {Release} of {Cloudy}},
	url = {https://ui.adsabs.harvard.edu/abs/2013RMxAA..49..137F},
	doi = {10.48550/arXiv.1302.4485},
	urldate = {2026-07-06},
	publisher = {arXiv},
	author = {Ferland, G. J. and Porter, R. L. and van Hoof, P. A. M. and Williams, R. J. R. and Abel, N. P. and Lykins, M. L. and Shaw, G. and Henney, W. J. and Stancil, P. C.},
	month = apr,
	year = {2013},
	note = {ISSN: 0185-1101
Volume: 49
ADS Bibcode: 2013RMxAA..49..137F},
}

@inproceedings{eisenhauer_sinfoni_2003,
	title = {{SINFONI}: integral field spectroscopy at 50-milli-arcsecond resolution with the {ESO} {VLT}},
	volume = {4841},
	shorttitle = {{SINFONI}},
	url = {https://www.spiedigitallibrary.org/conference-proceedings-of-spie/4841/1/SINFONI--integral-field-spectroscopy-at-50-milli-arcsecond-resolution/10.1117/12.459468},
	doi = {10.1117/12.459468},
	language = {en},
	urldate = {2026-07-06},
	booktitle = {Instrument {Design} and {Performance} for {Optical}/{Infrared} {Ground}-based {Telescopes}},
	publisher = {SPIE},
	author = {Eisenhauer, Frank and Abuter, Roberto and Bickert, Klaus and Biancat-Marchet, Fabio and Bonnet, Henri and Brynnel, Joar and Conzelmann, Ralf D. and Delabre, Bernard and Donaldson, Robert and Farinato, Jacopo and Fedrigo, Enrico and Genzel, Reinhard and Hubin, Norbert N. and Iserlohe, Christof and Kasper, Markus E. and Kissler-Patig, Markus and Monnet, Guy J. and Roehrle, Claudia and Schreiber, Juergen and Stroebele, Stefan and Tecza, Matthias and Thatte, Niranjan A. and Weisz, Harald},
	month = mar,
	year = {2003},
	pages = {1548},
}

@article{curtis-lake_jades_2026,
	title = {{JADES} data release 4 ─ {Paper} {I}. {Sample} selection, observing strategy and redshifts of the complete spectroscopic sample},
	volume = {549},
	issn = {0035-8711},
	url = {https://ui.adsabs.harvard.edu/abs/2026MNRAS.549ag836C},
	doi = {10.1093/mnras/stag836},
	urldate = {2026-07-06},
	journal = {Monthly Notices of the Royal Astronomical Society},
	publisher = {OUP},
	author = {Curtis-Lake, Emma and Cameron, Alex J. and Bunker, Andrew J. and Scholtz, Jan and Carniani, Stefano and Parlanti, Eleonora and D'Eugenio, Francesco and Jakobsen, Peter and Willmer, Christopher N. A. and Arribas, Santiago and Baker, William M. and Charlot, Stéphane and Chevallard, Jacopo and Circosta, Chiara and Curti, Mirko and Duan, Qiao and Eisenstein, Daniel J. and Hainline, Kevin and Ji, Zhiyuan and Johnson, Benjamin D. and Jones, Gareth C. and Maiolino, Roberto and Maseda, Michael V. and Perna, Michele and Pérez-González, Pablo G. and Rawle, Tim and Rieke, Marcia and Rinaldi, Pierluigi and Robertson, Brant and Rodríguez Del Pino, Bruno and Saxena, Aayush and Shivaei, Irene and Smit, Renske and Tacchella, Sandro and Übler, Hannah and Venturi, Giacomo and Williams, Christina C. and Willott, Chris},
	month = jul,
	year = {2026},
	note = {ADS Bibcode: 2026MNRAS.549ag836C},
	pages = {stag836},
}

@inproceedings{bacon_muse_2010,
	title = {The {MUSE} second-generation {VLT} instrument},
	volume = {7735},
	url = {https://www.spiedigitallibrary.org/conference-proceedings-of-spie/7735/1/The-MUSE-second-generation-VLT-instrument/10.1117/12.856027},
	doi = {10.1117/12.856027},
	language = {en},
	urldate = {2026-07-06},
	booktitle = {Ground-based and {Airborne} {Instrumentation} for {Astronomy} {III}},
	publisher = {SPIE},
	author = {Bacon, R. and Accardo, M. and Adjali, L. and Anwand, H. and Bauer, S. and Biswas, I. and Blaizot, J. and Boudon, D. and Brau-Nogue, S. and Brinchmann, J. and Caillier, P. and Capoani, L. and Carollo, C. M. and Contini, T. and Couderc, P. and Daguisé, E. and Deiries, S. and Delabre, B. and Dreizler, S. and Dubois, J. and Dupieux, M. and Dupuy, C. and Emsellem, E. and Fechner, T. and Fleischmann, A. and François, M. and Gallou, G. and Gharsa, T. and Glindemann, A. and Gojak, D. and Guiderdoni, B. and Hansali, G. and Hahn, T. and Jarno, A. and Kelz, A. and Koehler, C. and Kosmalski, J. and Laurent, F. and Floch, M. Le and Lilly, S. J. and Lizon, J.-L. and Loupias, M. and Manescau, A. and Monstein, C. and Nicklas, H. and Olaya, J.-C. and Pares, L. and Pasquini, L. and Pécontal-Rousset, A. and Pelló, R. and Petit, C. and Popow, E. and Reiss, R. and Remillieux, A. and Renault, E. and Roth, M. and Rupprecht, G. and Serre, D. and Schaye, J. and Soucail, G. and Steinmetz, M. and Streicher, O. and Stuik, R. and H, Valentin and Vernet, J. and Weilbacher, P. and Wisotzki, L. and Yerle, N.},
	month = jul,
	year = {2010},
	pages = {131},
}

@article{vanzella_pristine_2026,
	title = {A pristine, star-forming complex at z = 4.19},
	volume = {705},
	issn = {0004-6361},
	url = {https://ui.adsabs.harvard.edu/abs/2026A&A...705L..12V},
	doi = {10.1051/0004-6361/202557153},
	urldate = {2026-07-06},
	journal = {Astronomy and Astrophysics},
	publisher = {EDP},
	author = {Vanzella, E. and Messa, M. and Zanella, A. and Bolamperti, A. and Castellano, M. and Loiacono, F. and Bergamini, P. and Roberts Borsani, G. and Adamo, A. and Fontana, A. and Treu, T. and Calura, F. and Grillo, C. and Lombardi, M. and Rosati, P. and Gilli, R. and Meneghetti, M.},
	month = jan,
	year = {2026},
	note = {ADS Bibcode: 2026A\&A...705L..12V},
	pages = {L12},
}

@article{sun_galaxy_2026,
	title = {Galaxy {Metallicity} {Gradients} in the {Epoch} of {Reionization} from the {FIRE}-2 {Simulations}},
	volume = {1002},
	issn = {0004-637X},
	url = {https://ui.adsabs.harvard.edu/abs/2026ApJ..1002..148S},
	doi = {10.3847/1538-4357/ae5f78},
	urldate = {2026-07-06},
	journal = {The Astrophysical Journal},
	publisher = {IOP},
	author = {Sun, Xunda and Wang, Xin and Jiang, Fangzhou and Mo, Houjun and Ho, Luis C. and Zhou, Qianqiao and Ma, Xiangcheng and Zhan, Hu and Wetzel, Andrew and Graf, Russell L. and Hopkins, Philip F. and Kereš, Dušan and Stern, Jonathan},
	month = may,
	year = {2026},
	note = {ADS Bibcode: 2026ApJ..1002..148S},
	pages = {148},
}

@article{stanghellini_direct_2026,
	title = {Direct {Abundance} {Maps} and {Radial} {Metallicity} {Gradients} of {Two} {Galaxies} at z ∼ 4─5 in the {GARDEN} {Survey}},
	volume = {1004},
	issn = {0004-637X},
	url = {https://ui.adsabs.harvard.edu/abs/2026ApJ..1004..243S},
	doi = {10.3847/1538-4357/ae5a9f},
	urldate = {2026-07-06},
	journal = {The Astrophysical Journal},
	publisher = {IOP},
	author = {Stanghellini, Letizia and Kassin, Susan A. and Pacifici, Camilla and Morrison, Jane E. and Dickinson, Mark and Sukay, Ezra and Mulcahey, Celia R. and Bergeron, Louis E. and Regan, Michael W. and Willmer, Christopher N. A. and Weiner, Benjamin J. and Dencheva, Nadia and Law, David and de la Vega, Alexander and Koekemoer, Anton M. and Conselice, Christopher and Gardner, Jonathan P. and Guo, Yicheng and Hammer, Francois and Henry, Alina and Holwerda, Benne W. and Kartaltepe, Jeyhan and Lucas, Ray A. and Puech, Mathieu and Rafelski, Marc and Shivaei, Irene and Welker, Charlotte and Xu, Xinfeng and Yung, L. Y. Aaron},
	month = jun,
	year = {2026},
	note = {ADS Bibcode: 2026ApJ..1004..243S},
	pages = {243},
}

@article{simmonds_bursting_2025,
	title = {Bursting at the seams: the star-forming main sequence and its scatter at z = 3─9 using {NIRCam} photometry from {JADES}},
	volume = {544},
	issn = {0035-8711},
	shorttitle = {Bursting at the seams},
	url = {https://ui.adsabs.harvard.edu/abs/2025MNRAS.544.4551S},
	doi = {10.1093/mnras/staf1950},
	urldate = {2026-07-06},
	journal = {Monthly Notices of the Royal Astronomical Society},
	publisher = {OUP},
	author = {Simmonds, C. and Tacchella, S. and McClymont, W. and Curtis-Lake, E. and D'Eugenio, F. and Hainline, K. and Johnson, B. D. and Kravtsov, A. and Puskás, D. and Robertson, B. and Stoffers, A. and Willott, C. and Baker, W. M. and Belokurov, V. A. and Bhatawdekar, R. and Bunker, A. J. and Carniani, S. and Chevallard, J. and Curti, M. and Duan, Q. and Helton, J. M. and Ji, Z. and Looser, T. J. and Maiolino, R. and Maseda, M. V. and Shivaei, I. and Williams, C. C.},
	month = dec,
	year = {2025},
	note = {ADS Bibcode: 2025MNRAS.544.4551S},
	pages = {4551--4575},
}

@article{marconcini_ga-nifs_2025,
	title = {{GA}-{NIFS}: {Dissecting} the multiple sub-structures and probing their complex interactions in the {Lyα} emitter galaxy {CR7} at z = 6.6 with {JWST}/{NIRSpec}},
	volume = {699},
	issn = {0004-6361},
	shorttitle = {{GA}-{NIFS}},
	url = {https://ui.adsabs.harvard.edu/abs/2025A&A...699A.154M},
	doi = {10.1051/0004-6361/202452994},
	urldate = {2026-07-06},
	journal = {Astronomy and Astrophysics},
	publisher = {EDP},
	author = {Marconcini, C. and D'Eugenio, F. and Maiolino, R. and Arribas, S. and Bunker, A. and Carniani, S. and Charlot, S. and Perna, M. and Rodríguez Del Pino, B. and Übler, H. and Pérez-González, P. G. and Willott, C. J. and Böker, T. and Cresci, G. and Curti, M. and Lamperti, I. and Scholtz, J. and Parlanti, E. and Venturi, G.},
	month = jul,
	year = {2025},
	note = {ADS Bibcode: 2025A\&A...699A.154M},
	pages = {A154},
}

@article{faisst_alpinecristaljwst_2026,
	title = {The {ALPINE}─{CRISTAL}─{JWST} {Survey}: {The} {Fast} {Metal} {Enrichment} of {Massive} {Galaxies} at z ∼ 5},
	volume = {1004},
	issn = {0004-637X},
	shorttitle = {The {ALPINE}─{CRISTAL}─{JWST} {Survey}},
	url = {https://ui.adsabs.harvard.edu/abs/2026ApJ..1004...22F},
	doi = {10.3847/1538-4357/ae6645},
	urldate = {2026-07-06},
	journal = {The Astrophysical Journal},
	publisher = {IOP},
	author = {Faisst, Andreas L. and Liu, Lun-Jun and Dubois, Yohan and Nakazato, Yurina and Osman, Omima and Pallottini, Andrea and Vallini, Livia and Fujimoto, Seiji and Mobasher, Bahram and Wang, Wuji and Lin, Yu-Heng and Amorín, Ricardo O. and Aravena, Manuel and Assef, R. J. and Battisti, Andrew J. and Béthermin, Matthieu and Boquien, Médéric and Cassata, Paolo and da Cunha, Elisabete and Dam, Poulomi and de Lucia, Gabriella and De Looze, Ilse and Dessauges-Zavadsky, Miroslava and Ferrara, Andrea and Finner, Kyle and Fontanot, Fabio and Ginolfi, Michele and Gómez-Espinoza, Diego A. and Gruppioni, Carlotta and Gutiérrez-Vera, Nicol and Hadi, Ali and Herrera-Camus, Rodrigo and Hirschmann, Michaela and Ibar, Eduardo and Inami, Hanae and Kartaltepe, Jeyhan S. and Koekemoer, Anton M. and Kohandel, Mahsa and Lee, Lilian L. and Li, Yuan and Molina, Juan and Nanni, Ambra and Narayanan, Desika and Pozzi, Francesca and Relano, Monica and Romano, Michael and Sanders, David B. and Silverman, John D. and Sommovigo, Laura and Spilker, Justin and Tsujita, Akiyoshi and Übler, Hannah and G. C., Keerthi Vasan and Veraldi, Enrico and Villanueva, Vincente and Xie, Lizhi and Zamorani, Gianni},
	month = jun,
	year = {2026},
	note = {ADS Bibcode: 2026ApJ..1004...22F},
	pages = {22},
}

@article{garcia_metallicity_2026,
	title = {Metallicity {Gradients} in {Modern} {Cosmological} {Simulations}. {II}. {The} {Role} of {Bursty} versus {Smooth} {Feedback} at {High} {Redshift}},
	volume = {1001},
	issn = {0004-637X},
	url = {https://ui.adsabs.harvard.edu/abs/2026ApJ..1001..188G},
	doi = {10.3847/1538-4357/ae561e},
	urldate = {2026-04-30},
	journal = {The Astrophysical Journal},
	publisher = {IOP},
	author = {Garcia, Alex M. and Torrey, Paul and Bhagwat, Aniket and Shen, Xuejian and Vogelsberger, Mark and McClymont, William and Nagarajan-Swenson, Jaya and Ridolfo, Sophia G. and Zhu, Peixin and Zimmerman, Dhruv T. and Zier, Oliver and Biddle, Sarah and Sarkar, Arnab and Chakraborty, Priyanka and Wright, Ruby J. and Grasha, Kathryn and Costa, Tiago and Keating, Laura and Kannan, Rahul and Smith, Aaron and Garaldi, Enrico and Puchwein, Ewald and Ciardi, Benedetta and Hernquist, Lars and Kewley, Lisa J.},
	month = apr,
	year = {2026},
	note = {ADS Bibcode: 2026ApJ..1001..188G},
	pages = {188},
}

@article{garcia_does_2024,
	title = {Does the fundamental metallicity relation evolve with redshift? {I}: the correlation between offsets from the mass-metallicity relation and star formation rate},
	volume = {531},
	issn = {0035-8711},
	shorttitle = {Does the fundamental metallicity relation evolve with redshift?},
	url = {https://ui.adsabs.harvard.edu/abs/2024MNRAS.531.1398G},
	doi = {10.1093/mnras/stae1252},
	urldate = {2026-04-29},
	journal = {Monthly Notices of the Royal Astronomical Society},
	publisher = {OUP},
	author = {Garcia, Alex M. and Torrey, Paul and Ellison, Sara and Grasha, Kathryn and Hernquist, Lars and Zovaro, Henry R. M. and Chen, Qian-Hui and Hemler, Z. S. and Kewley, Lisa J. and Nelson, Erica J. and Wright, Ruby J.},
	month = jun,
	year = {2024},
	note = {ADS Bibcode: 2024MNRAS.531.1398G},
	pages = {1398--1408},
}

@misc{pollock_characterising_2026,
	title = {Characterising {Ly}\${\textbackslash}alpha\$ damping wings at the onset of reionisation: {Evidence} for highly efficient star formation driven by dense, neutral gas in {UV}-bright galaxies at \$z{\textgreater}9\$},
	shorttitle = {Characterising {Ly}\${\textbackslash}alpha\$ damping wings at the onset of reionisation},
	url = {https://ui.adsabs.harvard.edu/abs/2026arXiv260211783P},
	doi = {10.48550/arXiv.2602.11783},
	urldate = {2026-03-23},
	publisher = {arXiv},
	author = {Pollock, Clara L. and Heintz, Kasper E. and Witstok, Joris and Gottumukkala, Rashmi and Brammer, Gabriel and Bose, Sownak and Cameron, Alex J. and Dayal, Pratika and van Dokkum, Pieter and Fynbo, Johan and Gelli, Viola and Hayes, Matthew J. and Inoue, Akio K. and Lagos, Claudia del P. and Laursen, Peter and Meyer, Romain A. and Naidu, Rohan and Oesch, Pascal and Rowland, Lucie E. and Tanvir, Nial R. and Tacchella, Sandro and Terp, Chamilla and Valentino, Francesco and Walter, Fabian and Weaver, John and Witten, Callum},
	month = feb,
	year = {2026},
	note = {ADS Bibcode: 2026arXiv260211783P},
}

@article{dimitrijevic_flux_2007,
	title = {The flux ratio of the [{OIII}] λλ5007, 4959 lines in {AGN}: comparison with theoretical calculations},
	volume = {374},
	issn = {0035-8711},
	shorttitle = {The flux ratio of the [{OIII}] λλ5007, 4959 lines in {AGN}},
	url = {https://ui.adsabs.harvard.edu/abs/2007MNRAS.374.1181D},
	doi = {10.1111/j.1365-2966.2006.11238.x},
	urldate = {2025-05-28},
	journal = {Monthly Notices of the Royal Astronomical Society},
	publisher = {OUP},
	author = {Dimitrijević, M. S. and Popović, L. C. and Kovačević, J. and Dačić, M. and Ilić, D.},
	month = jan,
	year = {2007},
	note = {ADS Bibcode: 2007MNRAS.374.1181D},
	pages = {1181--1184},
}

@article{eisenstein_jades_2025,
	title = {The {JADES} {Origins} {Field}: {A} {New} {JWST} {Deep} {Field} in the {JADES} {Second} {NIRCam} {Data} {Release}},
	volume = {281},
	issn = {0067-0049},
	shorttitle = {The {JADES} {Origins} {Field}},
	url = {https://ui.adsabs.harvard.edu/abs/2025ApJS..281...50E},
	doi = {10.3847/1538-4365/ae1137},
	urldate = {2026-03-26},
	journal = {The Astrophysical Journal Supplement Series},
	publisher = {IOP},
	author = {Eisenstein, Daniel J. and Johnson, Benjamin D. and Robertson, Brant and Tacchella, Sandro and Hainline, Kevin and Jakobsen, Peter and Maiolino, Roberto and Bonaventura, Nina and Bunker, Andrew J. and Cameron, Alex J. and Cargile, Phillip A. and Curtis-Lake, Emma and Hausen, Ryan and Puskás, Dávid and Rieke, Marcia and Sun, Fengwu and Willmer, Christopher N. A. and Willott, Chris and Alberts, Stacey and Arribas, Santiago and Baker, William M. and Baum, Stefi and Bhatawdekar, Rachana and Carniani, Stefano and Charlot, Stephane and Chen, Zuyi and Chevallard, Jacopo and Curti, Mirko and DeCoursey, Christa and D'Eugenio, Francesco and de Graaff, Anna and Egami, Eiichi and Helton, Jakob M. and Ji, Zhiyuan and Jones, Gareth C. and Kumari, Nimisha and Lützgendorf, Nora and Laseter, Isaac and Looser, Tobias J. and Lyu, Jianwei and Maseda, Michael V. and Nelson, Erica and Parlanti, Eleonora and Rauscher, Bernard J. and Rawle, Tim and Rieke, George and Rix, Hans-Walter and Rujopakarn, Wiphu and Sandles, Lester and Saxena, Aayush and Scholtz, Jan and Sharpe, Katherine and Shivaei, Irene and Simmonds, Charlotte and Smit, Renske and Topping, Michael W. and Übler, Hannah and Venturi, Giacomo and Williams, Christina C. and Witstok, Joris and Woodrum, Charity},
	month = dec,
	year = {2025},
	note = {ADS Bibcode: 2025ApJS..281...50E},
	pages = {50},
}

@article{williams_panoramic_2025,
	title = {The {PANORAMIC} {Survey}: {Pure} {Parallel} {Wide} {Area} {Legacy} {Imaging} with {JWST}/{NIRCam}},
	volume = {979},
	issn = {0004-637X},
	shorttitle = {The {PANORAMIC} {Survey}},
	url = {https://doi.org/10.3847/1538-4357/ad97bc},
	doi = {10.3847/1538-4357/ad97bc},
	language = {en},
	number = {2},
	urldate = {2026-03-26},
	journal = {The Astrophysical Journal},
	publisher = {The American Astronomical Society},
	author = {Williams, Christina C. and Oesch, Pascal A. and Weibel, Andrea and Brammer, Gabriel and Cloonan, Aidan P. and Whitaker, Katherine E. and Barrufet, Laia and Bezanson, Rachel and Bowler, Rebecca A. A. and Dayal, Pratika and Franx, Marijn and Greene, Jenny E. and Hutter, Anne and Ji, Zhiyuan and Labbé, Ivo and Manning, Sinclaire M. and Maseda, Michael V. and Xiao, Mengyuan},
	month = jan,
	year = {2025},
	pages = {140},
}

@article{bagley_next_2024,
	title = {The {Next} {Generation} {Deep} {Extragalactic} {Exploratory} {Public} ({NGDEEP}) {Survey}},
	volume = {965},
	issn = {0004-637X},
	url = {https://ui.adsabs.harvard.edu/abs/2024ApJ...965L...6B},
	doi = {10.3847/2041-8213/ad2f31},
	urldate = {2026-03-26},
	journal = {The Astrophysical Journal},
	publisher = {IOP},
	author = {Bagley, Micaela B. and Pirzkal, Nor and Finkelstein, Steven L. and Papovich, Casey and Berg, Danielle A. and Lotz, Jennifer M. and Leung, Gene C. K. and Ferguson, Henry C. and Koekemoer, Anton M. and Dickinson, Mark and Kartaltepe, Jeyhan S. and Kocevski, Dale D. and Somerville, Rachel S. and Yung, L. Y. Aaron and Backhaus, Bren E. and Casey, Caitlin M. and Castellano, Marco and Chávez Ortiz, Óscar A. and Chworowsky, Katherine and Cox, Isabella G. and Davé, Romeel and Davis, Kelcey and Estrada-Carpenter, Vicente and Fontana, Adriano and Fujimoto, Seiji and Gardner, Jonathan P. and Giavalisco, Mauro and Grazian, Andrea and Grogin, Norman A. and Hathi, Nimish P. and Hutchison, Taylor A. and Jaskot, Anne E. and Jung, Intae and Kewley, Lisa J. and Kirkpatrick, Allison and Larson, Rebecca L. and Matharu, Jasleen and Natarajan, Priyamvada and Pentericci, Laura and Pérez-González, Pablo G. and Ravindranath, Swara and Rothberg, Barry and Ryan, Russell and Shen, Lu and Simons, Raymond C. and Snyder, Gregory F. and Trump, Jonathan R. and Wilkins, Stephen M.},
	month = apr,
	year = {2024},
	note = {ADS Bibcode: 2024ApJ...965L...6B},
	pages = {L6},
}

@article{williams_jems_2023,
	title = {{JEMS}: {A} {Deep} {Medium}-band {Imaging} {Survey} in the {Hubble} {Ultra} {Deep} {Field} with {JWST} {NIRCam} and {NIRISS}},
	volume = {268},
	issn = {0067-0049},
	shorttitle = {{JEMS}},
	url = {https://doi.org/10.3847/1538-4365/acf130},
	doi = {10.3847/1538-4365/acf130},
	language = {en},
	number = {2},
	urldate = {2026-03-26},
	journal = {The Astrophysical Journal Supplement Series},
	publisher = {The American Astronomical Society},
	author = {Williams, Christina C. and Tacchella, Sandro and Maseda, Michael V. and Robertson, Brant E. and Johnson, Benjamin D. and Willott, Chris J. and Eisenstein, Daniel J. and Willmer, Christopher N. A. and Ji, Zhiyuan and Hainline, Kevin N. and Helton, Jakob M. and Alberts, Stacey and Baum, Stefi and Bhatawdekar, Rachana and Boyett, Kristan and Bunker, Andrew J. and Carniani, Stefano and Charlot, Stephane and Chevallard, Jacopo and Curtis-Lake, Emma and de Graaff, Anna and Egami, Eiichi and Franx, Marijn and Kumari, Nimisha and Maiolino, Roberto and Nelson, Erica J. and Rieke, Marcia J. and Sandles, Lester and Shivaei, Irene and Simmonds, Charlotte and Smit, Renske and Suess, Katherine A. and Sun, Fengwu and Übler, Hannah and Witstok, Joris},
	month = oct,
	year = {2023},
	pages = {64},
}

@article{oesch_jwst_2023,
	title = {The {JWST} {FRESCO} survey: legacy {NIRCam}/grism spectroscopy and imaging in the two {GOODS} fields},
	volume = {525},
	issn = {0035-8711},
	shorttitle = {The {JWST} {FRESCO} survey},
	url = {https://ui.adsabs.harvard.edu/abs/2023MNRAS.525.2864O},
	doi = {10.1093/mnras/stad2411},
	urldate = {2026-03-26},
	journal = {Monthly Notices of the Royal Astronomical Society},
	publisher = {OUP},
	author = {Oesch, P. A. and Brammer, G. and Naidu, R. P. and Bouwens, R. J. and Chisholm, J. and Illingworth, G. D. and Matthee, J. and Nelson, E. and Qin, Y. and Reddy, N. and Shapley, A. and Shivaei, I. and van Dokkum, P. and Weibel, A. and Whitaker, K. and Wuyts, S. and Covelo-Paz, A. and Endsley, R. and Fudamoto, Y. and Giovinazzo, E. and Herard-Demanche, T. and Kerutt, J. and Kramarenko, I. and Labbe, I. and Leonova, E. and Lin, J. and Magee, D. and Marchesini, D. and Maseda, M. and Mason, C. and Matharu, J. and Meyer, R. A. and Neufeld, C. and Prieto Lyon, G. and Schaerer, D. and Sharma, R. and Shuntov, M. and Smit, R. and Stefanon, M. and Wyithe, J. S. B. and Xiao, M.},
	month = oct,
	year = {2023},
	note = {ADS Bibcode: 2023MNRAS.525.2864O},
	pages = {2864--2874},
}

@article{eisenstein_overview_2026,
	title = {Overview of the {JWST} {Advanced} {Deep} {Extragalactic} {Survey} ({JADES})},
	volume = {283},
	issn = {0067-0049},
	url = {https://ui.adsabs.harvard.edu/abs/2026ApJS..283....6E},
	doi = {10.3847/1538-4365/ae3163},
	urldate = {2026-03-26},
	journal = {The Astrophysical Journal Supplement Series},
	publisher = {IOP},
	author = {Eisenstein, Daniel J. and Willott, Chris and Alberts, Stacey and Arribas, Santiago and Bonaventura, Nina and Bunker, Andrew J. and Cameron, Alex J. and Carniani, Stefano and Charlot, Stephane and Curtis-Lake, Emma and D'Eugenio, Francesco and Ferruit, Pierre and Giardino, Giovanna and Hainline, Kevin and Hausen, Ryan and Jakobsen, Peter and Johnson, Benjamin D. and Maiolino, Roberto and Rauscher, Bernard J. and Rieke, Marcia and Rieke, George and Rix, Hans-Walter and Robertson, Brant and Stark, Daniel P. and Tacchella, Sandro and Williams, Christina C. and Willmer, Christopher N. A. and Baker, William M. and Baum, Stefi and Bhatawdekar, Rachana and Boyett, Kristan and Chen, Zuyi and Chevallard, Jacopo and Circosta, Chiara and Curti, Mirko and Danhaive, A. Lola and DeCoursey, Christa and Endsley, Ryan and de Graaff, Anna and Dressler, Alan and Egami, Eiichi and Helton, Jakob M. and Hviding, Raphael E. and Ji, Zhiyuan and Jones, Gareth C. and Kumari, Nimisha and Lützgendorf, Nora and Laseter, Isaac and Looser, Tobias J. and Lyu, Jianwei and Maseda, Michael V. and Nelson, Erica and Parlanti, Eleonora and Perna, Michele and Puskás, Dávid and Rawle, Tim and Rodríguez Del Pino, Bruno and Rujopakarn, Wiphu and Sandles, Lester and Saxena, Aayush and Scholtz, Jan and Sharpe, Katherine and Shivaei, Irene and Silcock, Maddie S. and Simmonds, Charlotte and Skarbinski, Maya and Smit, Renske and Stone, Meredith and Suess, Katherine A. and Sun, Fengwu and Tang, Mengtao and Topping, Michael W. and Übler, Hannah and Villanueva, Natalia C. and Wallace, Imaan E. B. and Whitler, Lily and Witstok, Joris and Woodrum, Charity},
	month = mar,
	year = {2026},
	note = {ADS Bibcode: 2026ApJS..283....6E},
	pages = {6},
}

@article{suess_medium_2024,
	title = {Medium {Bands}, {Mega} {Science}: {A} {JWST}/{NIRCam} {Medium}-band {Imaging} {Survey} of {A2744}},
	volume = {976},
	issn = {0004-637X},
	shorttitle = {Medium {Bands}, {Mega} {Science}},
	url = {https://ui.adsabs.harvard.edu/abs/2024ApJ...976..101S},
	doi = {10.3847/1538-4357/ad75fe},
	urldate = {2026-03-25},
	journal = {The Astrophysical Journal},
	publisher = {IOP},
	author = {Suess, Katherine A. and Weaver, John R. and Price, Sedona H. and Pan, Richard and Wang, Bingjie and Bezanson, Rachel and Brammer, Gabriel and Cutler, Sam E. and Labbé, Ivo and Leja, Joel and Williams, Christina C. and Whitaker, Katherine E. and Atek, Hakim and Dayal, Pratika and de Graaff, Anna and Feldmann, Robert and Franx, Marijn and Fudamoto, Yoshinobu and Fujimoto, Seiji and Furtak, Lukas J. and Goulding, Andy D. and Greene, Jenny E. and Khullar, Gourav and Kokorev, Vasily and Kriek, Mariska and Lorenz, Brian and Marchesini, Danilo and Maseda, Michael V. and Matthee, Jorryt and Miller, Tim B. and Mitsuhashi, Ikki and Mowla, Lamiya A. and Muzzin, Adam and Naidu, Rohan P. and Nanayakkara, Themiya and Nelson, Erica J. and Oesch, Pascal A. and Setton, David J. and Shipley, Heath and Smit, Renske and Spilker, Justin S. and van Dokkum, Pieter and Zitrin, Adi},
	month = nov,
	year = {2024},
	note = {ADS Bibcode: 2024ApJ...976..101S},
	pages = {101},
}

@article{morishita_beacon_2025,
	title = {{BEACON}: {JWST} {NIRCam} {Pure}-parallel {Imaging} {Survey}. {I}. {Survey} {Design} and {Initial} {Results}},
	volume = {983},
	issn = {0004-637X},
	shorttitle = {{BEACON}},
	url = {https://ui.adsabs.harvard.edu/abs/2025ApJ...983..152M},
	doi = {10.3847/1538-4357/adbbdc},
	urldate = {2026-03-25},
	journal = {The Astrophysical Journal},
	publisher = {IOP},
	author = {Morishita, Takahiro and Mason, Charlotte A. and Kreilgaard, Kimi C. and Trenti, Michele and Treu, Tommaso and Vulcani, Benedetta and Zhang, Yechi and {Abdurro'uf} and Alavi, Anahita and Atek, Hakim and Bahé, Yannick and Bradač, Maruša and Bradley, Larry D. and Bunker, Andrew J. and Coe, Dan and Colbert, James and Gelli, Viola and Hayes, Matthew J. and Jones, Tucker and Kodama, Tadayuki and Leethochawalit, Nicha and Liu, Zhaoran and Malkan, Matthew A. and Mehta, Vihang and Metha, Benjamin and Newman, Andrew B. and Rafelski, Marc and Roberts-Borsani, Guido and Rutkowski, Michael J. and Scarlata, Claudia and Stiavelli, Massimo and Sutanto, Ryo A. and Takahashi, Kosuke and Teplitz, Harry I. and Wang, Xin},
	month = apr,
	year = {2025},
	note = {ADS Bibcode: 2025ApJ...983..152M},
	pages = {152},
}

@misc{naidu_all_2024,
	title = {All the {Little} {Things} in {Abell} 2744: \${\textgreater}\$1000 {Gravitationally} {Lensed} {Dwarf} {Galaxies} at \$z=0-9\$ from {JWST} {NIRCam} {Grism} {Spectroscopy}},
	shorttitle = {All the {Little} {Things} in {Abell} 2744},
	url = {https://ui.adsabs.harvard.edu/abs/2024arXiv241001874N},
	doi = {10.48550/arXiv.2410.01874},
	urldate = {2026-03-25},
	publisher = {arXiv},
	author = {Naidu, Rohan P. and Matthee, Jorryt and Kramarenko, Ivan and Weibel, Andrea and Brammer, Gabriel and Oesch, Pascal A. and Lechner, Peter and Furtak, Lukas J. and Di Cesare, Claudia and Torralba, Alberto and Kotiwale, Gauri and Bezanson, Rachel and Bouwens, Rychard J. and Chandra, Vedant and Claeyssens, Adélaïde and Danhaive, A. Lola and Frebel, Anna and de Graaff, Anna and Greene, Jenny E. and Heintz, Kasper E. and Ji, Alexander P. and Kashino, Daichi and Katz, Harley and Labbe, Ivo and Leja, Joel and Li, Yijia and Maseda, Michael V. and Richard, Johan and Shivaei, Irene and Simcoe, Robert A. and Sobral, David and Suess, Katherine A. and Tacchella, Sandro and Williams, Christina C.},
	month = oct,
	year = {2024},
	note = {ADS Bibcode: 2024arXiv241001874N},
}

@article{paris_glass-jwst_2023,
	title = {The {GLASS}-{JWST} {Early} {Release} {Science} {Program}. {II}. {Stage} {I} {Release} of {NIRCam} {Imaging} and {Catalogs} in the {Abell} 2744 {Region}},
	volume = {952},
	issn = {0004-637X},
	url = {https://ui.adsabs.harvard.edu/abs/2023ApJ...952...20P},
	doi = {10.3847/1538-4357/acda8a},
	urldate = {2026-03-25},
	journal = {The Astrophysical Journal},
	publisher = {IOP},
	author = {Paris, Diego and Merlin, Emiliano and Fontana, Adriano and Bonchi, Andrea and Brammer, Gabriel and Correnti, Matteo and Treu, Tommaso and Boyett, Kristan and Calabrò, Antonello and Castellano, Marco and Chen, Wenlei and Yang, Lilan and Glazebrook, Karl and Kelly, Patrick and Koekemoer, Anton M. and Leethochawalit, Nicha and Mascia, Sara and Mason, Charlotte and Morishita, Takahiro and Nonino, Mario and Pentericci, Laura and Polenta, Gianluca and Roberts-Borsani, Guido and Santini, Paola and Trenti, Michele and Vanzella, Eros and Vulcani, Benedetta and Windhorst, Rogier A. and Nanayakkara, Themiya and Wang, Xin},
	month = jul,
	year = {2023},
	note = {ADS Bibcode: 2023ApJ...952...20P},
	pages = {20},
}

@article{merlin_early_2022,
	title = {Early {Results} from {GLASS}-{JWST}. {II}. {NIRCam} {Extragalactic} {Imaging} and {Photometric} {Catalog}},
	volume = {938},
	issn = {0004-637X},
	url = {https://ui.adsabs.harvard.edu/abs/2022ApJ...938L..14M},
	doi = {10.3847/2041-8213/ac8f93},
	urldate = {2026-03-25},
	journal = {The Astrophysical Journal},
	publisher = {IOP},
	author = {Merlin, Emiliano and Bonchi, Andrea and Paris, Diego and Belfiori, Davide and Fontana, Adriano and Castellano, Marco and Nonino, Mario and Polenta, Gianluca and Santini, Paola and Yang, Lilan and Glazebrook, Karl and Treu, Tommaso and Roberts-Borsani, Guido and Trenti, Michele and Birrer, Simon and Brammer, Gabriel and Grillo, Claudio and Calabrò, Antonello and Marchesini, Danilo and Mason, Charlotte and Mercurio, Amata and Morishita, Takahiro and Strait, Victoria and Boyett, Kristan and Leethochawalit, Nicha and Nanayakkara, Themiya and Vulcani, Benedetta and Bradac, Marusa and Wang, Xin},
	month = oct,
	year = {2022},
	note = {ADS Bibcode: 2022ApJ...938L..14M},
	pages = {L14},
}

@article{pontoppidan_jwst_2022,
	title = {The {JWST} {Early} {Release} {Observations}},
	volume = {936},
	issn = {0004-637X},
	url = {https://ui.adsabs.harvard.edu/abs/2022ApJ...936L..14P},
	doi = {10.3847/2041-8213/ac8a4e},
	urldate = {2026-03-25},
	journal = {The Astrophysical Journal},
	publisher = {IOP},
	author = {Pontoppidan, Klaus M. and Barrientes, Jaclyn and Blome, Claire and Braun, Hannah and Brown, Matthew and Carruthers, Margaret and Coe, Dan and DePasquale, Joseph and Espinoza, Néstor and Marin, Macarena Garcia and Gordon, Karl D. and Henry, Alaina and Hustak, Leah and James, Andi and Jenkins, Ann and Koekemoer, Anton M. and LaMassa, Stephanie and Law, David and Lockwood, Alexandra and Moro-Martin, Amaya and Mullally, Susan E. and Pagan, Alyssa and Player, Dani and Proffitt, Charles and Pulliam, Christine and Ramsay, Leah and Ravindranath, Swara and Reid, Neill and Robberto, Massimo and Sabbi, Elena and Ubeda, Leonardo and Balogh, Michael and Flanagan, Kathryn and Gardner, Jonathan and Hasan, Hashima and Meinke, Bonnie and Nota, Antonella},
	month = sep,
	year = {2022},
	note = {ADS Bibcode: 2022ApJ...936L..14P},
	pages = {L14},
}

@article{scholtz_ga-nifs_2025,
	title = {{GA}-{NIFS}: {ISM} properties and metal enrichment in a merger-driven starburst during the epoch of reionization probed with {JWST} and {ALMA}},
	volume = {539},
	issn = {0035-8711},
	shorttitle = {{GA}-{NIFS}},
	url = {https://ui.adsabs.harvard.edu/abs/2025MNRAS.539.2463S},
	doi = {10.1093/mnras/staf518},
	urldate = {2026-03-25},
	journal = {Monthly Notices of the Royal Astronomical Society},
	publisher = {OUP},
	author = {Scholtz, J. and Curti, M. and D'Eugenio, F. and Übler, H. and Maiolino, R. and Marconcini, C. and Smit, R. and Perna, M. and Witstok, J. and Arribas, S. and Böker, T. and Bunker, A. J. and Carniani, S. and Charlot, S. and Cresci, G. and Lamperti, I. and Parlanti, E. and Pérez-González, P. G. and Rodríguez Del Pino, B. and Venturi, G.},
	month = may,
	year = {2025},
	note = {ADS Bibcode: 2025MNRAS.539.2463S},
	pages = {2463--2484},
}

@article{feltre_nuclear_2016,
	title = {Nuclear activity versus star formation: emission-line diagnostics at ultraviolet and optical wavelengths},
	volume = {456},
	issn = {0035-8711},
	shorttitle = {Nuclear activity versus star formation},
	url = {https://ui.adsabs.harvard.edu/abs/2016MNRAS.456.3354F},
	doi = {10.1093/mnras/stv2794},
	urldate = {2026-03-25},
	journal = {Monthly Notices of the Royal Astronomical Society},
	publisher = {OUP},
	author = {Feltre, A. and Charlot, S. and Gutkin, J.},
	month = mar,
	year = {2016},
	note = {ADS Bibcode: 2016MNRAS.456.3354F},
	pages = {3354--3374},
}

@article{mazzolari_narrow-line_2025,
	title = {Narrow-line {AGN} selection in {CEERS}: {Spectroscopic} selection, physical properties, and {X}-ray and radio analysis},
	volume = {700},
	issn = {0004-6361},
	shorttitle = {Narrow-line {AGN} selection in {CEERS}},
	url = {https://ui.adsabs.harvard.edu/abs/2025A&A...700A..12M},
	doi = {10.1051/0004-6361/202451860},
	urldate = {2026-03-25},
	journal = {Astronomy and Astrophysics},
	publisher = {EDP},
	author = {Mazzolari, Giovanni and Scholtz, Jan and Maiolino, Roberto and Gilli, Roberto and Traina, Alberto and López, Ivan E. and Übler, Hannah and Trefoloni, Bartolomeo and D'Eugenio, Francesco and Ji, Xihan and Mignoli, Marco and Vito, Fabio and Vignali, Cristian and Brusa, Marcella},
	month = aug,
	year = {2025},
	note = {ADS Bibcode: 2025A\&A...700A..12M},
	pages = {A12},
}

@article{mascia_insights_2023,
	title = {Insights into the reionization epoch from cosmic-noon-{C} {IV} emitters in the {VANDELS} survey},
	volume = {674},
	issn = {0004-6361},
	url = {https://ui.adsabs.harvard.edu/abs/2023A&A...674A.221M},
	doi = {10.1051/0004-6361/202245152},
	urldate = {2026-03-25},
	journal = {Astronomy and Astrophysics},
	publisher = {EDP},
	author = {Mascia, S. and Pentericci, L. and Saxena, A. and Belfiori, D. and Calabrò, A. and Castellano, M. and Saldana-Lopez, A. and Talia, M. and Amorín, R. and Cullen, F. and Garilli, B. and Guaita, L. and LLerena, M. and McLure, R. J. and Moresco, M. and Santini, P. and Schaerer, D.},
	month = jun,
	year = {2023},
	note = {ADS Bibcode: 2023A\&A...674A.221M},
	pages = {A221},
}

@article{nagao_gas_2006,
	title = {Gas metallicity in the narrow-line regions of high-redshift active galactic nuclei},
	volume = {447},
	issn = {0004-6361},
	url = {https://ui.adsabs.harvard.edu/abs/2006A&A...447..863N},
	doi = {10.1051/0004-6361:20054127},
	urldate = {2026-03-25},
	journal = {Astronomy and Astrophysics},
	publisher = {EDP},
	author = {Nagao, T. and Maiolino, R. and Marconi, A.},
	month = mar,
	year = {2006},
	note = {ADS Bibcode: 2006A\&A...447..863N},
	pages = {863--876},
}

@article{guo_high-redshift_2020,
	title = {High-redshift {Extreme} {Variability} {Quasars} from {Sloan} {Digital} {Sky} {Survey} {Multiepoch} {Spectroscopy}},
	volume = {905},
	issn = {0004-637X},
	url = {https://ui.adsabs.harvard.edu/abs/2020ApJ...905...52G},
	doi = {10.3847/1538-4357/abc2ce},
	urldate = {2026-03-25},
	journal = {The Astrophysical Journal},
	publisher = {IOP},
	author = {Guo, Hengxiao and Peng, Jiacheng and Zhang, Kaiwen and Burke, Colin J. and Liu, Xin and Sun, Mouyuan and Wang, Shu and Kong, Minzhi and Sheng, Zhenfeng and Wang, Tinggui and He, Zhicheng and Gu, Minfeng},
	month = dec,
	year = {2020},
	note = {ADS Bibcode: 2020ApJ...905...52G},
	pages = {52},
}

@misc{witten_not_2025,
	title = {Not all protoclusters host evolved galaxies: {Evidence} for reduced environmental effects in a lower halo mass protocluster at \$z = 7.66\$},
	shorttitle = {Not all protoclusters host evolved galaxies},
	url = {https://ui.adsabs.harvard.edu/abs/2025arXiv251105647W},
	doi = {10.48550/arXiv.2511.05647},
	urldate = {2026-03-23},
	publisher = {arXiv},
	author = {Witten, Callum and Oesch, Pascal A. and Bennett, Jake S. and Meyer, Romain A. and Giovinazzo, Emma and Covelo-Paz, Alba and Baker, William M. and Ivey, Lucy R.},
	month = nov,
	year = {2025},
	note = {ADS Bibcode: 2025arXiv251105647W},
}

@article{bezanson_jwst_2024,
	title = {The {JWST} {UNCOVER} {Treasury} {Survey}: {Ultradeep} {NIRSpec} and {NIRCam} {Observations} before the {Epoch} of {Reionization}},
	volume = {974},
	issn = {0004-637X},
	shorttitle = {The {JWST} {UNCOVER} {Treasury} {Survey}},
	url = {https://ui.adsabs.harvard.edu/abs/2024ApJ...974...92B},
	doi = {10.3847/1538-4357/ad66cf},
	urldate = {2026-03-23},
	journal = {The Astrophysical Journal},
	publisher = {IOP},
	author = {Bezanson, Rachel and Labbe, Ivo and Whitaker, Katherine E. and Leja, Joel and Price, Sedona H. and Franx, Marijn and Brammer, Gabriel and Marchesini, Danilo and Zitrin, Adi and Wang, Bingjie and Weaver, John R. and Furtak, Lukas J. and Atek, Hakim and Coe, Dan and Cutler, Sam E. and Dayal, Pratika and van Dokkum, Pieter and Feldmann, Robert and Förster Schreiber, Natascha M. and Fujimoto, Seiji and Geha, Marla and Glazebrook, Karl and de Graaff, Anna and Greene, Jenny E. and Juneau, Stéphanie and Kassin, Susan and Kriek, Mariska and Khullar, Gourav and Maseda, Michael and Mowla, Lamiya A. and Muzzin, Adam and Nanayakkara, Themiya and Nelson, Erica J. and Oesch, Pascal A. and Pacifici, Camilla and Pan, Richard and Papovich, Casey and Setton, David J. and Shapley, Alice E. and Smit, Renske and Stefanon, Mauro and Taylor, Edward N. and Williams, Christina C.},
	month = oct,
	year = {2024},
	note = {ADS Bibcode: 2024ApJ...974...92B},
	pages = {92},
}

@article{ritter_metal_2015,
	title = {Metal transport and chemical heterogeneity in early star forming systems},
	volume = {451},
	issn = {0035-8711},
	url = {https://ui.adsabs.harvard.edu/abs/2015MNRAS.451.1190R},
	doi = {10.1093/mnras/stv982},
	urldate = {2026-03-23},
	journal = {Monthly Notices of the Royal Astronomical Society},
	publisher = {OUP},
	author = {Ritter, Jeremy S. and Sluder, Alan and Safranek-Shrader, Chalence and Milosavljević, Miloš and Bromm, Volker},
	month = aug,
	year = {2015},
	note = {ADS Bibcode: 2015MNRAS.451.1190R},
	pages = {1190--1198},
}

@article{deugenio_jades_2026,
	title = {{JADES} and {BlackTHUNDER}: rest-frame {Balmer}-line absorption and the local environment in a {Little} {Red} {Dot} at z = 5},
	volume = {545},
	issn = {0035-8711},
	shorttitle = {{JADES} and {BlackTHUNDER}},
	url = {https://ui.adsabs.harvard.edu/abs/2026MNRAS.545f2117D},
	doi = {10.1093/mnras/staf2117},
	urldate = {2026-03-23},
	journal = {Monthly Notices of the Royal Astronomical Society},
	publisher = {OUP},
	author = {D'Eugenio, Francesco and Juodžbalis, Ignas and Ji, Xihan and Scholtz, Jan and Maiolino, Roberto and Carniani, Stefano and Perna, Michele and Mazzolari, Giovanni and Übler, Hannah and Arribas, Santiago and Bhatawdekar, Rachana and Bunker, Andrew J. and Cresci, Giovanni and Curtis-Lake, Emma and Hainline, Kevin and Inayoshi, Kohei and Isobe, Yuki and Ji, Zhiyuan and Johnson, Benjamin D. and Jones, Gareth C. and Looser, Tobias J. and Nelson, Erica J. and Parlanti, Eleonora and Puskás, Dávid and Rinaldi, Pierluigi and Robertson, Brant and Rodríguez Del Pino, Bruno and Shivaei, Irene and Sun, Fengwu and Tacchella, Sandro and Venturi, Giacomo and Volonteri, Marta and Williams, Christina C. and Willmer, Christopher N. A. and Willott, Chris and Witstok, Joris},
	month = jan,
	year = {2026},
	note = {ADS Bibcode: 2026MNRAS.545f2117D},
	pages = {staf2117},
}

@article{ivey_exploring_2026,
	title = {Exploring spatially resolved metallicities, dynamics, and outflows in low-mass galaxies at z ∼ 7.6},
	volume = {546},
	issn = {0035-8711},
	url = {https://ui.adsabs.harvard.edu/abs/2026MNRAS.546ag094I},
	doi = {10.1093/mnras/stag094},
	urldate = {2026-03-23},
	journal = {Monthly Notices of the Royal Astronomical Society},
	publisher = {OUP},
	author = {Ivey, L. R. and Scholtz, J. and Danhaive, A. L. and Koudmani, S. and Jones, G. C. and Maiolino, R. and Curti, M. and D'Eugenio, F. and Tacchella, S. and Baker, W. M. and Arribas, S. and Charlot, S. and Eisenstein, D. and Ji, Z. and Koller, M. and Laporte, N. and Perna, M. and Puskás, D. and Robertson, B. and Sijacki, D. and Trussler, J. A. A. and Witten, C.},
	month = mar,
	year = {2026},
	note = {ADS Bibcode: 2026MNRAS.546ag094I},
	pages = {stag094},
}

@article{gillman_resolved_2022,
	title = {The resolved chemical abundance properties within the interstellar medium of star-forming galaxies at z≍ 1.5},
	volume = {512},
	issn = {0035-8711},
	url = {https://ui.adsabs.harvard.edu/abs/2022MNRAS.512.3480G},
	doi = {10.1093/mnras/stac580},
	urldate = {2026-03-20},
	journal = {Monthly Notices of the Royal Astronomical Society},
	publisher = {OUP},
	author = {Gillman, S. and Puglisi, A. and Dudzevičiūtė, U. and Swinbank, A. M. and Tiley, A. L. and Harrison, C. M. and Molina, J. and Sharples, R. M. and Bower, R. G. and Cirasuolo, M. and Ibar, Edo and Obreschkow, D.},
	month = may,
	year = {2022},
	note = {ADS Bibcode: 2022MNRAS.512.3480G},
	pages = {3480--3499},
}

@article{garcia_does_2025,
	title = {Does the fundamental metallicity relation evolve with redshift? - {II}. {The} evolution in normalization of the mass-metallicity relation},
	volume = {536},
	issn = {0035-8711},
	shorttitle = {Does the fundamental metallicity relation evolve with redshift?},
	url = {https://ui.adsabs.harvard.edu/abs/2025MNRAS.536..119G},
	doi = {10.1093/mnras/stae2587},
	urldate = {2026-03-19},
	journal = {Monthly Notices of the Royal Astronomical Society},
	publisher = {OUP},
	author = {Garcia, Alex M. and Torrey, Paul and Ellison, Sara L. and Grasha, Kathryn and Chen, Qian-Hui and Hemler, Z. S. and Zimmerman, Dhruv T. and Wright, Ruby J. and Zovaro, Henry R. M. and Nelson, Erica J. and Sanders, Ryan L. and Kewley, Lisa J. and Hernquist, Lars},
	month = jan,
	year = {2025},
	note = {ADS Bibcode: 2025MNRAS.536..119G},
	pages = {119--144},
}

@article{looser_recently_2024,
	title = {A recently quenched galaxy 700 million years after the {Big} {Bang}},
	volume = {629},
	issn = {0028-0836},
	url = {https://ui.adsabs.harvard.edu/abs/2024Natur.629...53L},
	doi = {10.1038/s41586-024-07227-0},
	urldate = {2026-03-19},
	journal = {Nature},
	author = {Looser, Tobias J. and D'Eugenio, Francesco and Maiolino, Roberto and Witstok, Joris and Sandles, Lester and Curtis-Lake, Emma and Chevallard, Jacopo and Tacchella, Sandro and Johnson, Benjamin D. and Baker, William M. and Suess, Katherine A. and Carniani, Stefano and Ferruit, Pierre and Arribas, Santiago and Bonaventura, Nina and Bunker, Andrew J. and Cameron, Alex J. and Charlot, Stephane and Curti, Mirko and de Graaff, Anna and Maseda, Michael V. and Rawle, Tim and Rix, Hans-Walter and Del Pino, Bruno Rodríguez and Smit, Renske and Übler, Hannah and Willott, Chris and Alberts, Stacey and Egami, Eiichi and Eisenstein, Daniel J. and Endsley, Ryan and Hausen, Ryan and Rieke, Marcia and Robertson, Brant and Shivaei, Irene and Williams, Christina C. and Boyett, Kristan and Chen, Zuyi and Ji, Zhiyuan and Jones, Gareth C. and Kumari, Nimisha and Nelson, Erica and Perna, Michele and Saxena, Aayush and Scholtz, Jan},
	month = may,
	year = {2024},
	note = {ADS Bibcode: 2024Natur.629...53L},
	pages = {53--57},
}

@misc{rodriguez_del_pino_ga-nifs_2026,
	title = {{GA}-{NIFS}: high prevalence of dusty and metal-enriched outflows in massive and luminous star-forming galaxies at \$z{\textbackslash}sim3-9\$},
	shorttitle = {{GA}-{NIFS}},
	url = {https://ui.adsabs.harvard.edu/abs/2026arXiv260106255R},
	doi = {10.48550/arXiv.2601.06255},
	urldate = {2026-03-11},
	publisher = {arXiv},
	author = {Rodríguez Del Pino, B. and Arribas, S. and Perna, M. and Lamperti, I. and Bunker, A. and Carniani, S. and Charlot, S. and D'Eugenio, F. and Maiolino, R. and Übler, H. and Bertola, E. and Böker, T. and Cresci, G. and Jones, G. C. and Marconcini, C. and Parlanti, E. and Scholtz, J. and Venturi, G. and Zamora, S.},
	month = jan,
	year = {2026},
	note = {ADS Bibcode: 2026arXiv260106255R},
}

@article{cai_metal-free_2025,
	title = {A {Metal}-free {Galaxy} at z = 3.19? {Evidence} of {Late} {Population} {III} {Star} {Formation} at {Cosmic} {Noon}},
	volume = {993},
	issn = {0004-637X},
	shorttitle = {A {Metal}-free {Galaxy} at z = 3.19?},
	url = {https://ui.adsabs.harvard.edu/abs/2025ApJ...993L..52C},
	doi = {10.3847/2041-8213/ae1608},
	urldate = {2026-03-11},
	journal = {The Astrophysical Journal},
	publisher = {IOP},
	author = {Cai, Sijia and Li, Mingyu and Cai, Zheng and Wu, Yunjing and Yu, Fujiang and Dickinson, Mark and Sun, Fengwu and Fan, Xiaohui and Wang, Ben and Cullen, Fergus and Bian, Fuyan and Lin, Xiaojing and Zou, Jiaqi},
	month = nov,
	year = {2025},
	note = {ADS Bibcode: 2025ApJ...993L..52C},
	pages = {L52},
}

@article{bhagwat_spice_2024,
	title = {{SPICE}: the connection between cosmic reionization and stellar feedback in the first galaxies},
	volume = {531},
	issn = {0035-8711},
	shorttitle = {{SPICE}},
	url = {https://doi.org/10.1093/mnras/stae1125},
	doi = {10.1093/mnras/stae1125},
	number = {3},
	urldate = {2026-03-10},
	journal = {Monthly Notices of the Royal Astronomical Society},
	author = {Bhagwat, Aniket and Costa, Tiago and Ciardi, Benedetta and Pakmor, Rüdiger and Garaldi, Enrico},
	month = jul,
	year = {2024},
	pages = {3406--3430},
}

@article{kannan_introducing_2025,
	title = {Introducing the {THESAN}-{ZOOM} project: radiation-hydrodynamic simulations of high-redshift galaxies with a multi-phase interstellar medium},
	volume = {8},
	issn = {2565-6120},
	shorttitle = {Introducing the {THESAN}-{ZOOM} project},
	url = {https://ui.adsabs.harvard.edu/abs/2025OJAp....8E.153K},
	doi = {10.33232/001c.145804},
	urldate = {2026-03-10},
	journal = {The Open Journal of Astrophysics},
	author = {Kannan, Rahul and Puchwein, Ewald and Smith, Aaron and Borrow, Josh and Garaldi, Enrico and Keating, Laura and Vogelsberger, Mark and Zier, Oliver and McClymont, William and Shen, Xuejian and Popovic, Filip and Tacchella, Sandro and Hernquist, Lars and Springel, Volker},
	month = oct,
	year = {2025},
	note = {ADS Bibcode: 2025OJAp....8E.153K},
	pages = {153},
}

@article{smith_thesan_2022,
	title = {The {THESAN} project: {Lyman}-α emission and transmission during the {Epoch} of {Reionization}},
	volume = {512},
	issn = {0035-8711},
	shorttitle = {The {THESAN} project},
	url = {https://ui.adsabs.harvard.edu/abs/2022MNRAS.512.3243S},
	doi = {10.1093/mnras/stac713},
	urldate = {2026-03-10},
	journal = {Monthly Notices of the Royal Astronomical Society},
	publisher = {OUP},
	author = {Smith, A. and Kannan, R. and Garaldi, E. and Vogelsberger, M. and Pakmor, R. and Springel, V. and Hernquist, L.},
	month = may,
	year = {2022},
	note = {ADS Bibcode: 2022MNRAS.512.3243S},
	pages = {3243--3265},
}

@article{garaldi_thesan_2022,
	title = {The {THESAN} project: properties of the intergalactic medium and its connection to reionization-era galaxies},
	volume = {512},
	issn = {0035-8711},
	shorttitle = {The {THESAN} project},
	url = {https://ui.adsabs.harvard.edu/abs/2022MNRAS.512.4909G},
	doi = {10.1093/mnras/stac257},
	urldate = {2026-03-10},
	journal = {Monthly Notices of the Royal Astronomical Society},
	publisher = {OUP},
	author = {Garaldi, E. and Kannan, R. and Smith, A. and Springel, V. and Pakmor, R. and Vogelsberger, M. and Hernquist, L.},
	month = jun,
	year = {2022},
	note = {ADS Bibcode: 2022MNRAS.512.4909G},
	pages = {4909--4933},
}

@article{kannan_introducing_2022,
	title = {Introducing the {THESAN} project: radiation-magnetohydrodynamic simulations of the epoch of reionization},
	volume = {511},
	issn = {0035-8711},
	shorttitle = {Introducing the {THESAN} project},
	url = {https://ui.adsabs.harvard.edu/abs/2022MNRAS.511.4005K},
	doi = {10.1093/mnras/stab3710},
	urldate = {2026-03-10},
	journal = {Monthly Notices of the Royal Astronomical Society},
	publisher = {OUP},
	author = {Kannan, R. and Garaldi, E. and Smith, A. and Pakmor, R. and Springel, V. and Vogelsberger, M. and Hernquist, L.},
	month = apr,
	year = {2022},
	note = {ADS Bibcode: 2022MNRAS.511.4005K},
	pages = {4005--4030},
}

@article{hopkins_fire-2_2018,
	title = {{FIRE}-2 simulations: physics versus numerics in galaxy formation},
	volume = {480},
	issn = {0035-8711},
	shorttitle = {{FIRE}-2 simulations},
	url = {https://doi.org/10.1093/mnras/sty1690},
	doi = {10.1093/mnras/sty1690},
	number = {1},
	urldate = {2026-03-10},
	journal = {Monthly Notices of the Royal Astronomical Society},
	author = {Hopkins, Philip F and Wetzel, Andrew and Kereš, Dušan and Faucher-Giguère, Claude-André and Quataert, Eliot and Boylan-Kolchin, Michael and Murray, Norman and Hayward, Christopher C and Garrison-Kimmel, Shea and Hummels, Cameron and Feldmann, Robert and Torrey, Paul and Ma, Xiangcheng and Anglés-Alcázar, Daniel and Su, Kung-Yi and Orr, Matthew and Schmitz, Denise and Escala, Ivanna and Sanderson, Robyn and Grudić, Michael Y and Hafen, Zachary and Kim, Ji-Hoon and Fitts, Alex and Bullock, James S and Wheeler, Coral and Chan, T K and Elbert, Oliver D and Narayanan, Desika},
	month = oct,
	year = {2018},
	pages = {800--863},
}

@article{hutter_astraeus_2021,
	title = {Astraeus {I}: the interplay between galaxy formation and reionization},
	volume = {503},
	issn = {0035-8711},
	shorttitle = {Astraeus {I}},
	url = {https://ui.adsabs.harvard.edu/abs/2021MNRAS.503.3698H},
	doi = {10.1093/mnras/stab602},
	urldate = {2026-03-10},
	journal = {Monthly Notices of the Royal Astronomical Society},
	publisher = {OUP},
	author = {Hutter, Anne and Dayal, Pratika and Yepes, Gustavo and Gottlöber, Stefan and Legrand, Laurent and Ucci, Graziano},
	month = may,
	year = {2021},
	note = {ADS Bibcode: 2021MNRAS.503.3698H},
	pages = {3698--3723},
}

@article{pillepich_simulating_2018,
	title = {Simulating galaxy formation with the {IllustrisTNG} model},
	volume = {473},
	issn = {0035-8711},
	url = {https://ui.adsabs.harvard.edu/abs/2018MNRAS.473.4077P},
	doi = {10.1093/mnras/stx2656},
	urldate = {2026-03-10},
	journal = {Monthly Notices of the Royal Astronomical Society},
	publisher = {OUP},
	author = {Pillepich, Annalisa and Springel, Volker and Nelson, Dylan and Genel, Shy and Naiman, Jill and Pakmor, Rüdiger and Hernquist, Lars and Torrey, Paul and Vogelsberger, Mark and Weinberger, Rainer and Marinacci, Federico},
	month = jan,
	year = {2018},
	note = {ADS Bibcode: 2018MNRAS.473.4077P},
	pages = {4077--4106},
}

@article{dave_simba_2019,
	title = {simba: {Cosmological} simulations with black hole growth and feedback},
	volume = {486},
	issn = {0035-8711},
	shorttitle = {simba},
	url = {https://doi.org/10.1093/mnras/stz937},
	doi = {10.1093/mnras/stz937},
	number = {2},
	urldate = {2026-03-10},
	journal = {Monthly Notices of the Royal Astronomical Society},
	author = {Davé, Romeel and Anglés-Alcázar, Daniel and Narayanan, Desika and Li, Qi and Rafieferantsoa, Mika H and Appleby, Sarah},
	month = jun,
	year = {2019},
	pages = {2827--2849},
}

@article{vogelsberger_introducing_2014,
	title = {Introducing the {Illustris} {Project}: simulating the coevolution of dark and visible matter in the {Universe}},
	volume = {444},
	issn = {0035-8711},
	shorttitle = {Introducing the {Illustris} {Project}},
	url = {https://doi.org/10.1093/mnras/stu1536},
	doi = {10.1093/mnras/stu1536},
	number = {2},
	urldate = {2026-03-10},
	journal = {Monthly Notices of the Royal Astronomical Society},
	author = {Vogelsberger, Mark and Genel, Shy and Springel, Volker and Torrey, Paul and Sijacki, Debora and Xu, Dandan and Snyder, Greg and Nelson, Dylan and Hernquist, Lars},
	month = oct,
	year = {2014},
	pages = {1518--1547},
}

@article{li_insights_2025,
	title = {Insights on metal enrichment and environmental effects at z ≈ 5─7 with {JWST} {ASPIRE}/{EIGER} and the chemical evolution model},
	volume = {703},
	issn = {0004-6361},
	url = {https://ui.adsabs.harvard.edu/abs/2025A&A...703A.106L},
	doi = {10.1051/0004-6361/202555372},
	urldate = {2026-03-10},
	journal = {Astronomy and Astrophysics},
	publisher = {EDP},
	author = {Li, Zihao and Kakiichi, Koki and Christensen, Lise and Cai, Zheng and Dekel, Avishai and Fan, Xiaohui and Farina, Emanuele Paolo and Jun, Hyunsung D. and Li, Zhaozhou and Li, Mingyu and Pudoka, Maria and Sun, Fengwu and Trebitsch, Maxime and Walter, Fabian and Wang, Feige and Yang, Jinyi and Zhang, Huanian and Zou, Siwei},
	month = nov,
	year = {2025},
	note = {ADS Bibcode: 2025A\&A...703A.106L},
	pages = {A106},
}

@article{tacchella_h_2022,
	title = {H α emission in local galaxies: star formation, time variability, and the diffuse ionized gas},
	volume = {513},
	issn = {0035-8711},
	shorttitle = {H α emission in local galaxies},
	url = {https://ui.adsabs.harvard.edu/abs/2022MNRAS.513.2904T},
	doi = {10.1093/mnras/stac818},
	urldate = {2026-03-09},
	journal = {Monthly Notices of the Royal Astronomical Society},
	publisher = {OUP},
	author = {Tacchella, Sandro and Smith, Aaron and Kannan, Rahul and Marinacci, Federico and Hernquist, Lars and Vogelsberger, Mark and Torrey, Paul and Sales, Laura and Li, Hui},
	month = jun,
	year = {2022},
	note = {ADS Bibcode: 2022MNRAS.513.2904T},
	pages = {2904--2929},
}

@misc{venturi_ga-nifs_2025,
	title = {{GA}-{NIFS}: {Powerful} and frequent outflows in moderate-luminosity {AGN} at \$z{\textbackslash}sim3-6\$},
	shorttitle = {{GA}-{NIFS}},
	url = {https://ui.adsabs.harvard.edu/abs/2025arXiv251209996V},
	doi = {10.48550/arXiv.2512.09996},
	urldate = {2026-03-09},
	publisher = {arXiv},
	author = {Venturi, Giacomo and Carniani, Stefano and Bertola, Elena and Circosta, Chiara and Parlanti, Eleonora and Perna, Michele and Arribas, Santiago and Böker, Torsten and Bunker, Andrew and Charlot, Stéphane and D'Eugenio, Francesco and Maiolino, Roberto and Rodríguez del Pino, Bruno and Übler, Hannah and Cresci, Giovanni and Jones, Gareth C. and Kumari, Nimisha and Lamperti, Isabella and Marshall, Madeline A. and Scholtz, Jan and Zamora, Sandra},
	month = dec,
	year = {2025},
	note = {ADS Bibcode: 2025arXiv251209996V},
}

@article{mingozzi_sdss_2020,
	title = {{SDSS} {IV} {MaNGA}: {Metallicity} and ionisation parameter in local star-forming galaxies from {Bayesian} fitting to photoionisation models},
	volume = {636},
	issn = {0004-6361},
	shorttitle = {{SDSS} {IV} {MaNGA}},
	url = {https://ui.adsabs.harvard.edu/abs/2020A&A...636A..42M},
	doi = {10.1051/0004-6361/201937203},
	urldate = {2026-03-06},
	journal = {Astronomy and Astrophysics},
	publisher = {EDP},
	author = {Mingozzi, M. and Belfiore, F. and Cresci, G. and Bundy, K. and Bershady, M. and Bizyaev, D. and Blanc, G. and Boquien, M. and Drory, N. and Fu, H. and Maiolino, R. and Riffel, R. and Schaefer, A. and Storchi-Bergmann, T. and Telles, E. and Tremonti, C. and Zakamska, N. and Zhang, K.},
	month = apr,
	year = {2020},
	note = {ADS Bibcode: 2020A\&A...636A..42M},
	pages = {A42},
}

@article{calzetti_dust_2000,
	title = {The {Dust} {Content} and {Opacity} of {Actively} {Star}-forming {Galaxies}},
	volume = {533},
	issn = {0004-637X},
	url = {https://ui.adsabs.harvard.edu/abs/2000ApJ...533..682C},
	doi = {10.1086/308692},
	urldate = {2026-03-05},
	journal = {The Astrophysical Journal},
	publisher = {IOP},
	author = {Calzetti, Daniela and Armus, Lee and Bohlin, Ralph C. and Kinney, Anne L. and Koornneef, Jan and Storchi-Bergmann, Thaisa},
	month = apr,
	year = {2000},
	note = {ADS Bibcode: 2000ApJ...533..682C},
	pages = {682--695},
}

@article{ibrahim_impact_2025,
	title = {The impact of supernova feedback on metallicity-gradient evolution in cosmological simulations},
	volume = {544},
	issn = {0035-8711},
	url = {https://ui.adsabs.harvard.edu/abs/2025MNRAS.544..815I},
	doi = {10.1093/mnras/staf1727},
	urldate = {2026-03-04},
	journal = {Monthly Notices of the Royal Astronomical Society},
	publisher = {OUP},
	author = {Ibrahim, Dyna and Kobayashi, Chiaki},
	month = nov,
	year = {2025},
	note = {ADS Bibcode: 2025MNRAS.544..815I},
	pages = {815--835},
}

@article{deepak_global_2025,
	title = {A {Global} {Census} of {Metals} in the {Universe}},
	volume = {987},
	issn = {0004-637X},
	url = {https://ui.adsabs.harvard.edu/abs/2025ApJ...987..199D},
	doi = {10.3847/1538-4357/add732},
	urldate = {2026-03-02},
	journal = {The Astrophysical Journal},
	publisher = {IOP},
	author = {Deepak, Saloni and Howk, J. Christopher and Lehner, Nicolas and Péroux, Céline},
	month = jul,
	year = {2025},
	note = {ADS Bibcode: 2025ApJ...987..199D},
	pages = {199},
}

@article{rodriguez_del_pino_ga-nifs_2024,
	title = {{GA}-{NIFS}: {Co}-evolution within a highly star-forming galaxy group at z ∼ 3.7 witnessed by {JWST}/{NIRSpec} {IFS}},
	volume = {684},
	copyright = {© The Authors 2024},
	issn = {0004-6361, 1432-0746},
	shorttitle = {{GA}-{NIFS}},
	url = {https://www.aanda.org/articles/aa/abs/2024/04/aa48057-23/aa48057-23.html},
	doi = {10.1051/0004-6361/202348057},
	language = {en},
	urldate = {2025-11-21},
	journal = {Astronomy \& Astrophysics},
	publisher = {EDP Sciences},
	author = {Rodríguez Del Pino, B. and Perna, M. and Arribas, S. and D’Eugenio, F. and Lamperti, I. and Pérez-González, P. G. and Übler, H. and Bunker, A. and Carniani, S. and Charlot, S. and Maiolino, R. and Willott, C. J. and Böker, T. and Chevallard, J. and Cresci, G. and Curti, M. and Jones, G. C. and Parlanti, E. and Scholtz, J. and Venturi, G.},
	month = apr,
	year = {2024},
	pages = {A187},
}

@misc{laseter_investigation_2025,
	title = {An {Investigation} into the {Low}-{Mass} {Fundamental} {Metallicity} {Relation} in the {Local} and {High}-z {Universe}},
	url = {https://ui.adsabs.harvard.edu/abs/2025arXiv251015024L},
	doi = {10.48550/arXiv.2510.15024},
	urldate = {2026-03-02},
	publisher = {arXiv},
	author = {Laseter, Isaac H. and Maseda, Michael V. and Bunker, Andrew J. and Cameron, Alex J. and Curti, Mirko and Simmonds, Charlotte},
	month = oct,
	year = {2025},
	note = {ADS Bibcode: 2025arXiv251015024L},
}

@article{tacchella_jwst_2023,
	title = {{JWST} {NIRCam} + {NIRSpec}: interstellar medium and stellar populations of young galaxies with rising star formation and evolving gas reservoirs},
	volume = {522},
	issn = {0035-8711},
	shorttitle = {{JWST} {NIRCam} + {NIRSpec}},
	url = {https://ui.adsabs.harvard.edu/abs/2023MNRAS.522.6236T},
	doi = {10.1093/mnras/stad1408},
	urldate = {2026-03-02},
	journal = {Monthly Notices of the Royal Astronomical Society},
	publisher = {OUP},
	author = {Tacchella, Sandro and Johnson, Benjamin D. and Robertson, Brant E. and Carniani, Stefano and D'Eugenio, Francesco and Kumari, Nimisha and Maiolino, Roberto and Nelson, Erica J. and Suess, Katherine A. and Übler, Hannah and Williams, Christina C. and Adebusola, Alabi and Alberts, Stacey and Arribas, Santiago and Bhatawdekar, Rachana and Bonaventura, Nina and Bowler, Rebecca A. A. and Bunker, Andrew J. and Cameron, Alex J. and Curti, Mirko and Egami, Eiichi and Eisenstein, Daniel J. and Frye, Brenda and Hainline, Kevin and Helton, Jakob M. and Ji, Zhiyuan and Looser, Tobias J. and Lyu, Jianwei and Perna, Michele and Rawle, Timothy and Rieke, George and Rieke, Marcia and Saxena, Aayush and Sandles, Lester and Shivaei, Irene and Simmonds, Charlotte and Sun, Fengwu and Willmer, Christopher N. A. and Willott, Chris J. and Witstok, Joris},
	month = jul,
	year = {2023},
	note = {ADS Bibcode: 2023MNRAS.522.6236T},
	pages = {6236--6249},
}

@article{brook_hierarchical_2012,
	title = {Hierarchical formation of bulgeless galaxies - {II}. {Redistribution} of angular momentum via galactic fountains},
	volume = {419},
	issn = {0035-8711},
	url = {https://ui.adsabs.harvard.edu/abs/2012MNRAS.419..771B},
	doi = {10.1111/j.1365-2966.2011.19740.x},
	urldate = {2026-03-02},
	journal = {Monthly Notices of the Royal Astronomical Society},
	publisher = {OUP},
	author = {Brook, C. B. and Stinson, G. and Gibson, B. K. and Roškar, R. and Wadsley, J. and Quinn, T.},
	month = jan,
	year = {2012},
	note = {ADS Bibcode: 2012MNRAS.419..771B},
	pages = {771--779},
}

@article{brook_hierarchical_2011,
	title = {Hierarchical formation of bulgeless galaxies: why outflows have low angular momentum},
	volume = {415},
	issn = {0035-8711},
	shorttitle = {Hierarchical formation of bulgeless galaxies},
	url = {https://ui.adsabs.harvard.edu/abs/2011MNRAS.415.1051B},
	doi = {10.1111/j.1365-2966.2011.18545.x},
	urldate = {2026-03-02},
	journal = {Monthly Notices of the Royal Astronomical Society},
	publisher = {OUP},
	author = {Brook, C. B. and Governato, F. and Roškar, R. and Stinson, G. and Brooks, A. M. and Wadsley, J. and Quinn, T. and Gibson, B. K. and Snaith, O. and Pilkington, K. and House, E. and Pontzen, A.},
	month = aug,
	year = {2011},
	note = {ADS Bibcode: 2011MNRAS.415.1051B},
	pages = {1051--1060},
}

@article{oppenheimer_feedback_2010,
	title = {Feedback and recycled wind accretion: assembling the z = 0 galaxy mass function},
	volume = {406},
	issn = {0035-8711},
	shorttitle = {Feedback and recycled wind accretion},
	url = {https://ui.adsabs.harvard.edu/abs/2010MNRAS.406.2325O},
	doi = {10.1111/j.1365-2966.2010.16872.x},
	urldate = {2026-03-02},
	journal = {Monthly Notices of the Royal Astronomical Society},
	publisher = {OUP},
	author = {Oppenheimer, Benjamin D. and Davé, Romeel and Kereš, Dušan and Fardal, Mark and Katz, Neal and Kollmeier, Juna A. and Weinberg, David H.},
	month = aug,
	year = {2010},
	note = {ADS Bibcode: 2010MNRAS.406.2325O},
	pages = {2325--2338},
}

@misc{langeroodi_ultraviolet_2023,
	title = {Ultraviolet {Compactness} of {High}-{Redshift} {Galaxies} as a {Tracer} of {Early}-{Stage} {Gas} {Infall}, {Bursty} {Star} {Formation}, and {Offset} from the {Fundamental} {Metallicity} {Relation}},
	url = {https://ui.adsabs.harvard.edu/abs/2023arXiv230706336L},
	doi = {10.48550/arXiv.2307.06336},
	urldate = {2026-02-18},
	publisher = {arXiv},
	author = {Langeroodi, Danial and Hjorth, Jens},
	month = jul,
	year = {2023},
	note = {ADS Bibcode: 2023arXiv230706336L},
}

@article{rusta_metal-polluted_2025,
	title = {Metal-polluted {Population} {III} {Galaxies} and {How} to {Find} {Them}},
	volume = {989},
	issn = {0004-637X},
	url = {https://ui.adsabs.harvard.edu/abs/2025ApJ...989L..32R},
	doi = {10.3847/2041-8213/adf4e3},
	urldate = {2026-02-10},
	journal = {The Astrophysical Journal},
	publisher = {IOP},
	author = {Rusta, Elka and Salvadori, Stefania and Gelli, Viola and Schaerer, Daniel and Marconi, Alessandro and Koutsouridou, Ioanna and Carniani, Stefano},
	month = aug,
	year = {2025},
	note = {ADS Bibcode: 2025ApJ...989L..32R},
	pages = {L32},
}

@article{oppenheimer_mass_2008,
	title = {Mass, metal, and energy feedback in cosmological simulations},
	volume = {387},
	issn = {0035-8711},
	url = {https://ui.adsabs.harvard.edu/abs/2008MNRAS.387..577O},
	doi = {10.1111/j.1365-2966.2008.13280.x},
	urldate = {2026-02-09},
	journal = {Monthly Notices of the Royal Astronomical Society},
	publisher = {OUP},
	author = {Oppenheimer, Benjamin D. and Davé, Romeel},
	month = jun,
	year = {2008},
	note = {ADS Bibcode: 2008MNRAS.387..577O},
	pages = {577--600},
}

@article{ubler_why_2014,
	title = {Why stellar feedback promotes disc formation in simulated galaxies},
	volume = {443},
	issn = {0035-8711},
	url = {https://ui.adsabs.harvard.edu/abs/2014MNRAS.443.2092U},
	doi = {10.1093/mnras/stu1275},
	urldate = {2026-02-09},
	journal = {Monthly Notices of the Royal Astronomical Society},
	publisher = {OUP},
	author = {Übler, Hannah and Naab, Thorsten and Oser, Ludwig and Aumer, Michael and Sales, Laura V. and White, Simon D. M.},
	month = sep,
	year = {2014},
	note = {ADS Bibcode: 2014MNRAS.443.2092U},
	pages = {2092--2111},
}

@misc{asada_glimpse-ddt_2026,
	title = {{GLIMPSE}-{DDT} spectroscopic properties of faint-end galaxies at \$z{\textbackslash}sim6\$: {Towards} first metal enrichment, dust production, and ionizing photon production},
	shorttitle = {{GLIMPSE}-{DDT} spectroscopic properties of faint-end galaxies at \$z{\textbackslash}sim6\$},
	url = {http://arxiv.org/abs/2601.20045},
	doi = {10.48550/arXiv.2601.20045},
	urldate = {2026-02-05},
	publisher = {arXiv},
	author = {Asada, Yoshihisa and Fujimoto, Seiji and Chisholm, John and Naidu, Rohan P. and Atek, Hakim and Brammer, Gabriel and Furtak, Lukas J. and Kokorev, Vasily and Pan, Richard and Basu, Arghyadeep and Bromm, Volker and Dessauges-Zavadsky, Miroslava and Hsiao, Tiger Yu-Yang and Jecmen, Michelle and Korber, Damien and Liu, Boyuan and McKinney, Jed and McQuinn, Kristen B. W. and Schaerer, Daniel},
	month = jan,
	year = {2026},
	note = {arXiv:2601.20045 [astro-ph]},
}

@article{cueto_astraeus_2024,
	title = {{ASTRAEUS}. {IX}. {Impact} of an evolving stellar initial mass function on early galaxies and reionisation},
	volume = {686},
	issn = {0004-6361},
	url = {https://ui.adsabs.harvard.edu/abs/2024A&A...686A.138C},
	doi = {10.1051/0004-6361/202349017},
	urldate = {2026-01-28},
	journal = {Astronomy and Astrophysics},
	publisher = {EDP},
	author = {Cueto, Elie R. and Hutter, Anne and Dayal, Pratika and Gottlöber, Stefan and Heintz, Kasper E. and Mason, Charlotte and Trebitsch, Maxime and Yepes, Gustavo},
	month = jun,
	year = {2024},
	note = {ADS Bibcode: 2024A\&A...686A.138C},
	pages = {A138},
}

@article{koller_both_2026,
	title = {Both stellar mass and gravitational potential shape the gas-phase metallicity},
	volume = {545},
	issn = {0035-8711},
	url = {https://ui.adsabs.harvard.edu/abs/2026MNRAS.545f2011K},
	doi = {10.1093/mnras/staf2011},
	urldate = {2026-01-22},
	journal = {Monthly Notices of the Royal Astronomical Society},
	publisher = {OUP},
	author = {Koller, Maria and Maiolino, Roberto and Baker, William M.},
	month = jan,
	year = {2026},
	note = {ADS Bibcode: 2026MNRAS.545f2011K},
	pages = {staf2011},
}

@misc{ferland_2017_2017,
	title = {The 2017 {Release} {Cloudy}},
	url = {https://ui.adsabs.harvard.edu/abs/2017RMxAA..53..385F},
	doi = {10.48550/arXiv.1705.10877},
	urldate = {2026-01-21},
	publisher = {arXiv},
	author = {Ferland, G. J. and Chatzikos, M. and Guzmán, F. and Lykins, M. L. and van Hoof, P. A. M. and Williams, R. J. R. and Abel, N. P. and Badnell, N. R. and Keenan, F. P. and Porter, R. L. and Stancil, P. C.},
	month = oct,
	year = {2017},
	note = {ISSN: 0185-1101
Volume: 53
ADS Bibcode: 2017RMxAA..53..385F},
}

@article{byler_nebular_2017,
	title = {Nebular {Continuum} and {Line} {Emission} in {Stellar} {Population} {Synthesis} {Models}},
	volume = {840},
	issn = {0004-637X},
	url = {https://ui.adsabs.harvard.edu/abs/2017ApJ...840...44B},
	doi = {10.3847/1538-4357/aa6c66},
	urldate = {2026-01-21},
	journal = {The Astrophysical Journal},
	publisher = {IOP},
	author = {Byler, Nell and Dalcanton, Julianne J. and Conroy, Charlie and Johnson, Benjamin D.},
	month = may,
	year = {2017},
	note = {ADS Bibcode: 2017ApJ...840...44B},
	pages = {44},
}

@article{rowland_rebels-ifu_2026,
	title = {{REBELS}-{IFU}: {Evidence} for metal-rich massive galaxies at z {\textasciitilde} 6 - 8},
	issn = {0035-8711},
	shorttitle = {{REBELS}-{IFU}},
	url = {https://ui.adsabs.harvard.edu/abs/2026MNRAS.tmp...14R},
	doi = {10.1093/mnras/staf2023},
	urldate = {2026-01-15},
	journal = {Monthly Notices of the Royal Astronomical Society},
	publisher = {OUP},
	author = {Rowland, Lucie E. and Stefanon, Mauro and Bouwens, Rychard and Hodge, Jacqueline and Algera, Hiddo and Fisher, Rebecca and Dayal, Pratika and Pallottini, Andrea and Stark, Daniel P. and Heintz, Kasper E. and Aravena, Manuel and Bowler, Rebecca A. A. and Cescon, Karin and Endsley, Ryan and Ferrara, Andrea and Fudamoto, Yoshinobu and Gonzalez, Valentino and Graziani, Luca and Gulis, Cindy and Herard-Demanche, Thomas and Inami, Hanae and Laza-Ramos, Andrès and van Leeuwen, Ivana and de Looze, Ilse and Nanayakkara, Themiya and Oesch, Pascal and Ormerod, Katherine and Palla, Marco and Sartorio, Nina S. and Schouws, Sander and Smit, Renske and Sommovigo, Laura and Toft, Sune and Weaver, John R. and van der Werf, Paul},
	month = jan,
	year = {2026},
	note = {ADS Bibcode: 2026MNRAS.tmp...14R},
}

@article{mcclymont_thesan-zoom_2026,
	title = {The {THESAN}-{ZOOM} project: {Mystery} {N}/{O} more - uncovering the origin of peculiar chemical abundances and a not-so-fundamental metallicity relation at 3 {\textless} z {\textless} 12},
	issn = {0035-8711},
	shorttitle = {The {THESAN}-{ZOOM} project},
	url = {https://ui.adsabs.harvard.edu/abs/2026MNRAS.tmp...20M},
	doi = {10.1093/mnras/stag016},
	urldate = {2026-01-14},
	journal = {Monthly Notices of the Royal Astronomical Society},
	publisher = {OUP},
	author = {McClymont, William and Tacchella, Sandro and Smith, Aaron and Kannan, Rahul and Garaldi, Enrico and Puchwein, Ewald and Isobe, Yuki and Ji, Xihan and Shen, Xuejian and Wang, Zihao and Belokurov, Vasily and Borrow, Josh and D'Eugenio, Francesco and Keating, Laura and Maiolino, Roberto and Monty, Stephanie and Vogelsberger, Mark and Zier, Oliver},
	month = jan,
	year = {2026},
	note = {ADS Bibcode: 2026MNRAS.tmp...20M},
}

@article{torrey_evolution_2019,
	title = {The evolution of the mass-metallicity relation and its scatter in {IllustrisTNG}},
	volume = {484},
	issn = {0035-8711},
	url = {https://ui.adsabs.harvard.edu/abs/2019MNRAS.484.5587T},
	doi = {10.1093/mnras/stz243},
	urldate = {2026-01-14},
	journal = {Monthly Notices of the Royal Astronomical Society},
	publisher = {OUP},
	author = {Torrey, Paul and Vogelsberger, Mark and Marinacci, Federico and Pakmor, Rüdiger and Springel, Volker and Nelson, Dylan and Naiman, Jill and Pillepich, Annalisa and Genel, Shy and Weinberger, Rainer and Hernquist, Lars},
	month = apr,
	year = {2019},
	note = {ADS Bibcode: 2019MNRAS.484.5587T},
	pages = {5587--5607},
}

@article{schaye_eagle_2015,
	title = {The {EAGLE} project: simulating the evolution and assembly of galaxies and their environments},
	volume = {446},
	issn = {0035-8711},
	shorttitle = {The {EAGLE} project},
	url = {https://ui.adsabs.harvard.edu/abs/2015MNRAS.446..521S},
	doi = {10.1093/mnras/stu2058},
	urldate = {2026-01-14},
	journal = {Monthly Notices of the Royal Astronomical Society},
	publisher = {OUP},
	author = {Schaye, Joop and Crain, Robert A. and Bower, Richard G. and Furlong, Michelle and Schaller, Matthieu and Theuns, Tom and Dalla Vecchia, Claudio and Frenk, Carlos S. and McCarthy, I. G. and Helly, John C. and Jenkins, Adrian and Rosas-Guevara, Y. M. and White, Simon D. M. and Baes, Maarten and Booth, C. M. and Camps, Peter and Navarro, Julio F. and Qu, Yan and Rahmati, Alireza and Sawala, Till and Thomas, Peter A. and Trayford, James},
	month = jan,
	year = {2015},
	note = {ADS Bibcode: 2015MNRAS.446..521S},
	pages = {521--554},
}

@article{klessen_first_2023,
	title = {The {First} {Stars}: {Formation}, {Properties}, and {Impact}},
	volume = {61},
	issn = {0066-4146},
	shorttitle = {The {First} {Stars}},
	url = {https://ui.adsabs.harvard.edu/abs/2023ARA&A..61...65K},
	doi = {10.1146/annurev-astro-071221-053453},
	urldate = {2026-01-08},
	journal = {Annual Review of Astronomy and Astrophysics},
	author = {Klessen, Ralf S. and Glover, Simon C. O.},
	month = aug,
	year = {2023},
	note = {ADS Bibcode: 2023ARA\&A..61...65K},
	pages = {65--130},
}

@article{wise_birth_2012,
	title = {The birth of a galaxy - {II}. {The} role of radiation pressure},
	volume = {427},
	issn = {0035-8711},
	url = {https://ui.adsabs.harvard.edu/abs/2012MNRAS.427..311W},
	doi = {10.1111/j.1365-2966.2012.21809.x},
	urldate = {2026-01-08},
	journal = {Monthly Notices of the Royal Astronomical Society},
	publisher = {OUP},
	author = {Wise, John H. and Abel, Tom and Turk, Matthew J. and Norman, Michael L. and Smith, Britton D.},
	month = nov,
	year = {2012},
	note = {ADS Bibcode: 2012MNRAS.427..311W},
	pages = {311--326},
}

@misc{nishigaki_dreamsii_2025,
	title = {{DREAMS}.{II}. {Galaxy} {Demographics} from {Direct} {Te}-{Based} {Metallicities} at z{\textasciitilde}2-10: {Tracing} the {Evolution} of the {Mass}-{Metallicity} and {Fundamental} {Relations}},
	shorttitle = {{DREAMS}.{II}. {Galaxy} {Demographics} from {Direct} {Te}-{Based} {Metallicities} at z{\textasciitilde}2-10},
	url = {https://ui.adsabs.harvard.edu/abs/2025arXiv251212983N},
	doi = {10.48550/arXiv.2512.12983},
	urldate = {2026-01-06},
	publisher = {arXiv},
	author = {Nishigaki, Moka and Nakajima, Kimihiko and Ouchi, Masami and Behroozi, Peter and Nakane, Minami and Takeda, Yui and Umeda, Hiroya and Yajima, Hidenobu and Yanagisawa, Hiroto},
	month = dec,
	year = {2025},
	note = {ADS Bibcode: 2025arXiv251212983N},
}

@article{kohandel_velocity_2020,
	title = {Velocity dispersion in the interstellar medium of early galaxies},
	volume = {499},
	issn = {0035-8711},
	url = {https://ui.adsabs.harvard.edu/abs/2020MNRAS.499.1250K},
	doi = {10.1093/mnras/staa2792},
	urldate = {2025-12-22},
	journal = {Monthly Notices of the Royal Astronomical Society},
	publisher = {OUP},
	author = {Kohandel, M. and Pallottini, A. and Ferrara, A. and Carniani, S. and Gallerani, S. and Vallini, L. and Zanella, A. and Behrens, C.},
	month = nov,
	year = {2020},
	note = {ADS Bibcode: 2020MNRAS.499.1250K},
	pages = {1250--1265},
}

@article{pallottini_survey_2022,
	title = {A survey of high-z galaxies: {SERRA} simulations},
	volume = {513},
	issn = {0035-8711},
	shorttitle = {A survey of high-z galaxies},
	url = {https://ui.adsabs.harvard.edu/abs/2022MNRAS.513.5621P},
	doi = {10.1093/mnras/stac1281},
	urldate = {2025-12-22},
	journal = {Monthly Notices of the Royal Astronomical Society},
	publisher = {OUP},
	author = {Pallottini, A. and Ferrara, A. and Gallerani, S. and Behrens, C. and Kohandel, M. and Carniani, S. and Vallini, L. and Salvadori, S. and Gelli, V. and Sommovigo, L. and D'Odorico, V. and Di Mascia, F. and Pizzati, E.},
	month = jul,
	year = {2022},
	note = {ADS Bibcode: 2022MNRAS.513.5621P},
	pages = {5621--5641},
}

@article{duan_galaxy_2025,
	title = {Galaxy mergers in the epoch of reionization – {I}. {A} {JWST} study of pair fractions, merger rates, and stellar mass accretion rates at z = 4.5–11.5},
	volume = {540},
	issn = {0035-8711},
	url = {https://ui.adsabs.harvard.edu/abs/2025MNRAS.540..774D},
	doi = {10.1093/mnras/staf638},
	urldate = {2025-12-22},
	journal = {Monthly Notices of the Royal Astronomical Society},
	publisher = {OUP},
	author = {Duan, Qiao and Conselice, Christopher J. and Li, Qiong and Austin, Duncan and Harvey, Thomas and Adams, Nathan J. and Duncan, Kenneth J. and Trussler, James and Ferreira, Leonardo and Westcott, Lewi and Harris, Honor and Windhorst, Rogier A. and Holwerda, Benne W. and Broadhurst, Thomas J. and Coe, Dan and Cohen, Seth H. and Du, Xiaojing and Driver, Simon P. and Frye, Brenda and Grogin, Norman A. and Hathi, Nimish P. and Jansen, Rolf A. and Koekemoer, Anton M. and Marshall, Madeline A. and Nonino, Mario and Ortiz, III, Rafael and Pirzkal, Nor and Robotham, Aaron and Ryan, Russell E. and Summers, Jake and D'Silva, Jordan C. J. and Willmer, Christopher N. A. and Yan, Haojing},
	month = jun,
	year = {2025},
	note = {ADS Bibcode: 2025MNRAS.540..774D},
	pages = {774--805},
}

@article{puskas_constraining_2025,
	title = {Constraining the major merger history of z {\textasciitilde} 3-9 galaxies using {JADES}: dominant in situ star formation},
	volume = {540},
	issn = {0035-8711},
	shorttitle = {Constraining the major merger history of z {\textasciitilde} 3-9 galaxies using {JADES}},
	url = {https://ui.adsabs.harvard.edu/abs/2025MNRAS.540.2146P},
	doi = {10.1093/mnras/staf813},
	urldate = {2025-12-22},
	journal = {Monthly Notices of the Royal Astronomical Society},
	publisher = {OUP},
	author = {Puskás, Dávid and Tacchella, Sandro and Simmonds, Charlotte and Hainline, Kevin and D'Eugenio, Francesco and Alberts, Stacey and Arribas, Santiago and Baker, William M. and Bunker, Andrew J. and Carniani, Stefano and Charlot, Stéphane and Duan, Qiao and Eisenstein, Daniel J. and Ji, Zhiyuan and Johnson, Benjamin D. and Jones, Gareth C. and Maiolino, Roberto and McClymont, William and Rieke, Marcia and Rinaldi, Pierluigi and Robertson, Brant and Übler, Hannah and Williams, Christina C. and Willmer, Christopher N. A. and Willott, Chris and Witstok, Joris},
	month = jul,
	year = {2025},
	note = {ADS Bibcode: 2025MNRAS.540.2146P},
	pages = {2146--2175},
}

@article{villar_martin_agn_2024,
	title = {{AGN} feedback can produce metal enrichment on galaxy scales},
	volume = {690},
	issn = {0004-6361},
	url = {https://ui.adsabs.harvard.edu/abs/2024A&A...690A.397V},
	doi = {10.1051/0004-6361/202449621},
	urldate = {2025-12-21},
	journal = {Astronomy and Astrophysics},
	publisher = {EDP},
	author = {Villar Martín, M. and López Cobá, C. and Cazzoli, S. and Pérez Montero, E. and Cabrera Lavers, A.},
	month = oct,
	year = {2024},
	note = {ADS Bibcode: 2024A\&A...690A.397V},
	pages = {A397},
}

@article{ma_origin_2016,
	title = {The origin and evolution of the galaxy mass-metallicity relation},
	volume = {456},
	issn = {0035-8711},
	url = {https://ui.adsabs.harvard.edu/abs/2016MNRAS.456.2140M},
	doi = {10.1093/mnras/stv2659},
	urldate = {2025-12-12},
	journal = {Monthly Notices of the Royal Astronomical Society},
	publisher = {OUP},
	author = {Ma, Xiangcheng and Hopkins, Philip F. and Faucher-Giguère, Claude-André and Zolman, Nick and Muratov, Alexander L. and Kereš, Dušan and Quataert, Eliot},
	month = feb,
	year = {2016},
	note = {ADS Bibcode: 2016MNRAS.456.2140M},
	pages = {2140--2156},
}

@article{langan_weak_2020,
	title = {Weak evolution of the mass-metallicity relation at cosmic dawn in the {FirstLight} simulations},
	volume = {494},
	issn = {0035-8711},
	url = {https://ui.adsabs.harvard.edu/abs/2020MNRAS.494.1988L},
	doi = {10.1093/mnras/staa880},
	urldate = {2025-12-12},
	journal = {Monthly Notices of the Royal Astronomical Society},
	publisher = {OUP},
	author = {Langan, Ivanna and Ceverino, Daniel and Finlator, Kristian},
	month = may,
	year = {2020},
	note = {ADS Bibcode: 2020MNRAS.494.1988L},
	pages = {1988--1993},
}

@article{nakazato_simulations_2023,
	title = {Simulations of {High}-redshift [{O} {III}] {Emitters}: {Chemical} {Evolution} and {Multiline} {Diagnostics}},
	volume = {953},
	issn = {0004-637X},
	shorttitle = {Simulations of {High}-redshift [{O} {III}] {Emitters}},
	url = {https://ui.adsabs.harvard.edu/abs/2023ApJ...953..140N},
	doi = {10.3847/1538-4357/ace25a},
	urldate = {2025-12-12},
	journal = {The Astrophysical Journal},
	publisher = {IOP},
	author = {Nakazato, Yurina and Yoshida, Naoki and Ceverino, Daniel},
	month = aug,
	year = {2023},
	note = {ADS Bibcode: 2023ApJ...953..140N},
	pages = {140},
}

@article{scholtz_jades_2025,
	title = {{JADES}: {A} large population of obscured, narrow-line active galactic nuclei at high redshift},
	volume = {697},
	issn = {0004-6361},
	shorttitle = {{JADES}},
	url = {https://ui.adsabs.harvard.edu/abs/2025A&A...697A.175S},
	doi = {10.1051/0004-6361/202348804},
	urldate = {2025-11-28},
	journal = {Astronomy and Astrophysics},
	publisher = {EDP},
	author = {Scholtz, Jan and Maiolino, Roberto and D'Eugenio, Francesco and Curtis-Lake, Emma and Carniani, Stefano and Charlot, Stephane and Curti, Mirko and Silcock, Maddie S. and Arribas, Santiago and Baker, William and Bhatawdekar, Rachana and Boyett, Kristan and Bunker, Andrew J. and Chevallard, Jacopo and Circosta, Chiara and Eisenstein, Daniel J. and Hainline, Kevin and Hausen, Ryan and Ji, Xihan and Ji, Zhiyuan and Johnson, Benjamin D. and Kumari, Nimisha and Looser, Tobias J. and Lyu, Jianwei and Maseda, Michael V. and Parlanti, Eleonora and Perna, Michele and Rieke, Marcia and Robertson, Brant and Del Pino, Bruno Rodríguez and Sun, Fengwu and Tacchella, Sandro and Übler, Hannah and Venturi, Giacomo and Williams, Christina C. and Willmer, Christopher N. A. and Willott, Chris and Witstok, Joris},
	month = may,
	year = {2025},
	note = {ADS Bibcode: 2025A\&A...697A.175S},
	pages = {A175},
}

@article{parlanti_ga-nifs_2025,
	title = {{GA}-{NIFS}: {Multiphase} analysis of a star-forming galaxy at z ∼ 5.5},
	volume = {695},
	issn = {0004-6361},
	shorttitle = {{GA}-{NIFS}},
	url = {https://ui.adsabs.harvard.edu/abs/2025A&A...695A...6P},
	doi = {10.1051/0004-6361/202451692},
	urldate = {2025-11-28},
	journal = {Astronomy and Astrophysics},
	publisher = {EDP},
	author = {Parlanti, Eleonora and Carniani, Stefano and Venturi, Giacomo and Herrera-Camus, Rodrigo and Arribas, Santiago and Bunker, Andrew J. and Charlot, Stéphane and D'Eugenio, Francesco and Maiolino, Roberto and Perna, Michele and Übler, Hannah and Böker, Torsten and Cresci, Giovanni and Curti, Mirko and Jones, Gareth C. and Lamperti, Isabella and Pérez-González, Pablo G. and Del Pino, Bruno Rodríguez and Zamora, Sandra},
	month = mar,
	year = {2025},
	note = {ADS Bibcode: 2025A\&A...695A...6P},
	pages = {A6},
}

@article{ubler_ga-nifs_2024,
	title = {{GA}-{NIFS}: {NIRSpec} reveals evidence for non-circular motions and {AGN} feedback in {GN20}},
	volume = {533},
	issn = {0035-8711},
	shorttitle = {{GA}-{NIFS}},
	url = {https://ui.adsabs.harvard.edu/abs/2024MNRAS.533.4287U},
	doi = {10.1093/mnras/stae1993},
	urldate = {2025-11-28},
	journal = {Monthly Notices of the Royal Astronomical Society},
	publisher = {OUP},
	author = {Übler, Hannah and D'Eugenio, Francesco and Perna, Michele and Arribas, Santiago and Jones, Gareth C. and Bunker, Andrew J. and Carniani, Stefano and Charlot, Stéphane and Maiolino, Roberto and Rodríguez del Pino, Bruno and Willott, Chris J. and Böker, Torsten and Cresci, Giovanni and Kumari, Nimisha and Lamperti, Isabella and Parlanti, Eleonora and Scholtz, Jan and Venturi, Giacomo},
	month = oct,
	year = {2024},
	note = {ADS Bibcode: 2024MNRAS.533.4287U},
	pages = {4287--4299},
}

@article{jones_ga-nifs_2024,
	title = {{GA}-{NIFS}: {JWST}/{NIRSpec} integral field unit observations of {HFLS3} reveal a dense galaxy group at z ∼ 6.3},
	volume = {682},
	issn = {0004-6361},
	shorttitle = {{GA}-{NIFS}},
	url = {https://ui.adsabs.harvard.edu/abs/2024A&A...682A.122J},
	doi = {10.1051/0004-6361/202347838},
	urldate = {2025-11-28},
	journal = {Astronomy and Astrophysics},
	publisher = {EDP},
	author = {Jones, Gareth C. and Übler, Hannah and Perna, Michele and Arribas, Santiago and Bunker, Andrew J. and Carniani, Stefano and Charlot, Stephane and Maiolino, Roberto and Del Pino, Bruno Rodríguez and Willott, Chris and Bowler, Rebecca A. A. and Böker, Torsten and Cameron, Alex J. and Chevallard, Jacopo and Cresci, Giovanni and Curti, Mirko and D'Eugenio, Francesco and Kumari, Nimisha and Saxena, Aayush and Scholtz, Jan and Venturi, Giacomo and Witstok, Joris},
	month = feb,
	year = {2024},
	note = {ADS Bibcode: 2024A\&A...682A.122J},
	pages = {A122},
}

@article{lamperti_ga-nifs_2024,
	title = {{GA}-{NIFS}: {JWST}/{NIRSpec} {IFS} view of the z ∼ 3.5 galaxy {GS5001} and its close environment at the core of a large-scale overdensity},
	volume = {691},
	issn = {0004-6361},
	shorttitle = {{GA}-{NIFS}},
	url = {https://ui.adsabs.harvard.edu/abs/2024A&A...691A.153L},
	doi = {10.1051/0004-6361/202451021},
	urldate = {2025-11-28},
	journal = {Astronomy and Astrophysics},
	publisher = {EDP},
	author = {Lamperti, Isabella and Arribas, Santiago and Perna, Michele and Rodríguez Del Pino, Bruno and Circosta, Chiara and Pérez-González, Pablo G. and Bunker, Andrew J. and Carniani, Stefano and Charlot, Stéphane and D'Eugenio, Francesco and Maiolino, Roberto and Übler, Hannah and Willott, Chris J. and Bertola, Elena and Böker, Torsten and Cresci, Giovanni and Curti, Mirko and Jones, Gareth C. and Kumari, Nimisha and Parlanti, Eleonora and Scholtz, Jan and Venturi, Giacomo},
	month = nov,
	year = {2024},
	note = {ADS Bibcode: 2024A\&A...691A.153L},
	pages = {A153},
}

@article{fujimoto_primordial_2025,
	title = {Primordial rotating disk composed of at least 15 dense star-forming clumps at cosmic dawn},
	volume = {9},
	issn = {2397-3366},
	url = {https://ui.adsabs.harvard.edu/abs/2025NatAs...9.1553F},
	doi = {10.1038/s41550-025-02592-w},
	urldate = {2025-11-28},
	journal = {Nature Astronomy},
	author = {Fujimoto, S. and Ouchi, M. and Kohno, K. and Valentino, F. and Giménez-Arteaga, C. and Brammer, G. B. and Furtak, L. J. and Kohandel, M. and Oguri, M. and Pallottini, A. and Richard, J. and Zitrin, A. and Bauer, F. E. and Boylan-Kolchin, M. and Dessauges-Zavadsky, M. and Egami, E. and Finkelstein, S. L. and Ma, Z. and Smail, I. and Watson, D. and Hutchison, T. A. and Rigby, J. R. and Welch, B. D. and Ao, Y. and Bradley, L. D. and Caminha, G. B. and Caputi, K. I. and Espada, D. and Endsley, R. and Fudamoto, Y. and González-López, J. and Hatsukade, B. and Koekemoer, A. M. and Kokorev, V. and Laporte, N. and Lee, M. and Magdis, G. E. and Ono, Y. and Rizzo, F. and Shibuya, T. and Shimasaku, K. and Sun, F. and Toft, S. and Umehata, H. and Wang, T. and Yajima, H.},
	month = aug,
	year = {2025},
	note = {ADS Bibcode: 2025NatAs...9.1553F},
	pages = {1553--1567},
}

@article{garcia_metallicity_2025,
	title = {Metallicity {Gradients} in {Modern} {Cosmological} {Simulations}. {I}. {Tension} between {Smooth} {Stellar} {Feedback} {Models} and {Observations}},
	volume = {989},
	issn = {0004-637X},
	url = {https://ui.adsabs.harvard.edu/abs/2025ApJ...989..147G},
	doi = {10.3847/1538-4357/adea51},
	urldate = {2025-11-28},
	journal = {The Astrophysical Journal},
	publisher = {IOP},
	author = {Garcia, Alex M. and Torrey, Paul and Bhagwat, Aniket and Wright, Ruby J. and Chen, Qian-Hui and Grasha, Kathryn and Ridolfo, Sophia and Hemler, Z. S. and Sarkar, Arnab and Chakraborty, Priyanka and Nelson, Erica J. and Sanders, Ryan L. and Costa, Tiago and Vogelsberger, Mark and Kewley, Lisa J. and Ellison, Sara L. and Hernquist, Lars},
	month = aug,
	year = {2025},
	note = {ADS Bibcode: 2025ApJ...989..147G},
	pages = {147},
}

@article{renzini_objective_2015,
	title = {An {Objective} {Definition} for the {Main} {Sequence} of {Star}-forming {Galaxies}},
	volume = {801},
	issn = {0004-637X},
	url = {https://ui.adsabs.harvard.edu/abs/2015ApJ...801L..29R},
	doi = {10.1088/2041-8205/801/2/L29},
	urldate = {2025-11-27},
	journal = {The Astrophysical Journal},
	publisher = {IOP},
	author = {Renzini, Alvio and Peng, Ying-jie},
	month = mar,
	year = {2015},
	note = {ADS Bibcode: 2015ApJ...801L..29R},
	pages = {L29},
}

@article{graf_spatial_2025,
	title = {Spatial {Variations} of {Stellar} {Elemental} {Abundances} in {FIRE} {Simulations} of {Milky} {Way}-mass {Galaxies}: {Patterns} {Today} {Mostly} {Reflect} {Those} at {Formation}},
	volume = {981},
	issn = {0004-637X},
	shorttitle = {Spatial {Variations} of {Stellar} {Elemental} {Abundances} in {FIRE} {Simulations} of {Milky} {Way}-mass {Galaxies}},
	url = {https://ui.adsabs.harvard.edu/abs/2025ApJ...981...47G},
	doi = {10.3847/1538-4357/adacd7},
	urldate = {2025-11-25},
	journal = {The Astrophysical Journal},
	publisher = {IOP},
	author = {Graf, Russell L. and Wetzel, Andrew and Bellardini, Matthew A. and Bailin, Jeremy},
	month = mar,
	year = {2025},
	note = {ADS Bibcode: 2025ApJ...981...47G},
	pages = {47},
}

@article{sharda_physics_2021,
	title = {The physics of gas phase metallicity gradients in galaxies},
	volume = {502},
	issn = {0035-8711},
	url = {https://ui.adsabs.harvard.edu/abs/2021MNRAS.502.5935S},
	doi = {10.1093/mnras/stab252},
	urldate = {2025-11-25},
	journal = {Monthly Notices of the Royal Astronomical Society},
	publisher = {OUP},
	author = {Sharda, Piyush and Krumholz, Mark R. and Wisnioski, Emily and Forbes, John C. and Federrath, Christoph and Acharyya, Ayan},
	month = apr,
	year = {2021},
	note = {ADS Bibcode: 2021MNRAS.502.5935S},
	pages = {5935--5961},
}

@article{bellardini_3d_2021,
	title = {{3D} gas-phase elemental abundances across the formation histories of {Milky} {Way}-mass galaxies in the {FIRE} simulations: initial conditions for chemical tagging},
	volume = {505},
	issn = {0035-8711},
	shorttitle = {{3D} gas-phase elemental abundances across the formation histories of {Milky} {Way}-mass galaxies in the {FIRE} simulations},
	url = {https://ui.adsabs.harvard.edu/abs/2021MNRAS.505.4586B},
	doi = {10.1093/mnras/stab1606},
	urldate = {2025-11-25},
	journal = {Monthly Notices of the Royal Astronomical Society},
	publisher = {OUP},
	author = {Bellardini, Matthew A. and Wetzel, Andrew and Loebman, Sarah R. and Faucher-Giguère, Claude-André and Ma, Xiangcheng and Feldmann, Robert},
	month = aug,
	year = {2021},
	note = {ADS Bibcode: 2021MNRAS.505.4586B},
	pages = {4586--4607},
}

@article{bellardini_3d_2022,
	title = {{3D} elemental abundances of stars at formation across the histories of {Milky} {Way}-mass galaxies in the {FIRE} simulations},
	volume = {514},
	issn = {0035-8711},
	url = {https://ui.adsabs.harvard.edu/abs/2022MNRAS.514.4270B},
	doi = {10.1093/mnras/stac1637},
	urldate = {2025-11-25},
	journal = {Monthly Notices of the Royal Astronomical Society},
	publisher = {OUP},
	author = {Bellardini, Matthew A. and Wetzel, Andrew and Loebman, Sarah R. and Bailin, Jeremy},
	month = aug,
	year = {2022},
	note = {ADS Bibcode: 2022MNRAS.514.4270B},
	pages = {4270--4289},
}

@article{mott_abundance_2013,
	title = {Abundance gradients in spiral discs: is the gradient inversion at high redshift real?},
	volume = {435},
	issn = {0035-8711},
	shorttitle = {Abundance gradients in spiral discs},
	url = {https://ui.adsabs.harvard.edu/abs/2013MNRAS.435.2918M},
	doi = {10.1093/mnras/stt1495},
	urldate = {2025-11-25},
	journal = {Monthly Notices of the Royal Astronomical Society},
	publisher = {OUP},
	author = {Mott, A. and Spitoni, E. and Matteucci, F.},
	month = nov,
	year = {2013},
	note = {ADS Bibcode: 2013MNRAS.435.2918M},
	pages = {2918--2930},
}

@article{poetrodjojo_sami_2021,
	title = {The {SAMI} {Galaxy} {Survey}: reconciling strong emission line metallicity diagnostics using metallicity gradients},
	volume = {502},
	issn = {0035-8711},
	shorttitle = {The {SAMI} {Galaxy} {Survey}},
	url = {https://doi.org/10.1093/mnras/stab205},
	doi = {10.1093/mnras/stab205},
	number = {3},
	urldate = {2025-11-24},
	journal = {Monthly Notices of the Royal Astronomical Society},
	author = {Poetrodjojo, Henry and Groves, Brent and Kewley, Lisa J and Sweet, Sarah M and Sanchez, Sebastian F and Medling, Anne M and López-Sánchez, Ángel R and Brough, Sarah and Cortese, Luca and van de Sande, Jesse and Vaughan, Sam and Richards, Samuel N and Bryant, Julia J and Croom, Scott M and Bland-Hawthorn, Joss and Goodwin, Michael and Lawrence, Jon S and Owers, Matt S and Scott, Nicholas},
	month = apr,
	year = {2021},
	pages = {3357--3373},
}

@article{khoram_direct-method_2025,
	title = {Direct-method metallicity gradients derived from spectral stacking with {SDSS}-{IV} {MaNGA}},
	volume = {693},
	copyright = {© The Authors 2025},
	issn = {0004-6361, 1432-0746},
	url = {https://www.aanda.org/articles/aa/abs/2025/01/aa51980-24/aa51980-24.html},
	doi = {10.1051/0004-6361/202451980},
	language = {en},
	urldate = {2025-11-24},
	journal = {Astronomy \& Astrophysics},
	publisher = {EDP Sciences},
	author = {Khoram, Amir H. and Belfiore, Francesco},
	month = jan,
	year = {2025},
	pages = {A150},
}

@article{li_negative_2025,
	title = {A negative stellar mass‑gaseous metallicity gradient relation of dwarf galaxies modulated by stellar feedback},
	volume = {698},
	issn = {0004-6361},
	url = {https://ui.adsabs.harvard.edu/abs/2025A&A...698A.208L},
	doi = {10.1051/0004-6361/202452978},
	urldate = {2025-11-24},
	journal = {Astronomy and Astrophysics},
	publisher = {EDP},
	author = {Li, Tie and Zhang, Hong-Xin and Lyu, Wenhe and Tang, Yimeng and Yao, Yao and Wang, Enci and Rong, Yu and Chen, Guangwen and Kong, Xu and Bian, Fuyan and Gu, Qiusheng and Johnston, Evelyn J. and Li, Xin and Mao, Shude and Shi, Yong and Wang, Junfeng and Wang, Xin and Yu, Xiaoling and Zheng, Zhiyuan},
	month = jun,
	year = {2025},
	note = {ADS Bibcode: 2025A\&A...698A.208L},
	pages = {A208},
}

@article{jones_resolved_2010,
	title = {Resolved spectroscopy of gravitationally lensed galaxies: recovering coherent velocity fields in subluminous z∼ 2–3 galaxies},
	volume = {404},
	issn = {0035-8711},
	shorttitle = {Resolved spectroscopy of gravitationally lensed galaxies},
	url = {https://doi.org/10.1111/j.1365-2966.2010.16378.x},
	doi = {10.1111/j.1365-2966.2010.16378.x},
	number = {3},
	urldate = {2025-11-24},
	journal = {Monthly Notices of the Royal Astronomical Society},
	author = {Jones, T. A. and Swinbank, A. M. and Ellis, R. S. and Richard, J. and Stark, D. P.},
	month = may,
	year = {2010},
	pages = {1247--1262},
}

@article{sharples_first_2013,
	title = {First {Light} for the {KMOS} {Multi}-{Object} {Integral}-{Field} {Spectrometer}},
	volume = {151},
	issn = {0722-6691},
	url = {https://ui.adsabs.harvard.edu/abs/2013Msngr.151...21S},
	urldate = {2025-11-24},
	journal = {The Messenger},
	author = {Sharples, R. and Bender, R. and Agudo Berbel, A. and Bezawada, N. and Castillo, R. and Cirasuolo, M. and Davidson, G. and Davies, R. and Dubbeldam, M. and Fairley, A. and Finger, G. and Förster Schreiber, N. and Gonte, F. and Hess, A. and Jung, I. and Lewis, I. and Lizon, J.-L. and Muschielok, B. and Pasquini, L. and Pirard, J. and Popovic, D. and Ramsay, S. and Rees, P. and Richter, J. and Riquelme, M. and Rodrigues, M. and Saviane, I. and Schlichter, J. and Schmidtobreick, L. and Segovia, A. and Smette, A. and Szeifert, T. and van Kesteren, A. and Wegner, M. and Wiezorrek, E.},
	month = mar,
	year = {2013},
	note = {ADS Bibcode: 2013Msngr.151...21S},
	pages = {21--23},
}

@article{wuyts_evolution_2016,
	title = {The {Evolution} of {Metallicity} and {Metallicity} {Gradients} from z = 2.7 to 0.6 with {KMOS3D}},
	volume = {827},
	issn = {0004-637X},
	url = {https://ui.adsabs.harvard.edu/abs/2016ApJ...827...74W},
	doi = {10.3847/0004-637X/827/1/74},
	urldate = {2025-11-24},
	journal = {The Astrophysical Journal},
	publisher = {IOP},
	author = {Wuyts, Eva and Wisnioski, Emily and Fossati, Matteo and Förster Schreiber, Natascha M. and Genzel, Reinhard and Davies, Ric and Mendel, J. Trevor and Naab, Thorsten and Röttgers, Bernhard and Wilman, David J. and Wuyts, Stijn and Bandara, Kaushala and Beifiori, Alessandra and Belli, Sirio and Bender, Ralf and Brammer, Gabriel B. and Burkert, Andreas and Chan, Jeffrey and Galametz, Audrey and Kulkarni, Sandesh K. and Lang, Philipp and Lutz, Dieter and Momcheva, Ivelina G. and Nelson, Erica J. and Rosario, David and Saglia, Roberto P. and Seitz, Stella and Tacconi, Linda J. and Tadaki, Ken-ichi and Übler, Hannah and van Dokkum, Pieter},
	month = aug,
	year = {2016},
	note = {ADS Bibcode: 2016ApJ...827...74W},
	pages = {74},
}

@article{leethochawalit_keck_2016,
	title = {A {Keck} {Adaptive} {Optics} {Survey} of a {Representative} {Sample} of {Gravitationally} {Lensed} {Star}-forming {Galaxies}: {High} {Spatial} {Resolution} {Studies} of {Kinematics} and {Metallicity} {Gradients}},
	volume = {820},
	issn = {0004-637X},
	shorttitle = {A {Keck} {Adaptive} {Optics} {Survey} of a {Representative} {Sample} of {Gravitationally} {Lensed} {Star}-forming {Galaxies}},
	url = {https://ui.adsabs.harvard.edu/abs/2016ApJ...820...84L},
	doi = {10.3847/0004-637X/820/2/84},
	urldate = {2025-11-24},
	journal = {The Astrophysical Journal},
	publisher = {IOP},
	author = {Leethochawalit, Nicha and Jones, Tucker A. and Ellis, Richard S. and Stark, Daniel P. and Richard, Johan and Zitrin, Adi and Auger, Matthew},
	month = apr,
	year = {2016},
	note = {ADS Bibcode: 2016ApJ...820...84L},
	pages = {84},
}

@article{swinbank_properties_2012,
	title = {The properties of the star-forming interstellar medium at z = 0.84-2.23 from {HiZELS}: mapping the internal dynamics and metallicity gradients in high-redshift disc galaxies},
	volume = {426},
	issn = {0035-8711},
	shorttitle = {The properties of the star-forming interstellar medium at z = 0.84-2.23 from {HiZELS}},
	url = {https://ui.adsabs.harvard.edu/abs/2012MNRAS.426..935S},
	doi = {10.1111/j.1365-2966.2012.21774.x},
	urldate = {2025-11-24},
	journal = {Monthly Notices of the Royal Astronomical Society},
	publisher = {OUP},
	author = {Swinbank, A. M. and Sobral, D. and Smail, Ian and Geach, J. E. and Best, P. N. and McCarthy, I. G. and Crain, R. A. and Theuns, T.},
	month = oct,
	year = {2012},
	note = {ADS Bibcode: 2012MNRAS.426..935S},
	pages = {935--950},
}

@article{queyrel_massiv_2012,
	title = {{MASSIV}: {Mass} {Assembly} {Survey} with {SINFONI} in {VVDS}. {III}. {Evidence} for positive metallicity gradients in z {\textasciitilde} 1.2 star-forming galaxies},
	volume = {539},
	issn = {0004-6361},
	shorttitle = {{MASSIV}},
	url = {https://ui.adsabs.harvard.edu/abs/2012A&A...539A..93Q},
	doi = {10.1051/0004-6361/201117718},
	urldate = {2025-11-24},
	journal = {Astronomy and Astrophysics},
	publisher = {EDP},
	author = {Queyrel, J. and Contini, T. and Kissler-Patig, M. and Epinat, B. and Amram, P. and Garilli, B. and Le Fèvre, O. and Moultaka, J. and Paioro, L. and Tasca, L. and Tresse, L. and Vergani, D. and López-Sanjuan, C. and Perez-Montero, E.},
	month = mar,
	year = {2012},
	note = {ADS Bibcode: 2012A\&A...539A..93Q},
	pages = {A93},
}

@article{yuan_metallicity_2011,
	title = {Metallicity {Gradient} of a {Lensed} {Face}-on {Spiral} {Galaxy} at {Redshift} 1.49},
	volume = {732},
	issn = {0004-637X},
	url = {https://ui.adsabs.harvard.edu/abs/2011ApJ...732L..14Y},
	doi = {10.1088/2041-8205/732/1/L14},
	urldate = {2025-11-24},
	journal = {The Astrophysical Journal},
	publisher = {IOP},
	author = {Yuan, T.-T. and Kewley, L. J. and Swinbank, A. M. and Richard, J. and Livermore, R. C.},
	month = may,
	year = {2011},
	note = {ADS Bibcode: 2011ApJ...732L..14Y},
	pages = {L14},
}

@article{sanchez_califa_2012,
	title = {{CALIFA}, the {Calar} {Alto} {Legacy} {Integral} {Field} {Area} survey. {I}. {Survey} presentation},
	volume = {538},
	issn = {0004-6361},
	url = {https://ui.adsabs.harvard.edu/abs/2012A&A...538A...8S},
	doi = {10.1051/0004-6361/201117353},
	urldate = {2025-11-24},
	journal = {Astronomy and Astrophysics},
	publisher = {EDP},
	author = {Sánchez, S. F. and Kennicutt, R. C. and Gil de Paz, A. and van de Ven, G. and Vílchez, J. M. and Wisotzki, L. and Walcher, C. J. and Mast, D. and Aguerri, J. A. L. and Albiol-Pérez, S. and Alonso-Herrero, A. and Alves, J. and Bakos, J. and Bartáková, T. and Bland-Hawthorn, J. and Boselli, A. and Bomans, D. J. and Castillo-Morales, A. and Cortijo-Ferrero, C. and de Lorenzo-Cáceres, A. and Del Olmo, A. and Dettmar, R.-J. and Díaz, A. and Ellis, S. and Falcón-Barroso, J. and Flores, H. and Gallazzi, A. and García-Lorenzo, B. and González Delgado, R. and Gruel, N. and Haines, T. and Hao, C. and Husemann, B. and Iglésias-Páramo, J. and Jahnke, K. and Johnson, B. and Jungwiert, B. and Kalinova, V. and Kehrig, C. and Kupko, D. and López-Sánchez, Á. R. and Lyubenova, M. and Marino, R. A. and Mármol-Queraltó, E. and Márquez, I. and Masegosa, J. and Meidt, S. and Mendez-Abreu, J. and Monreal-Ibero, A. and Montijo, C. and Mourão, A. M. and Palacios-Navarro, G. and Papaderos, P. and Pasquali, A. and Peletier, R. and Pérez, E. and Pérez, I. and Quirrenbach, A. and Relaño, M. and Rosales-Ortega, F. F. and Roth, M. M. and Ruiz-Lara, T. and Sánchez-Blázquez, P. and Sengupta, C. and Singh, R. and Stanishev, V. and Trager, S. C. and Vazdekis, A. and Viironen, K. and Wild, V. and Zibetti, S. and Ziegler, B.},
	month = feb,
	year = {2012},
	note = {ADS Bibcode: 2012A\&A...538A...8S},
	pages = {A8},
}

@article{croom_sydney-aao_2012,
	title = {The {Sydney}-{AAO} {Multi}-object {Integral} field spectrograph},
	volume = {421},
	issn = {0035-8711},
	url = {https://ui.adsabs.harvard.edu/abs/2012MNRAS.421..872C},
	doi = {10.1111/j.1365-2966.2011.20365.x},
	urldate = {2025-11-24},
	journal = {Monthly Notices of the Royal Astronomical Society},
	publisher = {OUP},
	author = {Croom, Scott M. and Lawrence, Jon S. and Bland-Hawthorn, Joss and Bryant, Julia J. and Fogarty, Lisa and Richards, Samuel and Goodwin, Michael and Farrell, Tony and Miziarski, Stan and Heald, Ron and Jones, D. Heath and Lee, Steve and Colless, Matthew and Brough, Sarah and Hopkins, Andrew M. and Bauer, Amanda E. and Birchall, Michael N. and Ellis, Simon and Horton, Anthony and Leon-Saval, Sergio and Lewis, Geraint and López-Sánchez, Á. R. and Min, Seong-Sik and Trinh, Christopher and Trowland, Holly},
	month = mar,
	year = {2012},
	note = {ADS Bibcode: 2012MNRAS.421..872C},
	pages = {872--893},
}

@article{berg_chaos_2015,
	title = {{CHAOS} {I}. {Direct} {Chemical} {Abundances} for {H} {II} {Regions} in {NGC} 628},
	volume = {806},
	issn = {0004-637X},
	url = {https://ui.adsabs.harvard.edu/abs/2015ApJ...806...16B},
	doi = {10.1088/0004-637X/806/1/16},
	urldate = {2025-11-21},
	journal = {The Astrophysical Journal},
	publisher = {IOP},
	author = {Berg, Danielle A. and Skillman, Evan D. and Croxall, Kevin V. and Pogge, Richard W. and Moustakas, John and Johnson-Groh, Mara},
	month = jun,
	year = {2015},
	note = {ADS Bibcode: 2015ApJ...806...16B},
	pages = {16},
}

@article{bresolin_abundance_2011,
	title = {The {Abundance} {Scatter} in {M33} from {H} {II} {Regions}: {Is} {There} {Any} {Evidence} for {Azimuthal} {Metallicity} {Variations}?},
	volume = {730},
	issn = {0004-637X},
	shorttitle = {The {Abundance} {Scatter} in {M33} from {H} {II} {Regions}},
	url = {https://ui.adsabs.harvard.edu/abs/2011ApJ...730..129B},
	doi = {10.1088/0004-637X/730/2/129},
	urldate = {2025-11-21},
	journal = {The Astrophysical Journal},
	publisher = {IOP},
	author = {Bresolin, Fabio},
	month = apr,
	year = {2011},
	note = {ADS Bibcode: 2011ApJ...730..129B},
	pages = {129},
}

@article{zaritsky_h_1994,
	title = {H {II} {Regions} and the {Abundance} {Properties} of {Spiral} {Galaxies}},
	volume = {420},
	issn = {0004-637X},
	url = {https://ui.adsabs.harvard.edu/abs/1994ApJ...420...87Z},
	doi = {10.1086/173544},
	urldate = {2025-11-21},
	journal = {The Astrophysical Journal},
	publisher = {IOP},
	author = {Zaritsky, Dennis and Kennicutt, Jr., Robert C. and Huchra, John P.},
	month = jan,
	year = {1994},
	note = {ADS Bibcode: 1994ApJ...420...87Z},
	pages = {87},
}

@article{magrini_metal_2010,
	title = {Metal production in {M} 33: space and time variations},
	volume = {512},
	issn = {0004-6361},
	shorttitle = {Metal production in {M} 33},
	url = {https://ui.adsabs.harvard.edu/abs/2010A&A...512A..63M},
	doi = {10.1051/0004-6361/200913564},
	urldate = {2025-11-21},
	journal = {Astronomy and Astrophysics},
	publisher = {EDP},
	author = {Magrini, L. and Stanghellini, L. and Corbelli, E. and Galli, D. and Villaver, E.},
	month = mar,
	year = {2010},
	note = {ADS Bibcode: 2010A\&A...512A..63M},
	pages = {A63},
}

@article{mckee_theory_2007,
	title = {Theory of {Star} {Formation}},
	volume = {45},
	issn = {0066-4146},
	url = {https://ui.adsabs.harvard.edu/abs/2007ARA&A..45..565M},
	doi = {10.1146/annurev.astro.45.051806.110602},
	urldate = {2025-11-21},
	journal = {Annual Review of Astronomy and Astrophysics},
	author = {McKee, Christopher F. and Ostriker, Eve C.},
	month = sep,
	year = {2007},
	note = {ADS Bibcode: 2007ARA\&A..45..565M},
	pages = {565--687},
}

@article{koeppen_evolution_1994,
	title = {The evolution of abundance gradients in spiral galaxies},
	volume = {281},
	issn = {0004-6361},
	url = {https://ui.adsabs.harvard.edu/abs/1994A&A...281...26K},
	urldate = {2025-11-21},
	journal = {Astronomy and Astrophysics},
	publisher = {EDP},
	author = {Koeppen, J.},
	month = jan,
	year = {1994},
	note = {ADS Bibcode: 1994A\&A...281...26K},
	pages = {26--34},
}

@article{keres_how_2005,
	title = {How do galaxies get their gas?},
	volume = {363},
	issn = {0035-8711},
	url = {https://doi.org/10.1111/j.1365-2966.2005.09451.x},
	doi = {10.1111/j.1365-2966.2005.09451.x},
	number = {1},
	urldate = {2025-11-21},
	journal = {Monthly Notices of the Royal Astronomical Society},
	author = {Kereš, Dušan and Katz, Neal and Weinberg, David H. and Davé, Romeel},
	month = oct,
	year = {2005},
	pages = {2--28},
}

@article{cresci_gas_2010,
	title = {Gas accretion as the origin of chemical abundance gradients in distant galaxies},
	volume = {467},
	copyright = {2010 Springer Nature Limited},
	issn = {1476-4687},
	url = {https://www.nature.com/articles/nature09451},
	doi = {10.1038/nature09451},
	language = {en},
	number = {7317},
	urldate = {2025-11-21},
	journal = {Nature},
	publisher = {Nature Publishing Group},
	author = {Cresci, G. and Mannucci, F. and Maiolino, R. and Marconi, A. and Gnerucci, A. and Magrini, L.},
	month = oct,
	year = {2010},
	pages = {811--813},
}

@article{ceverino_gas_2016,
	title = {Gas inflow and metallicity drops in star-forming galaxies},
	volume = {457},
	issn = {0035-8711},
	url = {https://ui.adsabs.harvard.edu/abs/2016MNRAS.457.2605C},
	doi = {10.1093/mnras/stw064},
	urldate = {2025-11-21},
	journal = {Monthly Notices of the Royal Astronomical Society},
	publisher = {OUP},
	author = {Ceverino, Daniel and Sánchez Almeida, Jorge and Muñoz Tuñón, Casiana and Dekel, Avishai and Elmegreen, Bruce G. and Elmegreen, Debra M. and Primack, Joel},
	month = apr,
	year = {2016},
	note = {ADS Bibcode: 2016MNRAS.457.2605C},
	pages = {2605--2612},
}

@article{torres-flores_star-forming_2014,
	title = {Star-forming regions and the metallicity gradients in the tidal tails: the case of {NGC} 92},
	volume = {438},
	issn = {0035-8711},
	shorttitle = {Star-forming regions and the metallicity gradients in the tidal tails},
	url = {https://ui.adsabs.harvard.edu/abs/2014MNRAS.438.1894T},
	doi = {10.1093/mnras/stt2340},
	urldate = {2025-11-20},
	journal = {Monthly Notices of the Royal Astronomical Society},
	publisher = {OUP},
	author = {Torres-Flores, S. and Scarano, S. and Mendes de Oliveira, C. and de Mello, D. F. and Amram, P. and Plana, H.},
	month = feb,
	year = {2014},
	note = {ADS Bibcode: 2014MNRAS.438.1894T},
	pages = {1894--1908},
}

@article{rich_integral_2012,
	title = {An {Integral} {Field} {Study} of {Abundance} {Gradients} in nearby {Luminous} {Infrared} {Galaxies}},
	volume = {753},
	issn = {0004-637X},
	url = {https://ui.adsabs.harvard.edu/abs/2012ApJ...753....5R},
	doi = {10.1088/0004-637X/753/1/5},
	urldate = {2025-11-20},
	journal = {The Astrophysical Journal},
	publisher = {IOP},
	author = {Rich, J. A. and Torrey, P. and Kewley, L. J. and Dopita, M. A. and Rupke, D. S. N.},
	month = jul,
	year = {2012},
	note = {ADS Bibcode: 2012ApJ...753....5R},
	pages = {5},
}

@article{rupke_gas-phase_2010,
	title = {Gas-phase {Oxygen} {Gradients} in {Strongly} {Interacting} {Galaxies}. {I}. {Early}-stage {Interactions}},
	volume = {723},
	issn = {0004-637X},
	url = {https://ui.adsabs.harvard.edu/abs/2010ApJ...723.1255R},
	doi = {10.1088/0004-637X/723/2/1255},
	urldate = {2025-11-20},
	journal = {The Astrophysical Journal},
	publisher = {IOP},
	author = {Rupke, David S. N. and Kewley, Lisa J. and Chien, L.-H.},
	month = nov,
	year = {2010},
	note = {ADS Bibcode: 2010ApJ...723.1255R},
	pages = {1255--1271},
}

@article{curti_klever_2020,
	title = {The {KLEVER} {Survey}: spatially resolved metallicity maps and gradients in a sample of 1.2 {\textless} z {\textless} 2.5 lensed galaxies},
	volume = {492},
	issn = {0035-8711},
	shorttitle = {The {KLEVER} {Survey}},
	url = {https://ui.adsabs.harvard.edu/abs/2020MNRAS.492..821C},
	doi = {10.1093/mnras/stz3379},
	urldate = {2025-11-20},
	journal = {Monthly Notices of the Royal Astronomical Society},
	publisher = {OUP},
	author = {Curti, Mirko and Maiolino, Roberto and Cirasuolo, Michele and Mannucci, Filippo and Williams, Rebecca J. and Auger, Matt and Mercurio, Amata and Hayden-Pawson, Connor and Cresci, Giovanni and Marconi, Alessandro and Belfiore, Francesco and Cappellari, Michele and Cicone, Claudia and Cullen, Fergus and Meneghetti, Massimo and Ota, Kazuaki and Peng, Yingjie and Pettini, Max and Swinbank, Mark and Troncoso, Paulina},
	month = feb,
	year = {2020},
	note = {ADS Bibcode: 2020MNRAS.492..821C},
	pages = {821--842},
}

@article{leja_how_2019,
	title = {How to {Measure} {Galaxy} {Star} {Formation} {Histories}. {II}. {Nonparametric} {Models}},
	volume = {876},
	issn = {0004-637X},
	url = {https://ui.adsabs.harvard.edu/abs/2019ApJ...876....3L},
	doi = {10.3847/1538-4357/ab133c},
	urldate = {2025-11-17},
	journal = {The Astrophysical Journal},
	publisher = {IOP},
	author = {Leja, Joel and Carnall, Adam C. and Johnson, Benjamin D. and Conroy, Charlie and Speagle, Joshua S.},
	month = may,
	year = {2019},
	note = {ADS Bibcode: 2019ApJ...876....3L},
	pages = {3},
}

@article{wang_inferring_2023,
	title = {Inferring {More} from {Less}: {Prospector} as a {Photometric} {Redshift} {Engine} in the {Era} of {JWST}},
	volume = {944},
	issn = {0004-637X},
	shorttitle = {Inferring {More} from {Less}},
	url = {https://ui.adsabs.harvard.edu/abs/2023ApJ...944L..58W},
	doi = {10.3847/2041-8213/acba99},
	urldate = {2025-11-17},
	journal = {The Astrophysical Journal},
	publisher = {IOP},
	author = {Wang, Bingjie and Leja, Joel and Bezanson, Rachel and Johnson, Benjamin D. and Khullar, Gourav and Labbé, Ivo and Price, Sedona H. and Weaver, John R. and Whitaker, Katherine E.},
	month = feb,
	year = {2023},
	note = {ADS Bibcode: 2023ApJ...944L..58W},
	pages = {L58},
}

@article{noll_analysis_2009,
	title = {Analysis of galaxy spectral energy distributions from far-{UV} to far-{IR} with {CIGALE}: studying a {SINGS} test sample},
	volume = {507},
	issn = {0004-6361},
	shorttitle = {Analysis of galaxy spectral energy distributions from far-{UV} to far-{IR} with {CIGALE}},
	url = {https://ui.adsabs.harvard.edu/abs/2009A&A...507.1793N},
	doi = {10.1051/0004-6361/200912497},
	urldate = {2025-11-17},
	journal = {Astronomy and Astrophysics},
	publisher = {EDP},
	author = {Noll, S. and Burgarella, D. and Giovannoli, E. and Buat, V. and Marcillac, D. and Muñoz-Mateos, J. C.},
	month = dec,
	year = {2009},
	note = {ADS Bibcode: 2009A\&A...507.1793N},
	pages = {1793--1813},
}

@article{charlot_simple_2000,
	title = {A {Simple} {Model} for the {Absorption} of {Starlight} by {Dust} in {Galaxies}},
	volume = {539},
	issn = {0004-637X},
	url = {https://ui.adsabs.harvard.edu/abs/2000ApJ...539..718C},
	doi = {10.1086/309250},
	urldate = {2025-11-17},
	journal = {The Astrophysical Journal},
	publisher = {IOP},
	author = {Charlot, Stéphane and Fall, S. Michael},
	month = aug,
	year = {2000},
	note = {ADS Bibcode: 2000ApJ...539..718C},
	pages = {718--731},
}

@article{sanchez-blazquez_medium-resolution_2006,
	title = {Medium-resolution {Isaac} {Newton} {Telescope} library of empirical spectra},
	volume = {371},
	issn = {0035-8711},
	url = {https://ui.adsabs.harvard.edu/abs/2006MNRAS.371..703S},
	doi = {10.1111/j.1365-2966.2006.10699.x},
	urldate = {2025-11-17},
	journal = {Monthly Notices of the Royal Astronomical Society},
	publisher = {OUP},
	author = {Sánchez-Blázquez, P. and Peletier, R. F. and Jiménez-Vicente, J. and Cardiel, N. and Cenarro, A. J. and Falcón-Barroso, J. and Gorgas, J. and Selam, S. and Vazdekis, A.},
	month = sep,
	year = {2006},
	note = {ADS Bibcode: 2006MNRAS.371..703S},
	pages = {703--718},
}

@article{choi_mesa_2016,
	title = {Mesa {Isochrones} and {Stellar} {Tracks} ({MIST}). {I}. {Solar}-scaled {Models}},
	volume = {823},
	issn = {0004-637X},
	url = {https://ui.adsabs.harvard.edu/abs/2016ApJ...823..102C},
	doi = {10.3847/0004-637X/823/2/102},
	urldate = {2025-11-17},
	journal = {The Astrophysical Journal},
	publisher = {IOP},
	author = {Choi, Jieun and Dotter, Aaron and Conroy, Charlie and Cantiello, Matteo and Paxton, Bill and Johnson, Benjamin D.},
	month = jun,
	year = {2016},
	note = {ADS Bibcode: 2016ApJ...823..102C},
	pages = {102},
}

@misc{brammer_grizli_2023,
	title = {grizli},
	url = {https://zenodo.org/records/8370018},
	doi = {10.5281/zenodo.8370018},
	urldate = {2025-10-28},
	publisher = {Zenodo},
	author = {Brammer, Gabriel},
	month = sep,
	year = {2023},
}


\appendix

\section{BLR and outflow in SMACS0723\_4590}\label{app:SMACS4590_BLR_outflow}

\begin{figure}
    \centering
    \includegraphics[width=0.4\paperwidth]{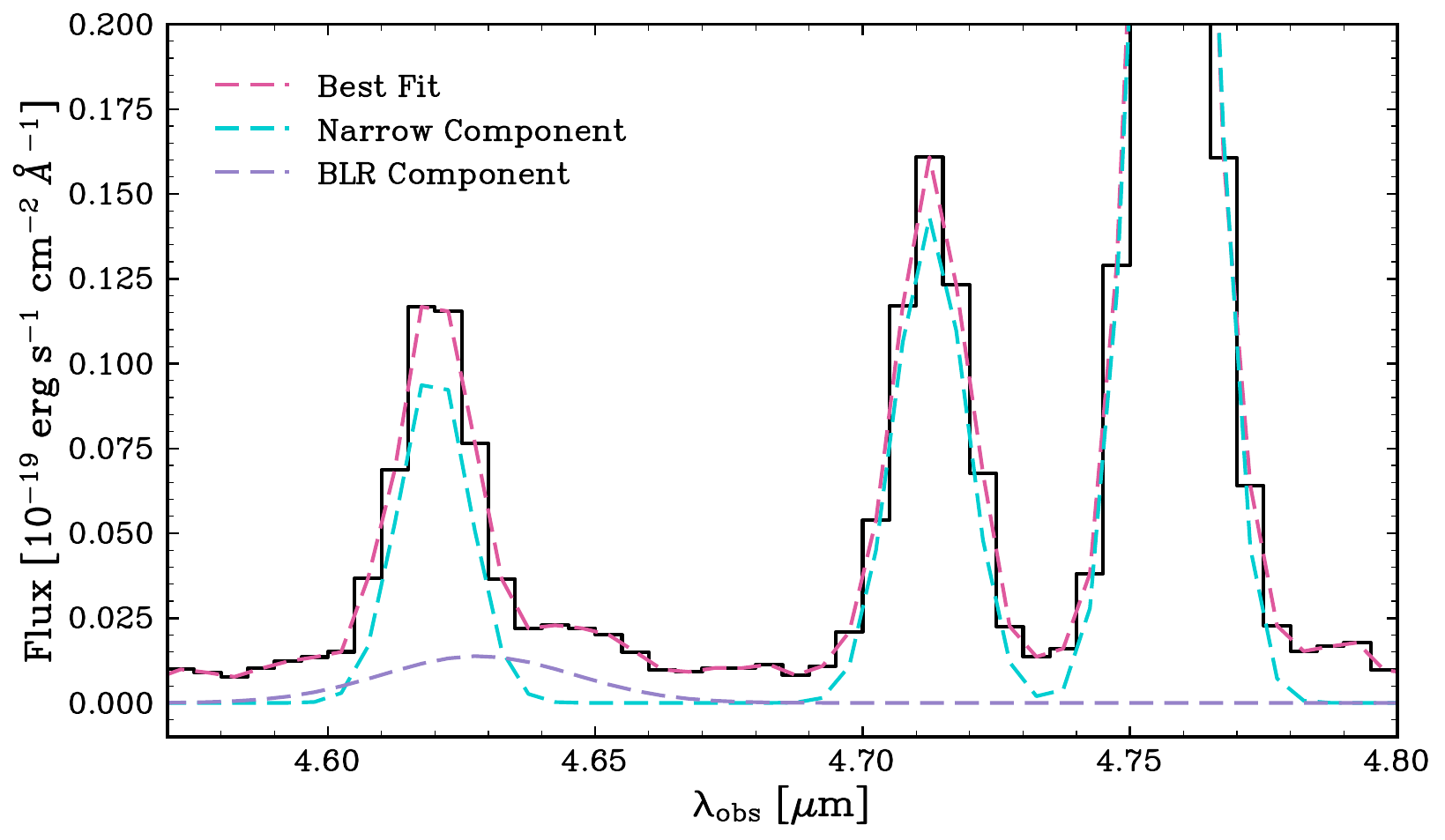}
    \includegraphics[width=0.4\paperwidth]{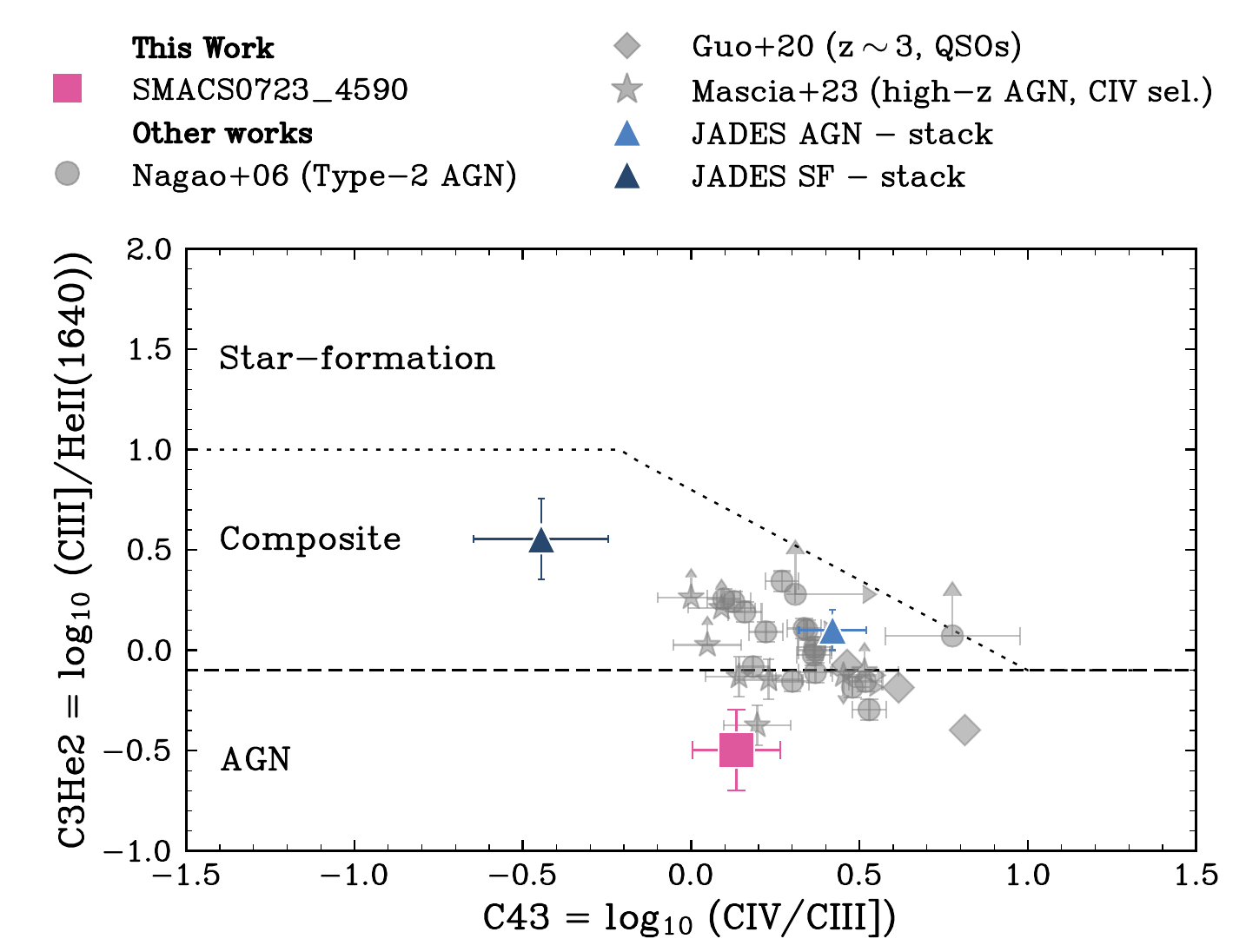}
    \includegraphics[width=0.4\paperwidth]{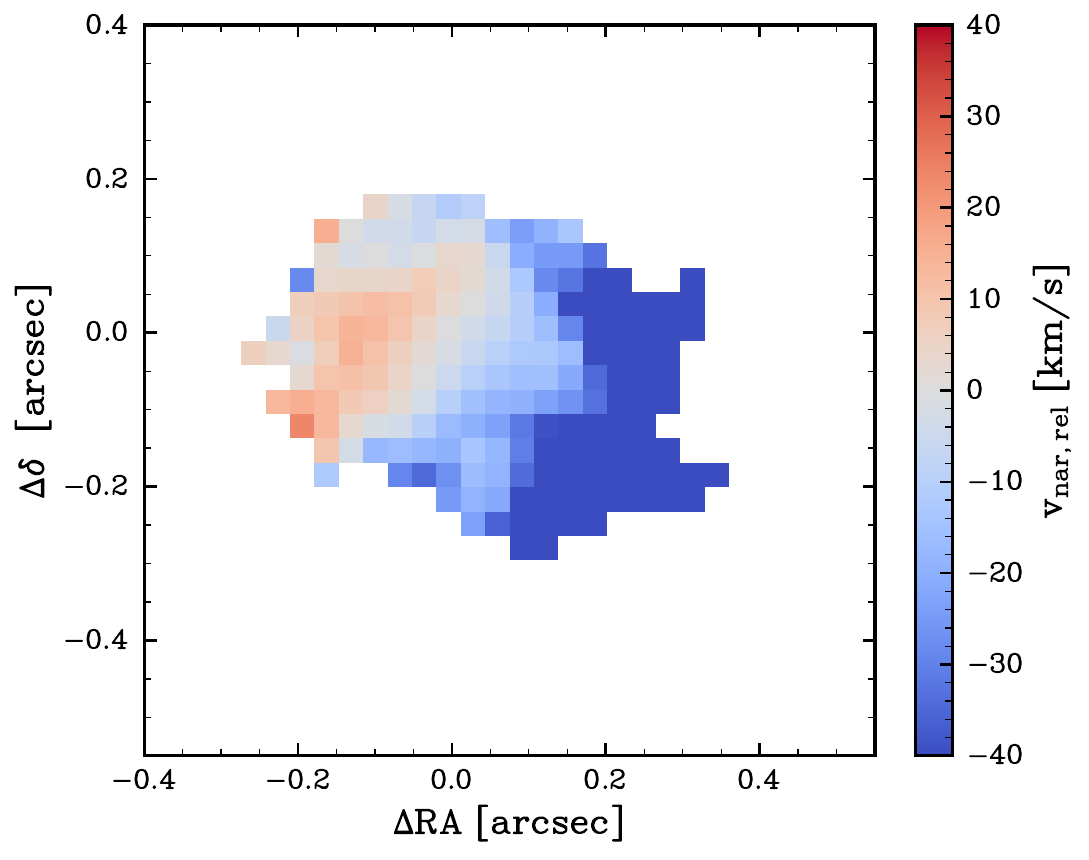}
    \caption{Evidence for AGN and outflow in SMACS0723\_4590. \textit{Top}: Detection of broad H$\beta$ component (purple dashed line) not seen in the $\rm [OIII]\lambda 5007$+H$\beta$. The data, the best-fit, and the narrow component in the lines are shown as black, pink, and blue lines, respectively. \textit{Middle}: UV emission line diagnostics based on CIV$\lambda1550$, CIII]$\lambda$1908, HeII$\lambda$1640 from the PRISM-IFS observations. The green squares, grey diamonds and blue diamonds show the type-2 AGN \citep{nagao_gas_2006},  CIV$\lambda1550$ detections from \citet{mascia_insights_2023} and quasars \citep{guo_high-redshift_2020}, respectively. The magenta and cyan squares show JADES stacks of AGN and star-forming galaxies from \citet{scholtz_jades_2025}. Based on this diagnostic, SMACS0723\_4590 is considered an AGN. \textit{Bottom}: 2D map the narrow line velocity, showing the increased offset, relative to the main galaxy, of the suspected outflow.}
    \label{fig:SMACS0723_4590_AGN}
\end{figure}

We detected a broad component in H$\beta$ in SMACS0723\_4590 in the integrated spectrum of the X target. We show the integrated spectrum highlighting the $\rm [OIII]\lambda 5007$+ H$\beta$ in the top panel of Figure~\ref{fig:SMACS0723_4590_AGN}. We find that the difference in the Bayesian information criterion (BIC) between the fits with and without a broad component in H$\beta$ is -20 in favour of the broad H$\beta$ component, showing a strong preference for the broad component in H$\beta$. We measure the FWHM of the broad H$\beta$ component of 2850$_{-290}^{350}$ km s$^{-1}$, with a velocity offset of 190$^{+30}_{-50}$ km s$^{-1}$ 

Presence of a broad component in permitted lines (such as H$\beta$) without any detection in forbidden lines (e.g. $\rm [OIII]\lambda 5007$) is a tell-tale sign of broad line region from an AGN \citep[e.g.][]{maiolino_jades_2024-1}. We further investigate the presence of AGN in this object based on the UV diagnostics in the bottom panel of Fig.~\ref{fig:SMACS0723_4590_AGN} - specifically using CIV$\lambda1550$, CIII]$\lambda$1908, HeII$\lambda$1640 used at high-z to identify AGN \citep{feltre_nuclear_2016, scholtz_jades_2025, mazzolari_narrow-line_2025}. The SMACS0723\_4590 is in the AGN part of this diagram. Based on this emission-line diagnostic and the presence of a broad component, we conclude that this object hosts an AGN.

We analysed the R2700 NIRSpec-IFS observations from a sister programme (PID: 2959, PI: J. Scholtz). The data were analysed using \texttt{QubeSpec} described in \citet[][]{scholtz_ga-nifs_2025}. We recover a velocity offset between the centre of SMACS0723\_4590 and the location of the possible outflow of roughly $\rm \delta v\sim 60 \ km \ s^{-1}$ (see bottom panel of Fig.~\ref{fig:SMACS0723_4590_AGN}). The velocity offset and enhanced metallicity in this region, together with the absence of any significant stellar continuum detection, suggest that this could indeed be an outflow driven by the suggested AGN within this source. In a similar way, \citet{deugenio_jades_2026} found evidence for an ionised outflow in a Little Red Dot (LRD) at $\rm z\sim 5$, which is also actively merging with another galaxy. This outflow exhibits a higher ionisation and dispersion compared to the main galaxy, which is in line with our results as indicated by the velocity offset and high $\rm R3$ measured within that area (see Fig.~\ref{fig:metal_grads_SMACS0723_4590}). 

Lastly, although we tentatively classify this source as an AGN, we infer metallicity gradients and other properties via calibrations based on star-forming galaxies. As reported in \citet{maiolino_black_2026} (Fig. B1), the harder ionising spectrum from an AGN primarily elevates the $\rm [OIII]/H\beta$ ratio; consequently, our derived metallicities would be only slightly lower than those inferred for a pure star-forming case. However, this discrepancy is minimal, and our results can therefore be treated as a robust characterisation of the galaxy's chemical enrichment.

\section{Compilation of spectra, NIRCam images, and emission line flux measurements of each galaxy}\label{app:spec_flux}

\begin{figure*}
    \centering
    \includegraphics[width=0.92\linewidth]{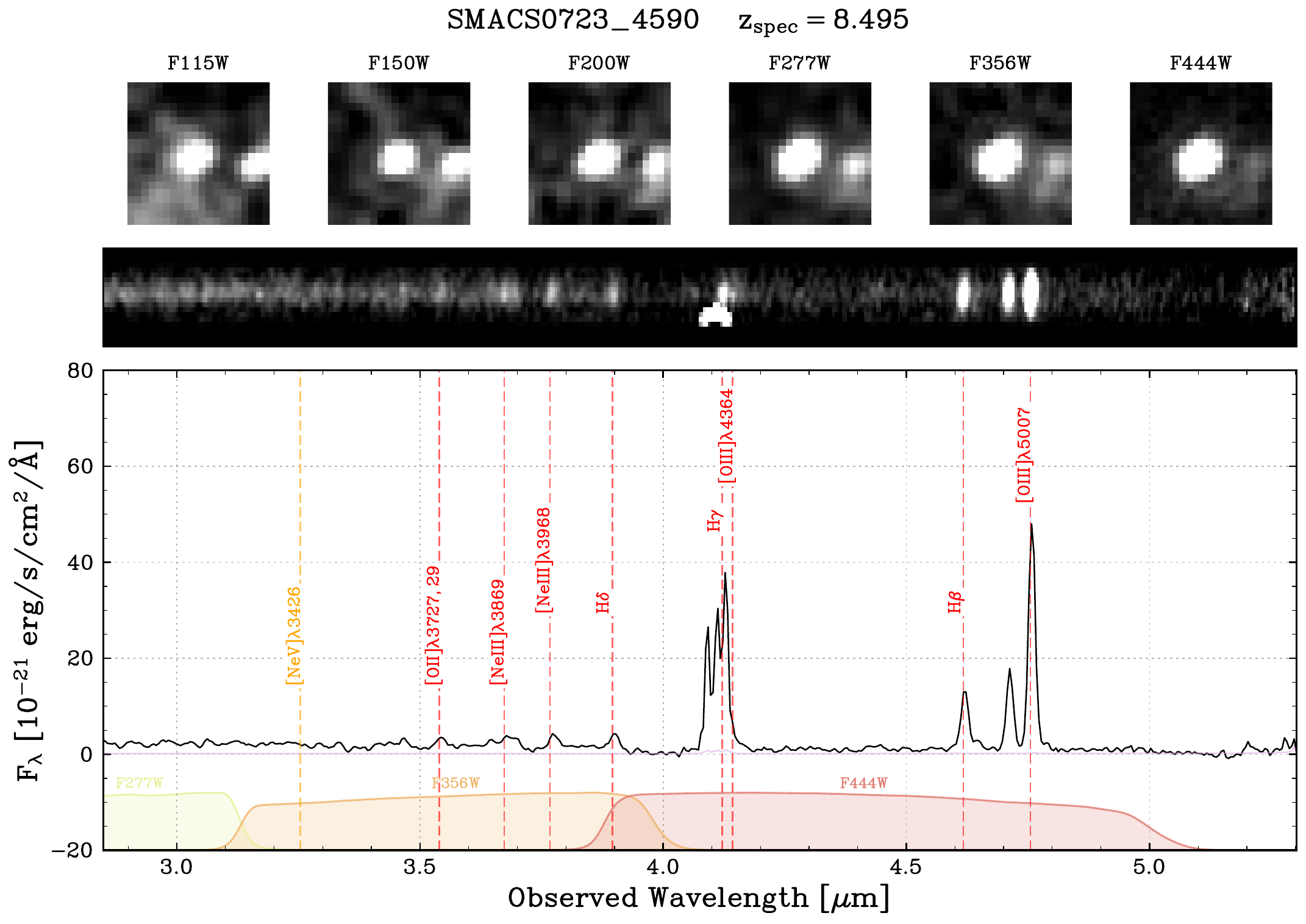}

    \vspace{0.2cm}

    \includegraphics[width=0.9\linewidth]{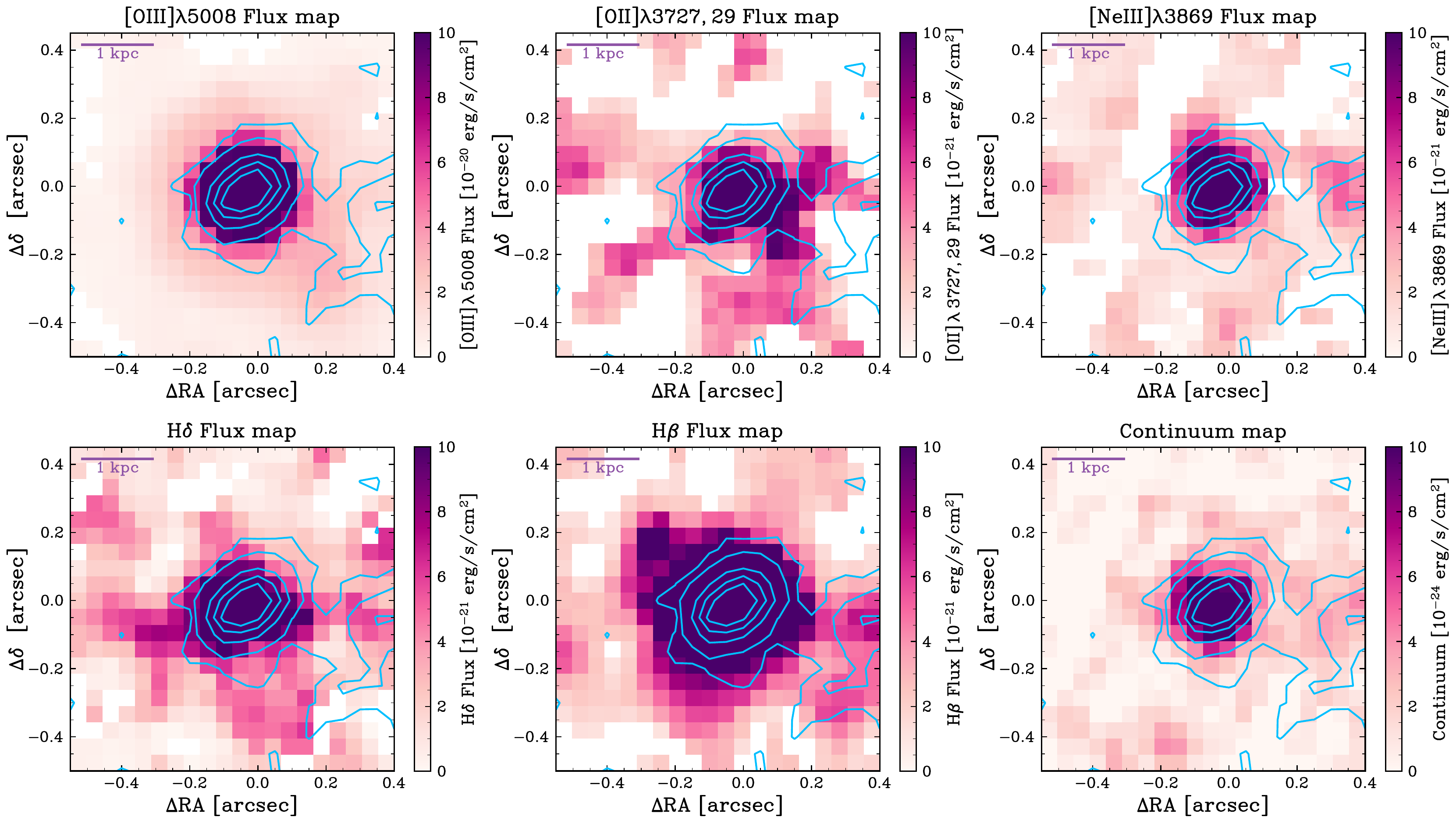}
    
    \caption{Integrated aperture spectrum and NIRCam filter (top) and emission line fluxes (bottom) measured via our Gaussian fitting routine (see section~\ref{sec:analysis-EL} for details) of SMACS0723\_4590. The first row depicts cutouts of the different NIRCam filters. For the spectrum, we present both the 2D spectrum at the top, obtained by collapsing the cube along only one spatial axis, and the traditional 1D spectrum below, which is obtained by collapsing along both spatial axes. The uncertainties of the 1D spectra are shown as a plum-coloured line. Both spectra are cut off at the lower limit of 3000\AA as the UV side is too noisy. Additionally, we highlight the different emission lines and corresponding NIRCam filter within the 1D spectrum. Here, we measured emission line fluxes using the $\rm 3 \times 3$-averaged spectra to provide smoothed views of the flux maps and display them for spaxels with $\rm S/N > 3$. The continuum map depicted in the bottom right is measured between $\rm H\gamma$ and $\rm H\beta$. Due to the inability to measure $\rm H\gamma$ and $\rm [OIII]\lambda 4364$ for this source, their two flux measurement subplots are not shown for this specific source. Blue contour lines trace the continuum measured via the F277W NIRCam filter for $\rm [3, 5, 10, 15, 20] \times RMS$ levels. The NIRSpec IFU cube was cut down to highlight the main galaxy.}
    \label{fig:spec_flux_SMACS0723_4590}
\end{figure*}

\begin{figure*}
    \centering
    \includegraphics[width=\linewidth]{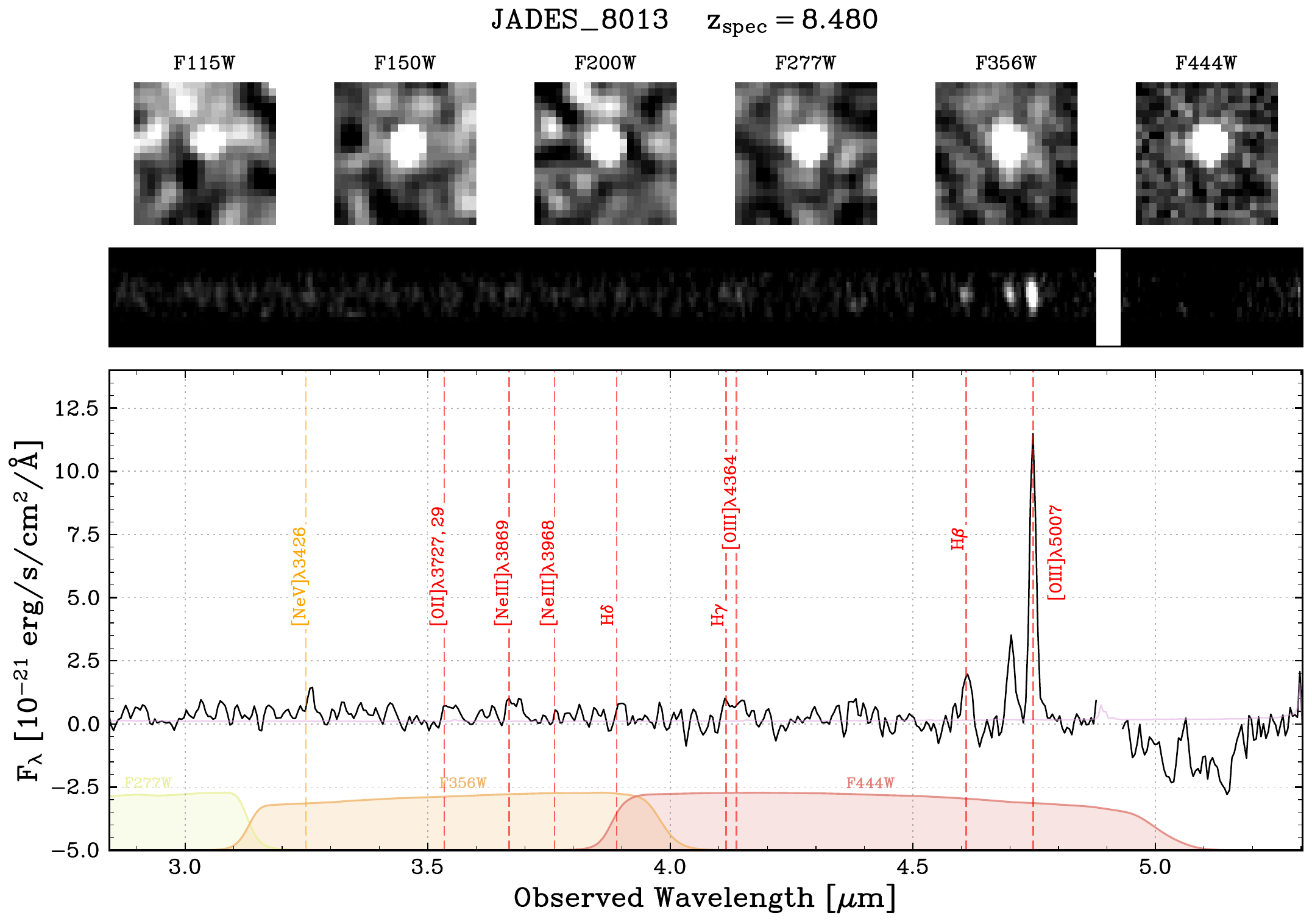}

    \vspace{0.3cm}

    \includegraphics[width=\linewidth]{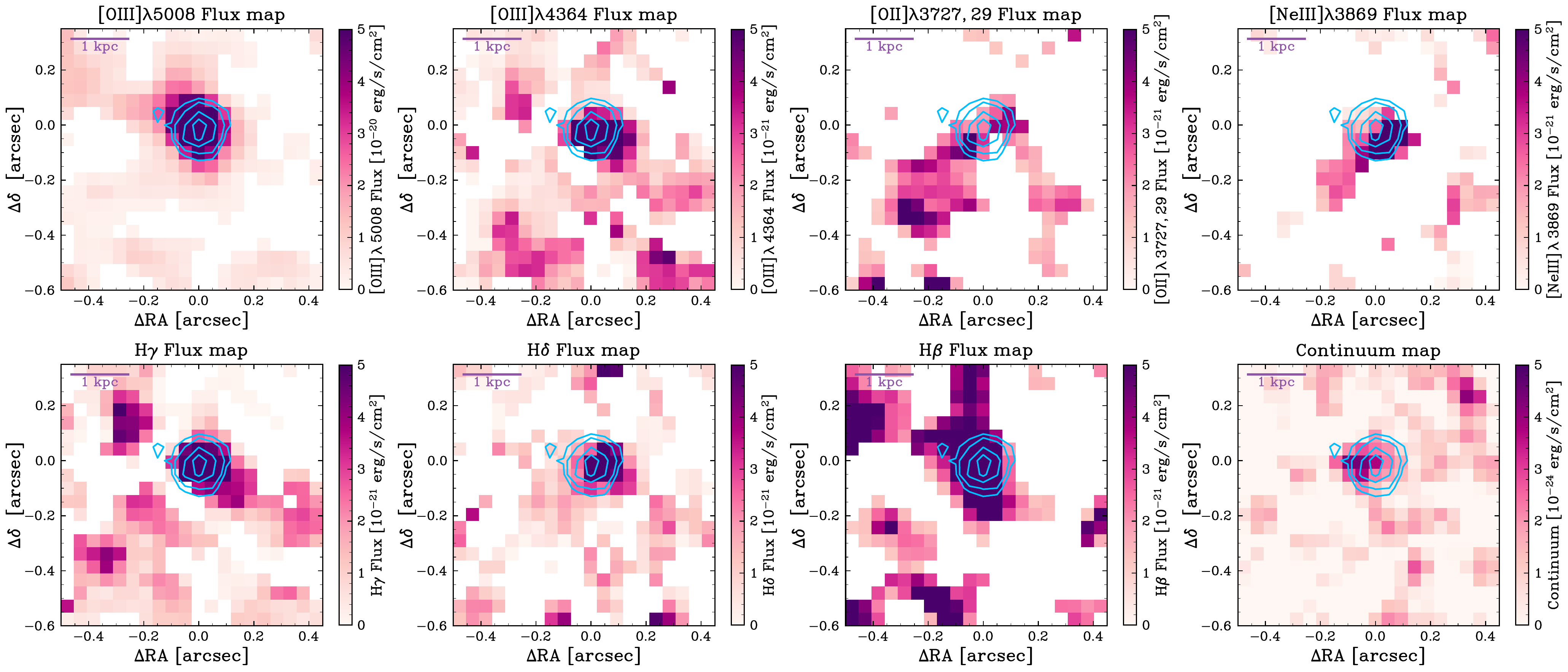}
    
    \caption{Same as Figure~\ref{fig:spec_flux_SMACS0723_4590} but for JADES\_8013.}
    \label{fig:spec_flux_JADES_8013}
\end{figure*}

\begin{figure*}
    \centering
    \includegraphics[width=\linewidth]{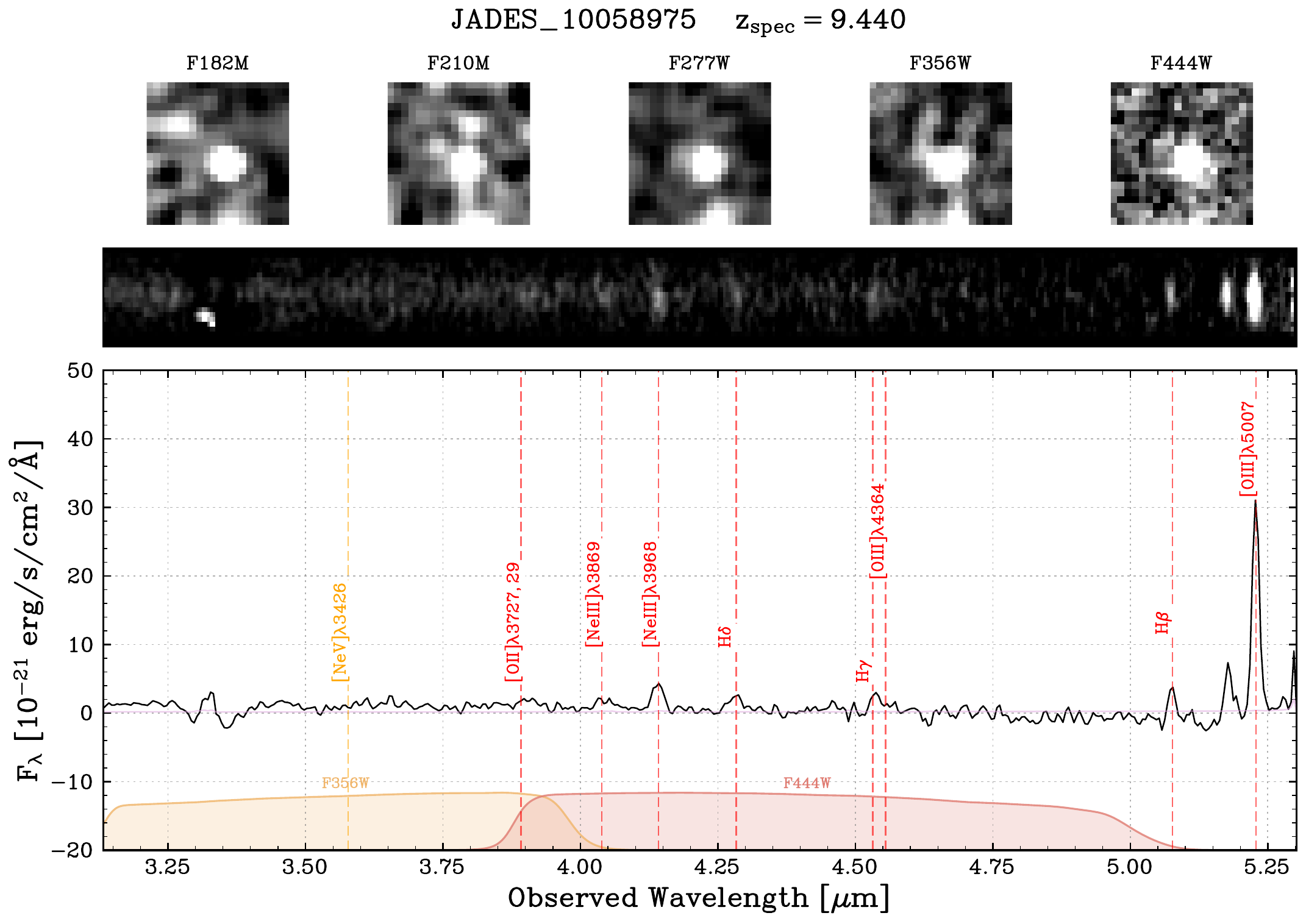}

    \vspace{0.3cm}

    \includegraphics[width=\linewidth]{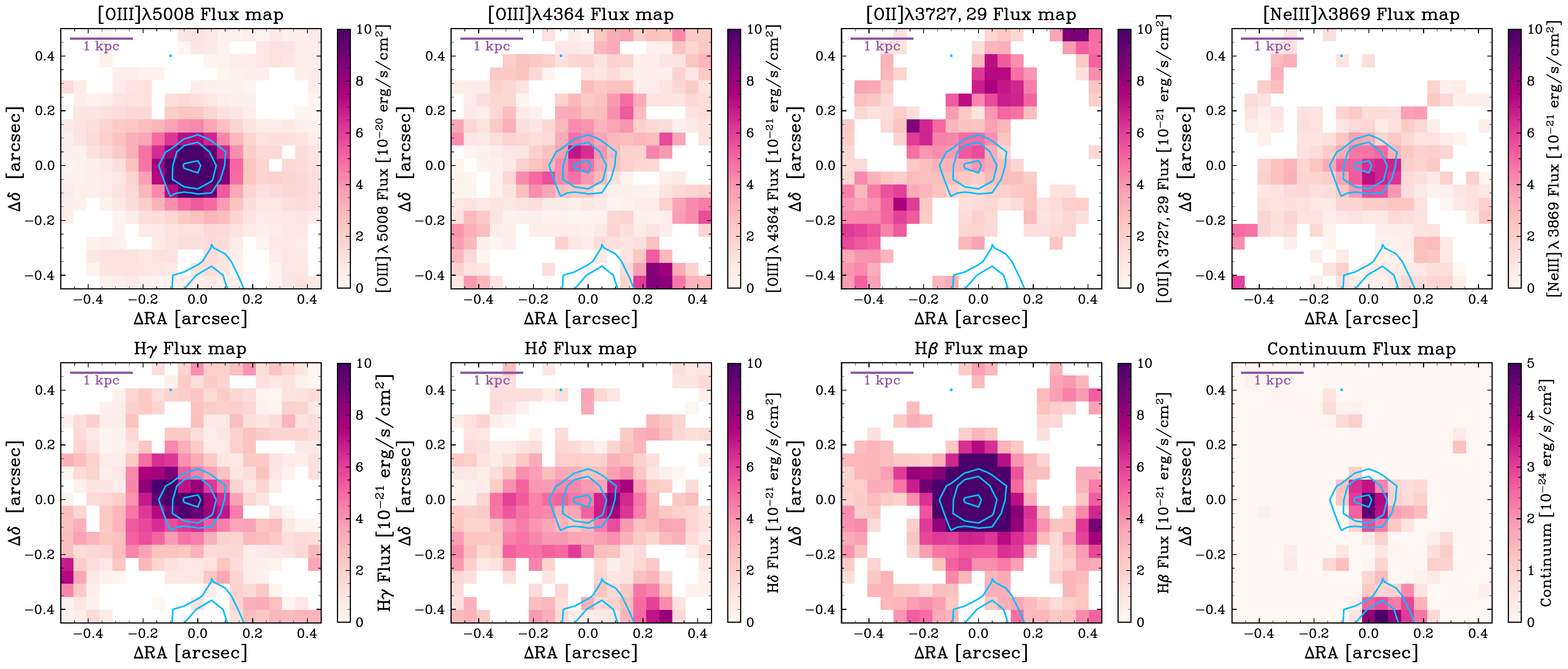}
    
    \caption{Same as Figure~\ref{fig:spec_flux_SMACS0723_4590} but for JADES\_10058975.}
    \label{fig:spec_flux_JADES_10058975}
\end{figure*}

\begin{figure*}
    \centering
    \includegraphics[width=\linewidth]{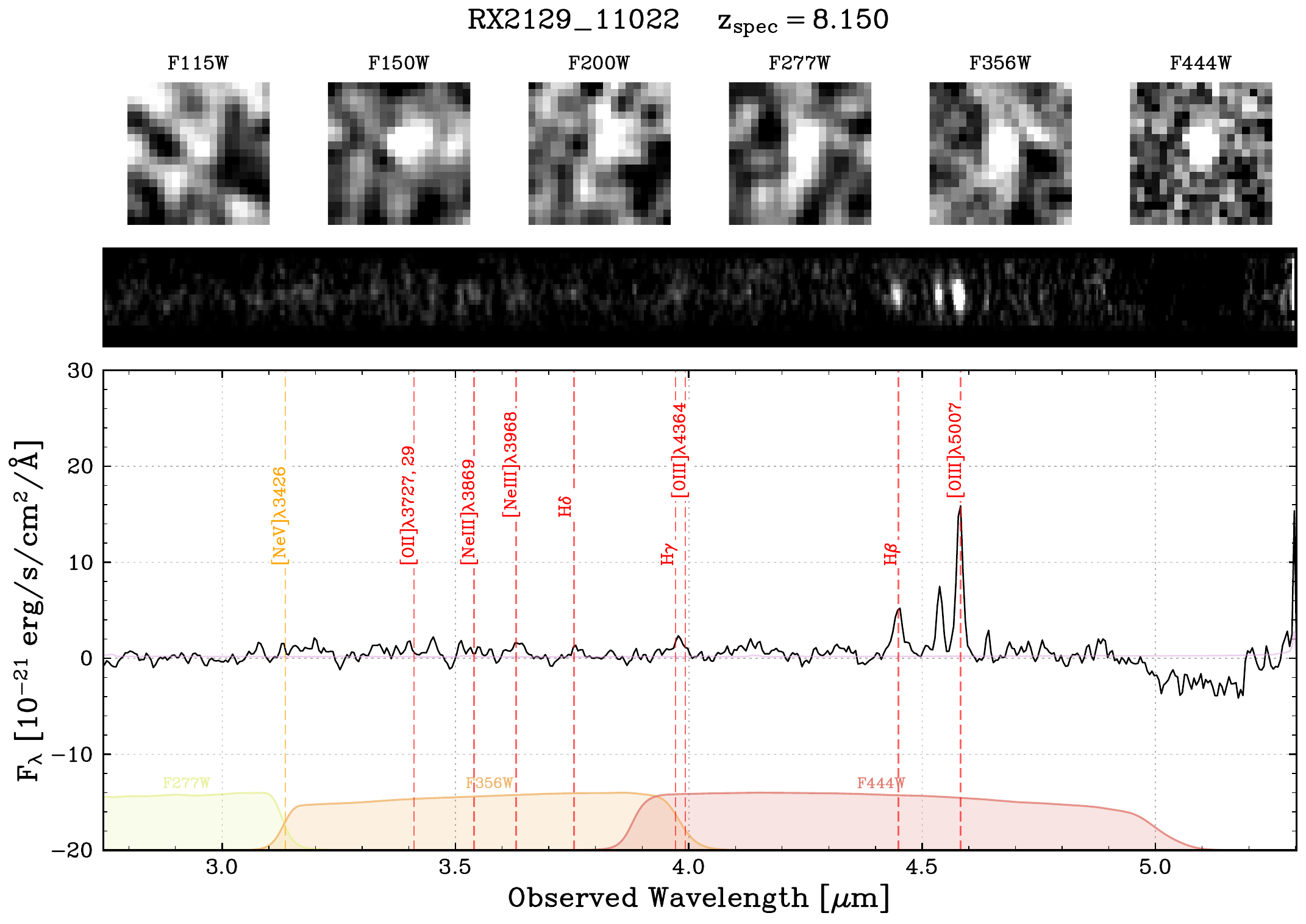}

    \vspace{0.3cm}

    \includegraphics[width=\linewidth]{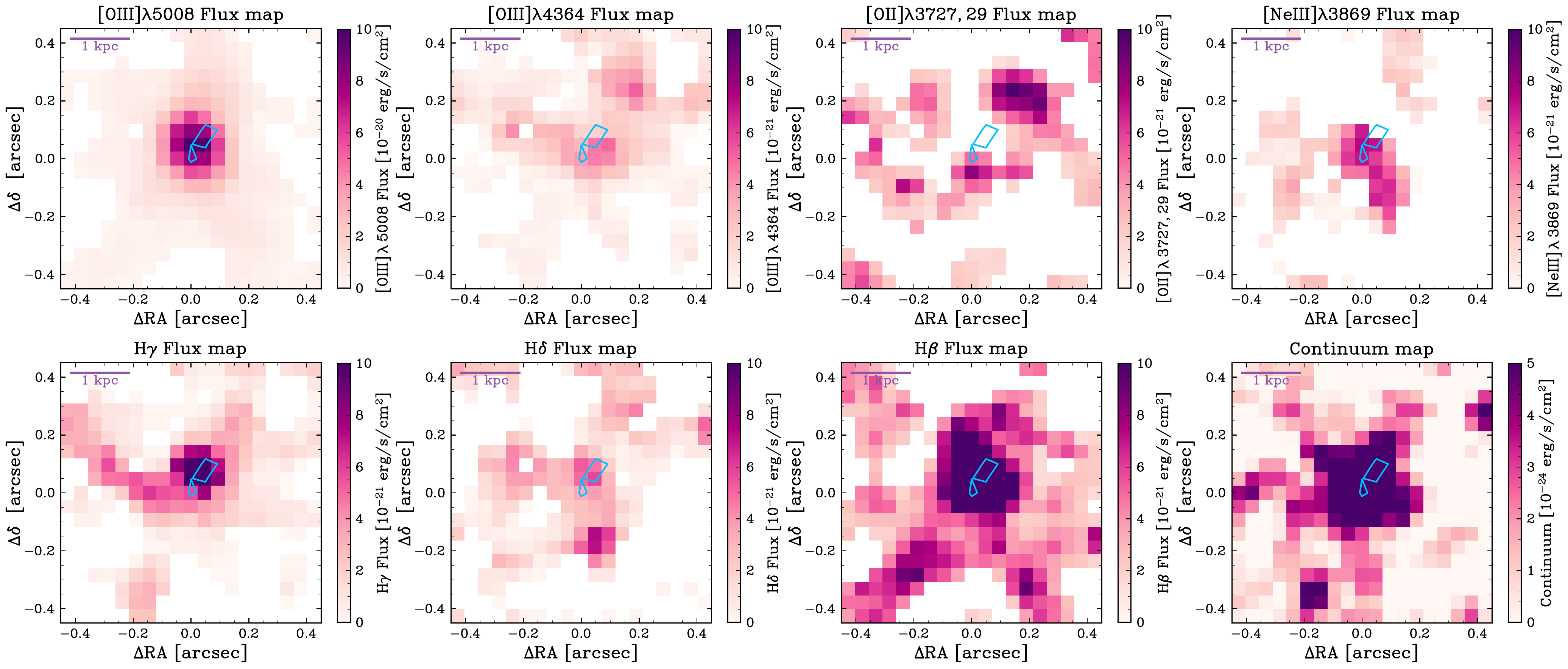}
    
    \caption{Same as Figure~\ref{fig:spec_flux_SMACS0723_4590} but for RX2129\_11022.}
    \label{fig:spec_flux_RX2129_11022}
\end{figure*}

\begin{figure*}
    \centering
    \includegraphics[width=\linewidth]{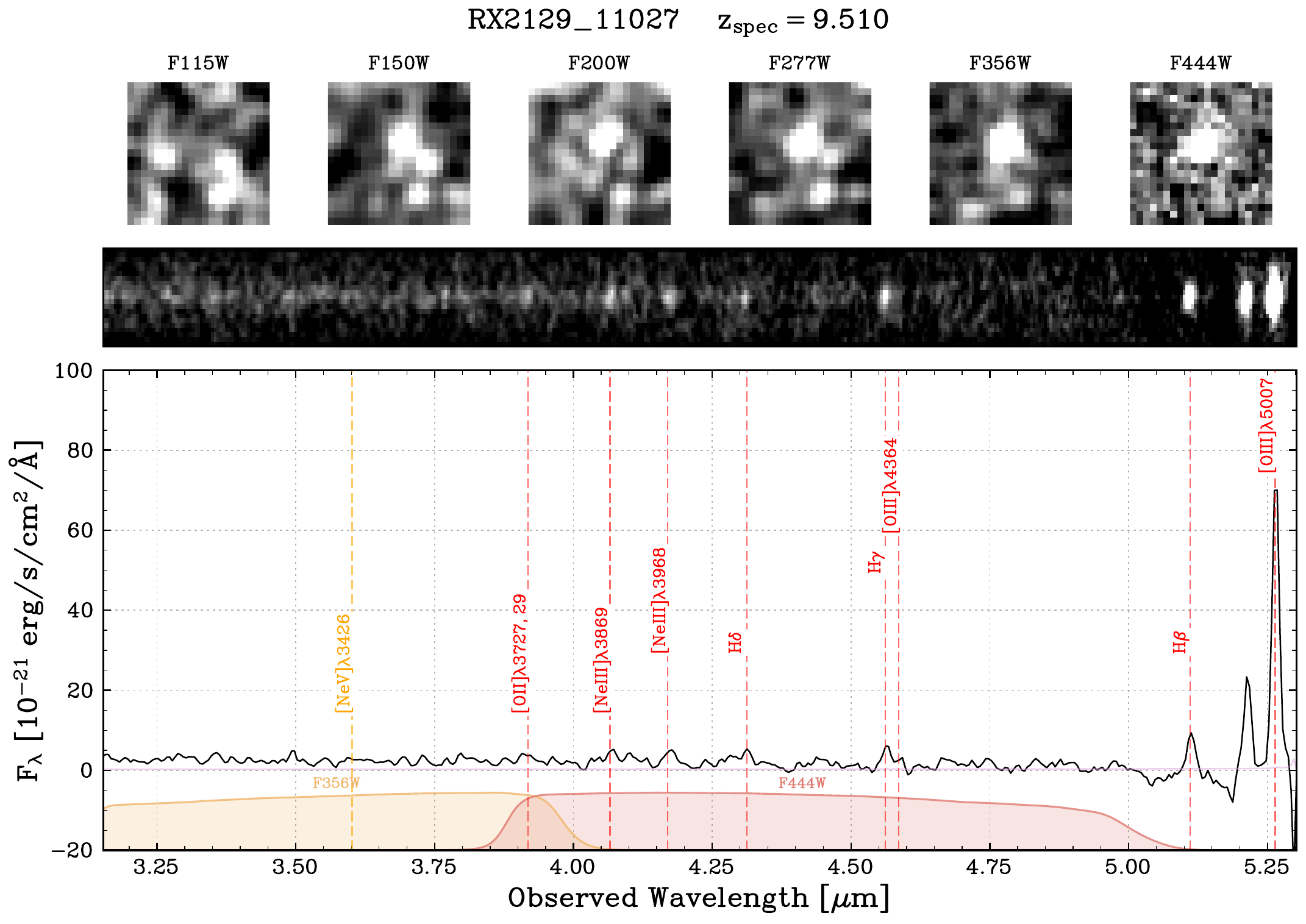}

    \vspace{0.3cm}

    \includegraphics[width=\linewidth]{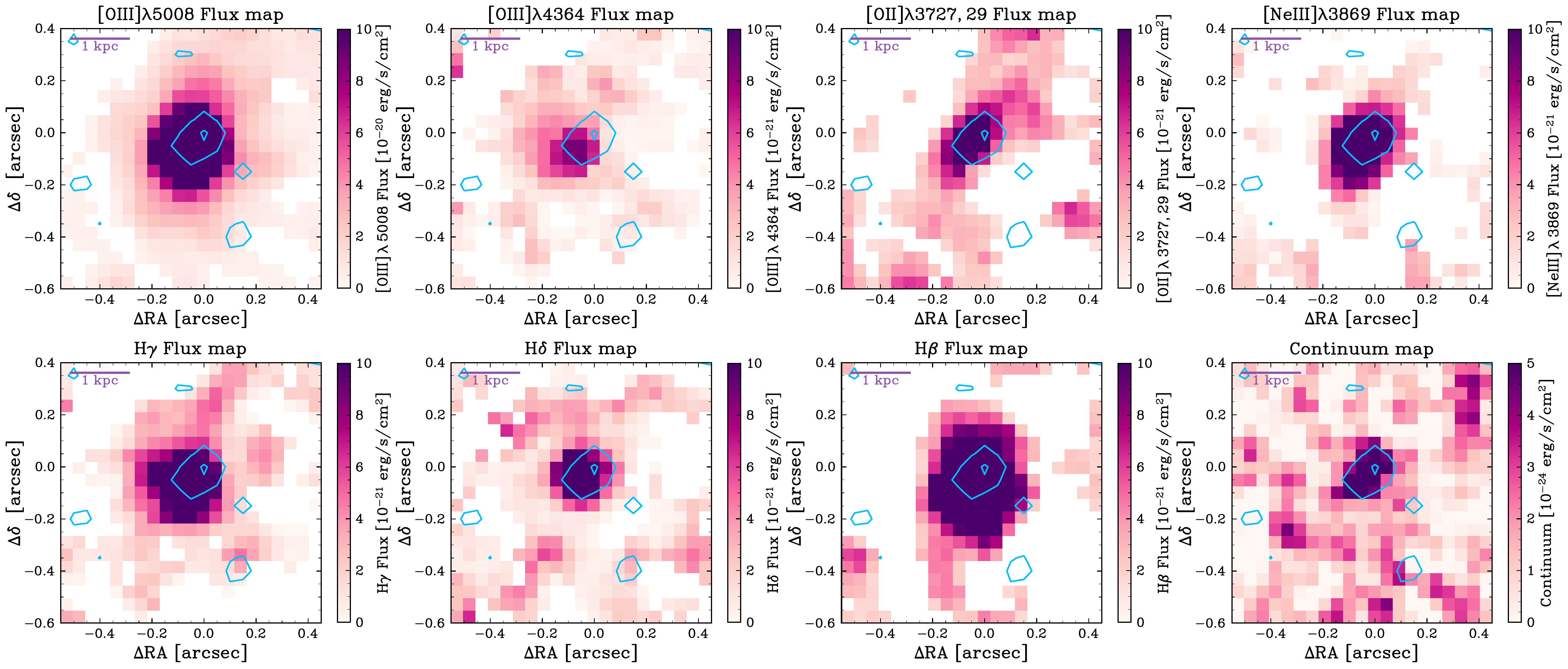}
    
    \caption{Same as Figure~\ref{fig:spec_flux_SMACS0723_4590} but for RX2129\_11027.}
    \label{fig:spec_flux_RX2129_11027}
\end{figure*}

\begin{figure*}
    \centering
    \includegraphics[width=\linewidth]{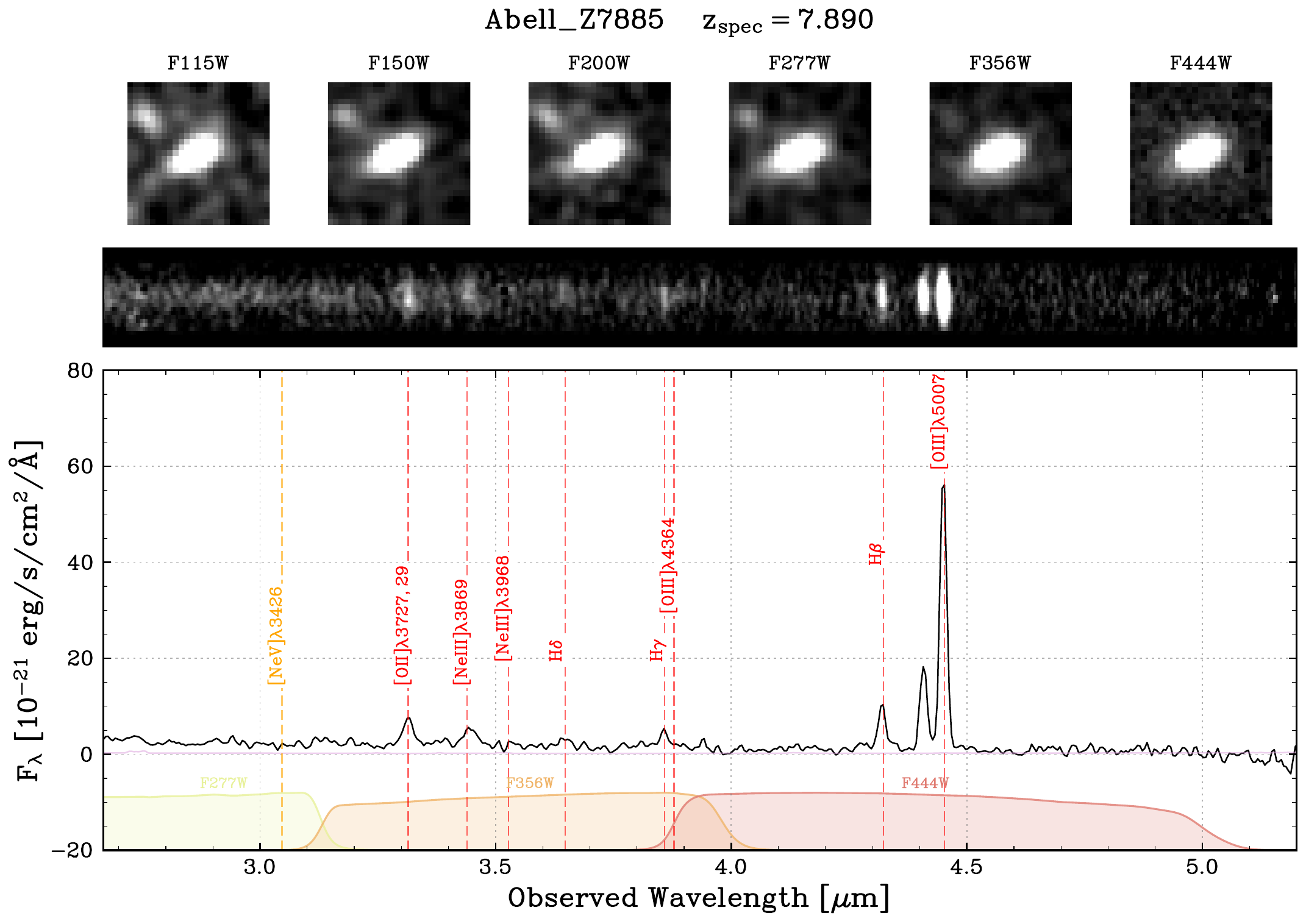}

    \vspace{0.3cm}

    \includegraphics[width=\linewidth]{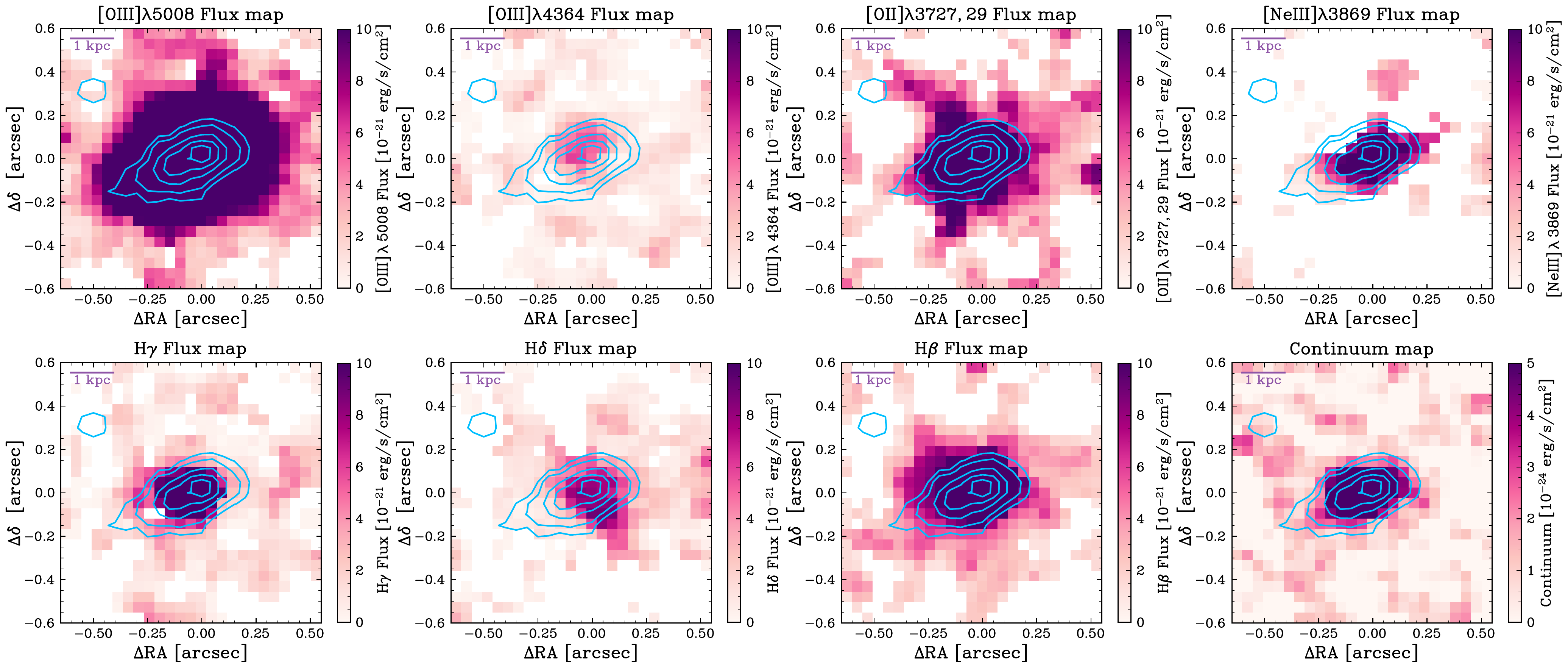}
    
    \caption{Same as Figure~\ref{fig:spec_flux_SMACS0723_4590} but for Abell\_Z7885.}
    \label{fig:spec_flux_Abell_Z7885}
\end{figure*}

\begin{figure*}
    \centering
    \includegraphics[width=\linewidth]{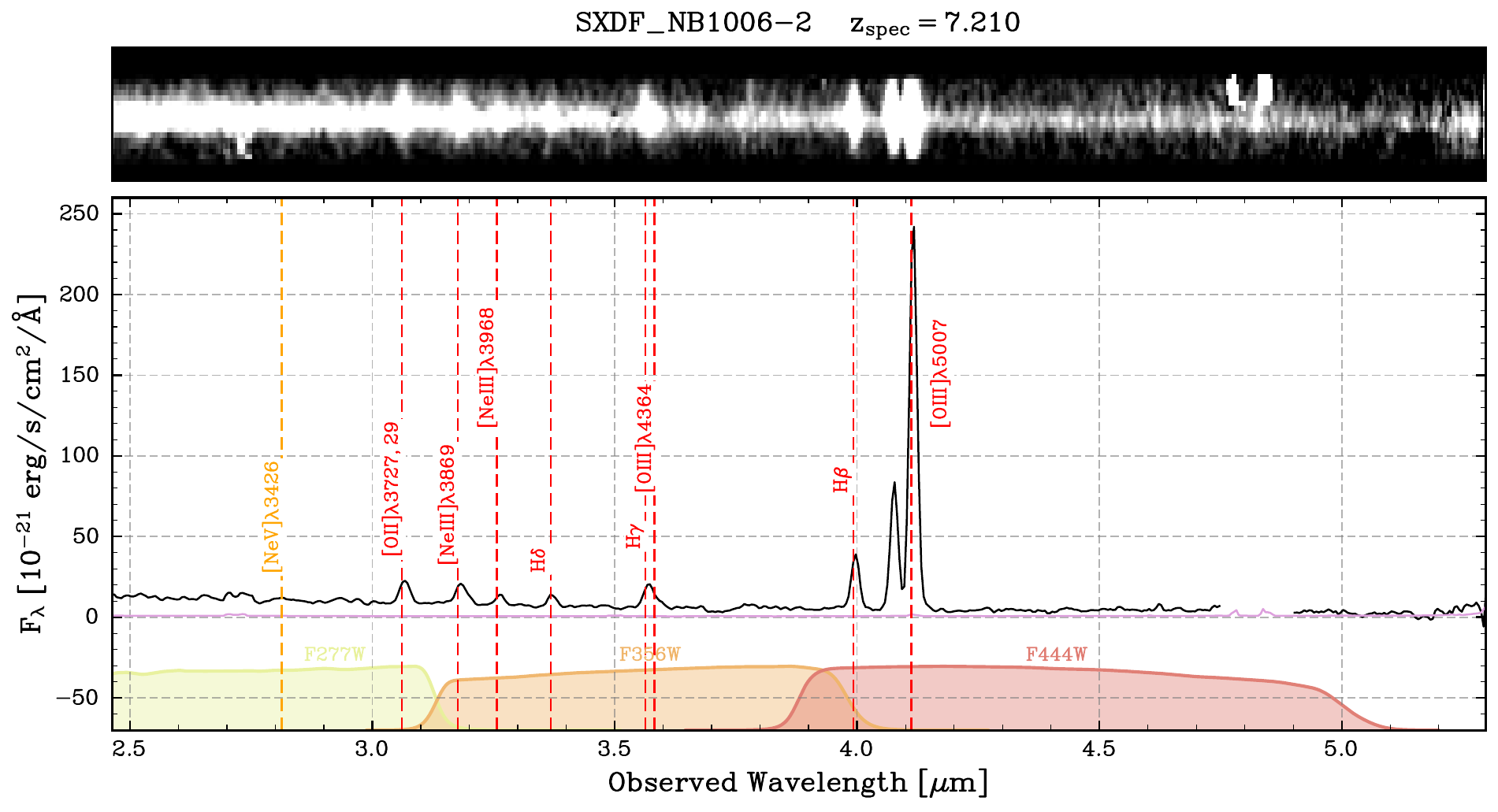}

    \vspace{0.3cm}

    \includegraphics[width=\linewidth]{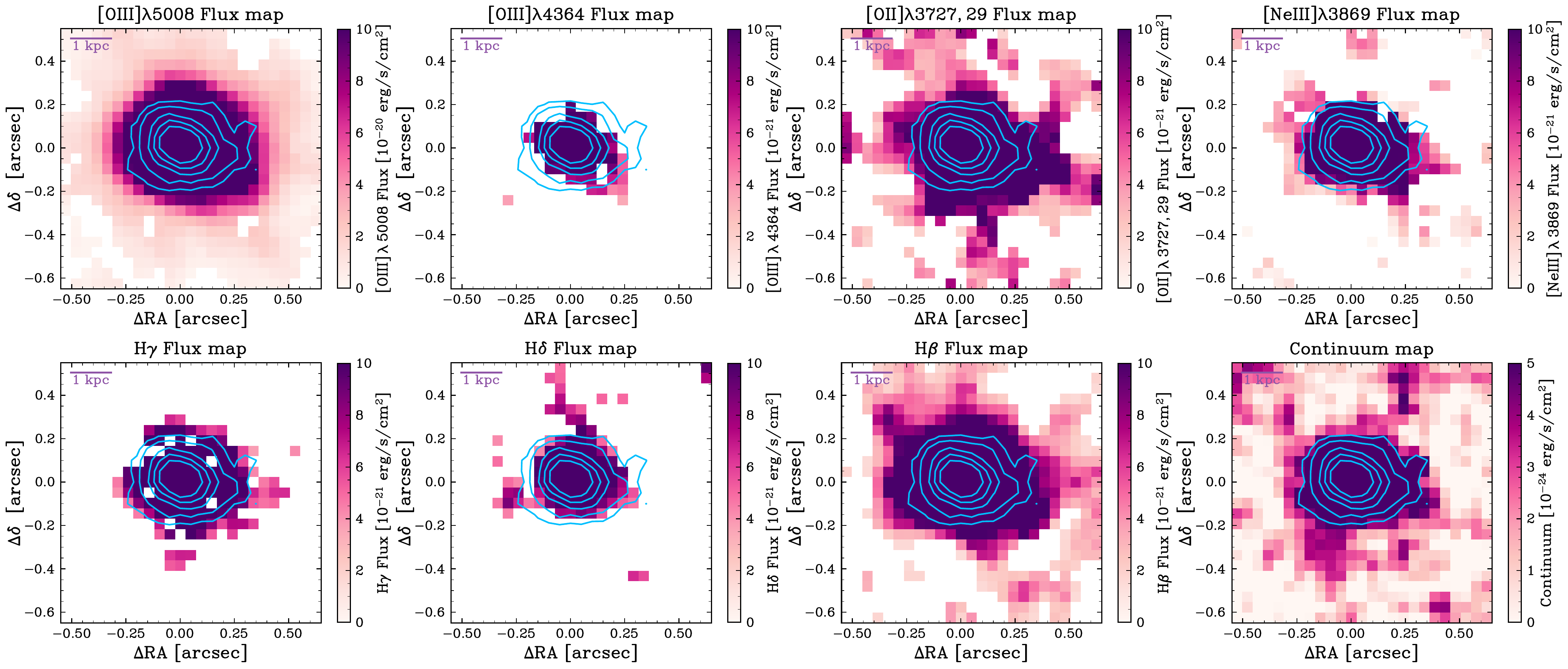}
    
    \caption{Same as Figure~\ref{fig:spec_flux_SMACS0723_4590} but for SXDF\_NB1006-2. As no NIRCam data is readily available for this object, we do not show any NIRCam cutouts and the blue contour lines represent the continuum measured between $\rm H\gamma$ and $\rm H\beta$ from the NIRSpec IFU data.}
    \label{fig:spec_flux_SXDF_NB1006-2}
\end{figure*}

\begin{figure*}
    \centering
    \includegraphics[width=\linewidth]{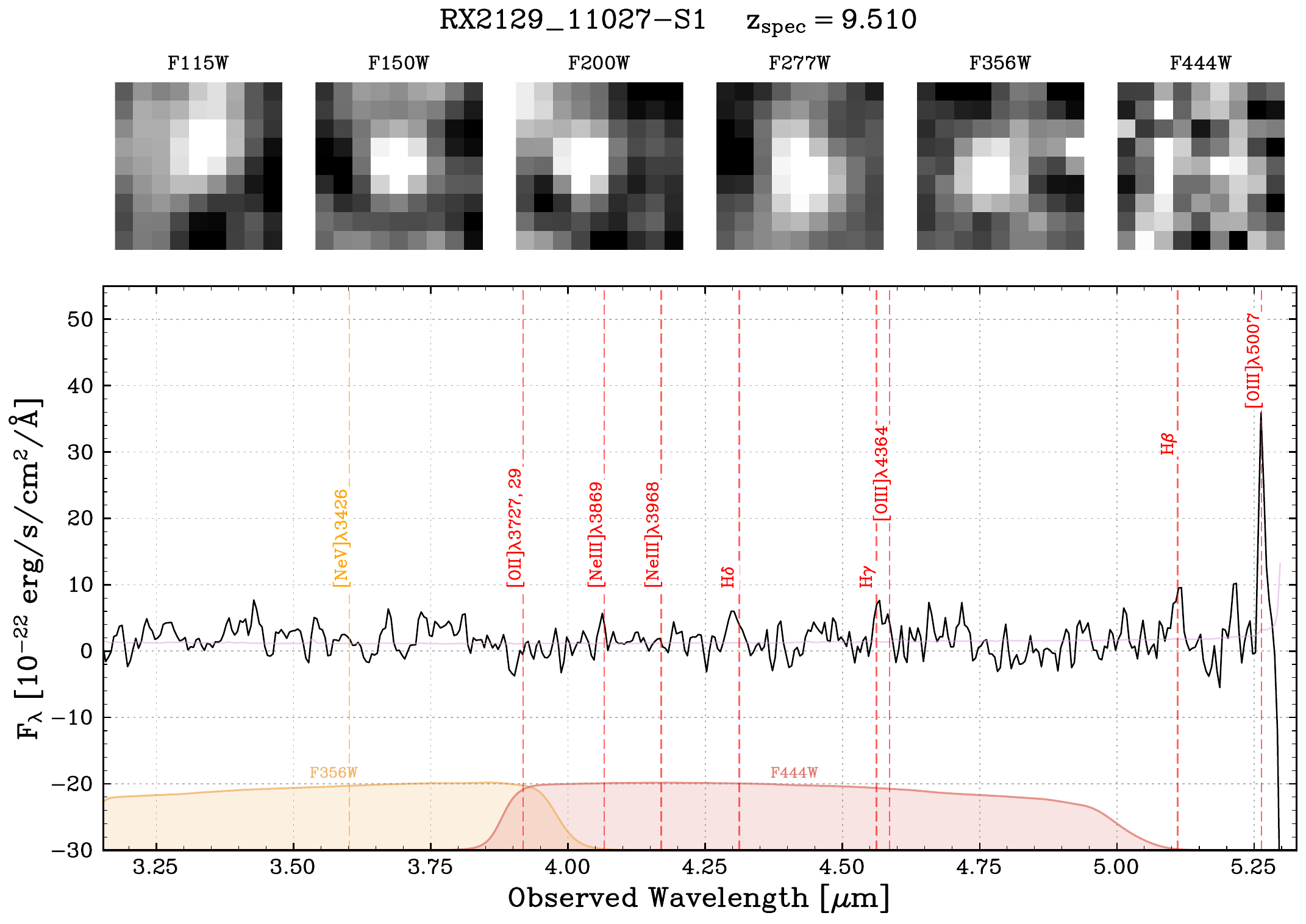}

    \caption{Same as Figure~\ref{fig:spec_flux_SMACS0723_4590} but for satellite RX2129\_11027-S1 and showing only the 1D spectrum and NIRCam filter images.}
    \label{fig:spec_RX2129_11027-S1}
\end{figure*}

\begin{figure*}
    \centering
    \includegraphics[width=\linewidth]{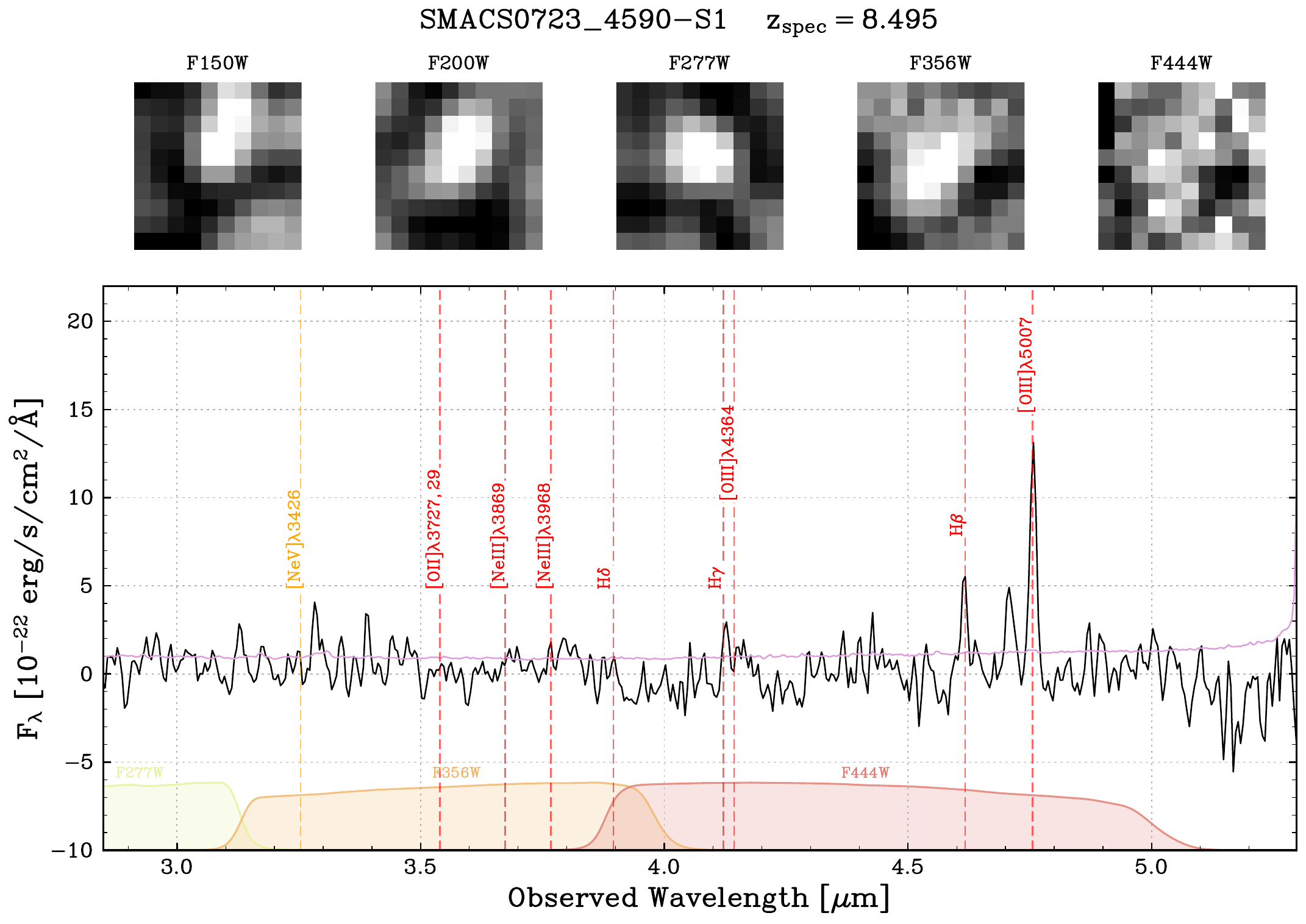}

    \caption{Same as Figure~\ref{fig:spec_flux_SMACS0723_4590} but for satellite SMACS0723\_4590-S1 and showing only the 1D spectrum and NIRCam filter images.}
    \label{fig:spec_SMACS0723_4590-S1}
\end{figure*}

\begin{figure*}
    \centering
    \includegraphics[width=\linewidth]{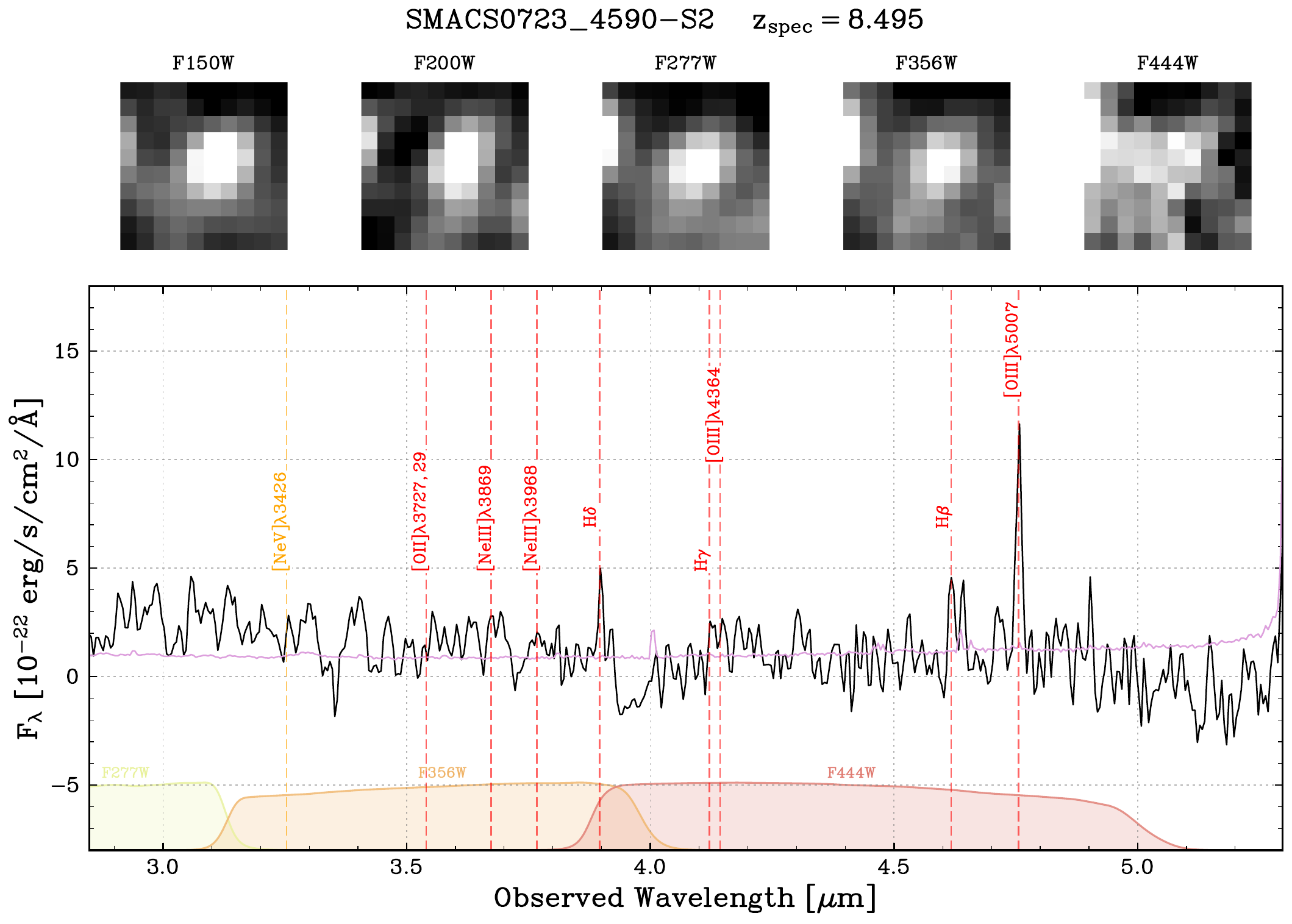}

    \caption{Same as Figure~\ref{fig:spec_flux_SMACS0723_4590} but for satellite SMACS0723\_4590-S2 and showing only the 1D spectrum and NIRCam filter images.}
    \label{fig:spec_SMACS0723_4590-S2}
\end{figure*}

In this appendix, we present the NIRCam filter images, 1D and 2D spectra, as well as running-median smoothed emission line flux maps of each galaxy in our sample. The results are shown in Figures~\ref{fig:spec_flux_SMACS0723_4590} to ~\ref{fig:spec_flux_SXDF_NB1006-2}. Additionally, we include the 1D spectra and NIRCam filter images of the three satellites, which are shown in Figures~\ref{fig:spec_RX2129_11027-S1} to ~\ref{fig:spec_SMACS0723_4590-S2}.

\section{Compilation of line flux ratios, metallicities and gradients of each galaxy}\label{app:metal_grads}

\begin{figure*}
    \centering

    \includegraphics[width=0.35\linewidth]{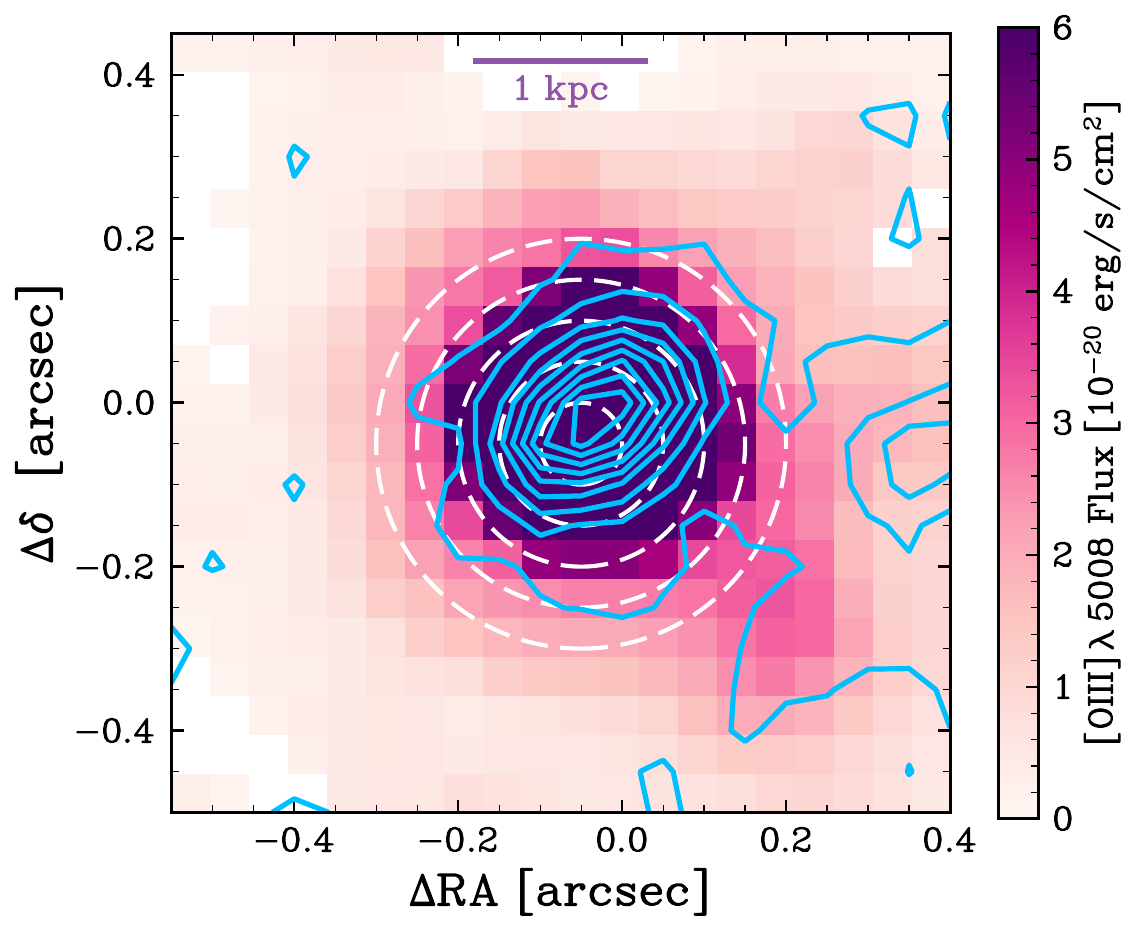}
    \hspace{0.8cm}
    \includegraphics[width=0.5\linewidth]{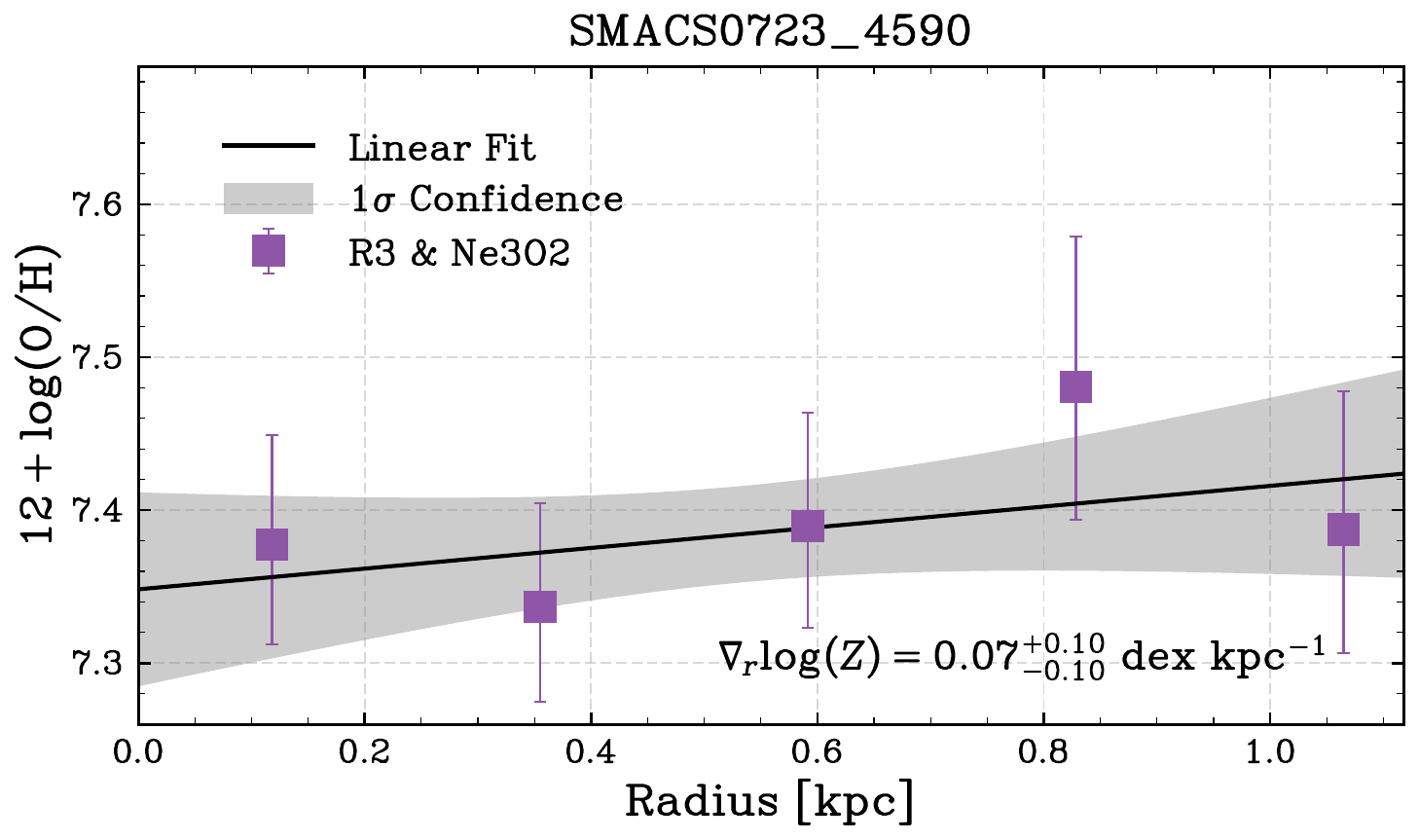}

    \vspace{0.3cm}
    
    \includegraphics[width=\linewidth]{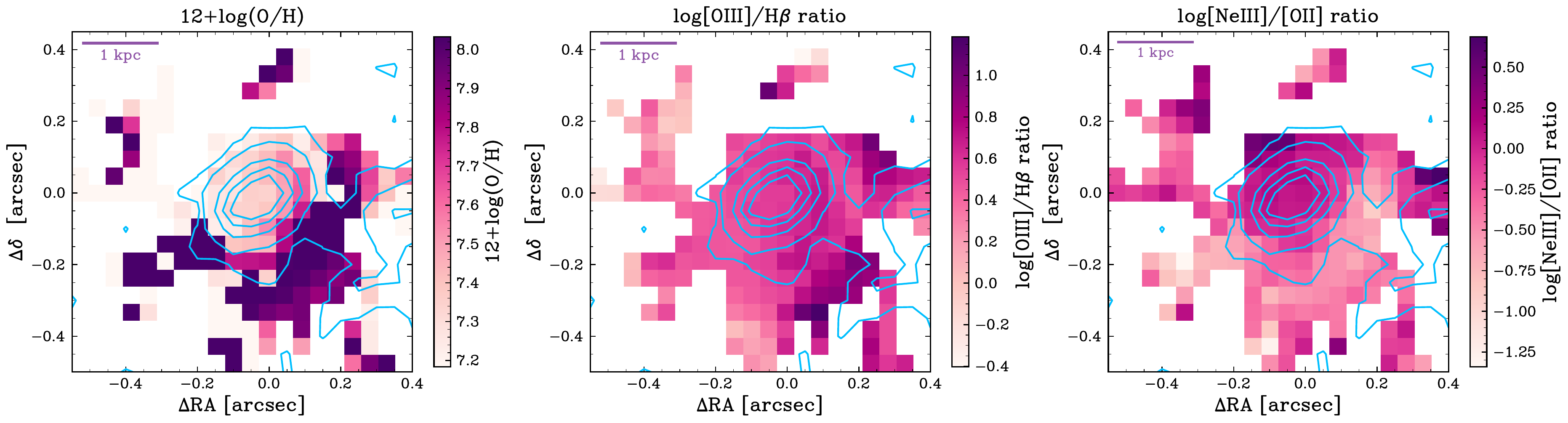}

    \vspace{0.3cm}

    \includegraphics[width=\linewidth]{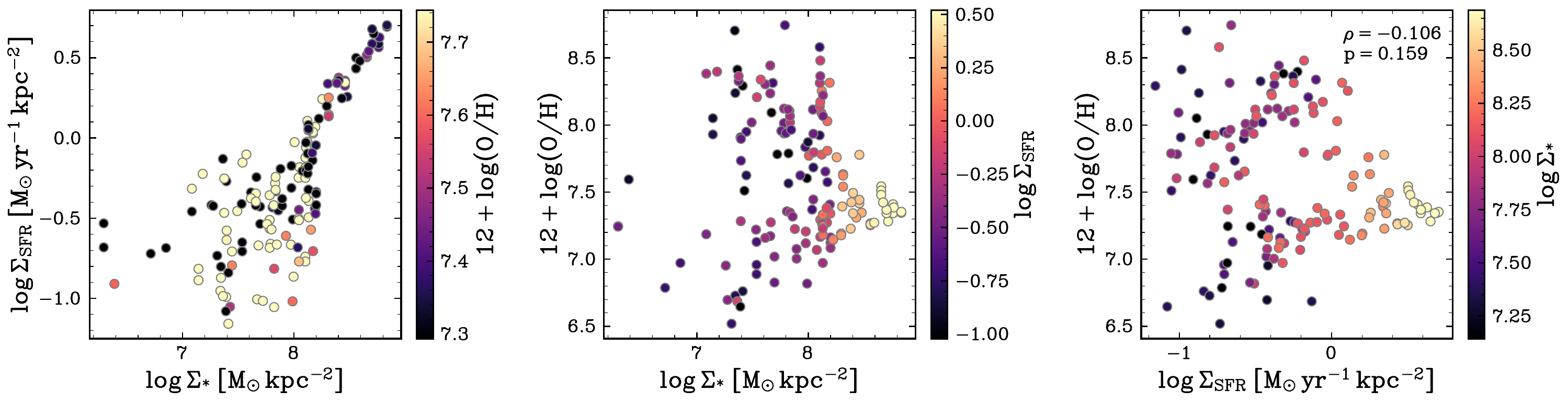}
    
    \caption{Spatially resolved metallicity, line flux ratios, and rFMR analysis of SMACS0723\_4590. \textit{Top row}: The measured $\rm [OIII]\,\lambda5008$ flux map derived from Gaussian fits to the $3 \times 3$ spaxel median-smoothed IFU cube. We overlay the annuli (white dashed rings) used to derive the gas-phase metallicity gradient, which is plotted in the adjacent panel. Blue contour lines trace the continuum measured via the F277W NIRCam filter for $\rm [3, 5, 10, 15, 20] \times RMS$ levels. \textit{Middle row}: Spatially resolved maps of the gas-phase metallicity (left), $\log(\rm [OIII]/H\beta)$ ratio (centre), and $\log(\rm [NeIII]/[OII])$ ratio (right). These maps are masked for $\rm S/N > 3$ in $\rm [OIII]$ and $\rm H\beta$, and $\rm S/N > 2$ for $\rm [NeIII]$ and $\rm [OII]$. \textit{Bottom row}: Three representations of the resolved Fundamental Metallicity Relation (rFMR). The right-most subplot also contains the Spearman coefficient and associated p-value between the metallicity and SFR surface density. All maps are generated from the spatially smoothed data cube.}
    \label{fig:metal_grads_SMACS0723_4590}
\end{figure*}

\begin{figure*}
    \centering

    \includegraphics[width=0.35\linewidth]{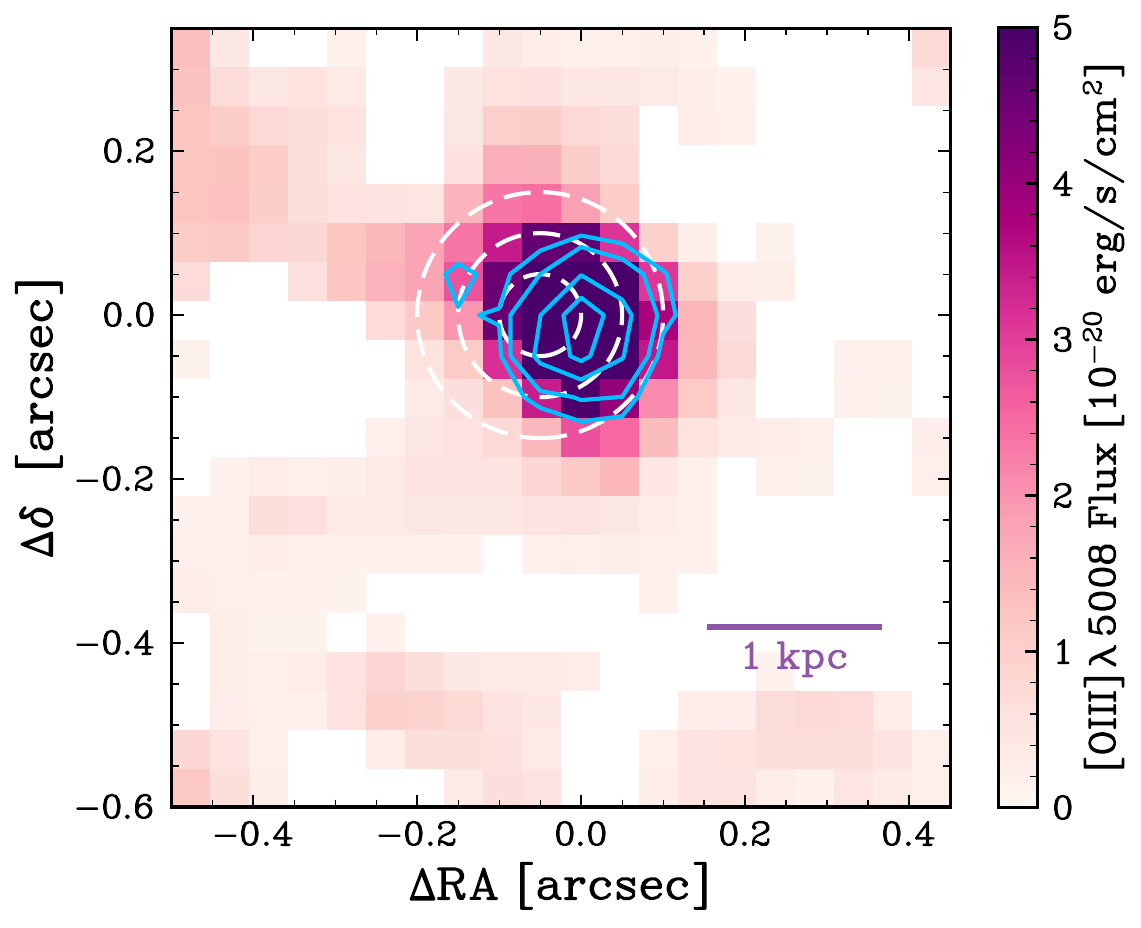}
    \hspace{0.8cm}
    \includegraphics[width=0.5\linewidth]{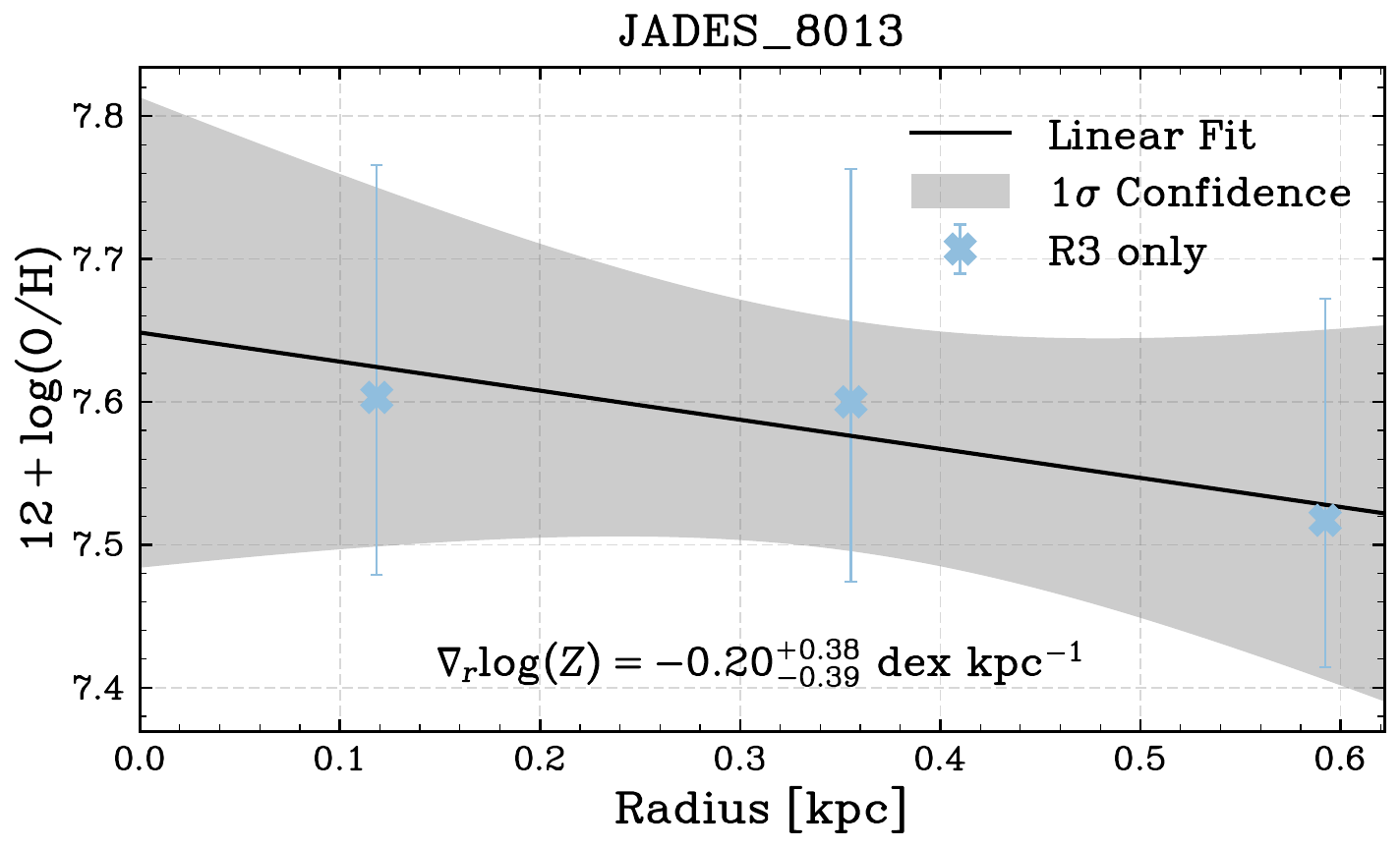}

    \vspace{0.3cm}
    
    \includegraphics[width=\linewidth]{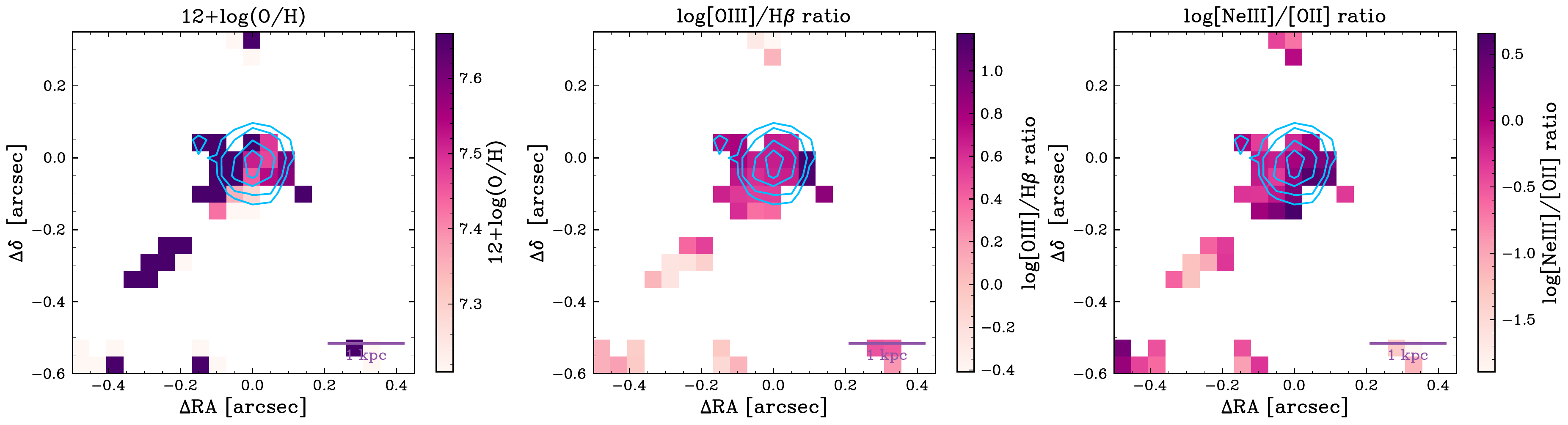}

    \vspace{0.3cm}

    \includegraphics[width=\linewidth]{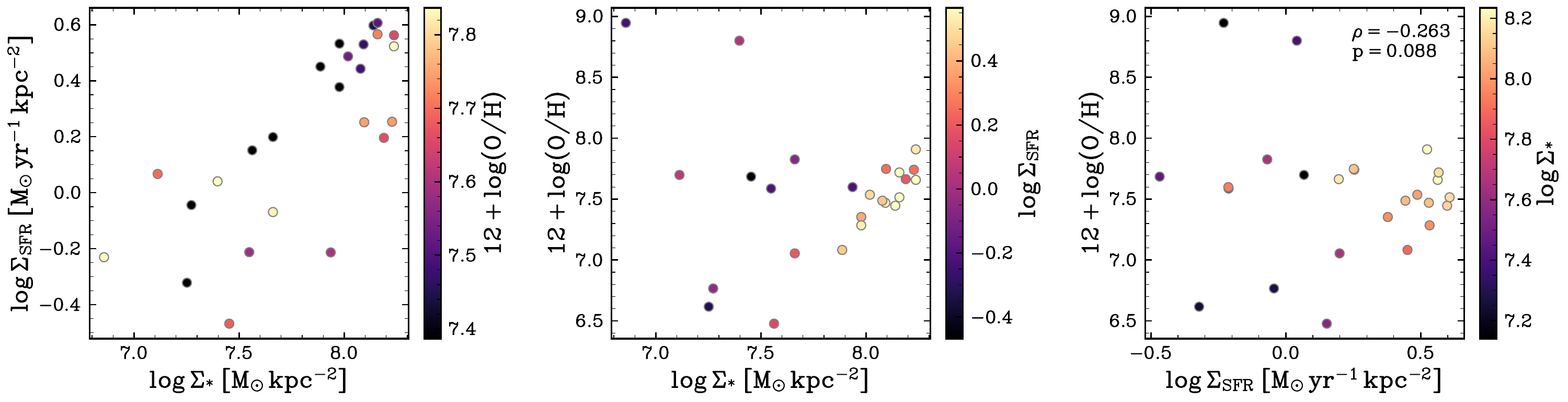}
    
    \caption{Same as Figure~\ref{fig:metal_grads_SMACS0723_4590} but for JADES\_8013.}
    \label{fig:metal_grads_JADES_8013}
\end{figure*}

\begin{figure*}
    \centering

    \includegraphics[width=0.35\linewidth]{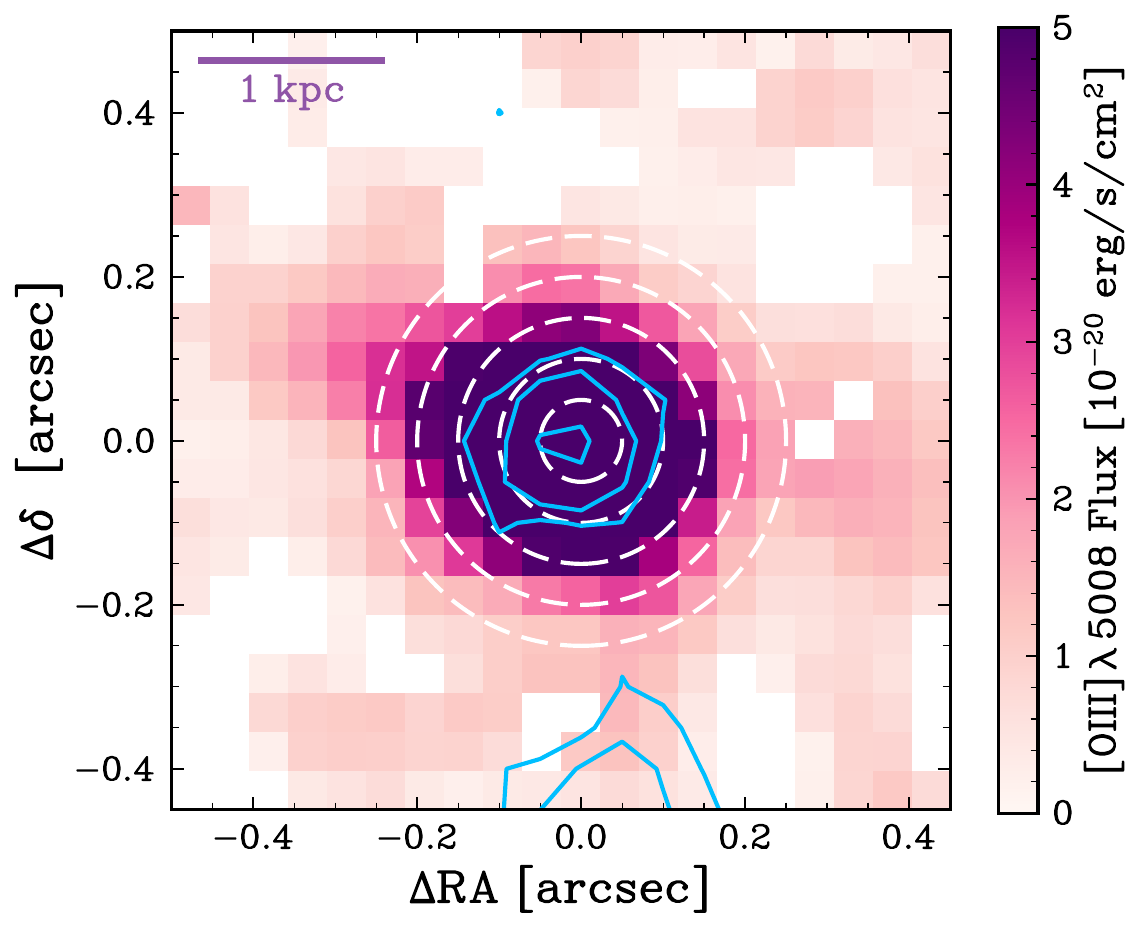}
    \hspace{0.8cm}
    \includegraphics[width=0.5\linewidth]{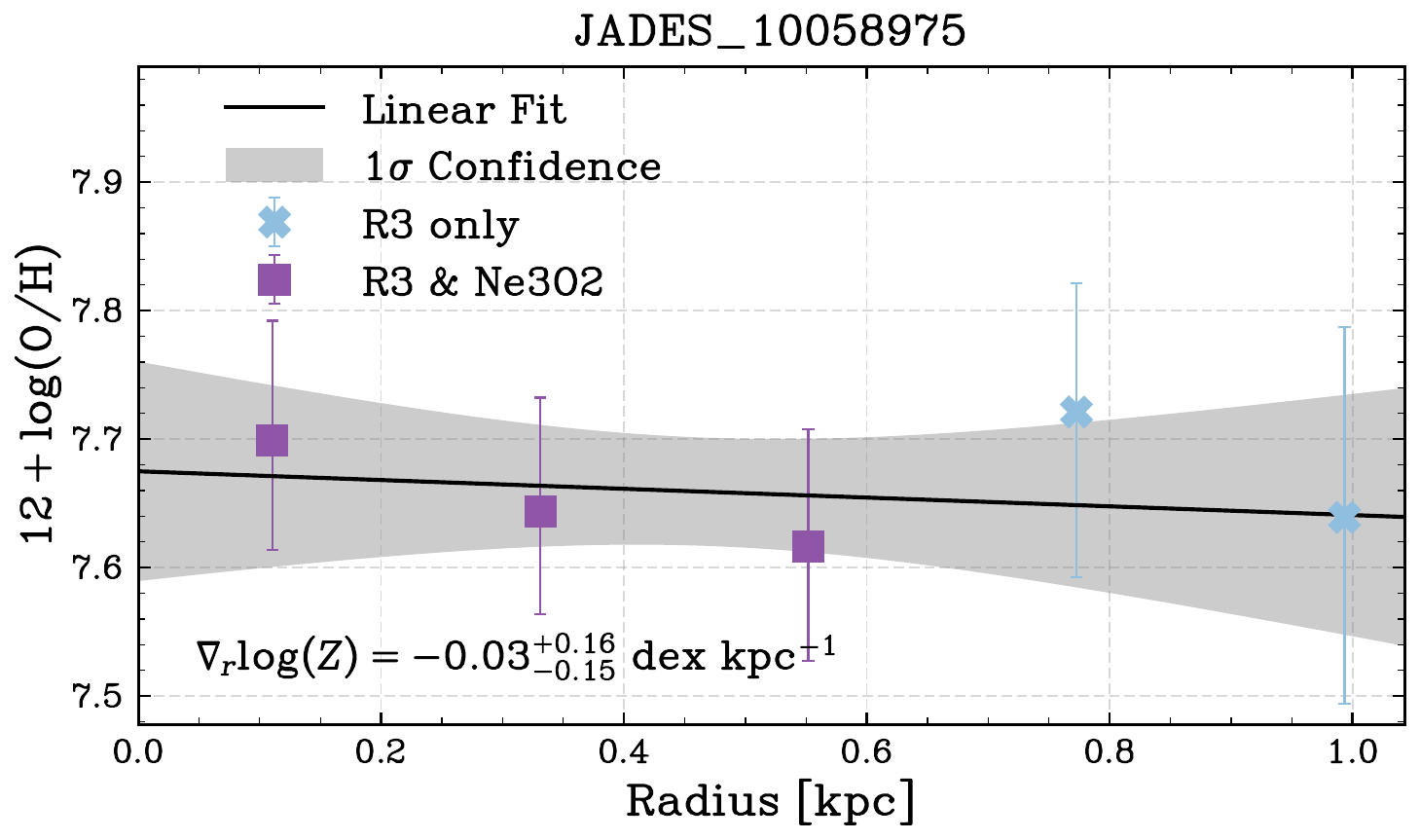}

    \vspace{0.3cm}
    
    \includegraphics[width=\linewidth]{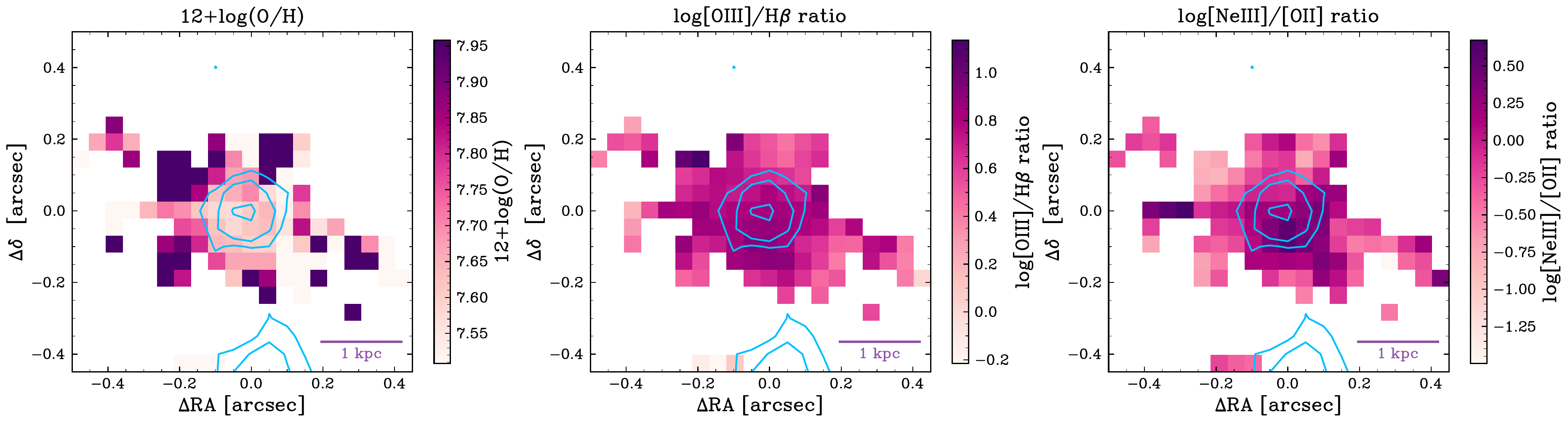}

    \vspace{0.3cm}

    \includegraphics[width=\linewidth]{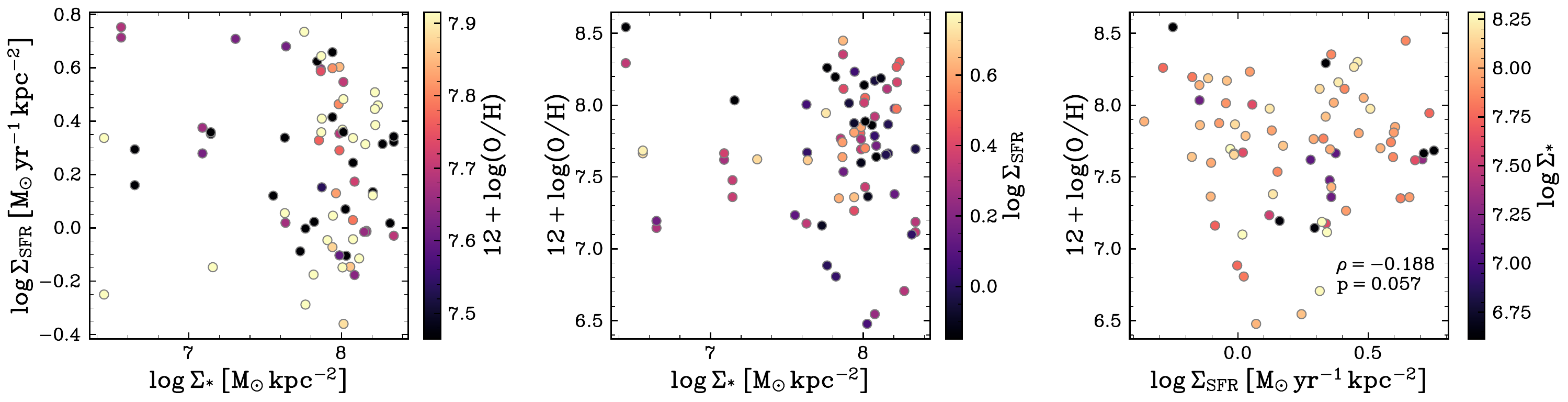}
    
    \caption{Same as Figure~\ref{fig:metal_grads_SMACS0723_4590} but for JADES\_10058975.}
    \label{fig:metal_grads_JADES_10058975}
\end{figure*}

\begin{figure*}
    \centering

    \includegraphics[width=0.35\linewidth]{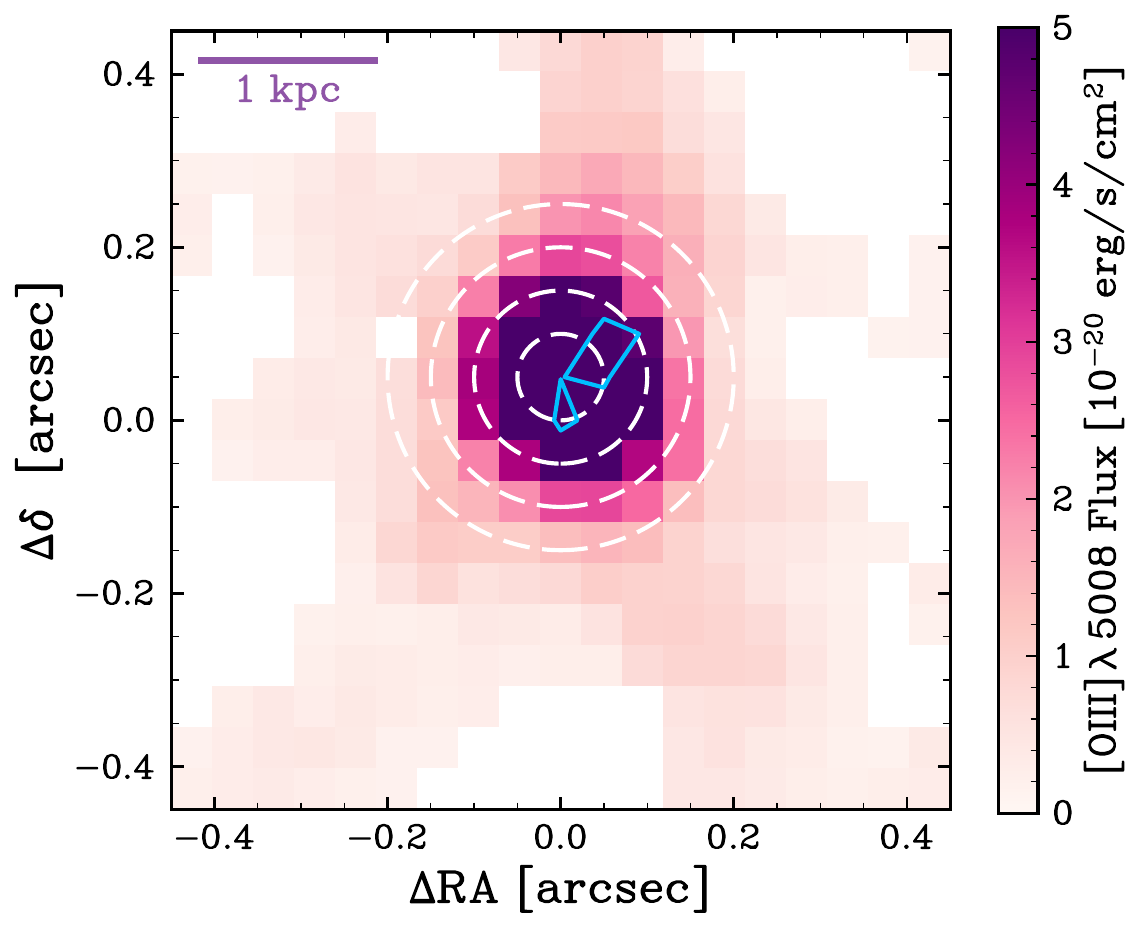}
    \hspace{0.8cm}
    \includegraphics[width=0.5\linewidth]{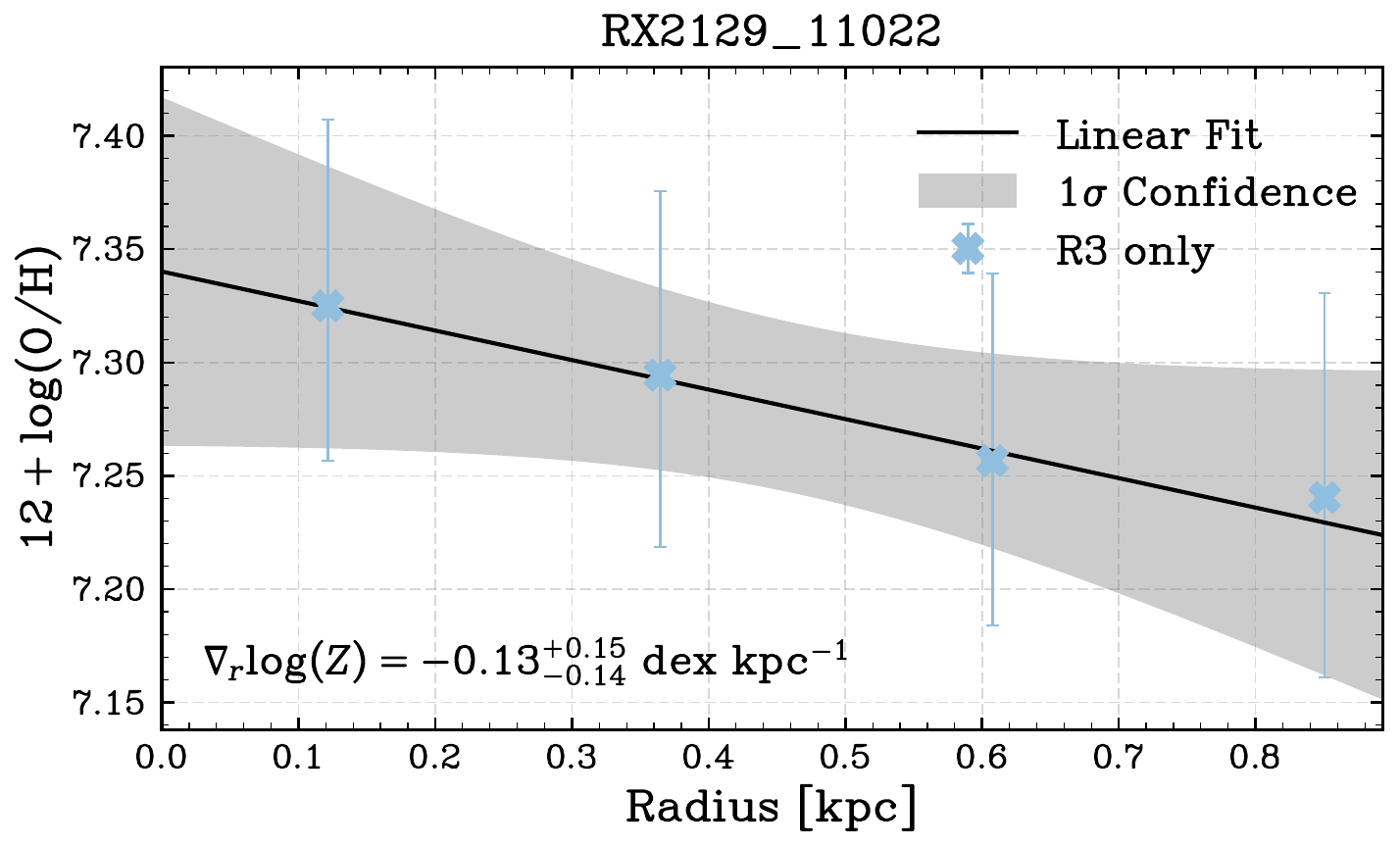}

    \vspace{0.3cm}
    
    \includegraphics[width=\linewidth]{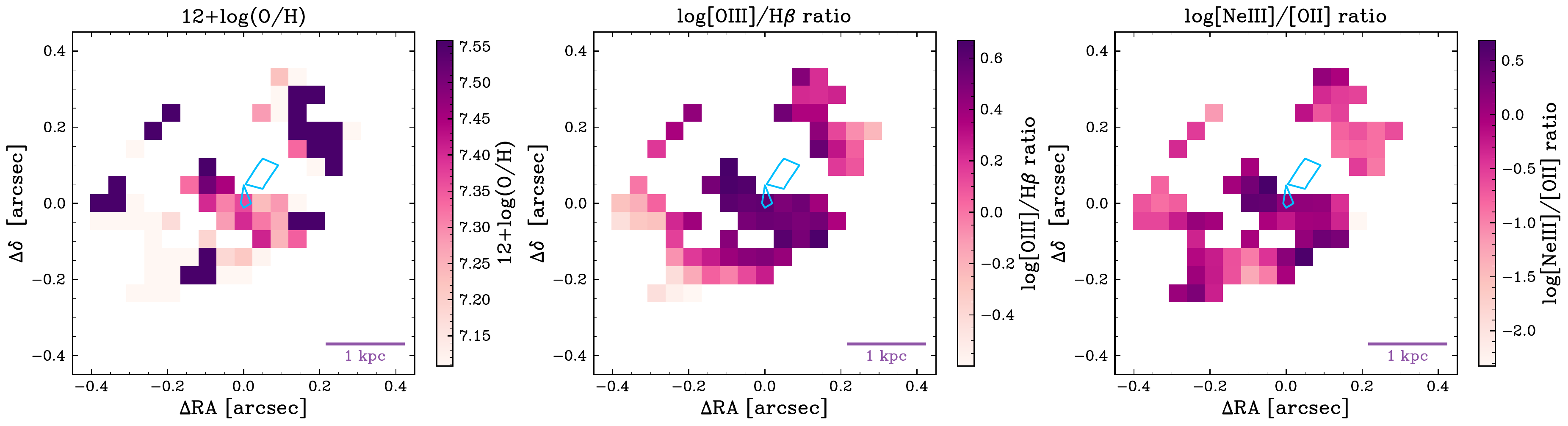}

    \vspace{0.3cm}

    \includegraphics[width=\linewidth]{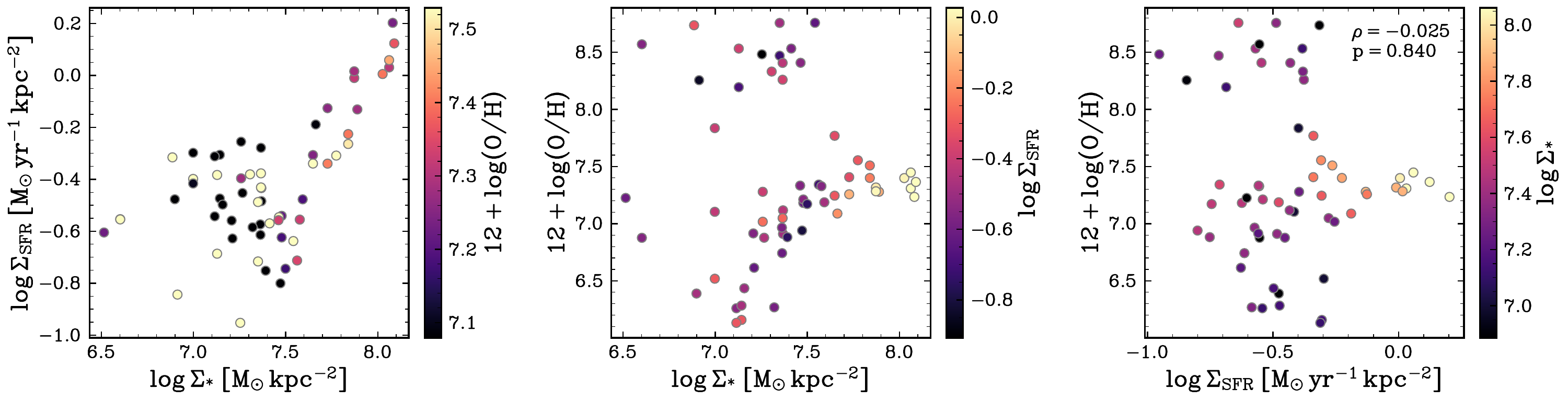}
    
    \caption{Same as Figure~\ref{fig:metal_grads_SMACS0723_4590} but for RX2129\_11022.}
    \label{fig:metal_grads_RX2129_11022}
\end{figure*}

\begin{figure*}
    \centering

    \includegraphics[width=0.35\linewidth]{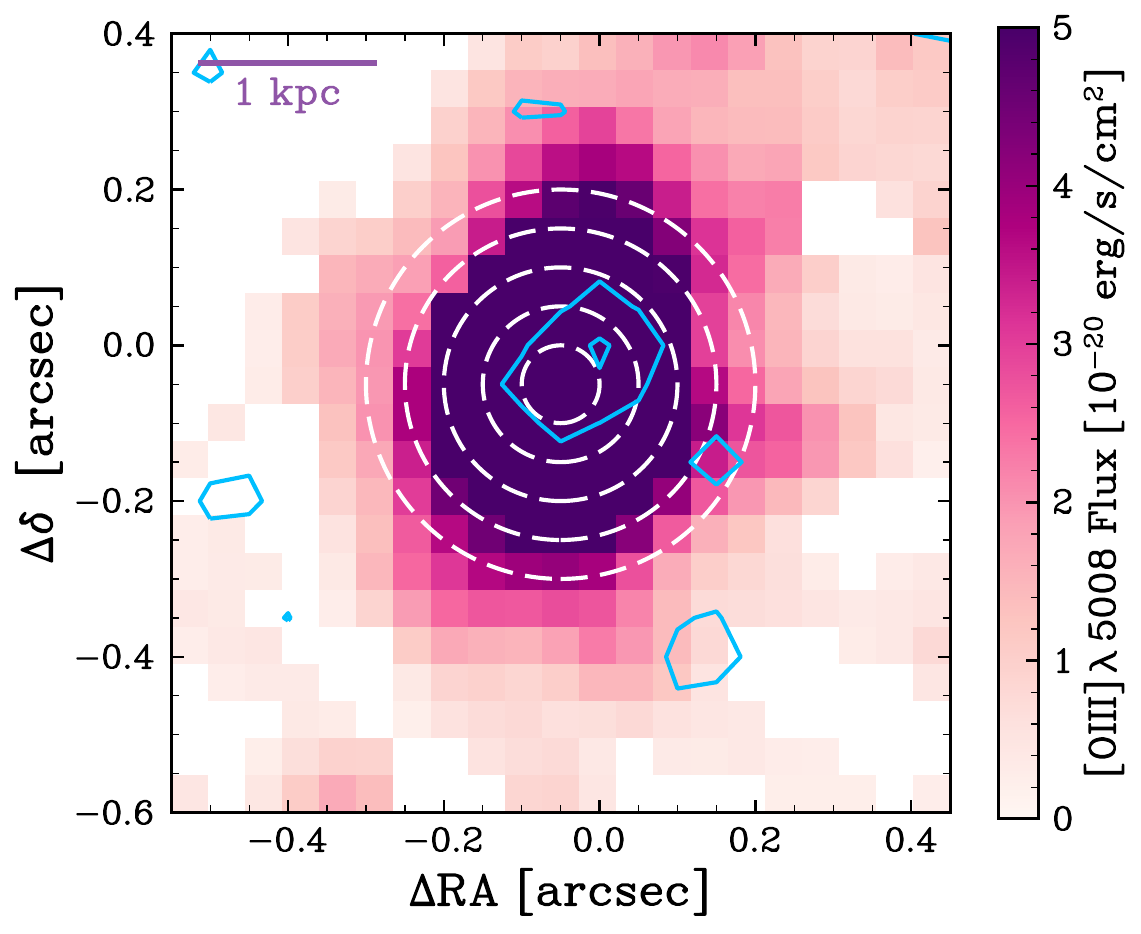}
    \hspace{0.8cm}
    \includegraphics[width=0.5\linewidth]{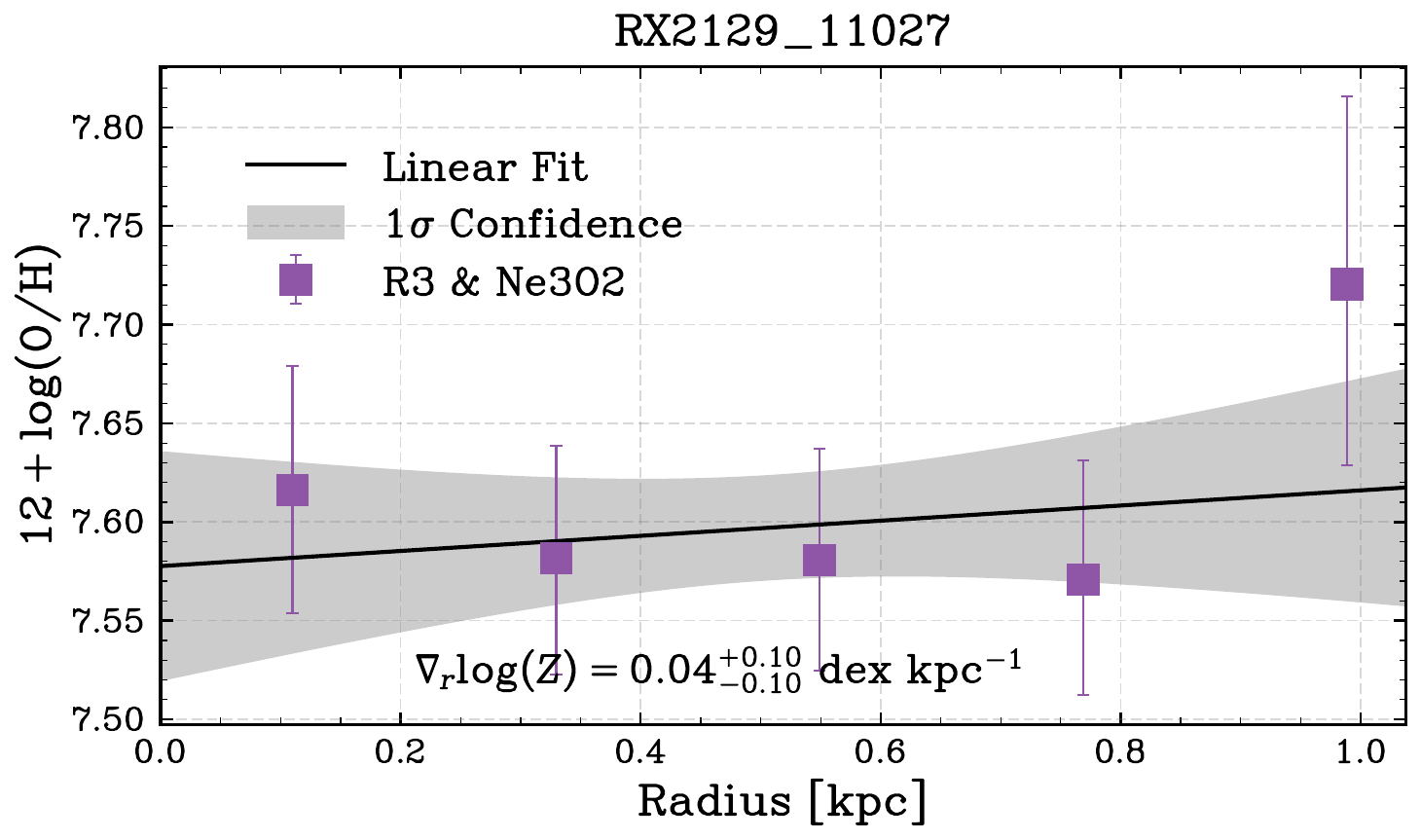}

    \vspace{0.3cm}
    
    \includegraphics[width=\linewidth]{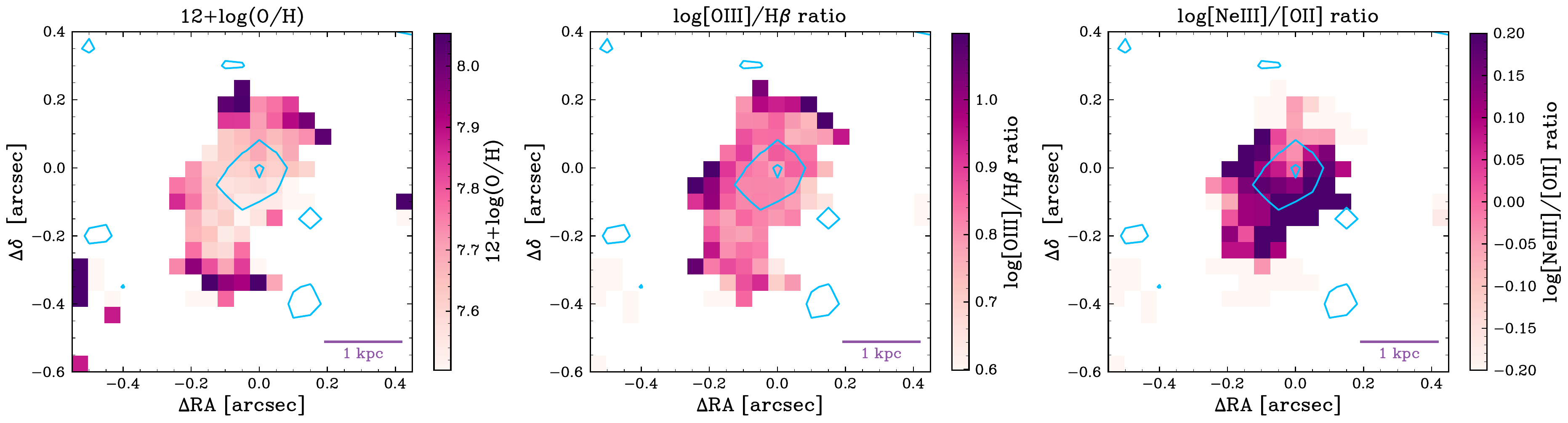}

    \vspace{0.3cm}

    \includegraphics[width=\linewidth]{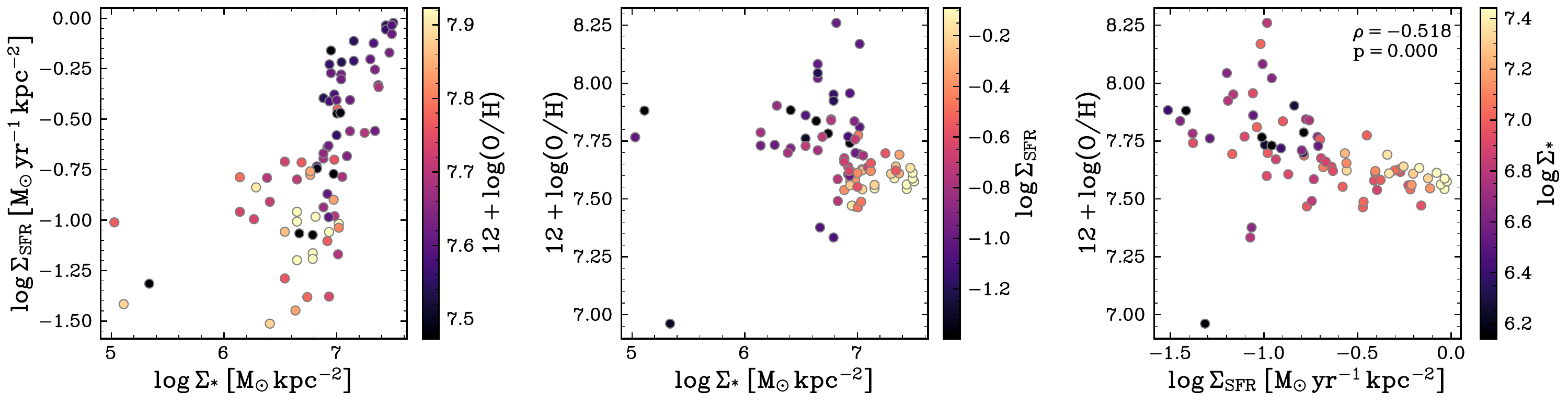}
    
    \caption{Same as Figure~\ref{fig:metal_grads_SMACS0723_4590} but for RX2129\_11027.}
    \label{fig:metal_grads_RX2129_11027}
\end{figure*}

\begin{figure*}
    \centering

    \includegraphics[width=0.35\linewidth]{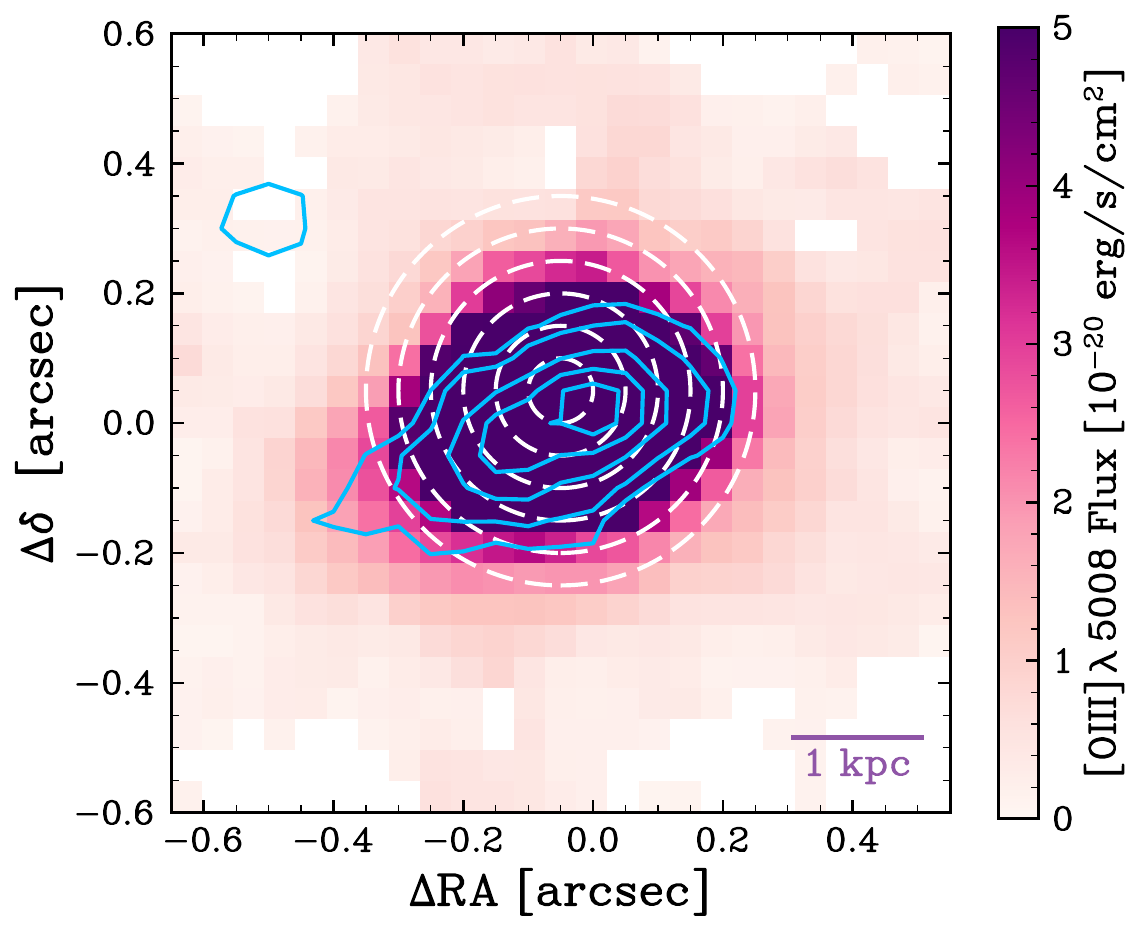}
    \hspace{0.8cm}
    \includegraphics[width=0.5\linewidth]{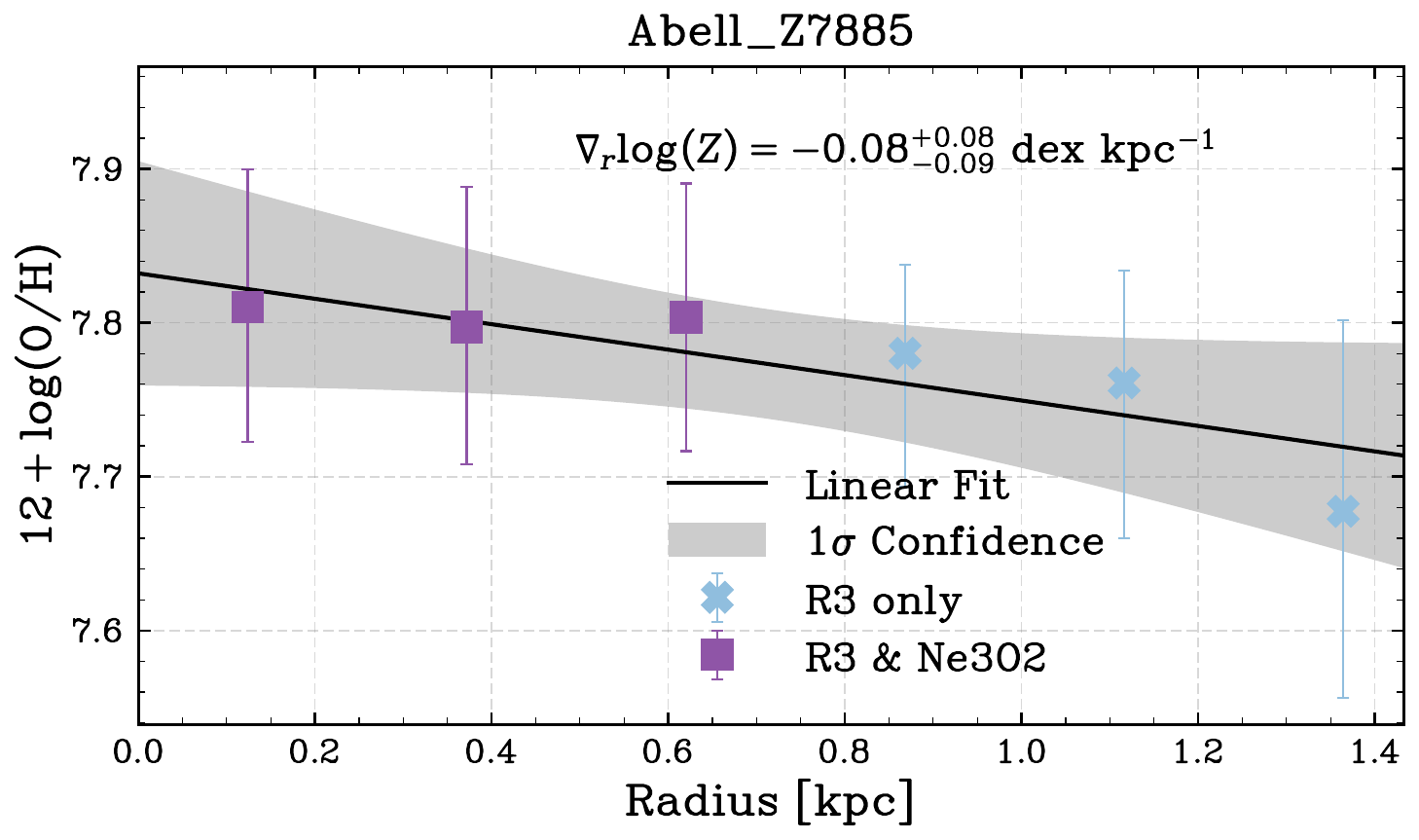}

    \vspace{0.3cm}
    
    \includegraphics[width=\linewidth]{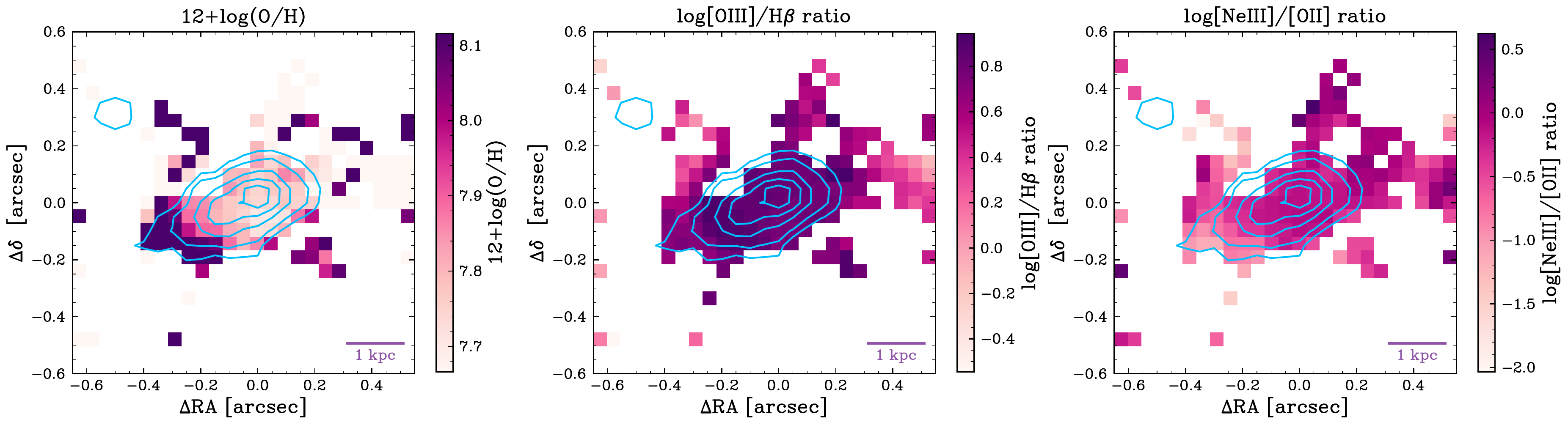}

    \vspace{0.3cm}

    \includegraphics[width=\linewidth]{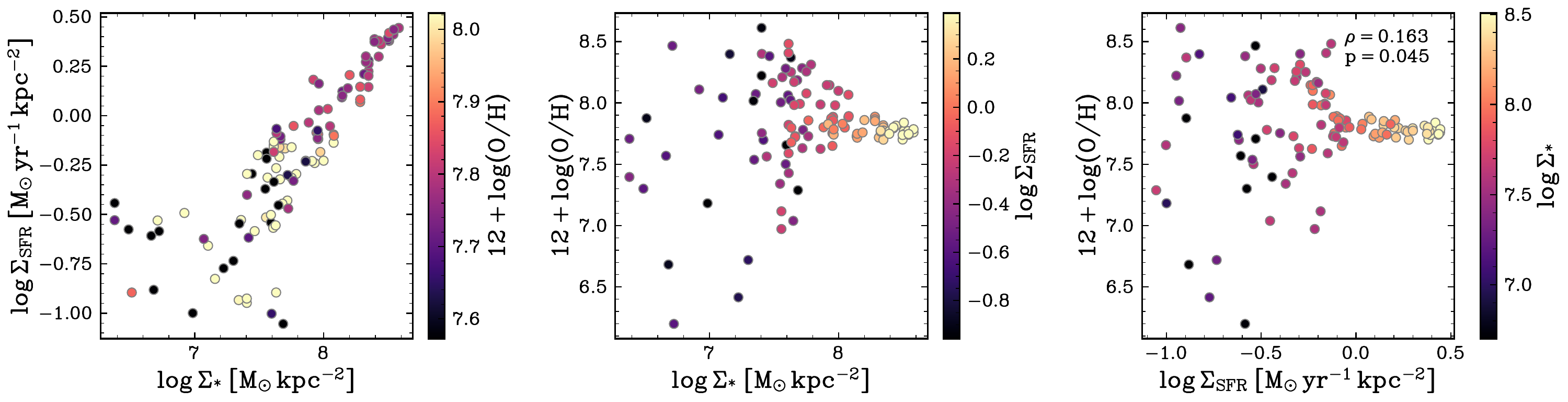}
    
    \caption{Same as Figure~\ref{fig:metal_grads_SMACS0723_4590} but for Abell\_Z7885.}
    \label{fig:metal_grads_Abell_Z7885}
\end{figure*}

\begin{figure*}
    \centering

    \includegraphics[width=0.35\linewidth]{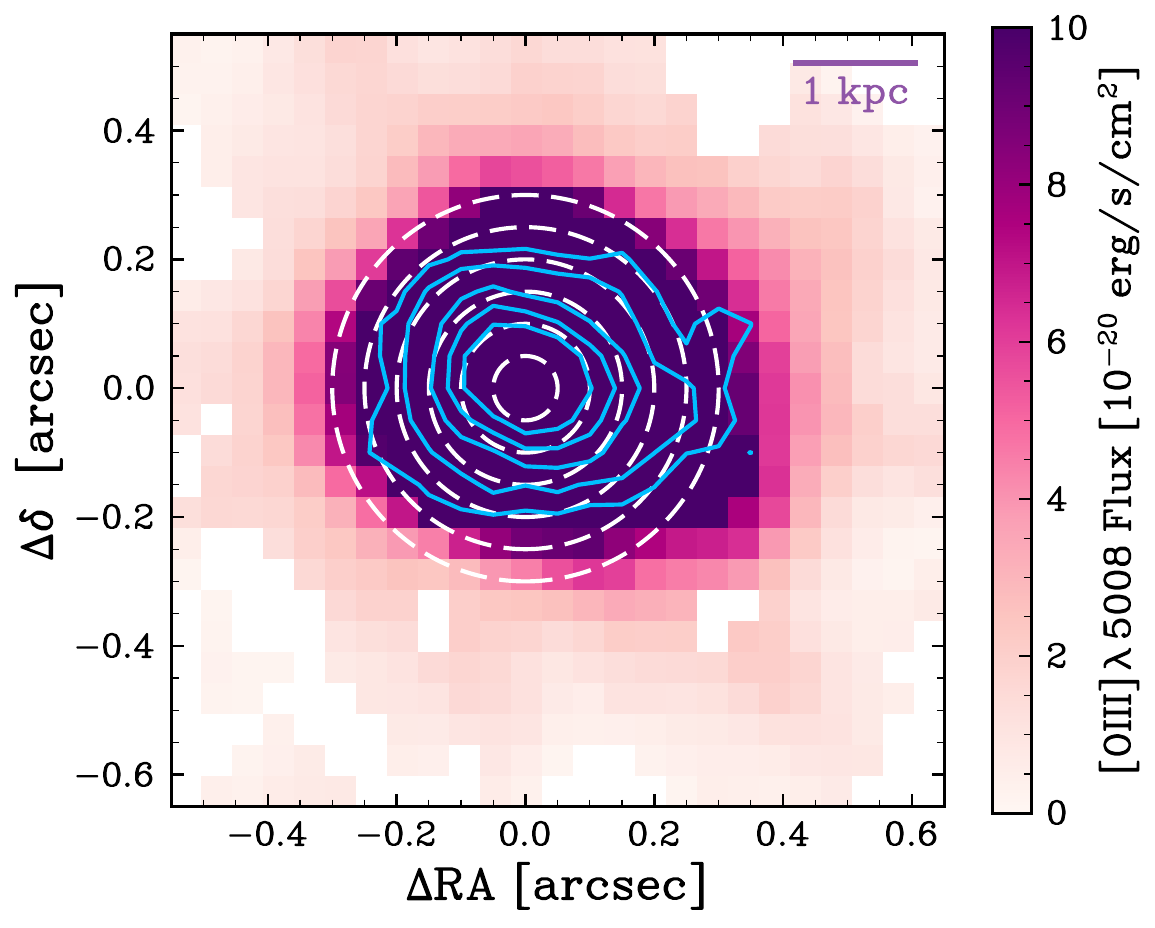}
    \hspace{0.8cm}
    \includegraphics[width=0.5\linewidth]{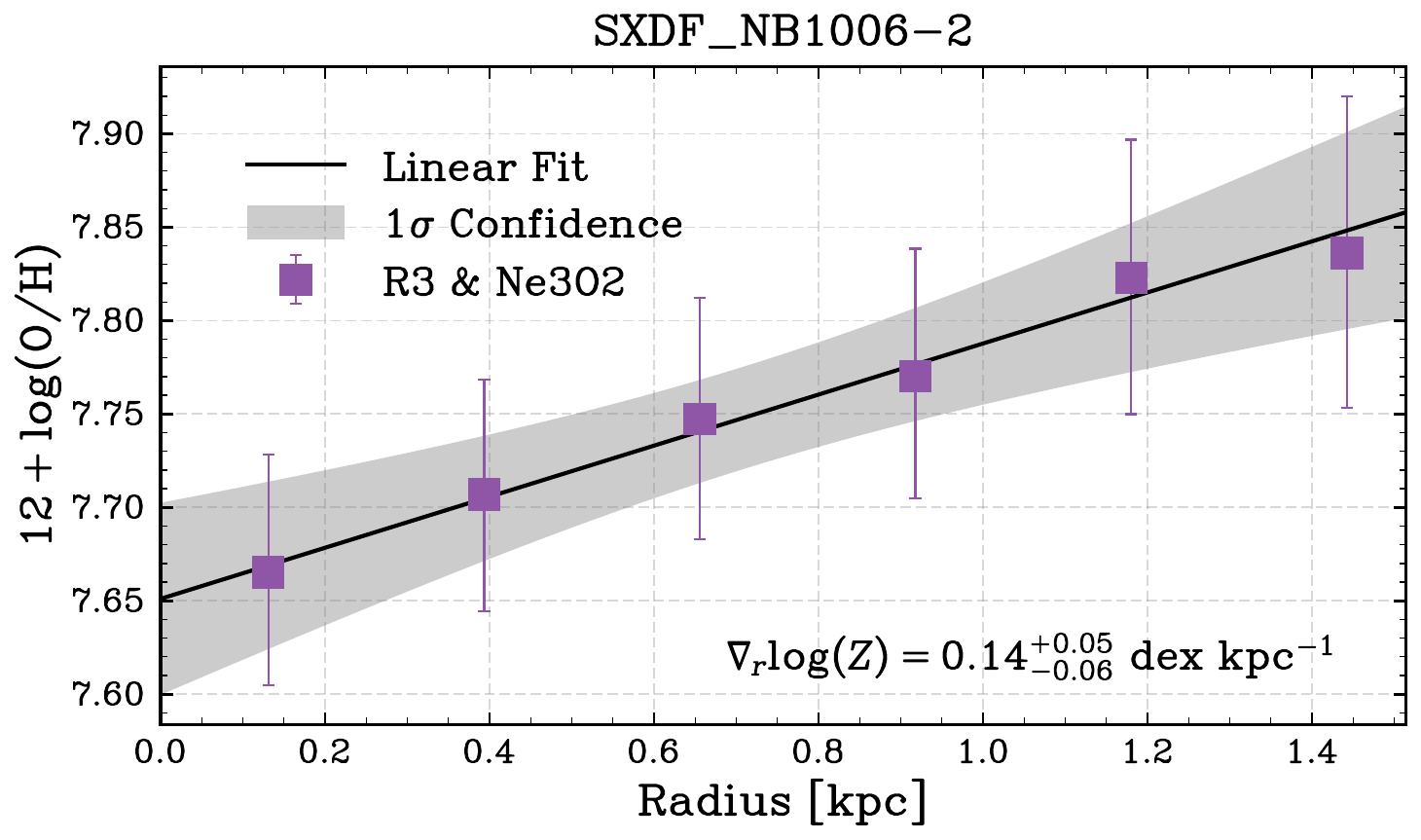}

    \vspace{0.3cm}
    
    \includegraphics[width=\linewidth]{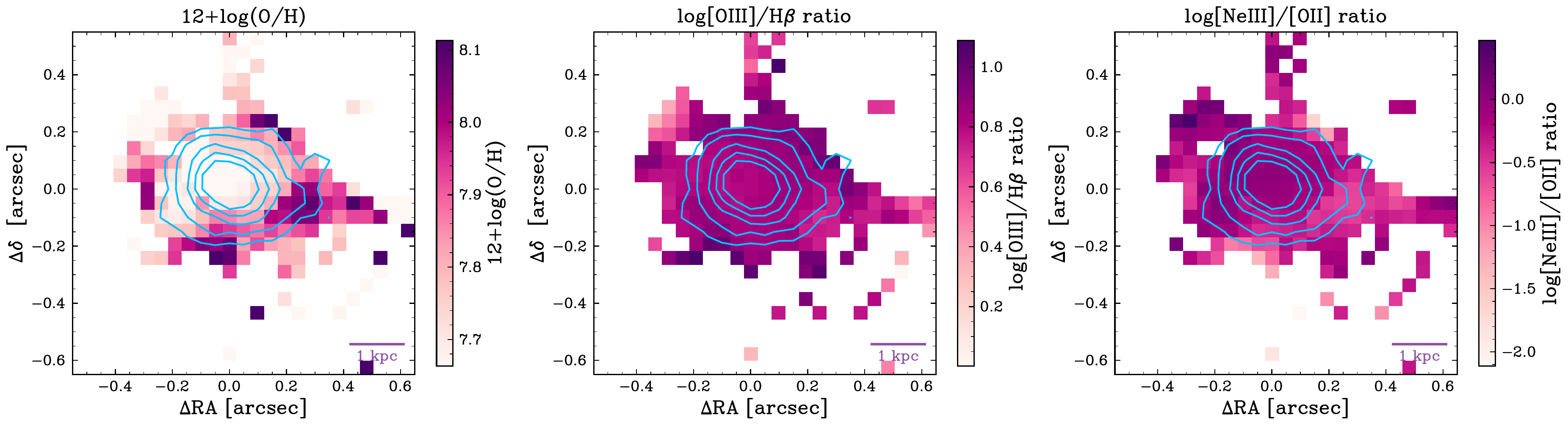}

    \vspace{0.3cm}

    \includegraphics[width=\linewidth]{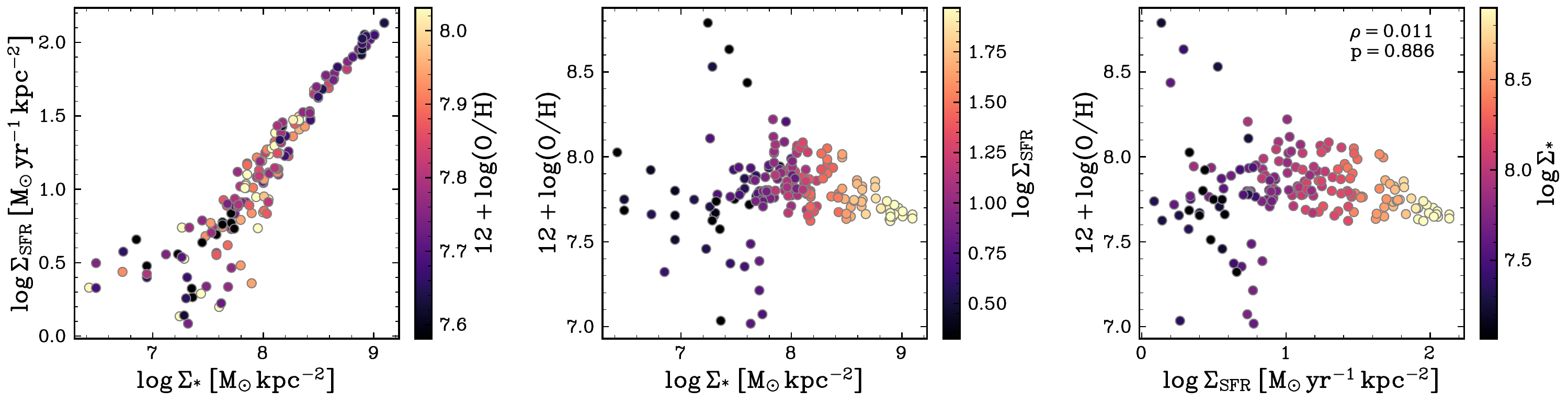}
    
    \caption{Same as Figure~\ref{fig:metal_grads_SMACS0723_4590} but for SXDF\_NB1006-2. As no NIRCam data is readily available for this object, the blue contour lines represent the continuum measured between $\rm H\gamma$ and $\rm H\beta$ from the NIRSpec IFU data.}
    \label{fig:metal_grads_SXDF_NB1006-2}
\end{figure*}

In this appendix, we provide a comprehensive compilation of NIRSpec IFU maps for the flux line ratios, metallicities, and metallicity gradients for each galaxy in our sample. These results encompass Figures~\ref{fig:metal_grads_SMACS0723_4590} to \ref{fig:metal_grads_SXDF_NB1006-2}.

\section{Stellar Mass Measurements using spectra only}

\begin{figure}
    \centering
    \includegraphics[width=\linewidth]{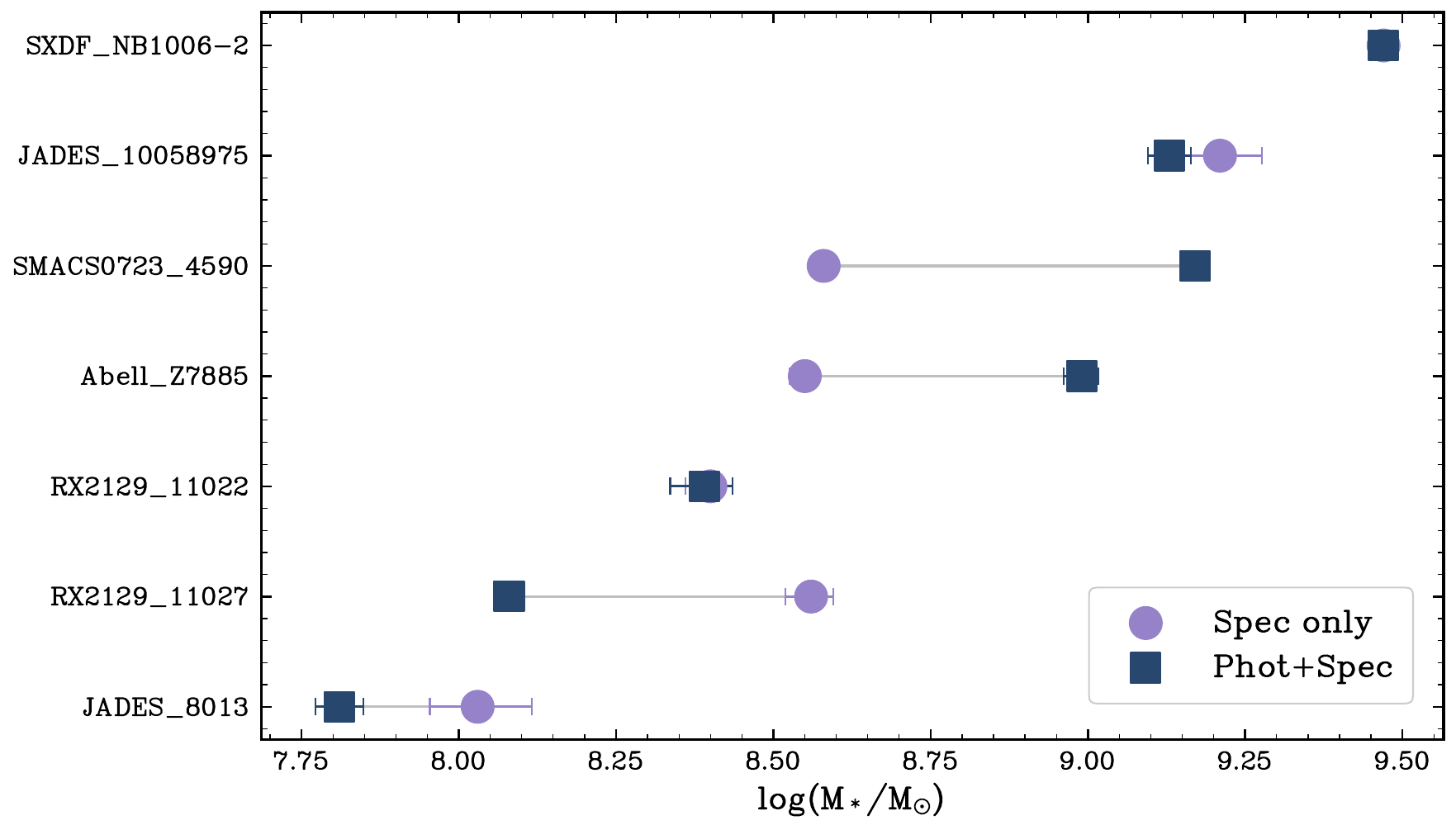}
    \caption{Comparison between stellar masses derived using NIRSpec IFU integrated spectra only (purple circles) and using these spectra and NIRCam photometry simultaneously (dark blue squares) during the SED fitting routine.}
    \label{fig:mass_comparison}
\end{figure}

In our main analysis, we derive stellar masses using Prospector, combining the derived NIRCam photometry with integrated spectra from the NIRSpec IFU data. The full details of this analysis are outlined in section~\ref{sec:sedfitting}. However, we also repeat the same analysis using only the integrated spectra from the NIRSpec IFU data, without the addition of photometry. Our results, shown in Figure~\ref{fig:mass_comparison}, differ depending on whether we include photometry in the SED fitting routine or not; nonetheless, no apparent trend is visible.

\bsp	
\label{lastpage}
\end{document}